%% file: ms.tex
\newif\ifreport
\reporttrue

\newcommand{\reportornot}[2]{\ifreport#1\else#2\fi}

\documentclass[format=acmsmall, review=false, screen=false, timestamp=false]{acmart}

\usepackage{multirow}
\usepackage{amsmath}
\usepackage{graphicx}
\usepackage{caption}
\usepackage{url}
\usepackage{titleref}
\usepackage{picins}
\usepackage{subfigure}
\usepackage{lscape}
\usepackage{array}
\usepackage{tabularx}
\usepackage{tabulary}
\usepackage{enumerate,letltxmacro}
\usepackage{enumitem}
\usepackage{longtable}
\usepackage{wrapfig}
\usepackage{setspace}
\usepackage{float}
\usepackage{multicol}
\usepackage{booktabs}
\usepackage{microtype}
\usepackage[figuresright]{rotating}
\usepackage[ansinew]{inputenc} 
\usepackage{cancel}
\usepackage{savesym}
\usepackage{textcomp}
\savesymbol{vec}
\usepackage{MnSymbol}
\usepackage{lineno,blindtext}
\usepackage{longtable}
\usepackage{scalerel}
\usepackage{everypage}
\usepackage{spverbatim}
\let\oldFootnote\footnote
\newcommand\nextToken\relax

\renewcommand\footnote[1]{%
    \oldFootnote{#1}\futurelet\nextToken\isFootnote}

\newcommand\isFootnote{%
    \ifx\footnote\nextToken\textsuperscript{,}\fi}
    
\newlist{compactitem}{itemize}{3} 
\newlist{compactenum}{itemize}{3} 
\setlist[compactitem]{label=$\circ$, nosep}
\setlist[compactenum]{label=$\circ$, nosep}

\setlist[compactitem]{label=$\circ$, leftmargin=1em, labelindent=0.1em, itemsep=0em, parsep=0em}
\setlist[compactenum]{label=$\circ$, leftmargin=1em, labelindent=0.1em, itemsep=0em, parsep=0em}

\usepackage[ruled]{algorithm2e} 
\renewcommand{\algorithmcfname}{ALGORITHM}
\SetAlFnt{\small}
\SetAlCapFnt{\small}
\SetAlCapNameFnt{\small}
\SetAlCapHSkip{0pt}
\IncMargin{-\parindent}

\graphicspath{{fig/}}

\input{artem.sty}

\newcommand{\MONTH}{%
  \ifcase\the\month
  \or January
  \or February
  \or March
  \or April
  \or May
  \or June
  \or July
  \or August
  \or September
  \or October
  \or November
  \or December
  \fi}

\renewcommand{\textrightarrow}{$\rightarrow$}

\received{April 2019}
\received[Revised]{\MONTH~\the\year}

\acmDOI{}
\acmPrice{}
\acmJournal{Preprint}
\acmVolume{1}
\acmNumber{1}
\acmArticle{2}
\acmYear{\the\year}
\acmMonth{\the\month}
\copyrightyear{\the\year}

\reportornot{
\setcopyright{none}

\usepackage[firstpage]{draftwatermark}
\SetWatermarkAngle{0}
\SetWatermarkFontSize{13pt}
\SetWatermarkHorCenter{131mm}
\SetWatermarkVerCenter{235mm}
\SetWatermarkLightness{0.6}
\SetWatermarkText{Preprint, September 2019}
}
{
\setcopyright{acmcopyright}
}

\begin{document}

\title[Process Query Language: Design, Implementation, and Evaluation]{Process Query Language:\\ Design, Implementation, and Evaluation}

\author{Artem Polyvyanyy}
\orcid{0000-0002-7672-1643}
\affiliation{%
  \institution{The University of Melbourne}
  \streetaddress{Level 8, Doug McDonell Building}
  \city{Parkville}
  \state{VIC}
  \postcode{3010}
  \country{Australia}}
\author{Arthur H. M. ter Hofstede}
\affiliation{%
  \institution{Queensland University of Technology}
  \streetaddress{2 George St.}
  \city{Brisbane City}
  \state{QLD}
  \postcode{4000}
  \country{Australia}}
\author{Marcello La Rosa}
\affiliation{%
  \institution{The University of Melbourne}
  \streetaddress{Level 8, Doug McDonell Building}
  \city{Parkville}
  \state{VIC}
  \postcode{3010}
  \country{Australia}}
\author{Chun Ouyang}
\affiliation{%
  \institution{Queensland University of Technology}
  \streetaddress{2 George St.}
  \city{Brisbane City}
  \state{QLD}
  \postcode{4000}
  \country{Australia}}
\author{Anastasiia Pika}
\affiliation{%
  \institution{Queensland University of Technology}
  \streetaddress{2 George St.}
  \city{Brisbane City}
  \state{QLD}
  \postcode{4000}
  \country{Australia}}

\begin{abstract}
Organizations can benefit from the use of practices, techniques, and tools from the area of business process management. 
Through the focus on processes, they create process models that require management, including support for versioning, refactoring and querying. 
Querying thus far has primarily focused on structural properties of models rather than on exploiting behavioral properties capturing aspects of model execution. 
While the latter is more challenging, it is also more effective, especially when models are used for auditing or process automation. 
The focus of this paper is to overcome the challenges associated with behavioral querying of process models in order to unlock its benefits. 
The first challenge concerns determining decidability of the building blocks of the query language, which are the possible behavioral relations between process tasks. 
The second challenge concerns achieving acceptable performance of query evaluation. 
The evaluation of a query may require expensive checks in all process models, of which there may be thousands. 
In light of these challenges, this paper proposes a special-purpose programming language, namely Process Query Language (PQL) for behavioral querying of process model collections. 
The language relies on a set of behavioral predicates between process tasks, whose usefulness has been empirically evaluated with a pool of process model stakeholders. 
This study resulted in a selection of the predicates to be implemented in PQL, whose decidability has also been formally proven. 
The computational performance of the language has been extensively evaluated through a set of experiments against two large process model collections. 

\end{abstract}



\thanks{Authors' addresses: A. Polyvyanyy {and} M. La Rosa, School of Computing and Information Systems, The University of Melbourne, Australia; 
A. H. M. ter Hofstede, C. Ouyang, {and} A. Pika, Business Process Management Discipline, Information Systems School, Queensland University of Technology, Brisbane, Australia.
}



\maketitle

\makeatletter
\DeclareFontShape{OT1}{cmtt}{bx}{n}{<5><6><7><8><9><10><10.95><12><14.4><17.28><20.74><24.88>cmttb10}{}
\DeclareFontShape{OT1}{cmtt}{m}{n}{<1><2><3><3><3><3><3><3><3><7><8><9>cmtt10}{}
\makeatother

\renewcommand{\shortauthors}{A. Polyvyanyy et al.}


\input{tex/introduction}
\input{tex/motivation}

\input{tex/preliminaries}
\input{tex/language}
\input{tex/implementation}
\input{tex/evaluation}
\input{tex/related}
\input{tex/conclusions}

\bibliographystyle{ACM-Reference-Format}
\bibliography{bibliography}

\newpage
\appendix
\input{tex/appendix_interview}
\input{tex/appendix_grammar}
\reportornot{}{\input{tex/appendix_predicate}}
\reportornot{}{\input{tex/appendix_implementation}}
\reportornot{}{\input{tex/appendix_evaluation}}
\input{tex/appendix_queries}

\end{document}

%% file: tex/introduction.tex
\newpage
\section{Introduction}
\label{sec:intro}

Through the application of methods and techniques from the field of business process management, organizations can identify, model, analyze, redesign, automate, monitor, and query their business processes~\cite{OuyangDAHM09,BusinessProcessManagement,DumasBPM13,PolyvyanyyOBA17}. 
This process-oriented thinking provides great benefits as making processes explicit through conceptual representations, \ie process models, allows organizations to subject these processes to various forms of analysis, to use them as the basis for automated support, and to adapt them more easily as well as more rapidly to continual changes imposed by the organization's environment, both internal and external.
As a consequence, some organizations have collected large numbers of process models. Examples of large process model collections reported in the literature are the SAP R/3 reference model (600+ models)~\cite{Curran98}, 
the IBM BIT collection in finance, telecommunication and other domains (700+ models)~\cite{FahlandFKLVW11}, 
a collection of healthcare process models from a German health insurance company (4,000+ models)~\cite{PolyvyanyySW08}, and a collection of insurance process models from Suncorp---one of the largest insurance groups in Australia (6,000+ models)~\cite{RosaDUD13}.

Process model collections evolve over time through model adaptations, mergers, and additions and their maintenance poses significant challenges. 
To support these activities, it should be possible to query a potentially large repository of process model to retrieve models with certain characteristics. 
\emph{Process querying} studies automated methods for managing, \eg retrieving or manipulating, process models and relationships between process models~\cite{PolyvyanyyOBA17}.
A process querying method is a technique that given a process model repository and a formal specification of an instruction to manage the repository, \ie a \emph{process query}, systematically implements the query in the repository.

Existing languages for specifying process queries, for example {BP-QL}~\cite{UniversityIEKM05} and {BPMN-Q}~\cite{Awad07}, predominantly focus on \emph{structural} aspects of process models~\cite{WangJWW14,PolyvyanyyOBA17}. 
Specifically, they treat process models as annotated directed graphs, where annotations are used to denote types of graph nodes, e.g., process tasks, activities, events, and decisions. 
The semantics of such queries is based on analysis of paths and patterns in the process model graphs. 
A more convenient and precise way to query a process model repository, though, is by using \emph{behavioral relations} between model elements, \eg relations which capture that process tasks can be performed in certain order, in parallel, or can never be performed as part of the same \emph{process instance}.\footnote{In this paper, we use the term process instance to refer to an artifact that encodes an execution of a process model, \eg an order of tasks that can be executed according to the semantics of the process model.}
Process querying methods based on process model graphs are not suitable to adequately analyze behavioral relations induced by process models for at least two reasons.
Firstly, behavioral relations are often defined over the collection of all process instances of a process model, which can be infinite. 
This poses a significant challenge to specify the corresponding query condition over the finite process model graph.
Secondly, it is well-known that infinitely many structurally different process models can describe the same behavioral relations~\cite{Polyvyanyy12,PolyvyanyyADG16}.
This poses a significant challenge to specify a finite query condition that caters for infinitely many structural patterns of process models.

\begin{wrapfigure}{r}{0.4\textwidth}
\flushright
\vspace{-2mm}
\includegraphics[width=0.427\textwidth]{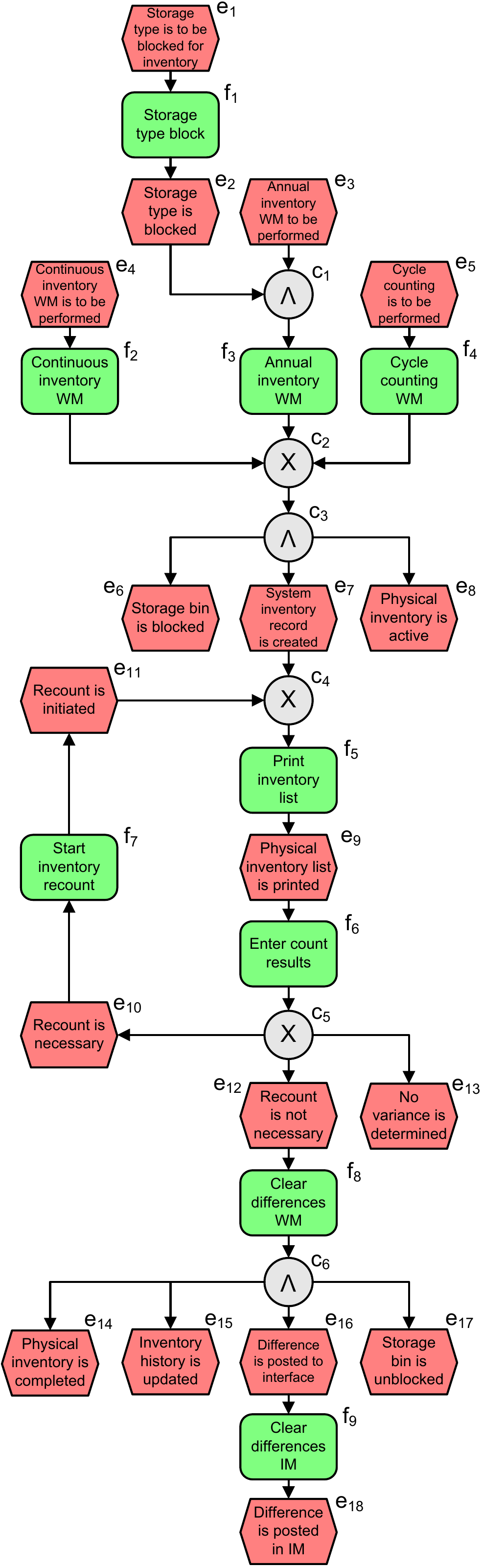}
\vspace{-7mm}
\caption{\label{fig:EPC}An EPC model from SAP R/3.}
\vspace{-4mm}
\end{wrapfigure}

The process model in \figurename~\ref{fig:EPC} was extracted from the SAP R/3 reference model~\cite{Curran98}.
It is captured using the EPC language~\cite{ScheerTA05}.
In this language, hexagons represent \emph{events}, rounded rectangles encode \emph{functions}, arrows capture the \emph{control flow}, and circles represent \emph{logical connectors}.
The model in \figurename~\ref{fig:EPC} should be retrieved as a response to the user's intent to discover all process models that describe executions in which event ``Physical inventory is active'' (denoted by $e_8$ in the figure) can be triggered concurrently (at the same time) with any occurrence of function ``Start inventory recount'' ($f_7$), and all occurrences of function $f_7$ precede, or are the cause for, all the subsequent occurrences of functions with a label that is similar to\, ``Clear differences'' (assuming that labels ``Clear differences WM'' ($f_8$) and ``Clear differences IM'' ($f_9$) are considered to be similar to ``Clear differences''). 
Due to the above outlined reasons, one cannot express the above\, intent to\, retrieve 

\noindent
process models using existing query languages based on process model structure, such as BP-QL or BPMN-Q. 

The added expressiveness of a query language grounded in behavioral relations comes at a price.
Behavioral relations cover a broad spectrum of inter-task dependencies that may be captured using special property specification languages, \eg temporal logics. 
Temporal logics are powerful enough to be able to express properties that are undecidable~\cite{EsparzaN94,EsparzaH08}. 
Hence, a query language that exploits behavioral relations needs to be carefully designed in terms of the behavioral inter-task dependencies that it supports.


While some relations are decidable, their use may not be very intuitive to the stakeholders of the language, \ie the business analysts that will end up formulating queries. 
Thus, it is important that a relation is likely to be frequently used in queries in practice to warrant support, and that its formal meaning is close to its perceived meaning. 
Another consideration is that query evaluations are performed in a ``reasonable'' amount of time. 
In fact, it is anticipated that process model stakeholders may wish to see the answers to their queries in (almost) real-time, so as to be able to quickly evaluate different scenarios when updating existing models or creating new ones.

This paper proposes Process Query Language (PQL)---a special-purpose programming language for querying repositories of process models based on information about executions, \ie \emph{process instances}, that these models describe. 
PQL programs are called \emph{queries}.
PQL allows formulating queries for retrieving models from repositories thereof using a selected number of behavioral inter-task relations, called \emph{predicates}.
The PQL predicates are built upon the \emph{4C spectrum}~\cite{PolyvyanyyWCRH14}---a systematic classification of possible behavioral relations between process model tasks according to four categories: conflict, co-occurrence, causality and concurrency.

PQL implements the \emph{process querying compromise}, refer to \sectionname~4.4 in~\cite{PolyvyanyyOBA17}, by supporting \emph{useful} and \emph{efficiently computable} process queries, whose\emph{practical relevance} has been validated with practitioners in terms of their perceived usefulness, importance, and intensity of use.
We demonstrate that the PQL predicates are \emph{decidable} over a given process model by reducing their computations to the reachability problem~\cite{Hack75}, the covering problem~\cite{Rackoff78}, or the problem of structural analysis over a complete prefix~\cite{McMillan92,EsparzaRV02} of the unfolding~\cite{NielsenPW81} of the model.
Despite the fact that the techniques for computing the PQL predicates are complex, \eg solving the reachability problem required exponential space~\cite{yale1976reachability}, the conducted experiments demonstrate the feasibility of using PQL in practical settings.

To facilitate query formulation, PQL is provided with the abstract syntax and a concrete syntax, the latter inspired by the SQL language. 
To tackle the performance problem typical for checking behavioral properties of process models, the implemented PQL runtime environment relies on the use of indexed behavioral relations, \ie behavioral relations get precomputed and reused during evaluation of PQL queries. 
The performance of PQL query evaluation is assessed through an extensive set of experiments with real-life and synthetic process model collections.

In summary, the contributions of this paper are as follows:

\begin{compactitem}
\item 
Empirical evidence of the appropriateness of behavioral process querying, \ie the quality of behavioral process querying to be a proper method for retrieving process models from repositories based on behavioral inter-task relations;
\item
A selection of empirically justified behavioral inter-task relations for behavioral process querying based on quantitative feedback from prospective users;
\item 
Design of a query language, \viz PQL, based on the selected behavioral inter-task relations and qualitative feedback from prospective users;
\item
An open-source implementation of the proposed query language;
\item
A performance evaluation of the PQL implementation that demonstrates the feasibility of running PQL queries in (almost) real-time over industrial process repositories;
\item
A procedure for deciding whether two tasks are in the \texttt{TotalConcurrent} behavioral relation, which is one of the empirically justified PQL predicates;
\item
Application of the label unification principle proposed in~\cite{PolyvyanyyWCRH14} to implement an approach for exploratory behavioral process querying.
\end{compactitem}

\noindent
These contributions build on and extend our prior work. 
For example, PQL adopts the abstract syntax of A Process-model Query Language (APQL)~\cite{HofstedeORS0P13} and proposes new concrete syntax and dynamic semantics. 
The dynamic semantics of PQL is grounded in the behavioral relations of the 4C spectrum and label unification principle~\cite{PolyvyanyyWCRH14}. 

The remainder of the paper is structured as follows.
The next section motivates PQL by discussing several example PQL queries.
\sectionname~\ref{sec:preliminaries} introduces basic notions that are used to convey the denotational semantics of PQL queries. 
\sectionname~\ref{sec:design} addresses the design of the PQL language. 
It starts by reporting on an empirical study with process model stakeholders that has led to the selection of behavioral predicates to implement in PQL and
discussing the principles that were followed in the design of PQL.
Then, it presents the abstract syntax, the concrete syntax, and the dynamic semantics of PQL. 
The section concludes by discussing techniques for computing the selected predicates.
\sectionname~\ref{sec:implementation} presents the software implementation of PQL, while \sectionname~\ref{sec:evaluation} reports results of an evaluation of this implementation using one industrial and one synthetic process model collection.
Finally, \sectionname~\ref{sec:related} talks about related work, whereas \sectionname~\ref{sec:conclusion} states concluding remarks.


%% file: tex/motivation.tex
\section{Motivating Examples}
\label{sec:examples}

\newcommand{\apqltext}[1]{\texttt{`{\hspace{-.2mm}#1\hspace{-.2mm}}'}}
\newcommand{\SELECT}{\small\texttt{\textbf{SELECT}}\normalsize}
\newcommand{\FROM}{\small\texttt{\textbf{FROM}}\normalsize}
\newcommand{\WHERE}{\small\texttt{\textbf{WHERE}}\normalsize}
\newcommand{\IN}{\small\texttt{\textbf{IN}}\normalsize}
\newcommand{\ANY}{\small\texttt{\textbf{ANY}}\normalsize}
\newcommand{\SOME}{\small\texttt{\textbf{SOME}}\normalsize}
\newcommand{\EACH}{\small\texttt{\textbf{EACH}}\normalsize}
\newcommand{\ALL}{\small\texttt{\textbf{ALL}}\normalsize}

\newcommand{\UNION}{\small\texttt{\textbf{UNION}}\normalsize}
\newcommand{\INTERSECT}{\small\texttt{\textbf{INTERSECT}}\normalsize}
\newcommand{\EXCEPT}{\small\texttt{\textbf{EXCEPT}}\normalsize}
\newcommand{\AND}{\small\texttt{\textbf{AND}}\normalsize}
\newcommand{\OR}{\small\texttt{\textbf{OR}}\normalsize}

\newcommand{\NOT}{\small\texttt{\textbf{NOT}}\normalsize}
\newcommand{\EQUALS}{\small\texttt{\textbf{EQUALS}}\normalsize}
\newcommand{\OVERLAPS}{\small\texttt{\textbf{OVERLAPS}}\normalsize}
\newcommand{\WITH}{\small\texttt{\textbf{WITH}}\normalsize}
\newcommand{\PROPER}{\small\texttt{\textbf{PROPER}}\normalsize}
\newcommand{\IS}{\small\texttt{\textbf{IS}}\normalsize}
\newcommand{\SUBSET}{\small\texttt{\textbf{SUBSET}}\normalsize}
\newcommand{\OF}{\small\texttt{\textbf{OF}}\normalsize}

\newcommand{\TRUE}{\small\texttt{\textbf{TRUE}}\normalsize}
\newcommand{\FALSE}{\small\texttt{\textbf{FALSE}}\normalsize}
\newcommand{\UNKNOWN}{\small\texttt{\textbf{UNKNOWN}}\normalsize}



This section introduces key elements of PQL via discussions of three example queries. 
The model in \figurename~\ref{fig:EPC} is used to set the context for the examples. 
\tablename~\ref{tab:predicate-scenarios} lists six scenarios of behavioral inter-task relations and the corresponding behavioral predicates to capture these relations.
These scenarios informally introduce behavioral predicates that specify typical behavioral relations between tasks and serve as underlying constructs of PQL. 
Precise definitions of these predicates and their computations are provided later in the paper. 
We assume that the model is stored in the ``/SAP-R3-EPC-Repo'' location of the process repository. 

\begin{table}[h!]
	\footnotesize
  \centering
	\vspace{-2mm}
  \caption{\small Scenarios and behavioral predicates of inter-task relations from the process in \figurename~\ref{fig:EPC}.}
	\vspace{-2mm}
	\begin{tabular}{| c | l | l |} 
		\hline
		{\scriptsize{\bf No.}} & \multicolumn{1}{c|}{\bf Scenario description} & \multicolumn{1}{c|}{\bf Behavioral predicate} \\ 
		\hline\hline
		1 & ``Start inventory recount'' ($f_7$) occurs in \emph{at least one} process instance & \texttt{CanOccur($f_7$)} \\ 
		\hline
		2 & ``Print inventory list'' ($f_5$) occurs in \emph{every} process instance & \texttt{AlwaysOccurs($f_5$)} \\ 
		\hline
		3 & ``Storage type is blocked'' ($e_2$) and ``Annual inventory WM to be performed'' ($e_3$) & \texttt{Cooccur($e_2$,$e_3$)} \\ 
		  & either \emph{both} occur or \emph{neither} occur in a process instance &  \\ 
		\hline
		4 & ``No variance is determined'' ($e_{13}$) and ``Clear differences WM'' ($f_8$) & \texttt{Conflict($e_{13}$,$f_8$)} \\ 
		  & \emph{never both} occur in a process instance &  \\ 
		\hline
		5 & \emph{All} occurrences of ``Start inventory recount'' ($f_7$) precede (i.e. \emph{cause}) \emph{all} those of & \texttt{TotalCausal($f_7$,$f_9$)} \\ 
			& ``Clear differences IM'' ($f_9$) in \emph{every} process instance where they \emph{both} occur & \\
		\hline
		6 & ``Physical inventory is active'' ($e_8$) occurs \emph{concurrently} with \emph{any} occurrence of  & \texttt{TotalConcurrent($e_8$,$f_7$)} \\ 
			& ``Start inventory recount'' ($f_7$) in \emph{every} process instance where they \emph{both} occur & \\
		\hline
	\end{tabular}
	\label{tab:predicate-scenarios}
	\vspace{-4mm}
\end{table}

\vspace{-2mm}
\paragraph{Example~1.} Recall the example query from the Introduction: The model in \figurename~\ref{fig:EPC} should be retrieved as a response to the user's intent to discover, from the repository, all process models that describe executions in which $e_8$ can be triggered concurrently with any occurrence of $f_7$, see Scenario~6 in Table~\ref{tab:predicate-scenarios}, and all occurrences of $f_7$ precede all occurrences of functions with a label that is similar to ``Clear differences'', such as the label of $f_8$ and that of $f_9$, see Scenario~5 in Table~\ref{tab:predicate-scenarios}. 


This user's intent can be captured in the following PQL query~(\emph{Q1}): 

\vspace*{.5\baselineskip}


\texttt{\SELECT~"ID" \FROM~"$/$SAP-R3-EPC-Repo"}

\texttt{\WHERE~TotalConcurrent("Physical inventory is active","Start inventory recount")}

\texttt{\ \ \ \ \ \ \AND~TotalCausal("Start inventory recount",$\sim$"Clear differences");}

\vspace*{.5\baselineskip}

\noindent 
This query expects to retrieve the process models and their IDs, where \texttt{"$/$SAP-R3-EPC-Repo"} is the location where SAP models are stored in the repository, and \texttt{$\sim$"Clear differences"} specifies a task (which is either an event or a function in EPC) with a label that is similar to ``Clear differences''. 


\vspace{-2mm}
\paragraph{Example~2.} The user's intent is to retrieve the models, with their IDs and titles, where at least one of the functions ``Continuous inventory WM'' ($f_2$), ``Annual inventory WM'' ($f_3$), and ``Print inventory list'' ($f_5$) occurs in every process instance, see Scenario~2 in Table~\ref{tab:predicate-scenarios}; or for events ``Storage type is blocked'' ($e_2$) and ``Annual inventory WM to be performed'' ($e_3$), either both or neither occur in a process instance, see Scenario~3 in Table~\ref{tab:predicate-scenarios}. A PQL query~(\emph{Q2}) to capture this intent follows. 

\vspace*{.5\baselineskip}


\texttt{\SELECT~"ID","Title" \FROM~"$/$SAP-R3-EPC-Repo"}

\texttt{\WHERE~AlwaysOccurs(\{"Continuous inventory WM","Annual inventory WM",}

\texttt{\hspace*{9.2em} "Print inventory list"\},\ANY)}

\texttt{\hspace*{2.6em} \OR~Cooccur("Storage type is blocked","Annual inventory WM to be performed");}

\vspace{.5\baselineskip}

\noindent 
Note that this query utilizes a well-known mechanism of macros for combining results of two or more predicate checks into a result of a single statement. 
Concretely, in query \emph{Q2}, three behavioral predicates connected via the logic OR operator: \texttt{AlwaysOccurs($f_2$)~\OR~AlwaysOccurs($f_3$)~\OR\linebreak~AlwaysOccurs($f_5$)}, are combined into one marco \texttt{AlwaysOccurs(\{$f_2$,$f_3$,$f_5$\},\ANY)}.

\vspace{-2mm}
\paragraph{Example~3.} The user's intent is to retrieve the process models, with all their attributes information (\eg ID, title, version, author, etc.), which satisfy all the four conditions listed below:

\vspace{-.5mm}
\begin{itemize}
	\item[(\emph{C1})] Function ``Start inventory recount'' ($f_7$), event ``No variance is determined'' ($e_{13}$), function or event having a label similar to ``Clear differences'' ($f_8$ or $f_9$), and function or event having a label similar to ``Difference is posted'' ($e_{16}$ or $e_{18}$), occur in at least one process instance;
	\item[(\emph{C2})] Inventory recount is optional, \ie $f_7$ does not occur in every process instance;
	\item[(\emph{C3})] None of the tasks with a label similar to ``Clear differences'' ($f_8$ and $f_9$) occurs if no variance is determined (\ie when $e_{13}$ takes place); and
	\item[(\emph{C4})] All occurrences of $f_7$ precede all occurrences of functions or events having a label similar to ``Clear differences'' ($f_8$ and $f_9$) and all occurrences of functions or events having a label similar to ``Difference is posted'' ($e_{16}$ and $e_{18}$).
\end{itemize}
\vspace{-.5mm}

The following PQL query~(\emph{Q3}) can be specified to capture the above user's intent. 

\vspace*{.5\baselineskip}


\texttt{$x=$ \{"Start inventory recount","No variance is determined"\};}

\texttt{$y=$ \{$\sim$"Clear differences"\};}

\texttt{$z=$ $y$~\UNION~\{$\sim$"Difference is posted"\};}

\texttt{$w=$ GetTasksAlwaysOccurs(GetTasks());}

\texttt{\SELECT~* \FROM~"$/$SAP-R3-EPC-Repo"}

\texttt{\WHERE~CanOccur($x$~\UNION~$z$,\ALL)~\AND~} \hspace*{3.0em}
\hspace*{8.4em} \mbox{\footnotesize{--~--\ \texttt{(C1)}}}

\texttt{\hspace*{3.1em}(\NOT~("Start inventory recount"~\IN~$w$))~\AND}
\hspace*{3.9em} \mbox{\footnotesize{--~--\ \texttt{(C2)}}}

\texttt{\hspace*{3.1em}Conflict("No variance is determined",$y$,\ALL)~\AND}
\hspace*{1.55em} \mbox{\footnotesize{--~--\ \texttt{(C3)}}}

\texttt{\hspace*{3.1em}TotalCausal("Start inventory recount",$z$,\ALL);}
\hspace*{2.5em} \mbox{\footnotesize{--~--\ \texttt{(C4)}}}

\vspace*{.5\baselineskip}
\noindent 
To facilitate the query definition, variables are used to store sets of tasks. 
Variable~$x$ stores tasks $f_7$ and $e_{13}$,
$y$ contains one task which refers to labels of $f_8$ and $f_9$, and 
$z$ stores a set of tasks as a result of combining the task in $y$ and tasks referring to labels of $e_{16}$ and $e_{18}$. 
Variable $w$ collects the set of tasks that occur in every process instance of a process model being examined. 
Note that \texttt{GetTasks()} retrieves all the tasks in a process model and \texttt{GetTasksAlwaysOccurs()} selects the tasks that occur in every process instance from an input set of tasks. 

The \texttt{\WHERE} clause captures the four conditions (\emph{C1} to \emph{C4}) of the above user's query intent, as marked in the comments (starting with `\mbox{\footnotesize{--~--}}'). 
Firstly, predicate macro \texttt{CanOccur($x$~\UNION~$z$,} \texttt{\ALL)} checks if every task in the set of tasks combined from $x$ and $z$ occurs in at least one process instance, \eg \texttt{CanOccur($f_7$)}, see Scenario~1 in Table~\ref{tab:predicate-scenarios}, is part of the check. 
Next, predicate \texttt{(\NOT~("Start inventory recount"~\IN~$w$))} checks if the occurrence of $f_7$ is optional. 
Then, predicate macro \texttt{Conflict("No variance is determined",$y$,\ALL)} checks if $e_{13}$ occurs in conflict with each of the tasks stored in variable~$y$, \eg it is necessary to check \texttt{Conflict($e_{13}$,$f_8$)}, see Scenario~4 in Table~\ref{tab:predicate-scenarios}. 
Finally, predicate macro \texttt{TotalCausal("Start inventory recount",$z$,\ALL)} checks if all occurrences of $f_7$ precede all occurrences of every task stored in variable~$z$, \eg \texttt{TotalCausal($f_7$,\linebreak$f_9$)} is one of the required checks, see Scenario~5 in Table~\ref{tab:predicate-scenarios}.

 

%% file: tex/preliminaries.tex
\section{Preliminaries}
\label{sec:preliminaries}

\reportornot{
This section introduces basic notions that are related to Petri net systems. 
These notions are used in the next section to support discussions on the dynamic semantics of PQL.
}{
This section introduces notions used in the next section to present the dynamic semantics of PQL.
}

\vspace{-1mm}
\subsection{Multisets, Sequences, and Strings}

\reportornot{A \emph{multiset}, or a \emph{bag}, is a generalization of the concept of a set that allows a multiset to contain multiple instances of the same element. }{}
Let $A$ be a set.
By $\mathcal{P}(A)$ and $\mathcal{B}(A)$\label{page:ref:2}, we denote the power set of $A$ and the set of all finite multisets over $A$, respectively.
For some multiset $B \in \mathcal{B}(A)$, $B(a)$ is the multiplicity of element $a$ in $B$, \ie the number of times element $a \in A$ appears in $B$;
we define a multiset $B \in \mathcal{B}(A)$ as a function $B : A \rightarrow \mathbb{N}_0$.\footnote{$\mathbb{N}_0$ denotes the set of all natural numbers including zero.}
For example, $B_1:=[]$, $B_2:=[a]$, $B_3:=[a,b,a]$, $B_4:=[a^2,b^1,a^0]$ are multisets over $A:=\{a,b\}$.
\reportornot{Multiset $B_1$ is \emph{empty}, \ie contains no elements. }{Multiset $B_1$ is \emph{empty}.}
Multiset $B_2$ contains a single element $a \in A$. 
Multiset $B_3$ contains three elements: two occurrences of element $a \in A$ and one occurrence of element $b \in A$. 
Multisets $B_3$ and $B_4$ are equal, \ie $B_3 = [a^2,b] = B_4$. 
The above examples demonstrate the notation for describing multisets and suggest that the ordering of elements in multisets is irrelevant.

\reportornot{The standard set operations have been extended to deal with multisets as follows.}{}
If element $a$ is a member of multiset $B$, this is denoted by $a \in B$, while if element $b$ is not a member of $B$, we write $b \not\in B$.
For example, for the multisets defined above, it holds that $a \in B_2$, $b \not\in B_2$, and $a,b \in B_3$. 
The union of two multisets $C$ and $D$, denoted by $C \cupplus D$, is the multiset that contains all elements of $C$ and $D$ such that the multiplicity of an element in the resulting multiset is equal to the sum of multiplicities of this element in $C$ and $D$.
For example, $[a,b,a^2] = B_2 \cupplus B_3 = [a^3,b]$.
The difference of two multisets $C$ and $D$, denoted by $C \setminus D$, is the multiset that for every element $x$ in $C$ contains $\mathit{max}(0,C(x)-D(x))$ occurrences of element $x$.
For example, it holds that $B_4 \setminus B_2 = [a,b]$ and $B_3 \setminus B_4 = []$.
Finally, the cardinality of a multiset $B$, denoted by $|B|$, is the sum of the occurrences of its members.
For example, it holds that $|B_1|=0$, $|B_2|=1$, $|B_3|=3=|B_4|$.


\reportornot{In mathematics, a \emph{sequence} is an ordered list of elements. 
Similar to multisets, in a sequence, instances of the same element can appear multiple times. 
However, unlike in multisets, positions of element occurrences in the sequence matter.}{}
By $\sigma:=\sequence{a_1,a_2,\ldots,a_n} \in A^*$, we denote a sequence of length $n \in \Nzero$ over a set $A$, $a_i \in A$, $i \in [1..\,n]$.\footnote{$A^*$ denotes the Kleene star operation on set $A$.}
\reportornot{The \emph{empty} sequence, \ie the sequence without elements, is denoted by $\sequence{}$.}{The \emph{empty} sequence is denoted by $\sequence{}$.}
By $|\sigma|$, we indicate the number of all occurrences of elements in $\sigma$.
By $\mathit{prefix}(\sigma,i)$, we denote the prefix of $\sigma$ up to but excluding position $i$, whereas
$\mathit{suffix}(\sigma,i)$ is the suffix of $\sigma$ starting from and including position $i$, $i \in \mathbb{N}$.
Let $\eta:=\sequence{\texttt{a},\texttt{b},\texttt{a},\texttt{b},\texttt{a},\texttt{h},\texttt{a},\texttt{l},\texttt{a},\texttt{m},\texttt{a},\texttt{h},\texttt{a}}$ be a sequence.
Then, $|\eta| = 13$, $\mathit{prefix}(\eta,6)=\sequence{\texttt{a},\texttt{b},\texttt{a},\texttt{b},\texttt{a}}$, and $\mathit{suffix}(\eta,6)=\sequence{\texttt{h},\texttt{a},\texttt{l},\texttt{a},\texttt{m},\texttt{a},\texttt{h},\texttt{a}}$. 


An \emph{alphabet} is any non-empty finite set. 
The elements of an alphabet are also called \emph{characters}. 
A \emph{character string} over some alphabet is a finite sequence of characters from that alphabet. 
The characters of a string are usually written next to one another. 
For example, $q:=\texttt{101011}$ is a character string over $\{\texttt{0},\texttt{1}\}$.
The character string of length zero is called the \emph{empty string} and is denoted by $\epsilon$.

By $\mathbb{C}$, we denote the universe of all finite character strings over letters of the English alphabet (both lower- and uppercase), numerals, punctuation, and whitespace characters.

\subsection{Petri Net Systems, Workflow Systems, and Soundness}
\label{sec:petri:net:systems}

A Petri net system is a model of a distributed system. 

\begin{define}{Petri net system}{def:system}{\quad\\}
A \emph{Petri net system}, or a \emph{system}, is a 5-tuple $S:=(P,T,F,\lambda,M)$, where 
$P$ and $T$ are finite disjoint sets of \emph{places} and \emph{transitions}, respectively, 
$F \subseteq (P \times T) \cup (T \times P)$ is the \emph{flow relation},
$\lambda : T \to \mathbb{C}$ is a \emph{labeling function} that assigns labels to transitions, and 
$M \in \mathcal{B}(P)$ is a \emph{marking} of $S$.
\end{define}

\noindent
Transitions of Petri net systems can be used to encode \emph{actions}. 
\reportornot{It is convenient to distinguish between \emph{observable} and \emph{silent} transitions to describe actions that have a well-defined meaning and those that have no domain interpretation, respectively. 
In addition, one may be willing to assign the same application domain semantics to several distinct transitions.}{}
If $\lambda(t) \neq \epsilon$, $t \in T$, then $t$ is \emph{observable}; otherwise, $t$ is \emph{silent}. 
An element $x \in P \cup T$ is a \emph{node} of $S$. 
A node $x$ is an \emph{input} of a node $y$ \ifaof $(x,y) \in F$, while a node $y$ is an \emph{output} of $x$. 
By $\pre{x}$, we denote the \emph{preset} of node $x$, \ie the set of all input nodes of $x$, while by $\post{x}$, we denote the \emph{postset} of $x$, \ie the set of all output nodes of $x$. 

The execution semantics of a Petri net system is defined in terms of possible states and state transitions.
A marking of a system encodes its state.
A marking $M$ is often interpreted as an assignment of \emph{tokens} to places, \ie marking $M$ `puts' $M(p)$ tokens at place $p$, $p \in P$.

Let $S:=(P,T,F,\lambda,M)$ be a system. 
A transition $t \in T$ is \emph{enabled} in $M$, denoted by $M[t\rangle$, \ifaof every input place of $t$ contains at least one token, \ie $\forall p \in \pre{t} : M(p) > 0$. 
If a transition $t \in T$ is enabled in $M$, then $t$ can \emph{occur}, which \emph{leads} to a fresh marking $M':=(M \setminus \pre{t}) \cupplus \post{t}$ of $S$, \ie transition $t$ `consumes' one token from every input place of $t$ and `produces' one token for every output place \reportornot{\linebreak

\begin{wrapfigure}{r}{0.38\textwidth}
\flushright
\includegraphics[width=0.385\textwidth]{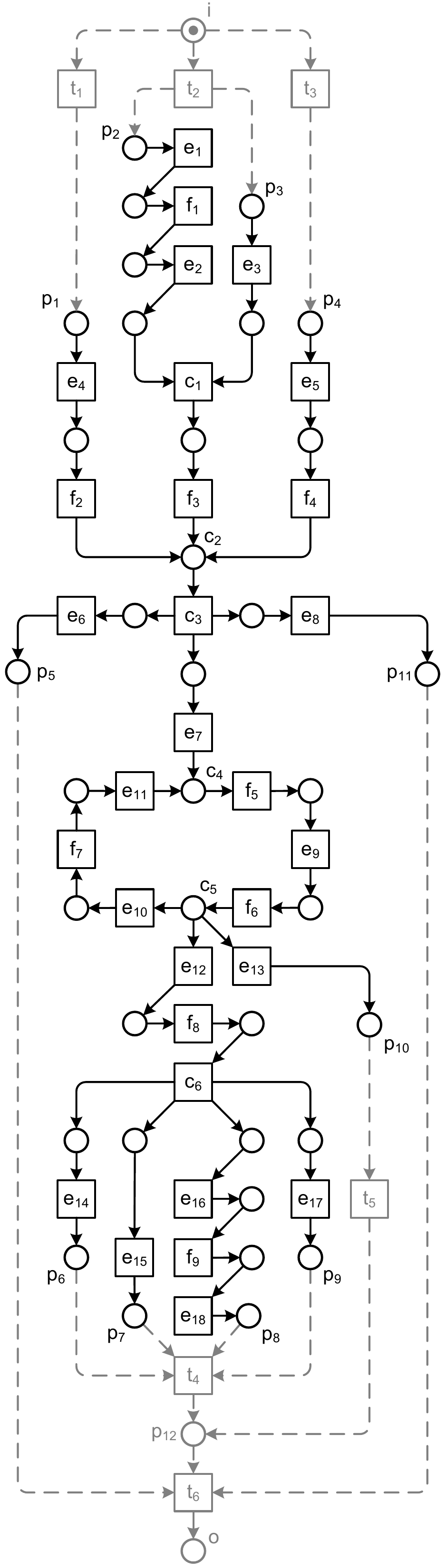}
\vspace{-7mm}
\caption{\label{fig:system}A workflow system.}
\vspace{-8mm}
\end{wrapfigure}

\noindent}{}of $t$.
By $M [t\rangle M'$, we denote the fact that an occurrence of $t$ leads from $M$ to $M'$.
A finite sequence of transitions $\sigma := \sequence{t_1,t_2,\ldots,t_n} \in T^*$, $n \in \Nzero$, is an \emph{occurrence sequence} of $S$ \ifaof 
$\sigma$ is empty or there exists a sequence of markings $\sequence{M_0,M_1,\ldots,M_n}$, such that $M_0=M$ and for every position $i \in [1..\,n]$ in $\sigma$ it holds that $M_{i-1}[t_i\rangle M_i$; we say that $\sigma$ \emph{leads} from $M_0$ to $M_n$ and denote this by $M_0[\sigma\rangle M_n$.


Petri net systems have a well-established graphical notation.
In this notation, places are visualized as circles, transitions are drawn as rectangles, where a label is depicted within the boundaries of the corresponding rectangle, every pair of nodes $(x,y)$ in the flow relation is drawn as a directed arc that leads from $x$ to $y$, and tokens induced by the marking of the Petri net system are depicted as black dots inside the assigned places.
\figurename~\ref{fig:system} shows a Petri net system that encodes the execution semantics of the model in \figurename~\ref{fig:EPC}.
It is a common practice to explain the execution semantics of\reportornot{}{\linebreak

\begin{wrapfigure}{r}{0.38\textwidth}
\flushright
\includegraphics[width=0.385\textwidth]{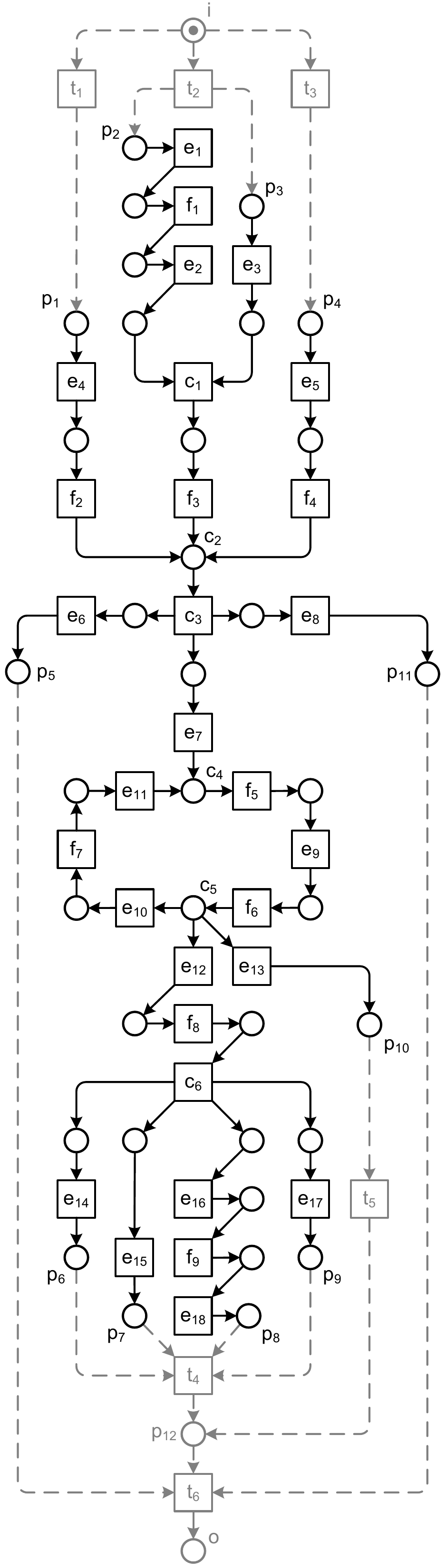}
\vspace{-5.5mm}
\caption{\label{fig:system}A workflow system.}
\vspace{-16mm}
\end{wrapfigure}

\noindent
}models captured using business process modelling languages through their translations to Petri nets.
Techniques for translating BPMN, BPEL, EPC, and YAWL models to Petri net systems are proposed in~\cite{DijkmanDO08}, \cite{LohmannVOS09}, \cite{Aalst99}, and~\cite{VerbeekAH07}, respectively.
The system in \figurename~\ref{fig:system} was obtained by first translating the model in \figurename~\ref{fig:EPC} into a Petri net system using the technique in~\cite{DongenJVA07} and then completing the system to a \emph{workflow system}~\cite{Aalst97ATPN}.
The fresh elements introduced during the completion are highlighted in gray, while the fresh arcs, in addition, are drawn using dashed lines.

In \figurename~\ref{fig:system}, transitions are labeled with short names while the full names are given in \figurename~\ref{fig:EPC}.
For example, transitions with labels $f_7$ and $e_9$ represent function ``\texttt{Start inventory recount}'' and event ``\texttt{Physical inventory list is printed}'' in the EPC model, respectively.
Transitions $t_1$, $t_2$, $t_3$, $t_4$, $t_5$, and $t_6$ introduced during the completion procedure, as well as transitions $c_1$, $c_3$ and $c_6$ that correspond to the logical AND connectors in \figurename~\ref{fig:EPC}, are silent, while all the other transitions in \figurename~\ref{fig:system} are observable. 

A \emph{workflow system} is a system with one \emph{source} place, one \emph{sink} place, every its node on a directed path from the source to the sink, and with marking that puts one token at the source place and no tokens elsewhere. 
The system in \figurename~\ref{fig:system} is a workflow system with source $i$ and sink $o$.
\reportornot{Workflow systems are used as abstract models for the explicit representation, validation, and verification of business procedures.}{}
Every execution of a workflow system that leads to the marking that puts one token at the sink place and no tokens elsewhere describes one business scenario~\cite{Aalst97ATPN}.

\reportornot{
\medskip
\begin{define}{Execution}{def:execution}{\quad\\}
An \emph{execution} of a workflow system $S:=(P,T,F,\lambda,[i])$ with the source place $i \in P$ is an occurrence sequence of $S$ that leads to $[o]$, where $o \in P$ is the sink place of $S$.
\end{define}
\smallskip
}{
\medskip
\begin{minipage}{8cm}
\begin{define}{Execution}{def:execution}{\quad\\}
An \emph{execution} of a workflow system $S:=(P,T,F,\lambda,[i])$ with the source place $i \in P$ is an occurrence sequence of $S$ that leads to $[o]$, where $o \in P$ is the sink place of $S$.
\end{define}
\end{minipage}
\smallskip
}

\noindent
Given a workflow system $S$, by $\mathbb{E}_S$\label{page:ref:7} we denote the set of all executions of $S$. 
Let $\sigma:=\sequence{t_1,t_2,\ldots,t_n} \in \mathbb{E}_S$, $n \in \Nzero$, be an execution of $S$.
Then, $\alpha := \sequence{\lambda(t_1),\lambda(t_2),\ldots,\lambda(t_n)}$ is the \emph{label execution} of $S$ induced by $\sigma$.
The sequences of transitions $\sequence{t_2,e_1,f_1,e_3}$ and $\sequence{t_3,e_5,f_4,c_3,e_7,f_5,e_9,e_8,f_6,e_6,e_{13},t_5,t_6}$ are two example occurrence sequences of the system $S$ in \figurename~\ref{fig:system}, whereas the latter is also an execution of $S$. 
Note that $\mathbb{E}_S$, where $S$ is the workflow system in \figurename~\ref{fig:system}, is infinite.

Petri net and workflow systems are subjects to semantic correctness constrains. 
One widely-used semantic correctness criterion for workflow systems is \emph{soundness}~\cite{Aalst97ATPN}.
Intuitively, every occurrence sequence of a sound workflow system can be extended (via occurrences of its enabled transitions) to an execution of the system, and every transition is an element of at least one execution of the system.
For example, the workflow system in \figurename~\ref{fig:system} is sound.

We define the dynamic semantics of PQL over sound workflow systems. 
This correctness requirement is rather technical and is imposed to simplify the definition and subsequent discussions of the proposed techniques for computing the PQL predicates, refer to~\sectionname~\ref{subsec:PQL:predicates:semantics}.


A Petri net system can often be transformed into a behaviorally equivalent workflow system. 
For example, one can use the technique from~\cite{PolyvyanyyGD12} to introduce the source place, while the technique from~\cite{KiepuszewskiHA03} can be applied to introduce the sink place.
These two techniques are computationally intensive.
One can often perform the completion efficiently.
For example, to obtain the workflow system in \figurename~\ref{fig:system} we used the generalized Refined Process Structure Tree (RPST)~\cite{PolyvyanyyVV10} of the underlying Petri net system. 
The RPST is the most fine-granular decomposition of a workflow graph into its single-entry-single-exit (SESE) components. 
In turn, the generalized version of the RPST is the most fine-granular decomposition of a workflow graph with multiple sources and multiple sinks into quasi SESE components, where quasi SESE components have multiple entries or multiple exits that correspond to sources or sinks of the underlying workflow graph.
The construction of the generalized RPST requires time that is linear in the size of the input graph~\cite{PolyvyanyyVV10}.

Among others, the generalized RPST of the graph in \figurename~\ref{fig:system} (without highlighted elements), identifies a quasi SESE component with entries $p_2,p_3$ and exit $c_1$, and a quasi SESE component with entries $p_1,p_2,p_3,p_4$ and exit $c_2$.
One can introduce single entry in each of these components.
Concretely, one can introduce transition $t_2$ as the counterpart to transition $c_1$ and place $i$ as the counterpart to place $c_2$.
Note that place $i$ is the fresh source of the system.
Similarly, one can introduce the sink in the system by introducing exits that match to entries of single entry quasi SESE components. 
In the example Petri net system, one can introduce transition $t_4$ as the counterpart of entry $c_6$ of the component with exits $p_6$, $p_7$, $p_8$, and $p_9$, 
place $p_{12}$ as the counterpart of entry $c_5$ of the component with exits $p_6$, $p_7$, $p_8$, $p_9$, and $p_{10}$ (note for fresh transition $t_5$), and
transition $t_6$ as the counterpart of entry $c_3$ of the component with exits $p_5$, $p_6$, $p_7$, $p_8$, $p_9$, $p_{10}$, and $p_{11}$.
To obtain a workflow system, one also needs to introduce the sink place $o$ as the only output of transition $t_6$.
This completion procedure runs in the time that is linear in the size of the input Petri net system.

%% file: tex/language.tex
\section{Design}
\label{sec:design}

This section discusses the design of PQL. 
It defines the main building blocks of the language, \ie its syntactic and semantic rules. 
\sectionname~\ref{sec:behavioral:relations} discusses the procedure we employed to select the basic behavioral predicates to include in the language.
Then, \sectionname~\ref{subsec:PQL:design:principles} summarizes the core principles followed in the design of PQL.
\sectionname~\ref{subsec:PQL:abstract:syntax} presents the abstract syntax of PQL; in an abstract syntax one can avoid committing to specific choices for keywords or to the order of various statements and concentrate on the design of the core structure of the language. 
\sectionname~\ref{subsec:PQL:concrete:syntax} is devoted to the discussion of a concrete syntax of the language, which constitutes its machine- and human-readable specification.
\sectionname~\ref{subsec:PQL:dynamic:semantics} breathes life into PQL queries by detailing their dynamic semantics, \ie the meaning of PQL queries.
\sectionname~\ref{subsec:PQL:predicates:semantics} states denotations of the PQL predicates and proposes techniques for computing them.
\sectionname~\ref{subsec:PQL:sample:queries} discusses sample PQL queries.
\reportornot{The reader can refer to \sectionname~\ref{sec:preliminaries} to get acquainted with some basic notions used in the subsequent discussions.}{}

\input{tex/predicates}
\subsection{Design Principles}
\label{subsec:PQL:design:principles}

PQL has been designed using the principles of suitability, simplicity, orthogonality, portability, decidability, and exploratory search support. 
Most of these principles are the standard principles of programming language design~\cite{Hoare73,Louden2011programming},
whereas ``exploratory search support'', motivated by the empirical evaluation reported in \sectionname~\ref{sec:behavioral:relations}, is borrowed from information retrieval.

\begin{compactitem}
\medskip
\item\textbf{Suitability.}
PQL queries should allow fulfilling practical tasks.
This is achieved by grounding the language in the behavioral predicates that are of practical relevance to process practitioners, refer to the previous section for details. 

\medskip
\item\textbf{Simplicity.}
PQL queries should allow capturing intents in short, succinct programs. 
They should be easy to read and comprehend.
The concrete syntax of PQL is inspired by SQL, which is a well-known language for querying relational databases, refer to~\sectionname~\ref{subsec:PQL:concrete:syntax} for further details including a justification for this decision. 
Out of 25 interviewees who participated in our empirical evaluation six have explicitly suggested that the invisaged process querying language should resemble SQL, with most of the participants being familiar with SQL.
Some direct quotes from the interviewees on the use of an SQL-like syntax include:
\smallskip
\begin{compactenum}
\item[$-$] ``\emph{SQL-like query language would be good ... people will adopt it if they can relate it so something familiar, like SQL.}'' (BPM consultant with 30 years experience in IT).
\item[$-$] ``\emph{SQL-like language is a science with which you can pose precise questions.}'' (Management consultant with 21 years of experience in design and setup of business processes).
\item[$-$] ``\emph{I think from an overall strategic level it'll bring a lot of benefits because different parts of the organization operate in different ways and being able to actually analyze it [process repository] through kind of a structured query language could be useful for an analyst.}'' (Business analyst with six months of experience).
\end{compactenum}
\smallskip

Finally, to keep queries short, PQL macros provide users with a mechanism to express several atomic statements using a single PQL construct. 

\medskip
\item\textbf{Orthogonality.}
PQL should be based on a small number of behavioral predicates that address orthogonal behavioral phenomena and allow combining them in many different ways to express complex queries.
PQL relies on the use of predicates grounded in the behavioral relations of the 4C spectrum~\cite{PolyvyanyyWCRH14}, which systematizes the four orthogonal behavioral relations of causality, conflict, concurrency and co-occurrence, refer to \sectionname~\ref{sec:predicates}.
Furthermore, PQL allows combining the predicates into propositional logic formulas to express complex query intents and supports set operations that can be used, for instance, to construct inputs to PQL macros.

\medskip
\item\textbf{Portability.}
PQL queries should be independent of implementation and execution environments, and data formats.
This is achieved by providing rigorous definitions of both the syntax and semantics of the language. 
Thus, one can implement PQL using different technologies that target various execution environments.
The semantics of PQL operates over Petri net systems, refer to \sectionname~\ref{subsec:PQL:predicates:definitions}. 
This allows using PQL over process models captured in a wide range of modeling languages, \eg BPMN, EPC, or YAWL, as models captured using most of the well-established process modeling languages can be translated to Petri net systems~\cite{LohmannVD09}.

\medskip
\item\textbf{Decidability.}
PQL queries should be decidable, \ie given a PQL query and a process model it should always be possible to decide if the model's behavior satisfies the query.
For each PQL predicate, it is either already known that it is decidable~\cite{PolyvyanyyWCRH14} or we show that it indeed can be computed, refer to \sectionname~\ref{app:PQL:predicates}. 

\medskip
\item\textbf{Exploratory search support.}
An exploratory search is an approach to information exploration which represents the activities carried out by users who are unfamiliar with the domain, or unsure about their goals and/or ways to achieve their goals~\cite{White2009}.
Often, these users apply querying to study the domain and/or foster learning. 

A user of PQL may be unfamiliar with process models stored in the repository or exact labels used to specify process tasks. 
Indeed, process models often suffer from the inconsistent usage of labels, even when developed for the same domain~\cite{Leopold13}. 
Consequently, a search procedure that relies on the exact comparison of task labels is likely to miss some important matches of similar tasks.
To address this issue in PQL, task labels can be expanded.
In information retrieval, a \emph{query expansion} is a process of reformulating the query to improve the effectiveness of search results~\cite{ManningRaghavanSchuetze08}. 
A task label can be reformulated into a similar label, \eg using the technique proposed in~\cite{AwadPW08}.
A fresh label can then be used to replace the original label in the seed query to obtain a new expanded query that can contribute the otherwise unanticipated relevant matches to the search procedure.
For example, the user may be inclined to accept that the label ``Print inventory list'' used to model a function in \figurename~\ref{fig:EPC} is similar to the label ``Produce inventory document'' that is used in a query.
Several interviewees, refer to \sectionname~\ref{subsec:empirical:evaluation:results}, suggested that the language for behavioral querying of process models should support the users in performing exploratory search, which lead to identification of the corresponding theme.
Some direct quotes of the participants of our study from this theme are listed below:

\smallskip
\begin{compactenum}
\item[$-$] ``\emph{It should be possible to specify activities where tasks similar to `Apply discount' can occur because having the exact string gets very difficult \ldots someone thinks `apply discount' \ldots someone calls it `discount the thing' or whatever. Is it something you also consider? I think this is something that makes such a query language quite powerful}'' (CEO of a company working in BPM and process mining consultancy)
\item[$-$] ``\emph{Sometimes the customers' language is different, but they mean the same thing \ldots that [support for label similarities] would be something helpful that I would definitely use.}'' (Senior consultant with three years experience)
\item[$-$] ``\emph{I think that [support for label similarities] will be interesting because people look for something at different angles to see the same thing.}'' (Business analyst with ten years experience)
\end{compactenum}

\end{compactitem}

\subsection{Abstract Syntax}
\label{subsec:PQL:abstract:syntax}

\reportornot{
This section discusses the syntax of PQL in the form of an \emph{abstract syntax}, which is also often referred to as an \emph{(abstract) grammar}. 
An example of a PQL query represented in its abstract syntax is provided at the end of the section. 

The grammar of PQL is defined using the notation introduced in~\cite{Meyer1990}. 
In this notation, the abstract grammar of a programming language consists of a finite set of names of \emph{constructs} and a finite set of \emph{productions}, each associated with a construct.
Each construct describes the structure of a set of objects, also called \emph{specimens} of the language, using productions of three types; these are the \emph{aggregate}, \emph{choice}, and \emph{list} productions.
}{}

The top construct of the PQL grammar is \texttt{Query}. 
It captures the core structure of all PQL programs. 
\small
\begin{eqnarray*}
\texttt{Query}							& \triangleq & \mathit{vars}:\texttt{Variables};\; \mathit{atts}:\texttt{Attributes};\; \mathit{locs}:\texttt{Locations};\; \mathit{pred}:\texttt{Predicate}
\end{eqnarray*}
\normalsize
The \texttt{Query} construct is defined as an \emph{aggregate} production composed of four components. 
\reportornot{
In general, an aggregate production defines a construct that is made of a fixed number of components. 
The components are separated by semicolons, each preceded by a \emph{tag} indicating its \emph{role} within the construct.
}{}
\reportornot{Thus, every PQL query is composed of variables, attributes, locations, and a predicate, which are distinguished via tags $\mathit{vars}$, $\mathit{atts}$, $\mathit{locs}$, and $\mathit{pred}$, respectively.}{Thus, every PQL query is composed of variables, attributes, locations, and a predicate.}
\reportornot{Intuitively, a PQL query specifies an intent to discover models, and their attributes, that are identified by the locations and satisfy the predicate, where the evaluation of the predicate relies on information stored in the variables; note that the detailed discussion of the precise meaning of PQL queries is postponed until \sectionname~\ref{subsec:PQL:dynamic:semantics}.}{Intuitively, a PQL query specifies an intent to discover models, and their attributes ($\mathit{atts}$), that are identified by the locations ($\mathit{locs}$) and satisfy the predicate ($\mathit{pred}$), where the evaluation of the predicate relies on information stored in the variables ($\mathit{vars}$).}
\reportornot{
The order in which the various specimens are listed in aggregate productions is irrelevant for the sake of the abstract grammar specification. 
This order is important in the context of the next section, in which one possible concrete syntax of PQL is proposed.

}{}
\reportornot{The \texttt{Query} construct defines the collection of all PQL queries.}{} 
One can capture a PQL query using abstract syntactic expressions. 
For example, the statement $\texttt{q} \;\triangleq\; \texttt{Query} (\mathit{vars}:\mathit{vs};\,\, \mathit{atts}:\mathit{as};\,\, \mathit{locs}:\mathit{ls};\,\, \mathit{pred}:p)$ defines a query having $vs$, $\mathit{as}$, $\mathit{ls}$, and $\mathit{p}$, as variables, attributes, locations, and a predicate, respectively (assuming that all the specimens, \ie $\mathit{vs}$, $\mathit{as}$, $\mathit{ls}$, and $p$, are provided). 

\reportornot{In PQL, variables, attributes, and locations are defined as \emph{list} productions, where a list production defines a sequence of zero, one, or more specimens of another construct.}{In PQL, variables, attributes, and locations are defined as \emph{list} productions.}
\small
\begin{eqnarray*}
\texttt{Variables}  & \triangleq & \texttt{Variable}^*\\
\texttt{Attributes}	& \triangleq & \texttt{Attribute}^+\\
\texttt{Locations}  & \triangleq & \texttt{Location}^+
\end{eqnarray*}
\normalsize
\noindent
\reportornot{Therefore, a PQL query defines a sequence of zero, one, or more variables, denoted by $\texttt{Variable}^*$; the asterisk symbol stands for the \emph{Kleene star}---its standard language theory meaning.}{A PQL query defines a sequence of zero, one, or more variables, denoted by $\texttt{Variable}^*$, while every sequence of attributes in a query must contain at least one attribute, denoted by $\texttt{Attribute}^+$.}
\reportornot{Every sequence of attributes must contain at least one attribute, denoted by $\texttt{Attribute}^+$; note that the asterisk symbol is replaced by a plus sign to signify that the list of locations cannot be empty. Similarly, every sequence of locations must contain at least one location specimen.}{}

PQL introduces a dedicated construct, denoted by \texttt{Variable}, to define \emph{variables}. 
\small
\begin{eqnarray*}
\texttt{Variable}  & \triangleq & \mathit{name}:\texttt{VariableName};\; \mathit{tasks}:\texttt{SetOfTasks}
\end{eqnarray*}
\normalsize
\noindent
\reportornot{A PQL variable associates a symbolic name with a set of tasks, or to be more precise, with a collection of \emph{PQL tasks}, \ie abstract concepts that represent tasks.}{A PQL variable associates a symbolic name with a set of \emph{PQL tasks}, \ie abstract concepts that represent tasks.} 
Tasks are introduced in the language to refer to atomic units of observable behavior captured in process models, \ie they are the smallest irreducible concepts that can be observed during execution of process models.
Each variable is an aggregate of two constructs: a variable name ($\mathit{name}:\texttt{VariableName}$) and a collection of tasks ($\mathit{tasks}:\texttt{SetOfTasks}$). 
Such a separation of the variable name from its associated content allows the name to be used independently of the exact information it represents. 
\reportornot{Thus, a variable name can be bound to a set of tasks during run time, and the content of the set may change during evaluation of the query.}{} 
When a predicate of some PQL query gets evaluated, every variable name that is mentioned in the predicate is replaced by the corresponding set of tasks.

PQL introduces the \texttt{Attribute} construct to allow specifying process model properties that must be retrieved in a response to a successful query matching exercise. 
\small
\begin{eqnarray*}
\texttt{Attribute}			& \triangleq & \texttt{Universe} \;\;|\;\; \texttt{AttributeName}
\end{eqnarray*}
\normalsize
A PQL attribute identifies a single property of a process model. 
The \texttt{Attribute} construct is specified as a \emph{choice} production with two alternatives. 
\reportornot{In general, a choice production defines a construct as a set of alternatives. The alternatives are separated by vertical bar symbols.}{} 
Hence, every PQL attribute is either the \emph{universe} attribute, denoted by \texttt{Universe}, or the \emph{name} attribute, denoted by \texttt{AttributeName}. 
The universe attribute refers to the list of all attributes of process models in the repository. 
The name attribute is introduced in PQL to allow searching for repository specific properties of process models, \eg unique identifier, creation date, author, version, title, description.
\reportornot{
The user can specify arbitrary name attributes in queries. 
However, only those supported by the repository will be returned in response to a successful query execution.
}{}

\reportornot{In a query, locations are used to refer to process models that should be matched with the query, \ie checked on whether they make the predicate of the query evaluate to \emph{true}.}{In a query, locations are used to refer to process models that should be matched with the query.}
\small
\begin{eqnarray*}
\texttt{Location}			& \triangleq & \texttt{Universe} \;\;|\;\; \texttt{LocationPath}
\end{eqnarray*}
\normalsize
\noindent 
\reportornot{A location identifies a single process model or a collection of process models. It is defined as a \emph{choice} production.}{}
A location is either the \emph{universe} location, denoted by \texttt{Universe}, or a \emph{path} location, denoted by \texttt{LocationPath}. 
The universe location is introduced in the language to refer to all process models in the scope of the query (usually, all process models in the repository). 
Path locations are introduced in PQL to allow fine-grained targeting of models based on paths in the repository. 
Here, we assume that repositories do indeed have mechanisms in place to address models via unique path identifiers, \eg using URIs~\cite{URI} or XPath expressions~\cite{XPath}. 

PQL provides several alternatives for specifying a set of tasks. 
A set of tasks can be defined as an enumeration of tasks, a result of standard operations on sets of tasks, information stored in a variable, a construction macro, or a dynamically-valued constant.
These various possibilities are captured in the \texttt{SetOfTasks} construct of the PQL grammar. 
\small
\begin{eqnarray*}
\texttt{SetOfTasks}		& \triangleq & \texttt{VariableName} \;\;|\;\; \texttt{SetOfAllTasks} \;\;|\;\; \texttt{SetOfTasksLiteral} \\
																				&											& |\;\; \texttt{SetOfTasksConstruction}  \;\;|\;\; \texttt{Union} \;\;|\;\; \texttt{Intersection} \;\;|\;\; \texttt{Difference}
\end{eqnarray*}
\normalsize
\noindent
\reportornot{The \texttt{SetOfTasks} construct is defined as a choice production.}{}
\reportornot{One can use the \texttt{VariableName} construct to refer to the set of tasks associated with (a name of) some variable.}{One can use the \texttt{VariableName} construct to refer to the set of tasks associated with some variable.} 
\reportornot{
Alternatively, one can specify a set of tasks using the $\texttt{SetOfAllTasks}$ construct. 
The $\texttt{SetOfAllTasks}$ construct constitutes a dynamically-valued constant that refers to the set of all tasks of the process model currently being matched to the query, refer to \sectionname~\ref{subsec:PQL:dynamic:semantics} for details.
}{
One can also specify a set of tasks using the $\texttt{SetOfAllTasks}$ construct, which constitutes a dynamically-valued constant that refers to the set of all tasks of the process model currently being matched to the query, refer to \sectionname~\ref{subsec:PQL:dynamic:semantics} for details.
} 

PQL proposes a notation to specify a set of tasks literal, \ie a notation for representing a set of tasks as a fixed value. 
\reportornot{A set of tasks literal can be defined using the $\texttt{SetOfTasksLiteral}$ construct, which is specified as a list production of zero, one, or more tasks.}{A set of tasks literal can be defined using the $\texttt{SetOfTasksLiteral}$ construct.} 
\small
\begin{eqnarray*}
\texttt{SetOfTasksLiteral}  & \triangleq & \texttt{Task}^*
\end{eqnarray*}
\normalsize
\reportornot{
As mentioned above, tasks are abstract representations of atomic units of observable behavior. 
PQL offers three ways to specify a task. 
These are captured in the definition of the \texttt{Task} construct below.
}{
PQL offers three ways to specify a task captured in the definition of the \texttt{Task} construct below.
}
\small
\begin{eqnarray*}
\texttt{Task}			  & \triangleq & \texttt{ExactTask} \;\;|\;\; \texttt{DefSimTask} \;\;|\;\; \texttt{SimTask}\\
\texttt{ExactTask}  & \triangleq & \mathit{label}:\texttt{Label}\\
\texttt{DefSimTask} & \triangleq & \mathit{label}:\texttt{Label}\\
\texttt{SimTask}    & \triangleq & \mathit{label}:\texttt{Label};\; \mathit{sim}:\texttt{Similarity}
\end{eqnarray*}
\normalsize
Intuitively, a PQL task is a collection of labels, \ie character strings, which are similar with a given label up to a given similarity degree threshold, refer to \sectionname~\ref{subsec:PQL:design:principles}.
\reportornot{
Note that the \texttt{ExactTask} construct and the \texttt{DefSimTask} construct, which are both defined by means of production $\mathit{label}:\texttt{Label}$, are distinguished at the level of concrete syntax, refer to \sectionname~\ref{subsec:PQL:concrete:syntax}.
}{}
The explanation of the differences in meanings of the three constructs that define a PQL task is proposed in \sectionname~\ref{subsec:PQL:dynamic:semantics}.

Another way to specify a set of tasks is to construct it.
\reportornot{To implement the construction, one can use the \texttt{SetOfTasksConstruction} construct, which is defined as a choice production below.}{} 
\small
\begin{eqnarray*}
\texttt{SetOfTasksConstruction}		& \triangleq & \texttt{UnaryPredicateConstruction} \;\;|\;\; \texttt{BinaryPredicateConstruction}
\end{eqnarray*}
\normalsize
\noindent
Given a set of tasks and a unary behavioral predicate, \texttt{UnaryPredicateConstruction} can be used to construct a set of tasks that contains every task from the given set for which the given behavioral predicate holds, and contains no other tasks.   
\reportornot{The given behavioral predicate must be evaluated in the context of the process model that is being matched to the query.}{}
Similarly, \texttt{BinaryPredicateConstruction} is introduced in PQL to allow selecting those tasks from a given set of tasks for which certain binary behavioral predicate holds, either with at least one or with all tasks taken from another given set of tasks. 
The choice of a quantifier type, either existential or universal, to be used during the above described selections is implemented via the \texttt{AnyAll} construct. 

\quad
\vspace{-9mm}

\begin{eqnarray*}
\texttt{UnaryPredicateConstruction}		& \triangleq & \mathit{name}:\texttt{UnaryPredicateName};\; \mathit{tasks}:\texttt{SetOfTasks}\\
\texttt{BinaryPredicateConstruction}  & \triangleq & \mathit{name}:\texttt{BinaryPredicateName};\; \mathit{tasks}_1:\texttt{SetOfTasks};\\
																																					&						 					& \mathit{tasks}_2:\texttt{SetOfTasks};\; q:\texttt{AnyAll}\\
\texttt{AnyAll}																								& \triangleq & \texttt{Any} \;\;|\;\; \texttt{All}
\end{eqnarray*}
\normalsize
\reportornot{\texttt{UnaryPredicateConstruction} and \texttt{BinaryPredicateConstruction} are associated with aggregate productions. 
The \texttt{AnyAll} construct is specified as a choice between the \texttt{Any} qualifier and the \texttt{All} qualifier, where \texttt{Any} and \texttt{All} stand for the existential quantifier type and the universal quantifier type, respectively.}{}
PQL uses \texttt{UnaryPredicateName} and \texttt{BinaryPredicateName} constructs to refer to unary and binary behavioral predicates, respectively.
PQL supports two unary and six binary behavioral predicates, refer to \sectionname~\ref{sec:behavioral:relations}.
\reportornot{These are the \texttt{CanOccur} and \texttt{AlwaysOccurs} unary behavioral predicates, and the \texttt{CanConflict}, \texttt{CanCooccur}, \texttt{Conflict}, \texttt{Cooccur}, \texttt{TotalCausal}, and \texttt{TotalConcurrent} binary behavioral predicates.}{} 
\small
\begin{eqnarray*}
\texttt{UnaryPredicateName}		& \triangleq & \texttt{CanOccur} \;\;|\;\; \texttt{AlwaysOccurs} \\
\texttt{BinaryPredicateName}	& \triangleq & \texttt{CanConflict} \;\;|\;\; \texttt{CanCooccur} \;\;|\;\; \texttt{Conflict}\\
																													& 										& \;\;|\;\; \texttt{Cooccur} \;\;|\;\; \texttt{TotalCausal} \;\;|\;\; \texttt{TotalConcurrent}
\end{eqnarray*}
\normalsize
Finally, a set of tasks can be constructed from other sets of tasks via the application of the fundamental set operations of \emph{union}, \emph{intersection}, and \emph{difference}, denoted by the $\texttt{Union}$, $\texttt{Intersection}$, and $\texttt{Difference}$ constructs, respectively. 

PQL proposes several ways to specify predicates in queries. 
\reportornot{All the options are captured in the choice production that is associated with the \texttt{Predicate} construct shown below.}{}
\small
\begin{eqnarray*}
\texttt{Predicate}	& \triangleq & \texttt{UnaryPredicate} \;\;|\;\; \texttt{BinaryPredicate} \;\;|\;\; \texttt{UnaryPredicateMacro}\\
																	& 					 					& \;\;|\;\; \texttt{BinaryPredicateMacro} \;\;|\;\; \texttt{SetPredicate} \;\;|\;\; \texttt{TruthValue} \;\;|\;\; \texttt{Negation}\\
																	&											& \;\;|\;\; \texttt{Conjunction} \;\;|\;\; \texttt{Disjunction} \;\;|\;\; \texttt{LogicalTest}
\end{eqnarray*}
\normalsize
For example, predicates can be captured as specimens of the \texttt{UnaryPredicate} construct or the \texttt{BinaryPredicate} construct.
\small
\begin{eqnarray*}
\texttt{UnaryPredicate}			& \triangleq & \mathit{name}:\texttt{UnaryPredicateName};\; \mathit{task}:\texttt{Task}\\
\texttt{BinaryPredicate}			& \triangleq & \mathit{name}:\texttt{BinaryPredicateName};\; \mathit{task}_1:\texttt{Task};\; \mathit{task}_2:\texttt{Task}
\end{eqnarray*}
\normalsize
\reportornot{
The \texttt{UnaryPredicate} construct and the \texttt{BinaryPredicate} construct are introduced in PQL to allow checking unary behavioral predicates and binary behavioral predicates, respectively. 
Both these constructs are aggregations of a name (specified by the \texttt{UnaryPredicateName} construct or the \texttt{BinaryPredicateName} construct) and a respective number of \texttt{Task} constructs; one for the \texttt{UnaryPredicate} construct and two for the \texttt{BinaryPredicate} construct.
}{
\noindent
}
PQL utilizes a well-known mechanism of macros for combining results of several \texttt{UnaryPredicate} or \texttt{BinaryPredicate} checks into a result of a single statement.
\small
\begin{eqnarray*}
\texttt{UnaryPredicateMacro}	& \triangleq & \mathit{name}:\texttt{UnaryPredicateName};\;  \mathit{tasks}:\texttt{SetOfTasks};\; q:\texttt{AnyAll}\\
\texttt{BinaryPredicateMacro}	& \triangleq & \texttt{BinaryPredicateMacroTaskSet} \;\;|\;\; \texttt{BinaryPredicateMacroSetSet}
\end{eqnarray*}
\normalsize
The aggregate production associated with the \texttt{UnaryPredicateMacro} construct is composed of a reference to a unary behavioral predicate ($name:\texttt{UnaryPredicateName}$), a set of tasks ($tasks:\texttt{SetOfTasks}$), and a quantifier ($q:\texttt{AnyAll}$). 
Intuitively, a single macro statement $p \triangleq \texttt{UnaryPredicateMacro}(\mathit{name}:n;\,\,\mathit{tasks}:\mathit{ts};\,\,\mathit{q}:x)$ is equivalent to a complex check of whether it holds that 
for at least one (if $x$ is set to \texttt{Any}) or for every (if $x$ is set to \texttt{All}) task $t$ in set of tasks $\mathit{ts}$ statement $\texttt{UnaryPredicate}(p.\mathit{name};\,\,  \mathit{task} : t)$ evaluates to \emph{true}.
Similarly, one can rely on the \texttt{BinaryPredicateMacro} construct to combine results of multiple \texttt{BinaryPredicate} checks. 

\quad
\vspace{-9mm}

\small
\begin{eqnarray*}
\texttt{BinaryPredicateMacroTaskSet}	& \triangleq & \mathit{name}:\texttt{BinaryPredicateName};\; \mathit{task}:\texttt{Task};\\
																			& 					 & \mathit{tasks}:\texttt{SetOfTasks};\; q:\texttt{AnyAll}\\
\texttt{BinaryPredicateMacroSetSet}		& \triangleq & \mathit{name}:\texttt{BinaryPredicateName};\; \mathit{tasks}_1:\texttt{SetOfTasks};\\
																			& 					 & \mathit{tasks}_2:\texttt{SetOfTasks};\; q:\texttt{AnySomeEachAll};\\
\texttt{AnySomeEachAll}								& \triangleq & \texttt{Any} \;\;|\;\; \texttt{Some} \;\;|\;\; \texttt{Each} \;\;|\;\; \texttt{All}
\end{eqnarray*}
\normalsize
\noindent
The \texttt{BinaryPredicateMacroTaskSet} construct is designed to allow checking whether a certain binary behavioral predicate ($\mathit{name}:\texttt{BinaryPredicateName}$) holds between a given task ($\mathit{task}:\texttt{Task}$) and either at least one (if the \texttt{AnyAll} construct is instantiated with the \texttt{Any} specimen) or every (if the \texttt{AnyAll} construct is instantiated with the \texttt{All} specimen) task in a given set of tasks ($\mathit{tasks}:\texttt{SetOfTasks}$).
Similarly, the \texttt{BinaryPredicateMacroSetSet} construct can be used to check whether a binary behavioral predicate of interest evaluates to \emph{true} for certain pairs of tasks in the Cartesian product of two given sets of tasks.

Note for the options to use \texttt{Some} and \texttt{Each} qualifier as a specimen of the \texttt{AnySomeEachAll} construct in the \texttt{AnySomeEachAll} production.  
When the \texttt{Some} option is used, one requests to check  whether for \emph{some} task in one given set of tasks the specified behavioral relation holds with each task from the other given set of tasks.
The \texttt{Each} option induces a check of whether for \emph{each} task in one set of tasks the specified behavioral relation holds with some task from the other set of tasks.

PQL supports checks of basic binary relations between (elements of) sets of tasks. 
These are captured by the choice production associated with the \texttt{SetPredicate} construct.
\small
\begin{eqnarray*}
\texttt{SetPredicate}					& \triangleq & \texttt{TaskInSetOfTasks} \;\;|\;\; \texttt{SetComparison} \\
\texttt{TaskInSetOfTasks}			& \triangleq & \mathit{task}:\texttt{Task};\; \mathit{tasks}:\texttt{SetOfTasks}\\
\texttt{SetComparison}				& \triangleq & \mathit{tasks}_1:\texttt{SetOfTasks};\; \mathit{oper}:\texttt{SetComparisonOperator};\; \mathit{tasks}_2:\texttt{SetOfTasks}\\
\texttt{SetComparisonOperator}	& \triangleq & \texttt{Identical} \;\;|\;\; \texttt{Different} \;\;|\;\; \texttt{OverlapsWith} \;\;|\;\; \texttt{SubsetOf} \;\;|\;\; \texttt{ProperSubsetOf}
\end{eqnarray*}
\normalsize
\noindent
\reportornot{
PQL can be used to check if a task is a member of a given set of tasks. 
This check can be accomplished using the \texttt{TaskInSetOfTasks} construct, which is specified as an aggregation of a task ($\mathit{task}:\texttt{Task}$) and a set of tasks ($\mathit{tasks}:\texttt{SetOfTasks}$). 
}{
PQL can be used to check if a task is a member of a given set of tasks using the \texttt{TaskInSetOfTasks} construct. 
}
\reportornot{
In addition, PQL can be used to compare sets of tasks using the $\texttt{SetComparison}$ construct. 
The $\texttt{SetComparison}$ construct is composed of two sets of tasks ($\mathit{tasks}_1:\texttt{SetOfTasks}$ and $\mathit{tasks}_2:\texttt{SetOfTasks}$) and a reference to a set comparison operator ($\mathit{oper}\,:\,\texttt{SetComparisonOperator}$).
}{
Also, PQL can be used to compare sets of tasks using the $\texttt{SetComparison}$ construct that consists of two sets of tasks ($\mathit{tasks}_1:\texttt{SetOfTasks}$ and $\mathit{tasks}_2:\texttt{SetOfTasks}$) and a reference to a set comparison operator ($\mathit{oper}\,:\,\texttt{SetComparisonOperator}$).
}
\reportornot{PQL supports five set comparison operations. 
These operations refer to checks of whether two sets of tasks are identical ($\texttt{Identical}$), different ($\texttt{Different}$), overlap ($\texttt{OverlapsWith}$), or
whether one set of tasks is a subset ($\texttt{SubsetOf}$) or a proper subset ($\texttt{ProperSubsetOf}$) of another set of tasks.
}{
PQL supports five set comparison operations that refer to checks of whether two sets of tasks are identical ($\texttt{Identical}$), different ($\texttt{Different}$), overlap ($\texttt{OverlapsWith}$), or
whether one set of tasks is a subset ($\texttt{SubsetOf}$) or a proper subset ($\texttt{ProperSubsetOf}$) of another set of tasks.
}

PQL operates with two truth values: \emph{true} and \emph{false}. 
This is reflected in the two literals of the choice production associated with the \texttt{TruthValue} construct, see below.
\small
\begin{eqnarray*}
\texttt{TruthValue}						& \triangleq & \texttt{True} \;\;|\;\; \texttt{False} 
\end{eqnarray*}
\normalsize
To allow complex logical statements, PQL supports standard logical operations. 
These are negation (\texttt{Negation}), conjunction (\texttt{Conjunction}), and disjunction (\texttt{Disjunction}).

PQL allows testing whether a given logic value is \emph{true} or \emph{false}. 
These checks are reflected in the options of the \texttt{LogicalTest} construct proposed below.
\small
\begin{eqnarray*}
\texttt{LogicalTest}	& \triangleq & \texttt{IsTrue} \;\;|\;\; \texttt{IsNotTrue} \;\;|\;\; \texttt{IsFalse} \;\;|\;\; \texttt{IsNotFalse}
\end{eqnarray*}
\normalsize
For a grammar of a language to be complete, all its constructs must be specified in terms of well-defined components, called the \emph{terminal} constructs. 
For example, the following constructs are the terminal constructs of the PQL grammar: \texttt{Any}, \texttt{Some}, \texttt{Each}, \texttt{All}, \texttt{Universe}, \texttt{Identical}, \texttt{Different}, \texttt{OverlapsWith}, \texttt{SubsetOf}, \texttt{ProperSubsetOf}, \texttt{True}, \texttt{False}, as well as all the constructs that are parts of the choice productions associated with the \texttt{UnaryPredicateName} and \texttt{BinaryPredicateName} constructs. 
All the above mentioned constructs do not have an internal structure and, thus, are atomic constructs of PQL.
Several of the proposed PQL constructs can be defined in terms of special sets. 
For example, PQL specifies 
\texttt{VariableName}, \texttt{AttributeName}, \texttt{LocationPath}, \texttt{Label}, and \texttt{Similarity}, as 
$\texttt{VariableName} \;\triangleq\; \mathit{id}:\mathbb{V}$, 
$\texttt{AttributeName} \;\triangleq\; \mathit{id}:\mathbb{C}$, 
$\texttt{LocationPath} \;\triangleq\; \mathit{id}:\mathbb{C}$, 
$\texttt{Label} \;\triangleq\; \mathit{value}:\mathbb{C}$, 
and $\texttt{Similarity} \;\triangleq\; \mathit{value}:[0 \, .. \, 1]$, respectively.
Recall that $\mathbb{C}$ is the universe of all finite character strings.
By $\mathbb{V}$, we denote the set of all legal PQL variable names.

\begin{figure}[t]
\centering
\includegraphics[width=1.02\linewidth, trim=5mm 0 0 0]{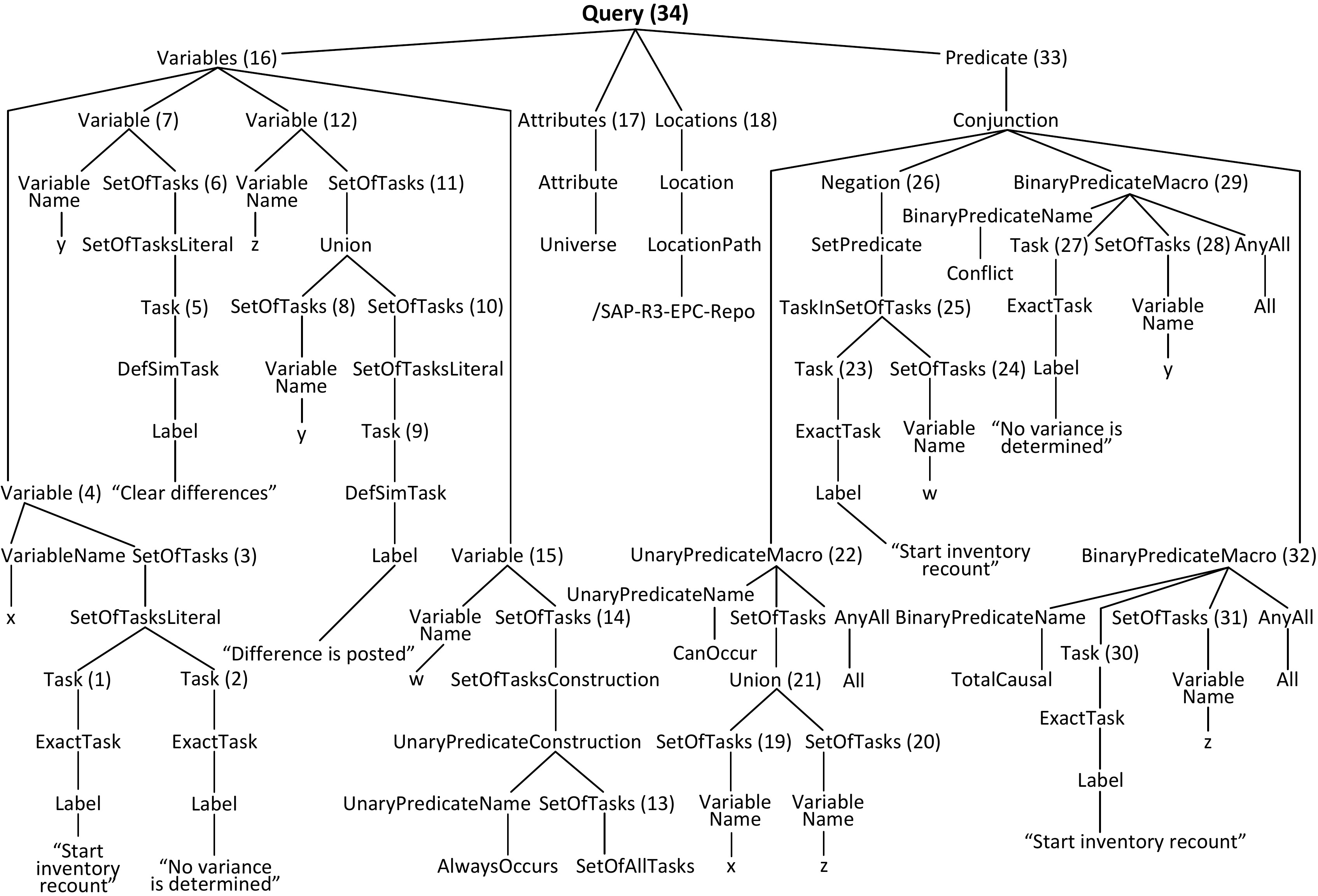}
\caption{\small Abstract syntax tree of the example PQL query~\emph{Q3} in \sectionname~\ref{sec:examples}. The numbers in parenthesis are used in \sectionname~\ref{subsec:PQL:dynamic:semantics} to explain the meaning of the query.}
\label{fig:syntaxtree}
\vspace{-4mm}
\end{figure}

PQL defines the \texttt{Negation} construct and all the four options associated with the \texttt{LogicalTest} construct in terms of a single \texttt{Predicate} component, \eg $\texttt{Negation} \triangleq \mathit{pred}:\texttt{Predicate}$, $\texttt{IsTrue} \triangleq \mathit{pred}:\texttt{Predicate}$, etc. 
For the sake of space considerations, at this stage we omit rigorous definitions of five PQL constructs: \texttt{Conjunction}, \texttt{Disjunction}, \texttt{Union}, \texttt{Intersection}, and \texttt{Difference}.
Intuitively, \texttt{Conjunction} and \texttt{Disjunction} can be defined as collections of predicates, whereas \texttt{Union}, \texttt{Intersection}, and \texttt{Difference} can be specified as collections of sets of tasks. 
\reportornot{
However, any definition of priorities for the operations that the above stated constructs represent in terms of grammar rules is rather lengthy and is driven by semantic, rather than syntactic, rules.
}{}

In the next section, we discuss priorities of various operations that are supported in PQL, whereas missing rigorous specifications of the five mentioned constructs can be found in \appendixname~\ref{app:PQL:grammar}.


\vspace*{.5\baselineskip}
\noindent 
\emph{An example of PQL's abstract syntax.} 
The abstract syntax of a PQL query can be represented as an abstract syntax tree. 
Each node of such an abstract syntax tree denotes a PQL construct. 
For example, \figurename~\ref{fig:syntaxtree} depicts the abstract syntax tree of sample PQL query~\emph{Q3} introduced in \sectionname~\ref{sec:examples}.

\subsection{Concrete Syntax}
\label{subsec:PQL:concrete:syntax}

\reportornot{The abstract syntax of PQL is independent of any particular representation.}{} 
This section proposes a specific encoding of the abstract syntax of PQL. 
This encoding constitutes one possible concrete syntax of PQL, \ie its machine- and human-readable representation.

The concrete syntax of PQL proposed in this section is inspired by SQL---a programming language for managing data stored in a relational database management system (DBMS)~\cite{DateD1997}. 
\reportornot{
Recall the three PQL query examples introduced in \sectionname~\ref{sec:examples}. 
All of them are specified using the concrete syntax of PQL, and it can be observed that they follow the concrete syntax of SQL. 
Being inspired by SQL, we keep the core structure of PQL queries as similar as possible to that of SQL queries and reuse SQL keywords in PQL, given that the contexts are similar. 
}{
We keep the core structure of PQL queries as similar as possible to that of SQL queries and reuse SQL keywords in PQL, given that the contexts are similar. 
}
The reason for this is threefold: 
\smallskip
\begin{compactitem}
\item
Despite addressing different domains, \ie dynamic processes versus static data, both languages serve the same purpose---the purpose of querying for information. 
\reportornot{Note that SQL was originally proposed to \emph{retrieve} data stored in quasi-relational DBMS~\cite{ChamberlinB1974}.}{}
\smallskip
\item
SQL is a widely used standard that is supported by just about every DBMS on the market. 
As a result, its syntax is well-recognized by technical specialists and analysts. 
By closely following the concrete syntax of SQL, PQL becomes readily usable by a wide range of stakeholders.
\smallskip
\item
As suggested by interviewees of the study reported in \sectionname~\ref{sec:behavioral:relations}, it would be beneficial for the syntax of the query language to resemble the syntax of SQL. One interviewee commented: 
\emph{``From an overall strategic point of view it'll bring a lot of benefits because different parts of the organization will be able to work together by using some kind of a structured query language''.}
\end{compactitem}
\smallskip

\noindent
Given a construct, one can specify its concrete syntax as a function that yields all its specific forms. 
\reportornot{PQL is defined as a textual language.}{}
\reportornot{Hence, for each PQL construct, its concrete syntax is given as a function that takes a specimen of the respective abstract construct as input and returns a collection of character strings that are accepted as its concrete encoding.}{For each PQL construct, its concrete syntax is given as a function that takes a specimen of the respective abstract construct as input and returns a collection of character strings that are accepted as its concrete encoding.}  
We denote such a function by the name of the construct with subscript \texttt{c}.

For example, the concrete syntax of a specimen of the \texttt{Query} construct is defined as follows.
\small
\begin{eqnarray*}
\texttt{Query}_\texttt{c}(\mathit{q}:\texttt{Query})	& \triangleq & \texttt{Variables}_\texttt{c}(\mathit{q.vars})\\
																																																	&											& \apqltext{\SELECT} \;\;\texttt{Attributes}_\texttt{c}(\mathit{q.atts})\\
																																																	&											& \apqltext{\FROM} \;\;\texttt{Locations}_\texttt{c}(\mathit{q.locs})\\
																																																	&											& (\apqltext{\WHERE} \;\; \texttt{Predicate}_\texttt{c}(\mathit{q.pred}))? \;\; \apqltext{;}
\end{eqnarray*}
\normalsize
\reportornot{We use regular expressions~\cite{AhoUllman1992} to define the concrete syntax of PQL specimens.}{}
\reportornot{Hence, as per the above definition, a PQL query is a character string that starts with a specification of variables, followed by the \SELECT~keyword, followed by a specification of attributes, followed by the \FROM~keyword, followed by a specification of locations, followed by the \WHERE~keyword, followed by a specification of a predicate, followed by the semicolon mark, \ie\apqltext{;}.}{Hence, a PQL query is a character string that starts with a specification of variables, followed by the \SELECT~keyword, followed by a specification of attributes, followed by the \FROM~keyword, followed by a specification of locations, followed by the \WHERE~keyword, followed by a specification of a predicate, followed by the semicolon mark, \ie\apqltext{;}.} 
There can be an arbitrary number of whitespace characters between any two subsequent components of a query string. 
The order of various components is fixed. 
\reportornot{Note that the presence of the \WHERE~clause in a PQL query is optional, \ie the \WHERE~keyword and the specification of the predicate can be skipped.}{Note that the presence of the \WHERE~clause in a PQL query is optional.}

\reportornot{
The reader might have already noticed that the core structure of a PQL query is similar to that of an SQL query signified with the declarative \SELECT~statement and used to formulate an intent for retrieving data from one or more database tables or expressions.
}{}

Specimens of PQL constructs that are associated with list productions are encoded as string concatenations of concrete forms of their components and whitespace characters. 
Often, we inject special symbols between every two subsequent components and/or at the beginning and end of the respective encodings. 
For example, the concrete syntax of a list of variables is defined as follows.
\small
\begin{eqnarray*}
\texttt{Variables}_\texttt{c}(\mathit{vs}:\texttt{Variables})	& \triangleq & \mathit{isEmpty(vs)} \;\; ? \;\; \apqltext{} : \texttt{Variable}_\texttt{c}(\mathit{vs}.\mathit{FIRST}) \;\; \texttt{Variables}_\texttt{c}(\mathit{vs}.\mathit{TAIL})
\end{eqnarray*}
\normalsize
That is, the encoding of the empty list of variables is the empty string. 
However, if a list of variables contains at least one element, its encoding is constructed as a concatenation of a concrete form of its first element, denoted by $\mathit{vs}.\mathit{FIRST}$, and an encoding of the list of its all other elements, denoted by $\mathit{vs}.\mathit{TAIL}$. 
The concrete syntax of a PQL variable is defined below.
\small
\begin{eqnarray*}
\texttt{Variable}_\texttt{c}(\mathit{v}:\texttt{Variable})	& \triangleq & \texttt{VariableName}_\texttt{c}(\mathit{v}.\mathit{name}) \;\; \apqltext{=} \;\; \texttt{SetOfTasks}_\texttt{c}(\mathit{v}.\mathit{tasks}) \;\; \apqltext{;}
\end{eqnarray*}
\normalsize
%
The concrete syntax of every other specimen of a PQL construct that is associated with a list production is defined similar to that of the \texttt{Variables} construct seen above.
However, every such encoding expects to include a special symbol between every two subsequent components. 
This is the comma symbol, \ie \apqltext{,}, for specimens of \texttt{Attributes}, \texttt{Locations}, and \texttt{SetOfTasksLiteral}, 
and the PQL keywords \UNION, \INTERSECT, \EXCEPT, \AND, and \OR, for specimens of \texttt{Union}, \texttt{Intersection}, \texttt{Difference}, \texttt{Conjunction}, and \texttt{Disjunction}, respectively. 
Additionally, every encoding of a specimen of the \texttt{SetOfTasksLiteral} construct must begin with the opening curly bracket, \ie \linebreak\apqltext{\{}, and end with the closing curly bracket, \ie \apqltext{\}}. 
\reportornot{For example, the character string \apqltext{\{"Buy item","Purchase product"\}} is a valid encoding of a specimen of the \texttt{SetOfTasksLiteral} construct that contains two elements; here, we follow the standard notation for specifying fixed sets.}{For example, the character string \apqltext{\{"Buy item","Purchase product"\}} is a valid encoding of a specimen of the \texttt{SetOfTasksLiteral} construct that contains two elements.} 
In the example , strings \texttt{"Buy item"} and \texttt{"Purchase product"} are valid encodings of tasks. 
PQL supports three alternative concrete encodings of a PQL task.
\small
\begin{eqnarray*}
\texttt{ExactTask}_\texttt{c}(\mathit{t}:\texttt{ExactTask})	& \triangleq & \apqltext{"} \;\; \texttt{Label}_\texttt{c}(\mathit{t}.\mathit{label}) \;\; \apqltext{"}\\
\texttt{DefSimTask}_\texttt{c}(\mathit{t}:\texttt{DefSimTask})	& \triangleq & \apqltext{$\sim$} \;\; \apqltext{"} \;\; \texttt{Label}_\texttt{c}(\mathit{t}.\mathit{label}) \;\; \apqltext{"}\\
\texttt{SimTask}_\texttt{c}(\mathit{t}:\texttt{SimTask})	& \triangleq & \apqltext{"} \;\; \texttt{Label}_\texttt{c}(\mathit{t}.\mathit{label}) \;\; \apqltext{"} \;\; \apqltext{[} \;\; \texttt{Similarity}_\texttt{c}(\mathit{t}.\mathit{sim}) \;\;\apqltext{]}
\end{eqnarray*}
\normalsize
Labels of PQL tasks must be enclosed in double quotes. 
A label can be preceded by the tilde symbol, \ie \apqltext{$\sim$}, or succeeded by an encoding of a similarity degree threshold enclosed in square brackets. 
The tilde symbol denotes that one is interested in all the tasks of which the label has a degree of similarity to the specified label that is equal or larger than some preconfigured value. 
A degree of similarity must be specified as a decimal representation of a real number greater or equal to zero and less than or equal to one, \eg $0.5$ or $.95$.
A specimen associated with a choice production is a specimen of one of the constructs from the list of alternatives of the production. 
\reportornot{Hence, a concrete syntax of an abstract grammar can (and often does) omit special encodings to signify choice productions, which is the case for the concrete syntax of PQL that is being proposed here.}{}
Thus, in the sequel, we only propose concrete encodings of the remaining aggregate productions of PQL.

A specimen of \texttt{Attribute} is a specimen of either the \texttt{Universe} or the \texttt{AttributeName} construct, whereas a specimen of the \texttt{Location} construct is a specimen of either the \texttt{Universe} or the \texttt{LocationPath} construct.
The \texttt{Universe} construct is denoted by \apqltext{*}.
The \texttt{AttributeName} and \texttt{LocationPath} specimens are denoted by character strings enclosed in double quotes.
\reportornot{However, we acknowledge that subsequent versions of PQL may adopt other concrete encodings for specimens of \texttt{AttributeName} and \texttt{LocationPath}, as well as for specimens of the \texttt{VariableName} construct discussed below.}{}

A concrete encoding of a specimen of the \texttt{VariableName} construct may contain lower case letters from the English alphabet, digits, and the underscore symbol, \ie \apqltext{\_}. 
\reportornot{It is necessary to use a letter or the underscore symbol at the start of a variable name; a digit at the start is not allowed. Subsequent characters may be letters, digits, or underscore symbols.}{It is necessary to use a letter or the underscore symbol at the start of a variable name.} 

Next, we define possible concrete encodings for specimens of the \texttt{SetOfAllTasks} construct, the \texttt{UnaryPredicateConstruction} construct, and the \texttt{BinaryPredicateConstruction} construct.
\small
\begin{eqnarray*}
\texttt{SetOfAllTasks}_\texttt{c}(\mathit{ts}:\texttt{SetOfAllTasks}) & \triangleq & \apqltext{GetTasks} \;\; \apqltext{(} \;\; \apqltext{)}
\end{eqnarray*}
\normalsize
\small
\begin{align*}
&\texttt{UnaryPredicateConstruction}_\texttt{c}(\mathit{upc}:\texttt{UnaryPredicateConstruction})	\;\; \triangleq\\
&\;\;\;\;\;\;\;\;\;\;\;\;\apqltext{GetTasks} \texttt{UnaryPredicateName}_\texttt{c}(\mathit{upc}.\mathit{name}) \;\; \apqltext{(} \;\; \texttt{SetOfTasks}_\texttt{c}(\mathit{upc}.\mathit{tasks}) \;\; \apqltext{)}
\end{align*}
\normalsize
\small
\begin{align*}
&\texttt{BinaryPredicateConstruction}_\texttt{c}(\mathit{bpc}:\texttt{BinaryPredicateConstruction})	\;\; \triangleq\\
&\;\;\;\;\;\;\;\;\;\;\;\;\apqltext{GetTasks} \texttt{BinaryPredicateName}_\texttt{c}(\mathit{bpc}.\mathit{name})\\ 
&\;\;\;\;\;\;\;\;\;\;\;\;\apqltext{(} \;\; \texttt{SetOfTasks}_\texttt{c}(\mathit{bpc}.\mathit{tasks}_1) \;\; \apqltext{,} \;\;  \texttt{SetOfTasks}_\texttt{c}(\mathit{bpc}.\mathit{tasks}_2) \;\; \apqltext{,} \;\; \texttt{AnyAll}_\texttt{c}(\mathit{bpc}.q) \;\; \apqltext{)}
\end{align*}
\normalsize
The concrete encodings of specimens of these constructs follow the syntax for specifying function calls used in many programming languages, \ie a name of a function to be called is followed by a comma-separated list of parameters which is enclosed in parentheses. 
For instance, one possible concrete encoding of a \texttt{SetOfAllTasks} specimen is \apqltext{GetTasks()}.
In the case of specimens of the \texttt{UnaryPredicateConstruction} construct and the \texttt{BinaryPredicateConstruction} construct, the names of functions are obtained by prefixing \apqltext{GetTasks} to names of unary and binary predicates, respectively. 
The remaining components are used as parameters of the corresponding functions. 
PQL exercises similar principles when specifying the concrete syntax of predicates and macros, both for the unary and binary cases. 
The concrete syntax for predicates proceeds as follows.
\small
\begin{align*}
&\texttt{UnaryPredicate}_\texttt{c}(\mathit{up}:\texttt{UnaryPredicate})	\;\; \triangleq\\ 
&\;\;\;\;\;\;\;\;\;\;\;\;\texttt{UnaryPredicateName}_\texttt{c}(\mathit{up}.\mathit{name}) \;\; \apqltext{(} \;\; \texttt{Task}_\texttt{c}(\mathit{up}.\mathit{task}) \;\; \apqltext{)}
\end{align*}
\normalsize
\small
\begin{align*}
&\texttt{BinaryPredicate}_\texttt{c}(\mathit{bp}:\texttt{BinaryPredicate})	\;\; \triangleq\\
&\;\;\;\;\;\;\;\;\;\;\;\;\texttt{BinaryPredicateName}_\texttt{c}(\mathit{bp}.\mathit{name}) \;\; \apqltext{(} \;\; \texttt{Task}_\texttt{c}(\mathit{bp}.\mathit{task}_1) \;\; \apqltext{,} \;\;  \texttt{Task}_\texttt{c}(\mathit{bp}.\mathit{task}_2) \;\; \apqltext{)}
\end{align*}
\normalsize
\reportornot{The only difference is that the names of these imitated function calls are not prefixed, but are solely composed of the concrete encodings of the respective predicate names.

}{

\noindent} 
The concrete syntax for denoting the PQL macros overloads the syntax for specifying function calls which encode the PQL predicates, \ie names of functions and types of outputs are the same, both for a given predicate and the corresponding macro. 
However, the types of inputs differ.
\small
\begin{align*}
&\texttt{UnaryPredicateMacro}_\texttt{c}(\mathit{upm}:\texttt{UnaryPredicateMacro})	\;\; \triangleq\\ 
&\;\;\;\;\;\;\;\;\;\;\;\;\texttt{UnaryPredicateName}_\texttt{c}(\mathit{upm}.\mathit{name}) \;\; \apqltext{(} \;\; \texttt{SetOfTask}_\texttt{c}(\mathit{upm}.\mathit{tasks}) \;\; \apqltext{,} \;\; \texttt{AnyAll}_\texttt{c}(\mathit{upm}.\mathit{q}) \;\;\apqltext{)}
\end{align*}
\normalsize
\small
\begin{align*}
&\texttt{BinaryPredicateMacroTaskSet}_\texttt{c}(\mathit{bpm}:\texttt{BinaryPredicateMacroTaskSet})	\;\; \triangleq\\ 
&\;\;\;\;\;\;\;\;\;\;\;\;\texttt{BinaryPredicateName}_\texttt{c}(\mathit{bpm}.\mathit{name})\\ 
&\;\;\;\;\;\;\;\;\;\;\;\;\apqltext{(} \;\; \texttt{Task}_\texttt{c}(\mathit{bpm}.\mathit{task}) \;\; \apqltext{,} \;\; \texttt{SetOfTask}_\texttt{c}(\mathit{bpm}.\mathit{tasks}) \;\; \apqltext{,} \;\; \texttt{AnyAll}_\texttt{c}(\mathit{bpm}.\mathit{q}) \;\;\apqltext{)}
\end{align*}
\normalsize
\small
\begin{align*}
&\texttt{BinaryPredicateMacroSetSet}_\texttt{c}(\mathit{bpm}:\texttt{BinaryPredicateMacroSetSet})	\;\; \triangleq\\ 
&\;\;\;\;\;\;\;\;\;\;\;\;\texttt{BinaryPredicateName}_\texttt{c}(\mathit{bpm}.\mathit{name})\\ 
&\;\;\;\;\;\;\;\;\;\;\;\;\apqltext{(} \;\; \texttt{SetOfTask}_\texttt{c}(\mathit{bpm}.\mathit{tasks}_1) \;\; \apqltext{,} \;\; \texttt{SetOfTask}_\texttt{c}(\mathit{bpm}.\mathit{tasks}_2) \;\; \apqltext{,} \;\; \texttt{AnySomeEachAll}_\texttt{c}(\mathit{bpm}.\mathit{q}) \;\;\apqltext{)}
\end{align*}
\normalsize
These syntax rules rely on the concrete encodings of \texttt{AnyAll} and \texttt{AnySomeEachAll}, which when instantiated are specified as a specimen of the \texttt{Any}, \texttt{Some}, \texttt{Each}, or \texttt{All} construct and are denoted by the PQL keywords \ANY, \SOME, \EACH, and \ALL, respectively.

A specimen of the \texttt{TaskInSetOfTasks} construct can be specified as follows.
\small
\begin{eqnarray*}
\texttt{TaskInSetOfTasks}_\texttt{c}(\mathit{in}:\texttt{TaskInSetOfTasks})	& \triangleq & \texttt{Task}_\texttt{c}(\mathit{in}.\mathit{task}) \;\; \apqltext{\IN} \;\; \texttt{SetOfTask}_\texttt{c}(\mathit{in}.\mathit{tasks})
\end{eqnarray*}
\normalsize
\reportornot{
That is, every character string that starts with an encoding of a task that is followed by the PQL keyword \IN~and ends with an encoding of a set of tasks specifies a specimen of  \texttt{TaskInSetOfTasks}.

}{\noindent}
A specimen of the \texttt{SetComparison} construct can be specified in the concrete syntax of PQL as two encodings of sets of tasks with a representation of a comparison operator in between. 
\small
\begin{align*}
&\texttt{SetComparison}_\texttt{c}(\mathit{comp}:\texttt{SetComparison})	\;\; \triangleq\\ 
&\;\;\;\;\;\;\;\;\;\;\;\; \texttt{SetOfTask}_\texttt{c}(\mathit{comp}.\mathit{tasks}_1) \;\; \texttt{SetComparisonOperator}_\texttt{c}(\mathit{comp}.\mathit{oper}) \;\; \texttt{SetOfTask}_\texttt{c}(\mathit{comp}.\mathit{tasks}_2)
\end{align*}
\normalsize
A set comparison operator is instantiated in PQL via a choice between specimens of terminal constructs \texttt{Identical}, \texttt{Different}, \texttt{OverlapsWith}, \texttt{SubsetOf}, and \texttt{ProperSubsetOf}, encoded using the keywords \EQUALS, \NOT~\EQUALS, \OVERLAPS~\WITH, \IS~\SUBSET~\OF, and, \IS~\PROPER~\SUBSET~\OF, respectively.
The terminal constructs \texttt{True} and \texttt{False} get encoded as the keywords \TRUE~and \FALSE, respectively.

Predicate names in PQL are encoded as names of the respective terminal constructs, \eg \texttt{CanOccur} and \texttt{TotalCausal} have concrete encodings \apqltext{CanOccur} and \apqltext{TotalCausal}, respectively.
Finally, the concrete encodings of \texttt{Negation} and the four logical test constructs are proposed below.
\small
\begin{eqnarray*}
\texttt{Negation}_\texttt{c}(\mathit{not}:\texttt{Negation})														& \triangleq & \apqltext{\NOT} \;\; \texttt{Predicate}_\texttt{c}(\mathit{not}.\mathit{pred}) \\
\texttt{IsTrue}_\texttt{c}(\mathit{test}:\texttt{IsTrue})																			& \triangleq & \texttt{Predicate}_\texttt{c}(\mathit{test}.\mathit{pred}) \;\; \apqltext{\IS} \;\; \apqltext{\TRUE}\\
\texttt{IsNotTrue}_\texttt{c}(\mathit{test}:\texttt{IsNotTrue})												& \triangleq & \texttt{Predicate}_\texttt{c}(\mathit{test}.\mathit{pred}) \;\; \apqltext{\IS} \;\; \apqltext{\NOT} \;\; \apqltext{\TRUE}\\
\texttt{IsFalse}_\texttt{c}(\mathit{test}:\texttt{IsFalse})																		& \triangleq & \texttt{Predicate}_\texttt{c}(\mathit{test}.\mathit{pred}) \;\; \apqltext{\IS} \;\; \apqltext{\FALSE}\\
\texttt{IsNotFalse}_\texttt{c}(\mathit{test}:\texttt{IsNotFalse})											& \triangleq & \texttt{Predicate}_\texttt{c}(\mathit{test}.\mathit{pred}) \;\; \apqltext{\IS} \;\; \apqltext{\NOT} \;\; \apqltext{\FALSE}
\end{eqnarray*}
\normalsize
\reportornot{These encodings rely on the use of the PQL keywords \NOT, \IS, \TRUE, \FALSE, and the concrete encoding of the \texttt{Predicate} construct, which is defined by concrete encodings of all the alternatives associated with the corresponding choice production.

}{\noindent}
As future work, we envision introduction of other specific encodings of the PQL grammar. 
We believe that availability of different concrete encodings will make PQL accessible to a wider audience.



\input{tex/semantics}

\input{tex/deciding}


\input{tex/sample_queries}

%% file: tex/predicates.tex
\subsection{Behavioral Querying and Basic Predicates}
\label{sec:behavioral:relations}

This section presents the results of an empirical study with process modeling experts that serve two purposes.\footnote{The study has been granted an approval on behalf of the University Human Research Ethics Committee, Queensland University of Technology, Australia, Ref. No.: 1000001158, and 
on behalf of the Human Ethics Advisory Group, the University of Melbourne, Australia, Ref. No.: 1851972.}
These results confirm that process querying based on behavioral inter-task relations is an appropriate method for retrieving process models from repositories. 
Moreover, they suggest a selection of basic behavioral predicates for inclusion in PQL.
The section starts by introducing a repertoire of behavioral predicates from our previous work~\cite{PolyvyanyyWCRH14}, refer to~\sectionname~\ref{sec:predicates}. 
Then, \sectionname~\ref{subsec:empirical:evaluation} presents the design of our empirical experiment.
Finally, \sectionname~\ref{subsec:empirical:evaluation:results} summarizes the results of the experiment.


\subsubsection{Behavioral Predicates}
\label{sec:predicates}

A \emph{behavioral relation} over process tasks in a model specifies an ordering constraint for occurrences of the tasks in the executions of the model. 
It has been shown that there are four fundamental categories of binary behavioral relations over process tasks: \emph{conflict}, \emph{causality}, \emph{concurrency} and \emph{co-occurrence}. 
These four categories of relations can be used to fully characterize any constraints over occurrences of process tasks~\cite{NielsenPW81,Engelfriet91,HaarKS13}. 
An occurrence of a process task implies that all its causally dependent tasks have already occurred and none of the conflicting tasks has been or will be observed, whereas two concurrent tasks can be enabled for simultaneous execution.
Finally, co-occurrence describes two process tasks that both occur in the same execution of a process model.\footnote{Note that the behavioral relations on tasks describe how they can be executed, and not how they are semantically related. For example, two tasks in the \emph{conflict} relation can never appear in the same execution of the model, not to be confused with semantic interference studied in Stroop experiments which look into how different concepts may conflict to hamper the understanding of the phenomenon~\cite{vanMaanen2009}.}

The \emph{4C spectrum}~\cite{PolyvyanyyWCRH14} is a systematic classification of behavioral relations grounded in the four categories of conflict, co-occurrence, causality, and concurrency. 
The spectrum proposes 18 basic relations over process tasks that can be combined using logical connectives into 318 predicates (63 conflict, 15 co-occurrence, 120 causality and 120 concurrency predicates). 
The basic relations are specified at different levels of `granularity', \ie given two process tasks they assess whether in all or some instances of the model, all or some occurrences of one task are in a relation with all or some occurrences of the other task.
For example, one of the 4C relations specifies that all occurrences of task \textbf{A} are concurrent to all occurrences of task \textbf{B} in all instances of a model, while another relation assesses whether at least one occurrence of \textbf{A} is concurrent to at least one occurrence of \textbf{B} in at least one instance of the process model.

We decided to use a subset of the 4C spectrum predicates to assess the relevance of using behavioral relations over process tasks for the purpose of querying process model collections.
Specifically, the selection of the predicates was driven by three factors: 
\begin{compactitem}
\item The selected predicates must cover all the four behavioral categories. 
\item The selected predicates must cover predicates of different granularity.
\item The list of selected predicates must be concise.
\end{compactitem}

\noindent
As a result, ten 4C predicates were selected: one conflict, one co-occurrence, four causality, and four concurrency predicates. 
In addition, in line with our previous work, we included two unary predicates that can be used to check if a given task can occur or always occurs in the executions of a process model~\cite{HofstedeORS0P13,PolyvyanyyRH14}. 
This led to the total of twelve predicates, which are reported with their names and text definitions in \tablename~\ref{tab:predicates}.
The reason for selecting only one conflict and one co-occurrence predicate is that all the 63 conflict and 15 co-occurrence predicates of the 4C spectrum stem from logical expressions over two basic relations, each addressing one of the respective behavioral categories. 
Four causal and four concurrent predicates were selected from the eight basic causal relations and the eight basic concurrency relations (as per the 4C spectrum), respectively. 
These predicates explore the different granularities of causality and concurrency.
For example, one can use \texttt{ExistCausal("Ship Order","Pay Order")} predicate to check if there exists an instance of a business process where shipment of order occurs before payment, which may signal a compliance issue.
Alternatively, one can use \texttt{TotalCausal("Pay Order","Ship Order")} to discover all the process models in which all the payments are finalized before all the shipments of the order take place and, thus, the aforementioned compliance issue does not manifest.
Finally, note that the ten 4C predicates assume an implicit condition that tasks \textbf{A} and \textbf{B} can occur in the model, \ie each of the tasks occurs in at least one possible execution of the model. 

\begin{table}[t]
\footnotesize
  \centering
  \caption{\small Twelve behavioral predicates.}
\begin{tabular}{|r|p{9.2cm}|}
\hline
\multicolumn{1}{ |r|  }{\multirow{1}{*}{\textbf{Behavioral predicate}} }  & \multicolumn{1}{ |l|  }{\multirow{1}{*}{\textbf{Definition}} } \\
\hline
\hline
\texttt{CanOccur(A)}\!\! & Find all process models where task \textbf{A} occurs in \emph{at least one} instance. \\
\hline
\texttt{AlwaysOccurs(A)}\!\! & Find all process models where task \textbf{A} occurs in \emph{every} instance. \\
\hline
\texttt{Cooccur(A,B)}\!\! & Find all process models where it holds that if task \textbf{A} occurs in \emph{some} instance then task \textbf{B} occurs in the \emph{same} instance, and \emph{vice versa}. \\
\hline
\texttt{Conflict(A,B)}\!\! & Find all process models where it holds that there is \emph{no} instance in which tasks \textbf{A} and \textbf{B} \emph{both} occur. \\
\hline
\texttt{ExistCausal(A,B)}\!\! & Find all process models where in \emph{at least one} instance it holds that \emph{some} occurrence of task \textbf{A} precedes \emph{some} occurrence of task \textbf{B}. \\
\hline
\texttt{ExistTotalCausal(A,B)}\!\! & Find all process models where in \emph{at least one} instance it holds that tasks \textbf{A} and \textbf{B} both occur and \emph{every} occurrence of task \textbf{A} precedes \emph{every} occurrence of task \textbf{B}. \\
\hline
\texttt{TotalExistCausal(A,B)}\!\! & Find all process models where for \emph{every} instance in which tasks \textbf{A} and \textbf{B} \emph{both} occur, it holds that \emph{some} occurrence of task \textbf{A} precedes \emph{some} occurrence of task \textbf{B}. \\
\hline
\texttt{TotalCausal(A,B)}\!\! & Find all process models where for \emph{every} instance in which tasks \textbf{A} and \textbf{B} \emph{both} occur, it holds that \emph{every} occurrence of task \textbf{A} precedes \emph{every} occurrence of task \textbf{B}. \\
\hline
\texttt{ExistConcurrent(A,B)}\!\! & Find all process models where in \emph{at least one} instance it holds that \emph{some} occurrence of task \textbf{A} can be executed at the same time with \emph{some} occurrence of task \textbf{B}. \\
\hline
\texttt{ExistTotalConcurrent(A,B)}\!\! & Find all process models where in \emph{at least one} instance it holds that tasks \textbf{A} and \textbf{B} \emph{both} occur and \emph{every} occurrence of task \textbf{A} can be executed at the same time with \emph{every} occurrence of task \textbf{B}. \\
\hline
\texttt{TotalExistConcurrent(A,B)}\!\! & Find all process models where for \emph{every} instance in which tasks \textbf{A} and \textbf{B} both occur, it holds that \emph{some} occurrence of task \textbf{A} can be executed at the same time with \emph{some} occurrence of task \textbf{B}. \\
\hline
\texttt{TotalConcurrent(A,B)}\!\! & Find all process models where for \emph{every} instance in which tasks \textbf{A} and \textbf{B} \emph{both} occur, it holds that \emph{every} occurrence of task \textbf{A} can be executed at the same time with  \emph{every} occurrence of task \textbf{B}. \\
\hline
\end{tabular}
\label{tab:predicates}
\vspace{-2mm}
\end{table}



\subsubsection{Experiment Design}
\label{subsec:empirical:evaluation}

Armed with the list of predicates from \tablename~\ref{tab:predicates}, we designed an experiment with two aims: 
(i) to gain understanding of the practical relevance of using behavioral predicates for querying process repositories, and (ii) to identify the most relevant predicates to implement in our query language. 
The experiment took the form of a one-hour semi-structured interview to seek expert opinions from practitioners that actively work with process models. 
The practitioners were contacted via public posting in dedicated Internet groups in the areas of business process management and process mining, or approached directly using our industry network.

As part of the interview, we first explained the rationale of the experiment and introduced the notion and main characteristics of the envisioned process query language. Next, we explained each of the twelve predicates in \tablename~\ref{tab:predicates} using a simple example, and conducted a short test to ensure that the interviewee had grasped the meaning of each predicate. 
Each predicate was presented in the form of a ``type of question'' that a process stakeholder, such as a business analyst, may need to get an answer to, about the process models that exist in their organization. 
For example, for the unary predicate \texttt{CanOccur}, we used the question ``Find all process models where task \textbf{A} occurs in \emph{at least one} instance'', while for the binary predicate \texttt{Conflict} we used the question ``Find all process models where it holds that there is \emph{no} instance in which tasks \textbf{A} and \textbf{B} both occur''.

We then proceeded with the actual questionnaire, which was divided into three parts. 
In the first part, we collected demographic information on the participants, such as role and experience with management of process models. 
The latter revolved around the number and type of process models managed, the key problems faced with these models, and the extent of process analysis conducted on these models on a daily basis.

Next, in the second part of the interview, we used four established metrics of data quality (usefulness, importance, likelihood, and frequency) as proxies for relevance, by asking the following four questions for each predicate, using 5-point Likert-type scales for the answers:

\begin{compactitem}
	\item How \textbf{useful} would an answer to such a question type be for your process analysis work?
	\item How \textbf{important} is such a question type to your process analysis work? 
	\item During process analysis, how \textbf{likely} does such a question type occur?
	\item During process analysis, how \textbf{frequently} does such a question type occur?
\end{compactitem}
	
\emph{Usefulness} and \emph{importance} are two external metrics on data quality~\cite{WandWang96}. 
They focus on the use and effect of an information system (in our case, of a given predicate in the PQL language) addressing the purpose and justification of the system, and its deployment in an organization. 
In the study, we adopted the definition of usefulness from~\cite{Davis89}, which states that perceived usefulness of a system is ``the degree to which a person believes that using a particular system would enhance his or her job performance''.
Importance of information is defined in~\cite{LarckerL80} as a degree to which information is a necessary input for task accomplishment.
Thus, usefulness and importance are measures of performance-enhancement and appropriateness, and are both task- and user-dependent.

While Wand and Wang present a range of external metrics (such as timeliness, flexibility, sufficiency, conciseness)~\cite{WandWang96}, we deemed usefulness and importance to be the most representative ones in our context (assessing the relevance of a given behavioral predicate), in light of the need to keep the interview brief. 
Internal metrics such as accuracy have been proven formally in this paper. 

We complemented usefulness and importance with \emph{likelihood} and \emph{frequency}, two other data quality metrics which act as proxy for the occurrence of a given predicate during process analysis. More specifically, these latter two metrics measure the \emph{intensity} of using a predicate in an organization, \ie the more likely and frequently a predicate occurs, the more intense is its use \cite{MendlingSR12}. 
Thus, likelihood and frequency are measures of significance and volume of a problem occurrence.
Likelihood is commonly understood as the condition of something being likely, or probable, while frequency is the rate at which something occurs over a period of time.

In the last part of the questionnaire, we allowed the interviewees to provide any additional comments on the above aspects of the envisioned query language. 

The complete interview instrument is provided in \appendixname~\ref{app:PQL:interview}. 
Presentation slides used to explain and test the understanding of the behavioral predicates are publicly available.\footnote{Presentation slides can be accessed here: \url{https://goo.gl/a9agBS}.}


\subsubsection{Experiment Results}
\label{subsec:empirical:evaluation:results}

We conducted the interviews with 25 practitioners. 
The results of two interviews were discarded, one because of inconsistency in the obtained feedback and the other because the interviewee background (CEO of a tool vendor with the experience of managing internal processes only), leading to a total of 23 interviews taken further to the analysis phase. 

All the interviewees work, or have worked in the past (for a significant time period), with process models and, hence, all of them are potential future users of PQL. 
Most of the participants of our study have a degree in Information Technology, Computer Science, Information Systems, Engineering, or Economics; at the undergraduate and/or graduate level. 
Many of the study participants hold dual degrees, including degrees in psychology, accounting, biology, political science, business and marketing, management, and business process management (BPM). 
Four participants received a Ph.D. degree. 
In their organizations, they have various roles, for example they are employed as business process analysts, business excellence managers, business architects, process architects, and BPM consultants. 
The professional experience of the participants ranges from half a year to over 40 years, with most of the interviewees having more than 7 years of professional experience (12.5 years on average). 
The interviewees reported that the number of process models they managed/analyzed in their practice varies from dozens to thousands (2,780 models on average).
These models belong to different domains: insurance, banking, investment and business recovery, HR, finance and budgeting, procurement, product lifecycle, IT and change management, media and healthcare. 
The type of models is also varied, ranging from simple and structured to large, complex, and unstructured models, captured using EPC and BPMN languages. 
The most recurring problems faced in managing process models are related to maintenance and understandability, validation, compliance management, standardization, and audit.

We analyzed the transcripts of the third parts of the interviews, refer to \appendixname~\ref{app:PQL:interview}, to 
identify themes/categories~\cite{Ryan2003} that relate to the motivation/design of a language for behavioral querying of process model repositories.
As a result, we identified these themes: relevance of behavioral querying, use cases of behavioral querying, relevance of behavioral predicates for querying, label similarity, and concrete syntax.
These themes informed the design of PQL, refer to~\sectionname~\ref{subsec:PQL:design:principles}.
In the end, 18 out of 23 interviewees have explicitly commented on the usefulness and importance of behavioral querying. 
Next, we list some direct quotes in support of this claim.
\smallskip
\begin{compactenum}
\item ``\emph{Yes, definitely, it would be a good idea to be able to analyze our process repository in a way like this [behavioral querying].}'' (Business analyst that manages 420 process models).
\item ``\emph{If you're trying to look for something that you can improve on, having these queries and trying to find processes would help, so rather than businesses coming to you, you can be a bit more proactive.}'' (Business excellence manager with 18 month experience in this role).
\item ``\emph{That [behavioral querying] would be extremely useful. It would save enormous amounts of time. Actually, it would enable us to undertake an analysis that we can't do at the moment. So, it would open more doors for us to actually be able to do different, potentially more valuable things.}'' (Senior business excellence manager with 20 years experience in analysis of process models).
\item ``\emph{This [behavioral querying] can be useful from a governance perspective, i.e., to be able to check process controls are in place. This also can assist with risks associated with the process.}'' (Business analyst with 15 years experience).
\item ``\emph{I've been looking for years for some solutions in this field [behavioral querying] ... it's for me very important and it is extremely useful to get this information}'' (BPM software product manager and business analyst with over 15 years experience).
\item ``\emph{I think from an overall strategic level it'll bring a lot of benefits because different parts of the organization operate in different ways and being able to actually analyze it through kind of a structured query language could be useful for an analyst. I can see it being quite highly useful because repositories right now are static and searching through can be time-consuming.}'' (Business analyst with six months experience).
\end{compactenum}
\smallskip

To identify relevant predicates to include in our query language, we analyzed the central tendency of the responses obtained in the second parts of the interviews and selected predicates with high scores. 
Because collected responses are ordinal, refer to \appendixname~\ref{app:PQL:interview}, for each combination of a question and predicate, we analyzed the median and mode of the responses. 
The results are reported in \tablename~\ref{tab:empirical_results:median:and:mode}.
Each cell of the table between rows two and five and between columns two and thirteen reports the median and mode (median/mode) of the 23 responses collected for the question indicated in the first column of the corresponding row and the predicated indicated in the first row of the corresponding column; for example the median and mode of the 23 collected responses on the usefulness for the \texttt{Conflict} predicate are 4 (very useful) and 5 (extremely useful), respectively.
The last row in \tablename~\ref{tab:empirical_results:median:and:mode} reports the sums of medians/modes over the four questions.

We decided to include six, \ie half of the tested, most relevant, \ie those with highest median and mode values, behavioral predicates in PQL.
Consequently, \texttt{CanOccur}, \texttt{AlwaysOccurs}, \texttt{Cooccur}, \texttt{Conflict}, \texttt{TotalCausal}, and \texttt{TotalConcurrent} predicates were selected for inclusion in PQL; the corresponding columns in the table are highlighted in bold font.
Each of these predicates has scored a sum of at least 15, both for median and mode, of perceived usefulness, importance, likelihood, and frequency of usage for the purpose of behavioral querying of process repositories.
Moreover, all the median and mode values for the selected predicates are at least 3, indicating that the respondents tended to rank these predicates as (at least) moderately useful, moderately important, occasionally frequent, and neutrally likely to be used for querying.
Remarkably, the selected predicates cover all the four behavioral categories, \ie conflict, causality, concurrency, and co-occurrence.
Interestingly, the causality and concurrency predicates with the total property were perceived as being more relevant than their existential counterparts (``Exist'', ``ExistTotal'' and ``TotalExist''). 
The ``total'' version is arguably simpler to understand and to relate to practice (\eg for compliance purposes), since it requires all instances of a process model to satisfy the behavioral relation captured by the predicate. 
Finally, because \texttt{Cooccur} and \texttt{Conflict} are defined as macros over \texttt{CanCooccur} and \texttt{CanConflict} predicates of the 4C spectrum, refer to~\cite{PolyvyanyyWCRH14} for details, we included \texttt{CanCooccur} and \texttt{CanConflict} in the selection to result in the repertoire of eight core PQL predicates.
\texttt{CanCooccur(A,B)} verifies if the model specifies at least one instance that contains tasks \textbf{A} and \textbf{B}, while \texttt{CanConflict(A,B)} checks if the model describes an instance that contains task \textbf{A} but does not contain task \textbf{B}; \sectionname~\ref{subsec:PQL:predicates:semantics} gives rigorous definitions of all the PQL predicates.

\begin{table}[htbp]
\footnotesize
  \centering
  \caption{\small Medians and modes of the responses to the four questions on the relevance of evaluated behavioral predicates obtained in 23 interviews with business analysts (Median/Mode).}
\begin{tabular}{r|c|c|c|c|c|c|c|c|c|c|c|c|}
\cline{2-13}
& \rotatebox[origin=l]{90}{\texttt{\textbf{CanOccur}}\;\;} & \rotatebox[origin=l]{90}{\texttt{\textbf{AlwaysOccurs}}\;\;} & \rotatebox[origin=l]{90}{\texttt{\textbf{Cooccur}}\;\;} & \rotatebox[origin=l]{90}{\texttt{\textbf{Conflict}}\;\;} & \rotatebox[origin=l]{90}{\texttt{ExistCausal}\;\;} & \rotatebox[origin=l]{90}{\texttt{ExistTotalCausal}\;\;} & \rotatebox[origin=l]{90}{\texttt{TotalExistCausal}\;\;} & \rotatebox[origin=l]{90}{\texttt{\textbf{TotalCausal}}\;\;} & \rotatebox[origin=l]{90}{\texttt{ExistConcurrent}\;\;} & \rotatebox[origin=l]{90}{\texttt{ExistTotalConcurrent}\;\;} & \rotatebox[origin=l]{90}{\texttt{TotalExistConcurrent}\;\;} & \rotatebox[origin=l]{90}{\texttt{\textbf{TotalConcurrent}}\;\;} \\
\hline
\multicolumn{1}{ |r|  }{\multirow{1}{*}{Useful} } & \textbf{4/5} & \textbf{4/4} &	\textbf{4/5} & \textbf{4/4} &	3/2 &	4/4 &	3/4 &	\textbf{4/5} & 3/4 & 3/4 & 4/4 & \textbf{4/4} \\
\hline
\multicolumn{1}{ |r|  }{\multirow{1}{*}{Important} } & \textbf{4/4} & \textbf{4/4} & \textbf{4/5} & \textbf{4/4} & 3/2 & 4/4 & 3/2 & \textbf{4/5} & 3/2 & 3/2 & 3/4 & \textbf{4/4} \\
\hline
\multicolumn{1}{ |r|  }{\multirow{1}{*}{Likely} } & \textbf{5/5} & \textbf{4/4} & \textbf{4/4} & \textbf{4/4} & 3/4 & 3/4 & 4/4 & \textbf{4/4} & 3/4 & 3/3 & 3/4 & \textbf{4/4} \\
\hline
\multicolumn{1}{ |r|  }{\multirow{1}{*}{Frequent} } & \textbf{4/3} & \textbf{3/3} & \textbf{3/3} & \textbf{3/3} & 3/3 & 3/3 & 3/3 & \textbf{3/3} & 3/3 & 3/3 & 3/3 & \textbf{3/3} \\
\hline
\hline
\multicolumn{1}{ |r|  }{\multirow{1}{*}{Total} } & \textbf{17/15} & \textbf{15/15} & \textbf{15/17} & \textbf{15/15} & 12/11 & 14/15 & 13/13 & \textbf{15/17} & 12/13 & 12/12 & 13/15 & \textbf{15/15} \\
\hline
    \end{tabular}
\label{tab:empirical_results:median:and:mode}
\end{table}

To generalize the results, we performed \emph{sign tests} to check if the medians of the responses are significantly greater than certain values. 
The sign test is a nonparametric test for hypotheses about a population median given a sample of observations from that population~\cite{Sprent2011}.
The decision to perform sign tests is justified by these three observations:
(i) the scales used to collect the responses, except for the likelihood scale, are not symmetric,
(ii) the distances between the answers are not always uniform, and
(iii) the collected responses are often not normally distributed.

For each pair of a question and behavioral predicate, we performed three one-tailed sign tests to test hypotheses of the form $H^x_0 : M \leq x$, where $M$ is the median response to the question w.r.t. the predicate and $x \in \{1.5, 2.5, 3.5\}$.
Thus, if one succeeds in rejecting, for example, hypothesis $H^{2.5}_0$, then the expected median response to the corresponding question w.r.t. the behavioral predicate is greater than 2.5.
\tablename~\ref{tab:empirical_results:expected} reports expected values of the answers to the four questions for the six selected predicates, obtained by rejecting the corresponding hypothesis based on p-values of at most 0.05. 
For example, the value of 4 for the usefulness of the \texttt{CanOccur} predicate reported in \tablename~\ref{tab:empirical_results:expected} indicates that we, based on the collected responses, were able to reject hypothesis $H^{3.5}_0$ for the corresponding combination of question and predicate.
Thus, if one repeats the study, we are at least 95\% confident that the median response on the usefulness of the \texttt{CanOccur} predicate will be at least 4.
Therefor, the expected responses to the four questions w.r.t. the six selected predicates are at least: 
`moderately useful' or `very useful' (usefulness),
`moderately important' or `very important' (importance),
`neutral' or `likely' (likelihood), and
`occasionally' or `almost every time' (frequency).

\begin{table}[htbp]
\footnotesize
  \centering
  \caption{\small Expected minimal medians of responses calculated using sign tests (p-value of 0.05).}
\begin{tabular}{r|c|c|c|c|c|c|}
\cline{2-7}
& \rotatebox[origin=l]{90}{\texttt{CanOccur}\;\;} & \rotatebox[origin=l]{90}{\texttt{AlwaysOccurs}\;\;} & \rotatebox[origin=l]{90}{\texttt{Cooccur}\;\;} & \rotatebox[origin=l]{90}{\texttt{Conflict}\;\;} & \rotatebox[origin=l]{90}{\texttt{TotalCausal}\;\;} & \rotatebox[origin=l]{90}{\texttt{TotalConcurrent}\;\;} \\
\hline
\multicolumn{1}{ |r|  }{\multirow{1}{*}{Useful} } & 4 &	3 &	4 &	3 &	4 &	4 \\
\hline
\multicolumn{1}{ |r|  }{\multirow{1}{*}{Important} } & 4 & 3 & 3 & 3 & 3 & 3 \\
\hline
\multicolumn{1}{ |r|  }{\multirow{1}{*}{Likely} } & 4 & 3 & 4 & 3 & 4 & 3 \\
\hline
\multicolumn{1}{ |r|  }{\multirow{1}{*}{Frequent} } & 3 & 3 & 3 & 4 & 3 & 3 \\
\hline
    \end{tabular}
\label{tab:empirical_results:expected}
\end{table}

To check whether the number of conducted interviews was sufficient, we estimated the statistical power of our study using G*Power 3.1~\cite{Faul2009}. 
Given a sample size of $n = 23$, and expecting a medium effect size (0.3) and an $\alpha$ error probability of 0.05, our experiment design achieves a statistical power of 0.93, which is well above the suggested threshold of 0.8.

%% file: tex/semantics.tex
\subsection{Dynamic Semantics}
\label{subsec:PQL:dynamic:semantics}

\reportornot{
This section specifies the dynamic semantics of PQL.
As an example, at the end of this section, the reader will find a detailed explanation of the dynamic semantics of query \emph{Q3} from \sectionname~\ref{sec:examples}. 
The dynamic semantics of PQL is proposed using \emph{meaning functions} that specify the effects of its valid constructs using mathematical denotations. 
These mathematical denotations are defined over the following domains:
}{
This section specifies the dynamic semantics of PQL using \emph{meaning functions} that describe the effects of valid constructs using mathematical denotations over the following domains:
}

\smallskip
\begin{compactitem}
\item $\mathbb{A}$, a universe of \emph{attribute names};
\item $\mathbb{B}$, a universe of \emph{attribute values};
\item $\mathbb{L}$, a universe of \emph{locations};
\item $\mathbb{S}$, a universe of \emph{Petri net systems}; 
\item $\mathbb{T}:=\powerset_{\geq 1}{(\mathbb{C})}$, the universe of all \emph{tasks} over the universe of character strings.\footnote{Given a set $A$, by $\powerset_{\geq 1}{(A)}$ we denote the set of all non-empty subsets of $A$, \ie the power set of $A$ without the empty set.} 
\end{compactitem}
\smallskip

\noindent
Let $\chi : \mathbb{A} \rightarrow \powerset_{\geq 1}{(\mathbb{B})}$ be a function that maps attribute names onto sets of permissible attribute values. 
A PQL query formulates a request to retrieve process models (and their attributes) from a given process model repository. 
Hence, every PQL query is executed in the context of a repository.
To permit subsequent formal discussions, we give a rigorous definition of a process model repository.

\begin{define}{Repository}{def:repository}{\quad\\}
A \emph{process model repository}, or a \emph{repository}, is a 6-tuple $R:=(S,A,L,\mathit{val},\mathit{loc},\precsim)$, where 
$S \subseteq \mathbb{S}$ is a finite set of \emph{systems}, 
$A \subseteq \mathbb{A}$ is a finite set of \emph{attribute names}, 
$L \subseteq \mathbb{L}$ is a set of \emph{locations}, 
$\mathit{val} : S \times A \rightarrow \mathbb{B}$ is the \emph{attribute value assignment} function, such that $\forall s \in S \;\forall a \in A : \mathit{val}(s,a) \in \chi(a)$, 
$\mathit{loc} : S \rightarrow L$ is the \emph{location assignment} function, and 
$\precsim$ is a reflexive binary relation over $L$, called \emph{location map}.
\end{define}

\reportornot{
\noindent
A system, refer to \definitionname~\ref{def:system}, is a specification of a process model that is stored in the repository.
Function $\mathit{val}$ assigns values to attributes of systems, \ie an attribute $a \in \mathbb{A}$ of a system $s \in \mathbb{S}$ has value $\mathit{val}(s,a)$. 
Function $\mathit{loc}$ assigns locations to systems, \ie a system $s \in \mathbb{S}$ is located at $\mathit{loc}(s)$.
Finally, the location map is introduced to implement nesting of locations, \ie a location $l_1 \in L$ is said to be nested in a location $l_2 \in L$ if and only if it holds that $l_1 \precsim l_2$.
}{

\noindent}
By $\texttt{Repository}$, we denote the universe of all possible repositories.
Let $k:=(k_1,k_2,\ldots,k_n) \in K_1 \times K_2 \times \ldots \times K_n$ be a point in $n$-dimensional space, where $K_1, K_2,\ldots, K_n$ are some sets. 
The \emph{projection function} $\pi_i(k)$, $i \in [1\,..\,n]$, is defined as $\pi_i(k):=k_i$, where $k_i$ is the $i$-th coordinate of $k$. 

The meaning function of the \texttt{Query} construct is defined as follows.
\small
\begin{eqnarray*}
M_{\texttt{Query}}							&:& \texttt{Query} \times \texttt{Repository} \rightarrow \mathbb{S} \times \powerset{(\mathbb{A} \times \mathbb{B})}\\
M_{\texttt{Query}}[q : \texttt{Query},(S,A,L,\mathit{val},\mathit{loc},\precsim) : \texttt{Repository}]							
                                & \triangleq & \{(s,x) \in \mathbb{S} \times \powerset{(\mathbb{A} \times \mathbb{B})} \divides \\
                                & & (\exists\; l \in M_{\texttt{Locations}}(q.locs) : loc(s) \precsim l) \;\land \\
                                & & (\pi_1(x)=M_{\texttt{Attributes}}(q.atts)) \;\land \\
                                & & (\forall (a,v) \in x : v = val(s,a)) \;\land \\
                                & & (M_{\texttt{Predicate}}(q.pred,s,M_{\texttt{Variables}}(q.vars,s)))\}
\end{eqnarray*}
\normalsize
\noindent
Similar as in~\cite{Meyer1990}, we denote the meaning function of a construct \texttt{c} by $M_\texttt{c}$.

\reportornot{The signature of the meaning function of the \texttt{Query} construct determines that its specimen defines a set of pairs, each composed of a system and a set of attribute-value pairs.}{}
The result of executing a query $q$ in the context of a repository consists of all the systems at locations nested in $q.\mathit{locs}$ that satisfy predicate $q.\mathit{pred}$, and attribute values of these systems requested in $q.\mathit{atts}$. 
The meaning of a specimen of the \texttt{Locations} construct is defined below.
\small
\begin{eqnarray*}
M_{\texttt{Locations}}					&:& \texttt{Locations} \rightarrow \powerset_{\geq 1}{(\mathbb{L})}\\ 
M_{\texttt{Locations}}[ls : \texttt{Locations}]							
                                & \triangleq & \bigcup_{i \,\in\, [1..|\mathit{ls}|]} M_{\texttt{Location}}(\mathit{ls}_i)\\
M_{\texttt{Location}}					  &:& \texttt{Location} \rightarrow \powerset_{\geq 1}{(\mathbb{L})}\\ 
M_{\texttt{Location}}[l : \texttt{Location}]							
                                & \triangleq & 
\begin{cases} 
    \;\mathbb{L} & l \,\,\,\textbf{\texttt{is}}\,\,\, \texttt{Universe} \\
    \;\{M_{\texttt{LocationPath}}(l)\} & \mathit{otherwise}
\end{cases}\\
M_{\texttt{LocationPath}}				&:& \texttt{LocationPath} \rightarrow \mathbb{L}
\end{eqnarray*}
\normalsize
\reportornot{\noindent
The denotation of a specimen of the \texttt{Locations} construct is a non-empty subset of $\mathbb{L}$, which results from the union of meanings of its elements.
The meaning of an element of a \texttt{Locations} specimen, \ie the meaning of a specimen of the \texttt{Location} construct, is, again, a non-empty set of locations.
If a location is a specimen of \texttt{Universe}, then its denotation is the universe of locations $\mathbb{L}$. 
Alternatively, if a location is a specimen of \texttt{LocationPath}, then its denotation is a singleton that contains a location that is associated with the specimen.
The association between specimens of \texttt{LocationPath} and locations is implemented in the meaning function of the \texttt{LocationPath} construct, see the signature above. 
Here, we refrain from giving a specific definition to the function, but assume the use of some such function definition.

}{\noindent} 
Note the use of the \textbf{\texttt{is}} operator above. 
We use this operator to specify a boolean expression that checks if a given specimen is a specimen of some given construct.
For example, the expression $l \,\,\,\textbf{\texttt{is}}\,\,\, \texttt{Universe}$ used above evaluates to \emph{true} if and only if $l$ is a specimen of the \texttt{Universe} construct.

The meaning of a specimen of the \texttt{Attributes} construct is defined similarly to that of the \texttt{Locations} construct and is proposed below.
\small
\begin{eqnarray*}
M_{\texttt{Attributes}}					&:& \texttt{Attributes} \rightarrow \powerset_{\geq 1}{(\mathbb{A})} \\ 
M_{\texttt{Attributes}}[as : \texttt{Attributes}]							
                                & \triangleq & \bigcup_{i \,\in\, [1..|\mathit{as}|]} M_{\texttt{Attribute}}(\mathit{as}_i)\\
M_{\texttt{Attribute}}					&:& \texttt{Attribute} \rightarrow \powerset_{\geq 1}{(\mathbb{A})} \\ 
M_{\texttt{Attribute}}[a : \texttt{Attribute}]							
                                & \triangleq & 
\begin{cases} 
    \;\mathbb{A} & a \,\,\,\textbf{\texttt{is}}\,\,\, \texttt{Universe} \\
    \;\{M_{\texttt{AttributeName}}(a)\} & \mathit{otherwise}
\end{cases}\\
M_{\texttt{AttributeName}}		  &:& \texttt{AttributeName} \rightarrow \mathbb{A}
\end{eqnarray*}
\normalsize
\reportornot{\noindent
Hence, the denotation of a specimen of the \texttt{Attributes} construct is a non-empty subset of $\mathbb{A}$.
In line with \texttt{LocationPath}, we assume the existence of some function that maps specimens of \texttt{AttributeName} onto $\mathbb{A}$ that is used as the meaning function of the \texttt{AttributeName} construct.

}{

\noindent}
The \emph{overriding union} of $f : X_1 \rightarrow Y_1$ by $g : X_2 \rightarrow Y_2$, denoted by $f \oplus g$, is defined by $g \cup \set{(x,f(x))}{x \in \mathit{dom}(f) \setminus \mathit{dom}(g)}$.
Given a sequence of functions $\mathit{fs}$, $|\mathit{fs}|=n$, $\bigoplus_{i=1}^{n} \mathit{fs}_i$ denotes the expression
$((\ldots((\mathit{fs}_1 \oplus\, \mathit{fs}_2) \oplus\, \mathit{fs}_3) \oplus \ldots \,\mathit{fs}_{n-1}) \oplus \,\mathit{fs}_n)$.

Next, we propose the denotation of a specimen of the \texttt{Variables} construct.
\small
\begin{eqnarray*}
M_{\texttt{Variables}}					&:& \texttt{Variables} \times \mathbb{S} \rightarrow \powerset{(\mathbb{V} \times \powerset{(\mathbb{T})})}\\ 
M_{\texttt{Variables}}[vs : \texttt{Variables}\!,\, s : \mathbb{S}]
                                & \triangleq & \bigoplus_{i=1}^{|\mathit{vs}|}\,\{M_{\texttt{Variable}}(\mathit{vs}_i,s,M_{\texttt{Variables}}(\mathit{prefix}(\mathit{vs},i),s))\}\\ 
M_{\texttt{Variable}}					  &:& \texttt{Variable} \times \mathbb{S} \times \powerset{(\mathbb{V} \times \powerset{(\mathbb{T})})} \rightarrow \mathbb{V} \times \powerset{(\mathbb{T})}\\
M_{\texttt{Variable}}[v:\texttt{Variable},\, s : \mathbb{S},\, vals: \powerset{(\mathbb{V} \times \powerset{(\mathbb{T})})}]	& \triangleq & 
                                               (v.\mathit{name}.\mathit{id},M_\texttt{SetOfTasks}(\mathit{v.\mathit{tasks},s,\mathit{vals}}))
\end{eqnarray*}
\normalsize
\noindent
Recall that $\mathbb{V}$ refers to the set of all legal PQL variable names.
The denotation of a specimen $\mathit{vs}$ of \texttt{Variables} is a set of variable-value pairs, where a value is a set of tasks, computed in the context of some process model $s \in \mathbb{S}$ as the overriding union of meanings of the individual elements of $\mathit{vs}$. 
Note that the overriding union is applied in the order the elements appear in $\mathit{vs}$. 
An element of $\mathit{vs}$ at position $i \in [1..|vs|]$ is a specimen $v$ of \texttt{Variable}.
The meaning of $v$ is derived in the context of $s$ and variable-value pairs $\mathit{vals}$ computed based on the prefix of $\mathit{vs}$ up to but excluding position $i$, \ie values of variables declared prior to $v$, and is defined as a pair composed of a legal PQL variable name associated with $v.\mathit{name}$ and a set of tasks that stems from the denotation of $v.\mathit{tasks}$.

Let $\mathit{labels} : \mathbb{S} \rightarrow \powerset{(\mathbb{C})}$, be a function that given a system results in the set of all the labels of its observable transitions.
Let $\mathit{similar} : \mathbb{C} \times [0\,..\,1] \rightarrow \powerset_{\geq 1}{(\mathbb{C})}$ be a function that given a label and a similarity level threshold results in the set of all labels in $\mathbb{C}$ that have similarity scores with the given label that are equal to or greater than the given threshold.
Note that different implementations of PQL may rely on different techniques for scoring label similarities. 
However, it must hold that $c \in \mathit{similar}(c,x)$, $c \in \mathbb{C}$ and $x \in [0\,..\,1]$.
Then, the denotation of the \texttt{SetOfTasks} construct is as follows.
\small
\begin{eqnarray*}
M_{\texttt{SetOfTasks}}   &:& \texttt{SetOfTasks} \times \mathbb{S} \times \powerset{(\mathbb{V} \times \powerset{(\mathbb{T})})} \rightarrow \powerset{(\mathbb{T})}\\ 
\begin{split}
M_{\texttt{SetOfTasks}}[\mathit{ts}:\texttt{SetOfTasks},\\ s : \mathbb{S},\, \mathit{vals}: \powerset{(\mathbb{V} \times \powerset{(\mathbb{T})})}]	
\end{split}
& \triangleq & 
\begin{cases} 
    \;\mathit{vals}(\mathit{ts}.\mathit{id}) & \mathit{ts}\,\,\,\textbf{\texttt{is}}\,\,\,\texttt{VariableName} \\
    \;\set{\{\lambda\} \in \mathbb{T}}{\lambda \in \mathit{labels}(s)} & \mathit{ts}\,\,\,\textbf{\texttt{is}}\,\,\,\texttt{SetOfAllTasks} \\
    \;\bigcup_{i \,\in\, [1..|\mathit{ts}|]} \{M_{\texttt{Task}}(\mathit{ts}_i)\} & \mathit{ts}\,\,\,\textbf{\texttt{is}}\,\,\,\texttt{SetOfTasksLiteral}\\
    \;M_{\texttt{SetOfTasksConstruction}}(\mathit{ts},s,\mathit{vals}) & \mathit{ts}\,\,\,\textbf{\texttt{is}}\,\,\,\texttt{SetOfTasksConstruction} \\
    \;\bigcup_{i \in [1..|\mathit{ts}|]} M_{\texttt{SetOfTasks}}(\mathit{ts}_i,s,\mathit{vals}) & \mathit{ts}\,\,\,\textbf{\texttt{is}}\,\,\,\texttt{Union}\\
    \;\bigcap_{i \in [1..|\mathit{ts}|]} M_{\texttt{SetOfTasks}}(\mathit{ts}_i,s,\mathit{vals}) & \mathit{ts}\,\,\,\textbf{\texttt{is}}\,\,\,\texttt{Intersection}\\
    \;M_{\texttt{Difference}}(\mathit{ts},s,\mathit{vals}) & \mathit{ts}\,\,\,\textbf{\texttt{is}}\,\,\,\texttt{Difference}
\end{cases}
\end{eqnarray*}
\normalsize
\reportornot{
\noindent
If a set of tasks is specified as a specimen $\mathit{ts}$ of \texttt{VariableName}, then its denotation is the value associated with $\mathit{ts}.\mathit{id}$, \ie $\mathit{vals}(\mathit{ts}.\mathit{id})$, where $\mathit{vals}$ maps variable names onto their values.
If a set of tasks is specified as a specimen of the \texttt{SetOfAllTasks} construct, then its denotation is the set of all `singleton' tasks of system $s$ that is currently being matched with the query, \ie the set that for every observable label $\lambda \in \mathit{labels(s)}$ of $s$ contains task $\{\lambda\}$ and does not contain any other tasks. 
If a set of tasks is given as a specimen of \texttt{SetOfTasksLiteral}, then its denotation is the set composed of meanings of its elements, where each element is a specimen of the \texttt{Task} construct.

}{

\noindent}
The denotation of the \texttt{Task} construct is as follows.
\small
\begin{eqnarray*}
M_{\texttt{Task}}         &:& \texttt{Task} \rightarrow \mathbb{T}\\
M_{\texttt{Task}}[t:\texttt{Task}] & \triangleq &
\begin{cases} 
    \; \{t.\mathit{label}\} & \mathit{t}\,\,\,\textbf{\texttt{is}}\,\,\,\texttt{ExactTask} \\
    \; \mathit{similar}(t.\mathit{label},\mathit{defSim}) & \mathit{t}\,\,\,\textbf{\texttt{is}}\,\,\,\texttt{DefSimTask}\\
    \; \mathit{similar}(t.\mathit{label},t.\mathit{sim}.\mathit{value}) & \mathit{t}\,\,\,\textbf{\texttt{is}}\,\,\,\texttt{SimTask} 
\end{cases}
\end{eqnarray*}
\normalsize
\noindent
A specimen of \texttt{Task} denotes a PQL task, \ie a non-empty set of character strings. 
The meaning of a specimen $t$ of \texttt{ExactTask} is a singleton that contains label $t.\mathit{label}$.
If $t$ is a specimen of \texttt{DefSimTask} or \texttt{SimTask}, then its denotation is the set of all labels in $\mathbb{C}$ that have similarity scores with $t.\mathit{label}$ greater than or equal to the default similarity threshold $\mathit{defSim}$ or $t.\mathit{sim}.\mathit{value}$, respectively.
Here, $\mathit{defSim} \in [0\,..\,1]$ is some global constant that specifies the default label similarity threshold.

A set of tasks can be constructed by selecting tasks from a given set of tasks, where the selection is implemented using a behavioral relation. 
To this end, one can use the \texttt{SetOfTasksConstruction} construct. 
For brevity and presentation considerations, we omit the precise definition of the meaning of \texttt{SetOfTasksConstruction} and revert to definitions of denotations of the two constructs that constitute the choice production associated with \texttt{SetOfTasksConstruction}.
\small
\begin{eqnarray*}
\begin{split}
M_{\texttt{UnaryPredicateConstruction}}[\mathit{upc}:\\\texttt{UnaryPredicateConstruction},\\ s : \mathbb{S},\, \mathit{vals}: \powerset{(\mathbb{V} \times \powerset{(\mathbb{T})})}]	
\end{split}
& \triangleq & 
\begin{cases} 
    \begin{split}
    \; \{t \in M_{\texttt{SetOfTasks}}(\mathit{upc}.\mathit{tasks},\\s,\mathit{vals}) \mid \mathit{canOccur}(s,t)\} 
    \end{split}
    & 
    \mathit{upc}.\mathit{name}\,\,\,\textbf{\texttt{is}}\,\,\,\texttt{CanOccur} \\
    & \\    
    \begin{split}
    \; \set{t \in M_{\texttt{SetOfTasks}}(\mathit{upc}.\mathit{tasks},s,\\\mathit{vals})}{\mathit{alwaysOccurs}(s,t)} 
    \end{split}
    &
    \mathit{upc}.\mathit{name}\,\,\,\textbf{\texttt{is}}\,\,\,\texttt{AlwaysOccurs}
\end{cases}
\end{eqnarray*}
\normalsize
\reportornot{\noindent
The denotation of a specimen $\mathit{upc}$ of \texttt{UnaryPredicateConstruction} is a set that contains every task from the set defined by $\mathit{upc}.\mathit{tasks}$ for which a behavioral relation denoted by $\mathit{upc}.\mathit{name}$ holds (in the system which is being matched to the query).
If $\mathit{upc}.\mathit{name}$ is of type \texttt{CanOccur}, the $\mathit{canOccur}$ relation is used.
Otherwise, the $\mathit{alwaysOccurs}$ relation is employed to select tasks.
}{

\noindent}
Note that denotations of $\mathit{canOccur}$ and $\mathit{alwaysOccurs}$ predicates are proposed in \sectionname~\ref{subsec:PQL:predicates:semantics}.

\reportornot{The meaning of a specimen of \texttt{BinaryPredicateConstruction} is defined similarly to the meaning of the \texttt{UnaryPredicateConstruction} construct. 
However, the selection of tasks is carried out based on a binary behavioral predicate.}{}
\small
\begin{eqnarray*}
\begin{split}
M_{\texttt{BinaryPredicateConstruction}}[\mathit{bpc}:\\\texttt{BinaryPredicateConstruction},\\ s : \mathbb{S},\, \mathit{vals}: \powerset{(\mathbb{V} \times \powerset{(\mathbb{T})})}]	
\end{split}
& \triangleq & 
\begin{cases} 
    \begin{split}
    \; \{t_1 \in M_{\texttt{SetOfTasks}}(\mathit{bpc}.\mathit{tasks}_1,s,\mathit{vals}) \mid\\ \exists\; t_2 \in M_{\texttt{SetOfTasks}}(\mathit{bpc}.\mathit{tasks}_2,s,\mathit{vals})\;:\\ \mathit{canConflict}(s,t_1,t_2)\} 
    \end{split}
    & 
    \begin{split}
    (\mathit{bpc}.\mathit{q}\,\,\,\textbf{\texttt{is}}\,\,\,\texttt{Any}) \,\land\,\\ (\mathit{bpc}.\mathit{name}\,\,\,\textbf{\texttt{is}}\\\texttt{CanConflict})
    \end{split}
    \\
    & \\    
    \begin{split}
    \; \{t_1 \in M_{\texttt{SetOfTasks}}(\mathit{bpc}.\mathit{tasks}_1,s,\mathit{vals}) \mid\\ \forall\; t_2 \in M_{\texttt{SetOfTasks}}(\mathit{bpc}.\mathit{tasks}_2,s,\mathit{vals})\;:\\ \mathit{canConflict}(s,t_1,t_2)\} 
    \end{split}
    & 
    \begin{split}
    (\mathit{bpc}.\mathit{q}\,\,\,\textbf{\texttt{is}}\,\,\,\texttt{All}) \,\land\,\\ (\mathit{bpc}.\mathit{name}\,\,\,\textbf{\texttt{is}}\\\texttt{CanConflict})
    \end{split}
    \\
    \;\;\;\ldots & 
\end{cases}
\end{eqnarray*}
\normalsize
\reportornot{\noindent
Thus, given a specimen $\mathit{bpc}$ of \texttt{BinaryPredicateConstruction}, its denotation is a set that contains every task from the set defined by $\mathit{bpc}.\mathit{tasks}_1$ which is in a binary behavioral relation with some or all tasks in the set defined by $\mathit{bpc}.\mathit{tasks}_2$. 
The choice of the quantifier is guided by the type of $\mathit{bpc}.q$, \ie existential if $\mathit{bpc}.q$ is of type \texttt{Any} and universal if $\mathit{bpc}.q$ is of type \texttt{All}.}{\noindent}
The use of a binary behavioral relation is determined by $\mathit{bpc}.\mathit{name}$. 
For example, $\mathit{bpc}.\mathit{name}$ of type $\texttt{CanConflict}$ calls for the use of the $\mathit{CanConflict}$ relation (see above). 
Similarly, a specimen $\mathit{bpc}.\mathit{name}$ of type \texttt{CanCooccur}, \texttt{Conflict}, \texttt{Cooccur}, \texttt{TotalCausal}, and \texttt{TotalConcurrent}, signifies the use of the $\mathit{canCooccur}$, $\mathit{conflict}$, $\mathit{cooccur}$, $\mathit{totalCausal}$, and $\mathit{totalConcurrent}$ relation, respectively (not shown because of space considerations).
Again, the denotations of all the binary predicates of PQL are proposed in \sectionname~\ref{subsec:PQL:predicates:semantics}.
A set of tasks can be defined as a result of well-known operations on other sets of tasks, such as union, intersection, and difference. 

As an example, the denotation of the \texttt{Difference} construct is proposed below.
\small
\begin{eqnarray*}
\begin{split}
M_{\texttt{Difference}}[\mathit{d}:\texttt{Difference},\\ s : \mathbb{S},\, \mathit{vals}: \powerset{(\mathbb{V} \times \powerset{(\mathbb{T})})}]
\end{split} 
& \triangleq &
\begin{cases}
M_{\texttt{SetOfTasks}}(\mathit{d}_1,s,\mathit{vals}) \setminus M_{\texttt{SetOfTasks}}(\mathit{d}_2,s,\mathit{vals}) & |d|=2\\
M_{\texttt{SetOfTasks}}(\mathit{d}_1,s,\mathit{vals}) \setminus M_{\texttt{Difference}}(\mathit{suffix}(d,2),s,\mathit{vals}) & |d|>2
\end{cases}
\end{eqnarray*}
\normalsize
\noindent
The PQL grammar in \appendixname~\ref{app:PQL:grammar} specifies that brackets () have the highest priority, then difference $\setminus$, then intersection $\cap$, and finally union $\cup$.
Hence, the PQL grammar in \appendixname~\ref{app:PQL:grammar} specifies that expression $A \cup B \,\cap\, C \setminus (D \cup E) \setminus F$ must be evaluated as $A \cup (B \,\cap\, (C \setminus ((D \cup E) \setminus F)))$.

The denotation of \texttt{Predicate} is as follows.
\small
\begin{eqnarray*}
M_{\texttt{Predicate}} &:& \texttt{Predicate} \times \mathbb{S} \times \powerset{(\mathbb{V} \times \powerset{(\mathbb{T})})} \rightarrow \{\mathit{true},\,\mathit{false}\}
\end{eqnarray*}
\normalsize
\reportornot{\noindent
The signature above specifies that \texttt{Predicate} denotes a boolean value and is evaluated in the context of a system and variable-value pairs.
}{}
The \texttt{Predicate} construct is defined as a choice production of ten alternatives.
\reportornot{In what follows, we define and discuss the meaning of each of the alternatives.}{In what follows, for illustration purpose, we define and discuss the meaning of six alternatives, whereas detailed discussions of the remaining four alternatives can be found in \appendixname~\ref{app:predicate}.}

The \texttt{TruthValue} construct is introduced in PQL to refer to values of $\mathit{true}$ and $\mathit{false}$ from classical logic.
Its denotation is given below.
\small
\begin{eqnarray*}
M_{\texttt{TruthValue}}[p:\texttt{TruthValue},s:\mathbb{S},\mathit{vals}:\powerset{(\mathbb{V} \times \powerset{(\mathbb{T})})}] & \triangleq & 
\begin{cases} 
    \;\mathit{true}   & \mathit{p}\,\,\,\textbf{\texttt{is}}\,\,\,\texttt{True}\\
    \;\mathit{false}  & \mathit{p}\,\,\,\textbf{\texttt{is}}\,\,\,\texttt{False}
\end{cases}
\end{eqnarray*}
\normalsize
\noindent
A predicate can be defined as a negation, conjunction, or disjunction of other predicates.
\small
\begin{eqnarray*}
M_{\texttt{Negation}}[p:\texttt{Negation},s:\mathbb{S},\mathit{vals}:\powerset{(\mathbb{V} \times \powerset{(\mathbb{T})})}] & \triangleq & \neg\, M_{\texttt{Predicate}}(p.\mathit{pred},s,\mathit{vals})\\
M_{\texttt{Conjunction}}[p:\texttt{Conjunction},s:\mathbb{S},\mathit{vals}:\powerset{(\mathbb{V} \times \powerset{(\mathbb{T})})}]
      & \triangleq & \bigwedge_{i \,\in\, [1..|\mathit{p}|]} M_{\texttt{Predicate}}(\mathit{p}_i,s,\mathit{vals})\\
M_{\texttt{Disjunction}}[p:\texttt{Disjunction},s:\mathbb{S},\mathit{vals}:\powerset{(\mathbb{V} \times \powerset{(\mathbb{T})})}]
      & \triangleq & \bigvee_{i \,\in\, [1..|\mathit{p}|]} M_{\texttt{Predicate}}(\mathit{p}_i,s,\mathit{vals})
\end{eqnarray*}
\normalsize
\noindent
The PQL grammar in \appendixname~\ref{app:PQL:grammar} specifies that brackets () have the highest priority, then negation $\neg$, then conjunction $\land$, and finally disjunction $\lor$. 
Thus, in PQL, expression $\neg (a \lor b \land c) \lor d \land e$ is evaluated as $(\neg (a \lor (b \land c))) \lor (d \land e)$.

One can test whether a predicate evaluates to \emph{true} or \emph{false} as follows.
\small
\begin{eqnarray*}
\begin{split}
M_{\texttt{LogicalTest}}[p:\texttt{LogicalTest},\\s:\mathbb{S},\mathit{vals}:\powerset{(\mathbb{V} \times \powerset{(\mathbb{T})})}] 
\end{split}
& \triangleq & 
\begin{cases} 
    \;M_{\texttt{Predicate}}(p.\mathit{pred},s,\mathit{vals})   & p \,\,\,\textbf{\texttt{is}}\,\,\, \texttt{IsTrue} \lor p \,\,\,\textbf{\texttt{is}}\,\,\, \texttt{IsNotFalse}\\
    \;\neg\,\,M_{\texttt{Predicate}}(p.\mathit{pred},s,\mathit{vals})   & p \,\,\,\textbf{\texttt{is}}\,\,\, \texttt{IsFalse} \lor p \,\,\,\textbf{\texttt{is}}\,\,\, \texttt{IsNotTrue}\\
\end{cases}
\end{eqnarray*}
\normalsize
\noindent
The \texttt{TaskInSetOfTasks} construct tests if a task is a member of a set of tasks.
\small
\begin{eqnarray*}
\begin{split}
M_{\texttt{TaskInSetOfTasks}}[p:\texttt{TaskInSetOfTasks},\\s:\mathbb{S},\mathit{vals}:\powerset{(\mathbb{V} \times \powerset{(\mathbb{T})})}] 
\end{split}
& \triangleq & 
M_{\texttt{Task}}(p.\mathit{task}) \;\in\; M_{\texttt{SetOfTasks}}(p.\mathit{tasks},s,\mathit{vals})
\end{eqnarray*}
\normalsize
\reportornot{\noindent
Every specimen of \texttt{SetComparison} denotes a set comparison operation.\footnote{As the meaning of a \texttt{SetOfTasks} construct is a set composed of the meanings of all and only its \texttt{Task} elements, each interpreted as a set of character strings, the identity of elements is defined as set equality over sets of character strings.}
\addtolength{\jot}{-.2em}
\small
\begin{eqnarray*}
\begin{split}
M_{\texttt{SetComparison}}\\ [p:\texttt{SetComparison},\\s:\mathbb{S},\mathit{vals}:\powerset{(\mathbb{V} \times \powerset{(\mathbb{T})})}] 
\end{split}
& \triangleq & 
\begin{cases} 
\;  \begin{split}
    M_{\texttt{SetOfTasks}}(p.\mathit{tasks}_1,s,\mathit{vals}) =\;\;\;\;\;\;\;\;\;\;\\ M_{\texttt{SetOfTasks}}(p.\mathit{tasks}_2,s,\mathit{vals})\;\;\;\;\;\;\;\;\;\;
    \end{split}
    & 
    \begin{split}
    p.\mathit{oper}\,\,\,\textbf{\texttt{is}}\;\;\;\\\texttt{Identical}
    \end{split}
    \\[1em] 
\;  \begin{split}
    M_{\texttt{SetOfTasks}}(p.\mathit{tasks}_1,s,\mathit{vals}) \neq\\ M_{\texttt{SetOfTasks}}(p.\mathit{tasks}_2,s,\mathit{vals}) 
    \end{split}
    & 
    \begin{split}
    p.\mathit{oper}\,\,\,\textbf{\texttt{is}}\;\;\;\\\texttt{Different}
    \end{split}
    \\[1em] 
\;  \begin{split}
    M_{\texttt{SetOfTasks}}(p.\mathit{tasks}_1,s,\mathit{vals})\,\, \cap\\ M_{\texttt{SetOfTasks}}(p.\mathit{tasks}_2,s,\mathit{vals}) \neq \emptyset 
    \end{split}
    & 
    \begin{split}
    p.\mathit{oper}\,\,\,\textbf{\texttt{is}}\;\;\;\;\;\;\;\;\;\;\;\\\texttt{OverlapsWith}
    \end{split}
    \\[1em] 
\;  \begin{split}
    M_{\texttt{SetOfTasks}}(p.\mathit{tasks}_1,s,\mathit{vals})\,\, \subseteq\\ M_{\texttt{SetOfTasks}}(p.\mathit{tasks}_2,s,\mathit{vals})
    \end{split}
    & 
    \begin{split}
    p.\mathit{oper}\,\,\,\textbf{\texttt{is}}\;\\\texttt{SubsetOf}
    \end{split}
    \\[1em] 
\;  \begin{split}
    M_{\texttt{SetOfTasks}}(p.\mathit{tasks}_1,s,\mathit{vals})\,\, \subset\\ M_{\texttt{SetOfTasks}}(p.\mathit{tasks}_2,s,\mathit{vals})
    \end{split}
    & 
    \begin{split}
    p.\mathit{oper}\,\,\,\textbf{\texttt{is}}\;\;\;\;\;\;\;\;\;\;\;\;\;\;\;\;\\\texttt{ProperSubsetOf}
    \end{split}
    \\ 
\end{cases}
\end{eqnarray*}
\normalsize
\noindent
The denotation of a specimen of \texttt{UnaryPredicate} is a unary behavioral relation.
\small
\begin{eqnarray*}
\begin{split}
M_{\texttt{UnaryPredicate}}[p:\texttt{UnaryPredicate},\\s:\mathbb{S},\mathit{vals}:\powerset{(\mathbb{V} \times \powerset{(\mathbb{T})})}]
\end{split}
& \triangleq &
\begin{cases} 
    \;\mathit{canOccur(s,M_\texttt{Task}(p.\mathit{task}))} & p.\mathit{name}\,\,\,\textbf{\texttt{is}}\,\,\,\texttt{CanOccur}\\
    \;\mathit{alwaysOccurs(s,M_\texttt{Task}(p.\mathit{task}))} & p.\mathit{name}\,\,\,\textbf{\texttt{is}}\,\,\,\texttt{AlwaysOccurs}
\end{cases}
\end{eqnarray*}
\normalsize
\noindent
Note that the respective behavioral relation is computed for the system that is being matched to the query.
Similarly, the meaning of a specimen of \texttt{BinaryPredicate} is a binary behavioral relation.
\small
\begin{eqnarray*}
\begin{split}
M_{\texttt{BinaryPredicate}}\\ [p:\texttt{BinaryPredicate},\\s:\mathbb{S},\mathit{vals}:\powerset{(\mathbb{V} \times \powerset{(\mathbb{T})})}]
\end{split}
& \triangleq &
\begin{cases}
\;  \begin{split}
    \mathit{canConflict}(s,M_\texttt{Task}(p.\mathit{task}_1),\;\;\;\;\;\;\;\;\;\;\;\;\;\;\;\;\; \\M_\texttt{Task}(p.\mathit{task}_2))\;\;\;\;\;\;\;\;\;
    \end{split}
    & 
    \begin{split}
    p.\mathit{name}\,\,\,\textbf{\texttt{is}}\;\;\;\;\;\;\;\\\texttt{CanConflict}
    \end{split}
    \\[1em] 
\;  \begin{split}
    \mathit{canCooccur}(s,M_\texttt{Task}(p.\mathit{task}_1),\;\;\;\;\;\;\, \\M_\texttt{Task}(p.\mathit{task}_2))
    \end{split}
    & 
    \begin{split}
    p.\mathit{name}\,\,\,\textbf{\texttt{is}}\;\;\;\;\\\texttt{CanCooccur}
    \end{split}
    \\[1em] 
\;  \begin{split}
    \mathit{conflict}(s,M_\texttt{Task}(p.\mathit{task}_1),\;\;\;\;\;\;\;\;\;\;\;\;\;\;\;\;\, \\M_\texttt{Task}(p.\mathit{task}_2))
    \end{split}
    & 
    \begin{split}
    p.\mathit{name}\,\,\,\textbf{\texttt{is}}\!\!\\\texttt{Conflict}
    \end{split}
    \\[1em] 
\;  \begin{split}
    \mathit{cooccur}(s,M_\texttt{Task}(p.\mathit{task}_1),\;\;\;\;\;\;\;\;\;\;\;\;\;\;\; \\M_\texttt{Task}(p.\mathit{task}_2))
    \end{split}
    & 
    \begin{split}
    p.\mathit{name}\,\,\,\textbf{\texttt{is}}\!\!\!\!\!\!\!\\\texttt{Cooccur}
    \end{split}
    \\[1em] 
\;  \begin{split}
    \mathit{totalCausal}(s,M_\texttt{Task}(p.\mathit{task}_1),\;\;\;\;\;\;\;\;\, \\M_\texttt{Task}(p.\mathit{task}_2))
    \end{split}
    & 
    \begin{split}
    p.\mathit{name}\,\,\,\textbf{\texttt{is}}\;\;\;\;\;\;\;\\\texttt{TotalCausal}
    \end{split}
    \\[1em] 
\;  \begin{split}
    \mathit{totalConcurrent}(s,M_\texttt{Task}(p.\mathit{task}_1),\\M_\texttt{Task}(p.\mathit{task}_2))
    \end{split}
    & 
    \begin{split}
    p.\mathit{name}\,\,\,\textbf{\texttt{is}}\;\;\;\;\;\;\;\;\;\;\;\;\;\;\;\;\;\\\texttt{TotalConcurrent}
    \end{split}
    \\ 
\end{cases}\\
\end{eqnarray*}
\normalsize
\noindent
As of today, PQL offers two unary and six binary behavioral predicates, refer to \sectionname~\ref{subsec:empirical:evaluation:results}. 
We expect that future studies will justify the use of other behavioral predicates for process querying, which consequently will result in the introduction of new predicates in PQL, ultimately improving its expressiveness. 
The reader can refer to \sectionname~\ref{subsec:PQL:predicates:semantics} to learn about methods for computing the PQL predicates (both unary and binary predicates).

As already mentioned, PQL utilizes a well-known mechanism of macros to combine results of several unary or binary relations.
A specimen of \texttt{UnaryPredicateMacro} combines several results of unary behavioral relations using existential and universal quantifiers as follows.

\small
\begin{eqnarray*}
\begin{split}
M_{\texttt{UnaryPredicateMacro}}\\ [p:\texttt{UnaryPredicateMacro},\\s:\mathbb{S},\mathit{vals}:\powerset{(\mathbb{V} \times \powerset{(\mathbb{T})})}] 
\end{split}
& \triangleq & 
\begin{cases}
    \begin{split}
    \;\exists\,\, t \in M_\texttt{SetOfTasks}(p.\mathit{tasks},s,\mathit{vals}) :\\ \mathit{canOccur}(s,t) 
    \end{split}
    & 
    \begin{split}
    (p.\mathit{name}\,\,\,\textbf{\texttt{is}}\,\,\,\texttt{CanOccur}) \,\land\\ (p.q\,\,\,\textbf{\texttt{is}}\,\,\,\texttt{Any})
    \end{split}
    \\[1em]
    \begin{split}
    \;\forall\, t \in M_\texttt{SetOfTasks}(p.\mathit{tasks},s,\mathit{vals}) :\\ \mathit{canOccur}(s,t) 
    \end{split}
    & 
    \begin{split}
    (p.\mathit{name}\,\,\,\textbf{\texttt{is}}\,\,\,\texttt{CanOccur}) \,\land\\ (p.q\,\,\,\textbf{\texttt{is}}\,\,\,\texttt{All})
    \end{split}
    \\[1em]
    \begin{split}
    \;\exists\,\, t \in M_\texttt{SetOfTasks}(p.\mathit{tasks},s,\mathit{vals}) :\\ \mathit{alwaysOccurs}(s,t) 
    \end{split}
    & 
    \begin{split}
    (p.\mathit{name}\,\,\,\textbf{\texttt{is}}\,\,\,\texttt{AlwaysOccurs}) \,\land\\ (p.q\,\,\,\textbf{\texttt{is}}\,\,\,\texttt{Any})
    \end{split}
    \\[1em]
    \begin{split}
    \;\forall\, t \in M_\texttt{SetOfTasks}(p.\mathit{tasks},s,\mathit{vals}) :\\ \mathit{alwaysOccurs}(s,t) 
    \end{split}
    & 
    \begin{split}
    (p.\mathit{name}\,\,\,\textbf{\texttt{is}}\,\,\,\texttt{AlwaysOccurs}) \,\land\\ (p.q\,\,\,\textbf{\texttt{is}}\,\,\,\texttt{All})
    \end{split}
    \\[1em]
\end{cases}
\end{eqnarray*}
\normalsize
\noindent
For example, a specimen $p$ of \texttt{UnaryPredicateMacro} such that $p.\mathit{name}$ is of type \texttt{AlwaysOccurs} and $p.q$ is of type \texttt{All} denotes the truth value of \emph{true} in the context of system $s$ \ifaof for every task $t$ in the set of tasks defined by $p.\mathit{tasks}$ it holds that $t$ occurs in every instance of $s$.

PQL offers two types of binary predicate macros: macros that relate a task to a set of tasks, and macros that relate two sets of tasks.
One can use a specimen of the \texttt{BinaryPredicateMacroTaskSet} construct to test if a task is in a binary behavioral relation with some (or all) tasks in a given set.
\small
\begin{eqnarray*}
\begin{split}
M_{\texttt{BinaryPredicateMacroTaskSet}}\\ [p:\texttt{BinaryPredicateMacroTaskSet},\\s:\mathbb{S},\mathit{vals}:\powerset{(\mathbb{V} \times \powerset{(\mathbb{T})})}] 
\end{split}
& \triangleq & 
\begin{cases}
    \begin{split}
    \;\exists\,\, t \in M_\texttt{SetOfTasks}(p.\mathit{tasks},s,\mathit{vals}) :\;\;\;\;\;\;\;\\ \mathit{canConflict}(s,M_\texttt{Task}(p.\mathit{task}),t)\;\;\;\;\;\;\;
    \end{split}
    & 
    \begin{split}
    (p.\mathit{name}\,\,\,\textbf{\texttt{is}}\\\texttt{CanConflict}) \,\land\\ (p.q\,\,\,\textbf{\texttt{is}}\,\,\,\texttt{Any})
    \end{split}
    \\[2.5em]
    \begin{split}
    \;\forall\, t \in M_\texttt{SetOfTasks}(p.\mathit{tasks},s,\mathit{vals}) :\\ \mathit{canConflict}(s,M_\texttt{Task}(p.\mathit{task}),t) 
    \end{split}
    & 
    \begin{split}
    (p.\mathit{name}\,\,\,\textbf{\texttt{is}}\\\texttt{CanConflict}) \,\land\\ (p.q\,\,\,\textbf{\texttt{is}}\,\,\,\texttt{All})
    \end{split}\\
    \;\;\;\ldots &
\end{cases}
\end{eqnarray*}
\normalsize
\noindent
We only specify the denotation of a specimen $p$ for which it holds that $p.\mathit{name}$ is of type $\texttt{CanConflict}$. 
Note that the denotations of specimens for other binary predicate names are defined analogously.
E.g., one can use a specimen $p$ of \texttt{BinaryPredicateMacroTaskSet} such that $p.\mathit{name}$ is of type \texttt{TotalCausal} and $p.q$ is of type \texttt{All} to check if the task defined by $p.\mathit{task}$ is in the \texttt{TotalCausal} behavioral relation with every task in the set of tasks defined by $p.\mathit{tasks}$, etc.

Finally, the denotation of a specimen of \texttt{BinaryPredicateMacroSetSet} is as follows.
\small
\begin{eqnarray*}
\begin{split}
M_{\texttt{BinaryPredicateMacroSetSet}}\\ [p:\texttt{BinaryPredicateMacroSetSet},\\s:\mathbb{S},\mathit{vals}:\powerset{(\mathbb{V} \times \powerset{(\mathbb{T})})}] 
\end{split}
& \triangleq & 
\begin{cases}
    \begin{split}
    \;\exists\, t_1 \in M_\texttt{SetOfTasks}(p.\mathit{tasks}_1,s,\mathit{vs})\;\;\;\;\;\;\;\\ \exists\, t_2 \in M_\texttt{SetOfTasks}(p.\mathit{tasks}_2,s,\mathit{vs}):\;\;\;\;\;\;\;\\ \mathit{canConflict}(s,t_1,t_2)\;\;\;\;\;\;\;
    \end{split}
    & 
    \begin{split}
    (p.\mathit{name}\,\,\,\textbf{\texttt{is}}\\\texttt{CanConflict}) \,\land\\ (p.q\,\,\,\textbf{\texttt{is}}\,\,\,\texttt{Any})
    \end{split}
    \\[2.5em]
    \begin{split}
    \;\exists\, t_1 \in M_\texttt{SetOfTasks}(p.\mathit{tasks}_1,s,\mathit{vs})\;\;\;\;\;\;\;\\ \forall\, t_2 \in M_\texttt{SetOfTasks}(p.\mathit{tasks}_2,s,\mathit{vs}):\;\;\;\;\;\;\;\\ \mathit{canConflict}(s,t_1,t_2)\;\;\;\;\;\;\;
    \end{split}
    & 
    \begin{split}
    (p.\mathit{name}\,\,\,\textbf{\texttt{is}}\\\texttt{CanConflict}) \,\land\\ (p.q\,\,\,\textbf{\texttt{is}}\,\,\,\texttt{Some})
    \end{split}
    \\[2.5em]
    \begin{split}
    \;\forall\, t_1 \in M_\texttt{SetOfTasks}(p.\mathit{tasks}_1,s,\mathit{vs})\\ \exists\, t_2 \in M_\texttt{SetOfTasks}(p.\mathit{tasks}_2,s,\mathit{vs}):\\ \mathit{canConflict}(s,t_1,t_2) 
    \end{split}
    & 
    \begin{split}
    (p.\mathit{name}\,\,\,\textbf{\texttt{is}}\\\texttt{CanConflict}) \,\land\\ (p.q\,\,\,\textbf{\texttt{is}}\,\,\,\texttt{Each})
    \end{split}
    \\[2.5em]
    \begin{split}
    \;\forall\, t_1 \in M_\texttt{SetOfTasks}(p.\mathit{tasks}_1,s,\mathit{vs})\\ \forall\, t_2 \in M_\texttt{SetOfTasks}(p.\mathit{tasks}_2,s,\mathit{vs}):\\ \mathit{canConflict}(s,t_1,t_2) 
    \end{split}
    & 
    \begin{split}
    (p.\mathit{name}\,\,\,\textbf{\texttt{is}}\\\texttt{CanConflict}) \,\land\\ (p.q\,\,\,\textbf{\texttt{is}}\,\,\,\texttt{All})\\
    \end{split}\\
    \;\;\ldots &
\end{cases}
\end{eqnarray*}
\normalsize
\noindent
Again, for space considerations, we omit the precise definitions of denotations of all specimens of the \texttt{BinaryPredicateMacroSetSet} construct. 
Note that the missing definitions are similar to those shown above for a specimen $p$ for which $p.\mathit{name}$ is of type \texttt{CanConflict}.
For example, one can use a specimen $p$ of \texttt{BinaryPredicateMacroSetSet} such that $p.\mathit{name}$ is of type \texttt{conflict} and $p.q$ is of type \texttt{Any} to check if some task in the set of tasks defined by $p.\mathit{tasks}_1$ is in the \texttt{Conflict} behavioral relation with some task in the set of tasks defined by $p.\mathit{tasks}_2$, etc.

}{\noindent}
In \tablename~\ref{tab:semantics:Q3}, we detail the meaning of query \emph{Q3} from \sectionname~\ref{sec:examples} matched to the model in \figurename~\ref{fig:EPC}.
In the table, the numbers in the first column refer to nodes in the abstract syntax tree in \figurename~\ref{fig:syntaxtree}, \eg number 1 in the first column refers to node ``Task (1)'' in the figure. 
The second column contains denotations of the corresponding constructs, while the third column contains further comments.

\small
\setlength\extrarowheight{2pt}
\begin{longtable}{| p{.025\textwidth} | p{.45\textwidth} | p{.45\textwidth} |}
\caption{\small Meaning of the example PQL query \emph{Q3} in \sectionname~\ref{sec:examples} explained using \figurename~\ref{fig:syntaxtree}.}
\vspace{-2mm}
\label{tab:semantics:Q3}\\
\hline
\textbf{No.} & \textbf{Denotation} & \textbf{Comment} \\ 
\hline 
\hline 
1 & $\{$``Start inventory recount''$\}$ & This is a specimen of \texttt{ExactTask} and, thus, it denotes a singleton of its label. \\ 
\hline 
2 & $\{$``No variance is determined''$\}$ & This is a specimen of \texttt{ExactTask} and, thus, it denotes a singleton of its label. \\ 
\hline 
3 & $\{\{$``Start inventory recount''$\}$, $\{$``No variance is determined''$\}\}$ & This set of tasks is specified as a fixed value composed of tasks denoted by 1 and 2. \\ 
\hline 
4 & $(x,\{\{$``Start inventory recount''$\}$, $\{$``No variance is determined''$\}\})$ & This variable associates symbolic name $x$ with the set of tasks denoted by $3$. \\ 
\hline 
5 & $\{$``Clear differences'', ``Clear differences WM'', ``Clear differences IM''$\}$ & This is a specimen of \texttt{DefSimTask} and, thus, denotes the set of all labels in the repository that are similar to its label. \\ 
\hline 
6 & $\{\{$``Clear differences'', ``Clear differences WM'', ``Clear differences IM''$\}\}$ & This set of tasks is specified as a fixed value composed of the task denoted by 5. \\ 
\hline 
7 & $(y,\{\{$``Clear differences'', ``Clear differences WM'', ``Clear differences IM''$\}\})$ & This variable associates symbolic name $y$ with the set of tasks denoted by $6$. \\ 
\hline 
8 & $\{\{$``Clear differences'', ``Clear differences WM'', ``Clear differences IM''$\}\}$ & This set of tasks is specified as the value associated by 7 with symbolic name $y$. \\ 
\hline 
9 & $\{$``Difference is posted'', ``Difference is posted to interface'', ``Difference is posted to IM''$\}$ & This is a specimen of \texttt{DefSimTask} and, thus, denotes the set of all labels in the repository that are similar to its label. \\ 
\hline 
10 & $\{\{$``Difference is posted'', ``Difference is posted to interface'', ``Difference is posted to IM''$\}\}$ & This set of tasks is specified as a fixed value composed of the task denoted by 9. \\ 
\hline 
11 & $\{\{$``Clear differences'', ``Clear differences WM'', ``Clear differences IM''$\}$, $\{$``Difference is posted'', ``Difference is posted to interface'', ``Difference is posted to IM''$\}\}$ & This set of tasks is computed as the union of sets denoted by 8 and 10. \\ 
\hline 
12 & $(z,\{\{$``Clear differences'', ``Clear differences WM'', ``Clear differences IM''$\}$, $\{$``Difference is posted'', ``Difference is posted to interface'', ``Difference is posted to IM''$\}\})$ & This variable associates symbolic name $z$ with the set of tasks denoted by $11$. \\ 
\hline 
13 & $\{\{$``Storage type is to be blocked for inventory''$\}$, $\{$``Storage type block''$\}$, $\{$``Storage type is blocked''$\}$, $...\,\,\}$ & This set of tasks is defined by all the labels in the model from \figurename~\ref{fig:EPC}; each function and event defines one PQL task in the set. \\ 
\hline 
14 & $\{\{$``Storage bin is blocked''$\}$, $\{$``System inventory record is created''$\}$, $\{$``Physical inventory is active''$\}$, $\{$``Print inventory list''$\}$, $\{$``Physical inventory list is printed''$\}$, $\{$``Enter count results''$\}\}$ & The set of tasks from the set denoted by 13 that occur in every instance of the model in \figurename~\ref{fig:EPC}; the checks are performed on the workflow system in \figurename~\ref{fig:system} using the result of \lemmaname~\ref{lem:always:occurs:transition}. \\ 
\hline 
15 & $(w,\{\{$``Storage bin is blocked''$\}$, $\{$``System inventory record is created''$\}$, $\{$``Physical inventory is active''$\}$, $\{$``Print inventory list''$\}$, $\{$``Physical inventory list is printed''$\}$, $\{$``Enter count results''$\}\})$ & This variable associates symbolic name $z$ with the set of tasks denoted by $14$. \\ 
\hline 
16 & $\{(x,\{\{$``Start inventory recount''$\}$, $\{$``No variance is determined''$\}\})$, ... $\}$ & The set contains values of all the variables used in the query, i.e., it contains denotations of 4, 7, 12, and 15. \\ 
\hline 
17 & $\mathbb{A}$ & All the attribute names used in the repository. \\ 
\hline 
18 & $\{/\texttt{SAP-R3-EPC-Repo}\}$ & The set composed of one location specified by the location path \texttt{"/SAP-R3-EPC-Repo"}. \\ 
\hline 
19 & $\{\{$``Start inventory recount''$\}$, $\{$``No variance is determined''$\}\}$ & This set of tasks is specified as the value associated by 4 with symbolic name $x$. \\ 
\hline 
20 & $\{\{$``Clear differences'', ``Clear differences WM'', ``Clear differences IM''$\}$, $\{$``Difference is posted'', ``Difference is posted to interface'', ``Difference is posted to IM''$\}\}$ & This set of tasks is specified as the value associated by 12 with symbolic name $z$. \\ 
\hline 
21 & $\{\{$``Start inventory recount''$\}$, $\{$``No variance is determined''$\}$, $\{$``Clear differences'', ``Clear differences WM'', ``Clear differences IM''$\}$, $\{$``Difference is posted'', ``Difference is posted to interface'', ``Difference is posted to IM''$\}\}$ & This set of tasks is computed as the union of sets denoted by 19 and 20. \\ 
\hline 
22 & $\mathit{true}$; every task in the set denoted by 21 occurs in at least one instance of the model in \figurename~\ref{fig:EPC}. & The checks are performed on the workflow system in \figurename~\ref{fig:system} using the result of \lemmaname~\ref{lem:can:occur:transition}. \\ 
\hline 
23 & $\{$``Start inventory recount''$\}$ & This is a specimen of \texttt{ExactTask} and, thus, it denotes a singleton of its label. \\ 
\hline 
24 & $\{\{$``Storage bin is blocked''$\}$, $\{$``System inventory record is created''$\}$, $\{$``Physical inventory is active''$\}$, $\{$``Print inventory list''$\}$, $\{$``Physical inventory list is printed''$\}$, $\{$``Enter count results''$\}\}$ & This set of tasks is specified as the value associated by 15 with symbolic name $w$. \\ 
\hline 
25 & $\mathit{false}$ & The task denoted by 23 is not a member of the set of tasks denoted by 24. \\ 
\hline 
26 & $\mathit{true}$ & Negation of denotation of 25. \\ 
\hline 
27 & $\{$``No variance is determined''$\}$ & This is a specimen of \texttt{ExactTask} and, thus, it denotes a singleton of its label. \\ 
\hline 
28 & $\{\{$``Clear differences'', ``Clear differences WM'', ``Clear differences IM''$\}\}$ & This set of tasks is specified as the value associated by 8 with symbolic name $y$. \\ 
\hline 
29 & $\mathit{true}$; the task denoted by 27 is in conflict with every task in the set denoted by 28. & The checks are performed on the workflow system in \figurename~\ref{fig:system} based on \definitionname~\ref{def:PQL:predicates} using the technique proposed in~\cite{PolyvyanyyWCRH14}. \\ 
\hline 
30 & $\{$``Start inventory recount''$\}$ & This is a specimen of \texttt{ExactTask} and, thus, it denotes a singleton of its label. \\ 
\hline 
31 & $\{\{$``Clear differences'', ``Clear differences WM'', ``Clear differences IM''$\}$, $\{$``Difference is posted'', ``Difference is posted to interface'', ``Difference is posted to IM''$\}\}$ & This set of tasks is specified as the value associated by 12 with symbolic name $z$. \\ 
\hline 
32 & $\mathit{true}$; the task denoted by 30 is in the total causal relation with every task in the set denoted by 31. & The checks are performed on the workflow system in \figurename~\ref{fig:system} based on \definitionname~\ref{def:PQL:predicates} using the technique proposed in~\cite{PolyvyanyyWCRH14}. \\ 
\hline 
33 & $\mathit{true}$ & Conjunction of denotations of 22, 26, 29, and 32. \\ 
\hline 
34 & $\{(s, \{($Author, SAP$), ...\}), ...\}$ & A result of query \emph{Q3} from \sectionname~\ref{sec:examples}. \\ 
\hline
\end{longtable}
\normalsize

\noindent
Thus, the meaning of \emph{Q3} is a set of pairs, where the first element of each pair is a model nested in a location from the set denoted by 18 and the second element is the set of attribute-value pairs for the attribute names contained in the set denoted by 17.
In \tablename~\ref{tab:semantics:Q3}, $s$ in row 34 is the system in \figurename~\ref{fig:system}.

%% file: tex/deciding.tex
\subsection{Denotations of Basic Predicates} 
\label{subsec:PQL:predicates:semantics}


This section presents denotations of the basic PQL predicates and techniques for computing them. 

\subsubsection{Definitions of the PQL Predicates}
\label{subsec:PQL:predicates:definitions}

\reportornot{Note that all the subsequent discussions are restricted to workflow systems.}{}
Before providing denotations of the predicates over PQL tasks, as demanded by the dynamic semantics of PQL, refer to \sectionname~\ref{subsec:PQL:dynamic:semantics}, we give definitions of the PQL predicates over character strings.

\mypar{Predicates over character strings} 
Given a workflow system, the $\mathit{canOccur}$ and $\mathit{alwaysOccurs}$ unary predicates on an input character string are defined below.

\begin{ncdefine}{Can occur and always occurs}{def:can:occur:and:always:occurs}{\quad\\}
Let $S:=(P,T,F,\lambda,M)$ be a workflow system and let $x \in \mathbb{C}$ be a non-empty character string.
\begin{compactitem} 
  \item $x$ \emph{can occur} in $S$, denoted by $\mathit{canOccur}(S,x)$, \ifaof there is a label execution $\eta$ of $S$ such that $x \in \eta$.
  \item $x$ \emph{always occurs} in $S$, denoted by $\mathit{alwaysOccurs}(S,x)$, \ifaof for every label execution $\eta$ of $S$ it holds that $x \in \eta$.\footnote{Given a sequence $\sigma$, $x \in \sigma$ denotes the fact that $x$ is an element of $\sigma$.}
\hfill\ensuremath{\lrcorner}
\end{compactitem}
\end{ncdefine}

\noindent
Hence, a character string $x \in \mathbb{C}$ can occur in a workflow system $S$ \ifaof one can observe $x$ in some execution of $S$, \ie an activity denoted by $x$ can be performed in some business scenario captured in $S$. 
In turn, $x$ always occurs in $S$, \ifaof it is observed in every execution of $S$.

As already mentioned, the binary predicates of PQL are grounded in the behavioral relations of the 4C spectrum.
In~\cite{PolyvyanyyWCRH14}, the authors define the 4C spectrum on systems in which transitions have distinct labels and then extend those relations to the general case when the same label can be assigned to several transitions in the system. 
The PQL predicates are grounded in these generalized relations. 
Thus, the $\mathit{canConflict}$ and $\mathit{canCooccur}$ predicates are defined as follows.

\begin{ncdefine}{Basic conflict and co-occurrence}{def:can:conflict:can:co:occur}{\quad\\}
Let $S:=(P,T,F,\lambda,M)$ be a workflow system and let $x,y \in \mathbb{C}$ be two non-empty character strings.
\begin{compactitem} 
  \item $x$ \emph{can conflict with} $y$ in $S$, denoted by $\mathit{canConflict}(S,x,y)$, \ifaof there is a label execution $\eta$ of $S$ such that $x \in \eta$ and $y \not\in \eta$.
  \item $x$ and $y$ \emph{can co-occur} in $S$, denoted by $\mathit{canCooccur}(S,x,y)$, \ifaof there is a label execution $\eta$ of $S$ such that $x \in \eta$ and $y \in \eta$.
\hfill\ensuremath{\lrcorner}
\end{compactitem}
\end{ncdefine}

\noindent
Intuitively, $x$ can conflict with $y$ \ifaof there exists an execution of $S$ which performs $x$ but not $y$, whereas $x$ and $y$ can co-occur \ifaof they both can be observed in some execution of $S$.

The 4C spectrum uses the basic conflict and co-occurrence relations as building blocks to define several other relations, two of which are among the selected PQL predicates.

\begin{ncdefine}{Conflict and co-occurrence}{def:conflict:co:occur}{\quad\\}
Let $S:=(P,T,F,\lambda,M)$ be a workflow system and let $x,y \in \mathbb{C}$ be two non-empty character strings.
\begin{compactitem}
  \item $x$ and $y$ are in \emph{conflict} in $S$, denoted by $\mathit{conflict}(S,x,y)$, \ifaof\\ 
  $\mathit{canConflict}(S,x,y) \;\land\; \mathit{canConflict}(S,y,x) \;\land\; \neg \;\mathit{canCooccur}(S,x,y)$.
  \item $x$ and $y$ \emph{co-occur} in $S$, denoted by $\mathit{cooccur}(S,x,y)$, \ifaof\\ 
  $\neg\;\mathit{canConflict}(S,x,y) \;\land\; \neg\;\mathit{canConflict}(S,y,x) \;\land\; \mathit{canCooccur}(S,x,y)$.
\hfill\ensuremath{\lrcorner}
\end{compactitem}
\end{ncdefine}

\noindent
Accordingly, $x$ and $y$ are in conflict in $S$ \ifaof they cannot co-occur but can conflict with each other, \ie they can be observed in some executions of $S$, but never together in the same execution.
In contrast, $x$ and $y$ co-occur in $S$ \ifaof they can co-occur but cannot conflict, \ie they can be observed together in some executions of $S$, $x$ is never observed in an execution that does not include an occurrence of $y$, and $y$ is never observed in an execution that does not include an occurrence of $x$.

\begin{figure}[t]
\vspace{-3mm}
\centering
\includegraphics[scale=.6]{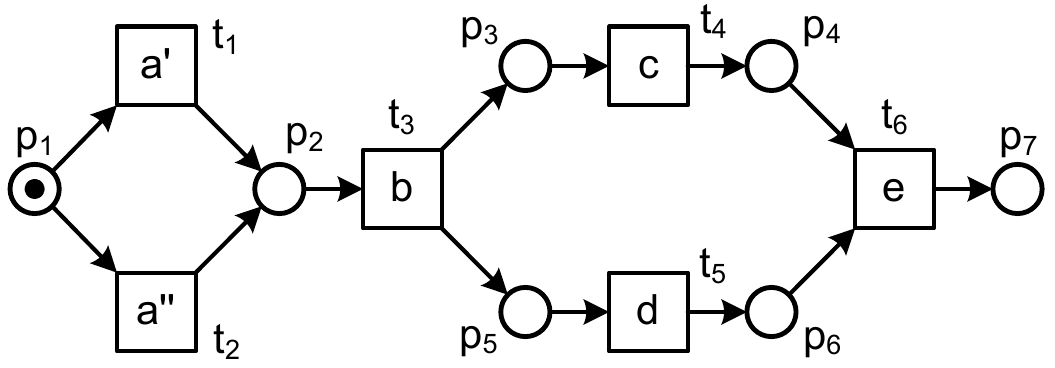}
\vspace{-2mm}
\caption{\small A workflow system.}
\label{fig:to:unify}
\vspace{-2mm}
\end{figure}

To give an example, consider the workflow system $S$ in \figurename~\ref{fig:to:unify}.
For this system, the expression $\mathit{canOccur(S,\texttt{a}')} \land \neg\;\mathit{alwaysOccurs(S,\texttt{a}')}$ evaluates to \emph{true}.
Indeed, there exists a label execution $\eta_1:=\sequence{\texttt{a}',\texttt{b},\texttt{c},\texttt{d},\texttt{e}}$ of $S$ that justifies the fact that $\texttt{a}'$ can occur in $S$, and a label execution $\eta_2:=\sequence{\texttt{a}'',\texttt{b},\texttt{c},\texttt{d},\texttt{e}}$ of $S$ that justifies the fact that $\texttt{a}'$ does not always occur in $S$.
Note that strings $\texttt{b}$, $\texttt{c}$, $\texttt{d}$, and $\texttt{e}$, always occur in $S$ as they are present in all of the four label executions of $S$.
Furthermore, both $\mathit{canConflict}(S,\texttt{b},\texttt{a}')$ and $\mathit{canCooccur}(S,\texttt{b},\texttt{a}')$ evaluate to \emph{true}, because $\texttt{b}$ can be observed without $\texttt{a}'$, for example in $\eta_2$, and with $\texttt{a}'$, for example in $\eta_1$.
Note that $\texttt{a}$ and $\texttt{a}'$ are in conflict, \ie $\mathit{conflict}(S,\texttt{a},\texttt{a}')$ evaluates to \emph{true}, as these two strings are never observed together.
Finally, \texttt{b} and \texttt{e} are example strings that co-occur in $S$, \ie $\mathit{cooccur}(S,\texttt{b},\texttt{e})$ evaluates to \emph{true}.


Executions of workflow systems capture orderings of transition occurrences. 
One can rely on \emph{processes} to adequately represent \emph{causality} and \emph{concurrency} relations on transition occurrences~\cite{GoltzR83}.


\begin{ncdefine}{Process}{def:process}{\quad\\}
A \emph{process} of a system $S:=(P,T,F,\lambda,M)$ is a 4-tuple $\pi:=(B,E,G,\rho)$, where 
$B$ is a set of \emph{conditions}, 
$E$ is a set of \emph{events}, 
$G \subseteq (B \times E) \cup (E \times B)$ is the \emph{flow relation}, 
such that $G^+$ is irreflexive and 
$\forall\, b \in B : |\set{e \in E}{(e,b) \in G}| \leq 1 \;\land\; |\set{e \in E}{(b,e) \in G}| \leq 1$, and 
$\rho : B \cup E \rightarrow P \cup T$ is such that:
\begin{compactitem}
  \item $\rho(B) \subseteq P$ and $\rho(E) \subseteq T$, \ie $\rho$ preserves the nature of nodes,
	\item 
  $\forall\, b_1,b_2 \in \Min{\pi} : (b_1,b_2) \not\in G^+ \land (b_2,b_1) \not\in G^+$, 
  $\forall\, b_1 \in B \setminus \Min{\pi} \exists\, b_2 \in \Min{\pi} : (b_2,b_1) \!\in G^+$, 
  and $\forall p \in P : M(p)=|\rho^{-1}(p) \cap \Min{\pi}|$, \ie $\pi$ starts at $M$, and 
	\item for every event $e \in E$ and for every place $p \in P$ it holds that\\ 
  $|\set{(p,t) \in F}{t = \rho(e)}|=|\rho^{-1}(p) \cap \pre{e}\,|$ and $|\set{(t,p) \in F}{t = \rho(e)}|=|\rho^{-1}(p) \cap \post{e\!}|$,\\ 
  \ie $\rho$ respects the environment of transitions.\footnote{$R^+$ denotes the transitive closure of a binary relation $R$.}\footnote{$f(X):=\set{f(x)}{x \in X}$ and $f^{-1}(z):=\set{y \in Y}{f(y)=z}$, where $X$ is a subset of $f$'s domain $Y$.}\footnote{$\Min{\pi}:=\set{b \in B}{\forall e \in E : (e,b) \not\in G}$.}\footnote{$\pre{e} := \set{b \in B}{(b,e) \in G}$ and $\post{e} := \set{b \in B}{(e,b) \in G}$.}
  \hfill\ensuremath{\lrcorner}
\end{compactitem}
\end{ncdefine}

\noindent
A process of a Petri net system is an acyclic bipartite graph, in which conditions and events are two disjoint sets of nodes, and a mapping from nodes of the process to nodes of the system.
\figurename~\ref{fig:processes} shows three processes of the system in \figurename~\ref{fig:to:unify}.
Conditions and events of a process are drawn as places and transitions, respectively.
The labels in the figures encode mappings of nodes of processes to nodes of the system.
In particular, each condition $b_i$ maps to place $p_i$, $i \in [1..7]$, and each event $e_j$ maps to transition $t_j$, $j \in [1..6]$.
In general, several conditions (events) of a process can refer to the same place (transition) of the corresponding system.
For example, there exist processes of the system in \figurename~\ref{fig:system} in which several conditions and several events refer to the same place and transition, respectively.
These are the processes that describe multiple occurrences of transitions in the loop with entry place $c_4$ and exit place $c_5$.

\begin{figure}[t]
\vspace{-3mm}
\begin{center}
  \subfigure[]{\label{fig:pi:1}\includegraphics[scale = .6]{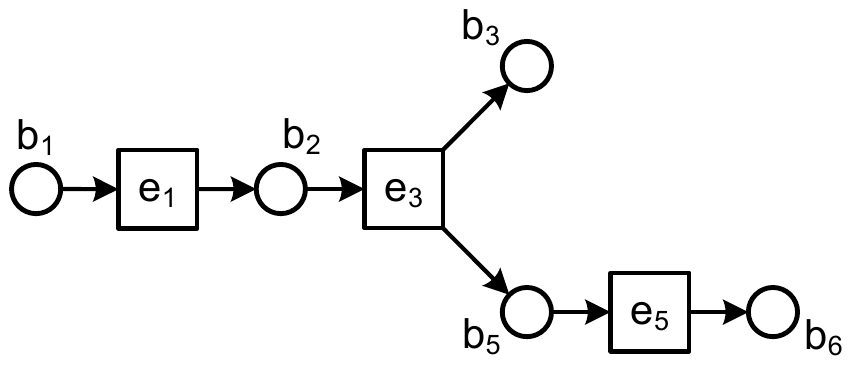}}
  \hspace{10mm}
  \subfigure[]{\label{fig:pi:2}\includegraphics[scale = .6]{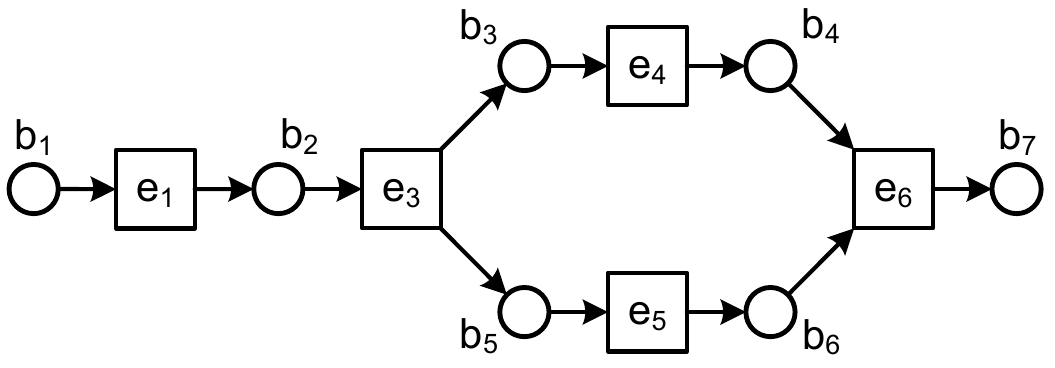}}
  \subfigure[]{\label{fig:pi:3}\includegraphics[scale = .6, trim = 0 0 0 3mm]{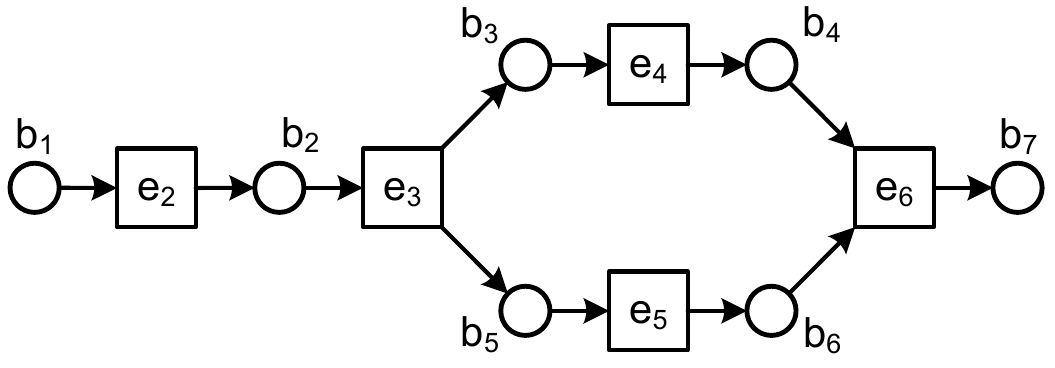}}
\end{center}
\vspace{-5mm}
\caption{\small Three processes of the system in \figurename~\ref{fig:to:unify}.}
\label{fig:processes}
\vspace{-4mm}
\end{figure}

Given a process $\pi$ of a system $S$, by $E_\pi$ and $\rho_\pi$ we refer to events of $\pi$ and the function that maps nodes of $\pi$ to nodes of $S$, respectively. 
A process $\pi$ of $S$ can be interpreted as a collection of occurrence sequences of $S$, where every event $e \in E_\pi$ describes an occurrence of transition $\rho_\pi(e)$. 
For example, the process in \figurename~\ref{fig:pi:1} describes a single occurrence sequence of the system in \figurename~\ref{fig:to:unify}, namely $\sequence{t_1,t_3,t_5}$, whereas 
the process in \figurename~\ref{fig:pi:2} describes two occurrence sequences of the system in \figurename~\ref{fig:to:unify}, namely $\sequence{t_1,t_3,t_4,t_5,t_6}$ and $\sequence{t_1,t_3,t_5,t_4,t_6}$.
Given a workflow system $S$, by $\Pi_S$ we denote the set that contains all and only processes of $S$ (up to isomorphism) that describe all the executions of $S$.
Note that the set of all processes of a workflow system can be infinite.
For example, the set of all processes of the system in \figurename~\ref{fig:system} that describe all its executions is infinite.
Let $S$ be the workflow system in \figurename~\ref{fig:to:unify}.
Then, $\Pi_S$ is composed of the two processes shown in \figurenames~\ref{fig:pi:2} and~\ref{fig:pi:3}; these two processes describe all the four executions of the system.
Note that the process from \figurename~\ref{fig:pi:1} is not in $\Pi_S$ because the occurrence sequence it describes is not an execution of $S$.
Finally, by $\Delta_S(x,y)$, where $x,y \in \mathbb{C}$, we refer to the set $\set{\pi \in \Pi_S}{\exists\, e_1,e_2 \in E_\pi : \lambda(\rho_\pi(e_1))=x \;\land\; \lambda(\rho_\pi(e_2))=y}$, \ie the set that consists of every process from $\Pi_S$ which contains an event that describes an occurrence of a transition with label $x$ and an event that describes an occurrence of a transition with label $y$.

Processes of systems can be characterized by the causality and concurrency relations over their nodes~\cite{NielsenPW81,GoltzR83}. 
An event $e_1 \in E_\pi$ \emph{causes} event $e_2 \in E_\pi$ of a process $\pi$, denoted by $e_1 \rightarrowtail_\pi e_2$, \ifaof $(e_1,e_2) \in G^+$. 
Two events $e_1 \in E_\pi$ and $e_2 \in E_\pi$ are \emph{concurrent} in $\pi$, denoted by $e_1 \;||_\pi\; e_2$, \ifaof $(e_1,e_2) \not\in G^+$ and $(e_2,e_1) \not\in G^+$. 
Intuitively, the fact that event $e_1$ is a cause for event $e_2$ means that one has to observe an occurrence of transition described by $e_1$ prior to observing an occurrence of transition described by $e_2$.
The fact that two events are concurrent means that the corresponding transitions can be enabled simultaneously in some occurrence sequence of the system and be performed one after another in any order.
For process $\pi$ in \figurename~\ref{fig:pi:3} it holds that $e_2 \rightarrowtail_\pi e_6$, $e_3 \rightarrowtail_\pi e_4$, and $e_5 \;||_\pi\; e_4$.

The total causal and total concurrent relations are now defined as follows.

\begin{ncdefine}{Total causality and total concurrency}{def:total:causality:and:concurrency}{\quad\\}
Let $S:=(P,T,F,\lambda,M)$ be a workflow system and let $x,y \in \mathbb{C}$ be two non-empty character strings.
\begin{compactitem}
  \item $x$ and $y$ are \emph{total causal} in $S$, denoted by $\mathit{totalCausal}(S,x,y)$, \ifaof\\
	$\forall\, \pi \in \Delta_S(x,y) \; \forall\, e_1 \in E_\pi \; \forall\, e_2 \in E_\pi : (e_1 \neq e_2 \;\land\; \lambda(\rho_\pi(e_1))=x \;\land\; \lambda(\rho_\pi(e_2))=y) \Rightarrow e_1 \; \rightarrowtail_\pi \; e_2$.
  \item $x$ and $y$ are \emph{total concurrent} in $S$, denoted by $\mathit{totalConcurrent}(S,x,y)$, \ifaof\\
	$\forall\, \pi \in \Delta_S(x,y) \; \forall\, e_1 \in E_\pi \; \forall\, e_2 \in E_\pi : (e_1 \neq e_2 \;\land\; \lambda(\rho_\pi(e_1))=x \;\land\; \lambda(\rho_\pi(e_2))=y) \Rightarrow e_1 \; ||_\pi \; e_2$.
\hfill\ensuremath{\lrcorner}
\end{compactitem}
\end{ncdefine}

\noindent
For example, $\mathit{totalCausal(S,\texttt{a}',\texttt{c})}$ holds for workflow system $S$ in \figurename~\ref{fig:to:unify}; in every execution of $S$ in which $\texttt{a}'$ and $\texttt{c}$ both occur it holds that every occurrence of $\texttt{a}'$ precedes every occurrence of $\texttt{c}$.
As another example, $\mathit{totalConcurrent(S,\texttt{c},\texttt{d})}$ holds for the same system.

\mypar{Predicates over PQL tasks}
Next, we lift the PQL predicates from individual labels to PQL tasks, \ie to sets of character strings.
To this end, we employ the label unification principle proposed in~\cite{PolyvyanyyWCRH14}.
Given a system $S$ and a character string $x \in \mathbb{C}$, \emph{label unification} of $x$ in $S$ is a transformation of $S$ into a fresh system $S'$ that is behaviorally equivalent to $S$ but in which every occurrence of $x$ is guaranteed to be triggered by a (fresh) dedicated transition, refer to \definitionname~6.2 and \lemmaname~6.5 in~\cite{PolyvyanyyWCRH14}.
Let $X$ be the set of all transitions of $S$ that have label $x$ such that $|X|>1$. 
The label unification of $x$ in $S$ augments $S$ to result in a system $S'$ that is behaviorally equivalent to $S$ but `forbids' occurrences of transitions in $X$ and
ensures that tokens from the input places of transitions in $X$ are `rerouted' to the only input place of the dedicated transition $\hat{t}$ that has label $x$.
Note that if $X$ is a singleton, then all occurrences of $x$ generated by $S$ are guaranteed to be triggered by transition $t \in X$ and, hence, there is no need to transform $S$, \ie it holds that $S'=S$.

\begin{figure}[t]
\vspace{-2mm}
\centering
\includegraphics[scale=.6]{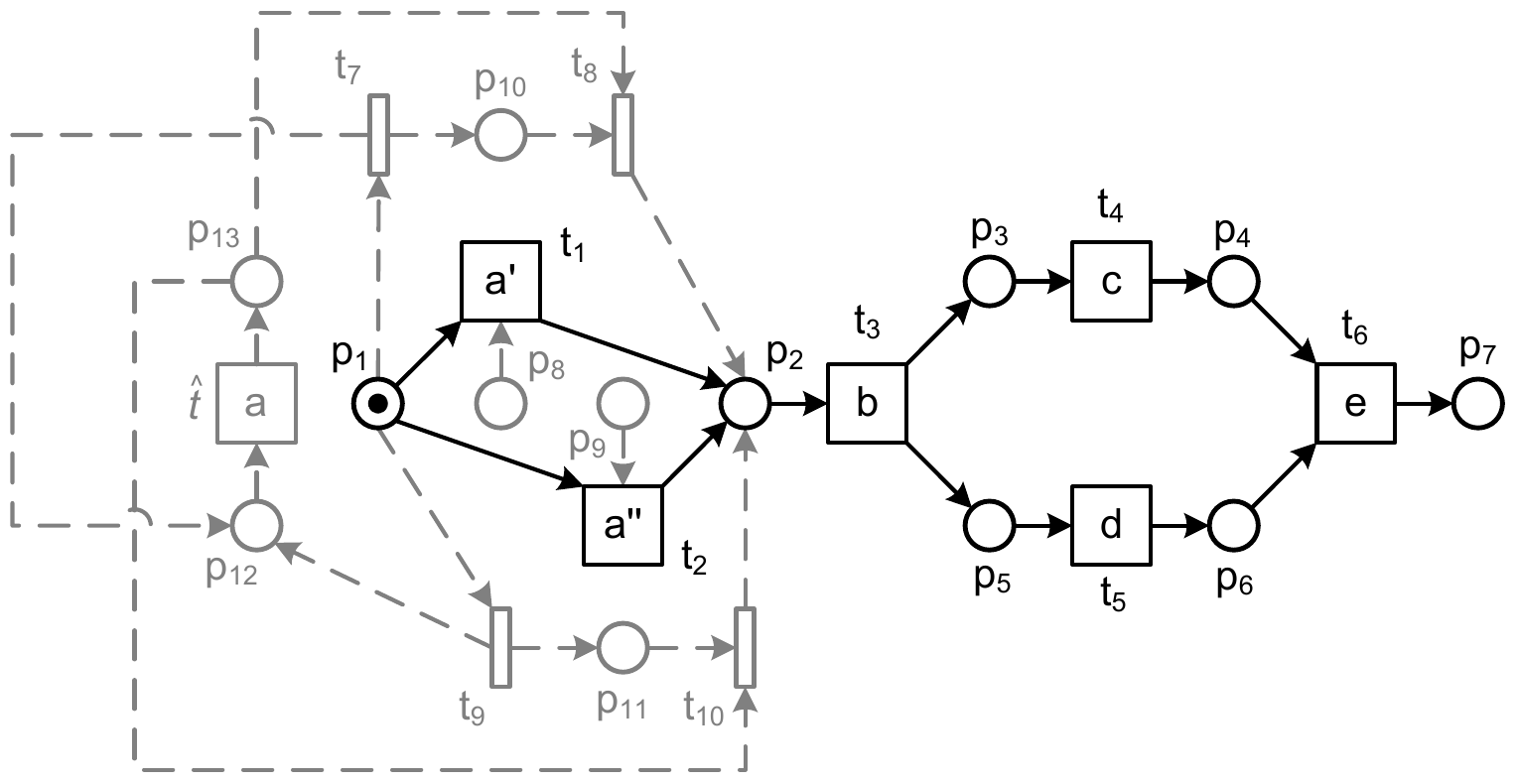}
\vspace{-2mm}
\caption{\small A system obtained via unification of the set of labels $\{\texttt{a},\texttt{a}'\}$ in the workflow system in \figurename~\ref{fig:to:unify}.} 
\label{fig:unified}
\vspace{-2mm}
\end{figure}

The concept of a PQL task is introduced in the language to allow handling several distinct labels as if they all represent the same activity.
For example, two character strings $\texttt{a}':=$``\texttt{process payment by cash}'' and $\texttt{a}'':=$``\texttt{process payment by check}'' may be seen as sufficiently similar to represent activity $\texttt{a}:=$``\texttt{process payment}''. 
PQL adapts the label unification principle to operationalize the above sketched intuition. 
\figurename~\ref{fig:unified} shows the result of performing label unification for labels $\texttt{a}'$ and $\texttt{a}''$ in the system in \figurename~\ref{fig:to:unify}.
The fresh elements introduced during the unification are highlighted in gray, while the fresh arcs, in addition, are depicted using the dashed lines.
Unlike in the system in \figurename~\ref{fig:to:unify}, transitions with labels $\texttt{a}'$ and $\texttt{a}''$
cannot occur in executions of the system in \figurename~\ref{fig:unified}. 
However, in the transformed system, the tokens from the input places of the ``forbidden'' transitions $t_1$ and $t_2$ can be rerouted to enable the fresh transition $\hat{t}$ that has label $\texttt{a}$ and can occur instead of both forbidden transitions; transition $\hat{t}$ is called the \emph{solitary} transition for labels $\texttt{a}'$ and $\texttt{a}''$.

By $\mathit{unify}(S,X)$, $X \in \powerset_{\geq 1}{(\mathbb{C})}$, we denote the result of the adapted label unification of $X$ in $S$. 
This result is a pair $(S',\hat{t})$, where $S'$ is the transformed system and $\hat{t}$ is the fresh solitary transition for labels in $X$.
For example, it holds that $\mathit{unify}(S,X) = (S',\hat{t})$, where $S$ is the system in \figurename~\ref{fig:to:unify}, $S'$ is the system in \figurename~\ref{fig:unified}, and $X = \{\texttt{a}',\texttt{a}''\}$.
\reportornot{

It is easy to see that the order of applying different label unifications has no impact on the resulting system as long as fresh solitary transitions have unique labels.}{The order of applying different label unifications has no impact on the resulting system.}
Finally, the predicates over PQL tasks are defined as follows.

\reportornot{\begin{ncdefine}{Predicates over PQL tasks}{def:PQL:predicates}{\quad\\}}{\begin{ncdefine}{Predicates over PQL tasks}{def:PQL:predicates}}
Let $S:=(P,T,F,\lambda,M)$ be a workflow system and let $X,Y \in \powerset_{\geq 1}{(\mathbb{C})}$ be two non-empty sets of non-empty character strings.
\reportornot{}{\vspace{-0.8mm}}
\begin{compactitem} 
  \item $X$ \emph{can occur} in $S$, denoted by $\mathit{canOccur}(S,X)$, \ifaof\\ it holds that $\mathit{canOccur}(S',x)$, where $\mathit{unify}(S,X)=(S',x)$.
  \item $X$ \emph{always occurs} in $S$, denoted by $\mathit{alwaysOccurs}(S,X)$, \ifaof\\ it holds that $\mathit{alwaysOccurs}(S',x)$, where $\mathit{unify}(S,X)=(S',x)$.
  \item $X$ \emph{can conflict with} $Y$ in $S$, denoted by $\mathit{canConflict}(S,X,Y)$, \ifaof it holds that $\mathit{canConflict}(S''\!,x,y)$, where $\mathit{unify}(S,X)=(S',x)$ and $\mathit{unify}(S',Y)=(S''\!,y)$.
  \item $X$ and $Y$ \emph{can co-occur} in $S$, denoted by $\mathit{canCooccur}(S,X,Y)$, \ifaof it holds that $\mathit{canCooccur}(S''\!,x,y)$, where $\mathit{unify}(S,X)=(S',x)$ and $\mathit{unify}(S',Y)=(S''\!,y)$.
  \item $X$ and $Y$ are in \emph{conflict} in $S$, denoted by $\mathit{conflict}(S,X,Y)$, \ifaof\\ it holds that $\mathit{conflict}(S''\!,x,y)$, where $\mathit{unify}(S,X)=(S',x)$ and $\mathit{unify}(S',Y)=(S''\!,y)$.
  \item $X$ and $Y$ \emph{co-occur} in $S$, denoted by $\mathit{cooccur}(S,X,Y)$, \ifaof\\ it holds that $\mathit{cooccur}(S''\!,x,y)$, where $\mathit{unify}(S,X)=(S',x)$ and $\mathit{unify}(S',Y)=(S''\!,y)$.
  \item $X$ and $Y$ are \emph{total causal} in $S$, denoted by $\mathit{totalCausal}(S,X,Y)$, \ifaof\\ it holds that $\mathit{totalCausal}(S''\!,x,y)$, where $\mathit{unify}(S,X)=(S',x)$ and $\mathit{unify}(S',Y)=(S''\!,y)$.
  \item $X$ and $Y$ are \emph{total concurrent} in $S$, denoted by $\mathit{totalConcurrent}(S,X,Y)$, \ifaof\\ it holds that $\mathit{totalConcurrent}(S''\!,x,y)$, where $\mathit{unify}(S,X)=(S',x)$ and $\mathit{unify}(S',Y)=(S''\!,y)$.
\hfill\ensuremath{\lrcorner}
\end{compactitem}
\end{ncdefine}

\noindent
Using the proposed denotations, one can verify that, for instance, $\mathit{alwaysOccurs}(S,\{\texttt{a}',\texttt{a}''\})$ evaluates to \emph{true}, where $S$ is the system in \figurename~\ref{fig:to:unify}.
Hence, for example, one can retrieve the system in \figurename~\ref{fig:to:unify} by issuing the query ``\texttt{\SELECT~* \FROM~* \WHERE~AlwaysOccurs($\sim$\texttt{a});}'', given that $\sim$\texttt{a} evaluates to $\{\texttt{a}',\texttt{a}''\}$.
One can use queries like the one exemplified to perform exploratory querying or address the problem of the inconsistent usage of labels in process repositories, where label $\texttt{a}$ from the user specified query gets replaced by labels $\texttt{a}'$ and $\texttt{a}''$ from the process repository.

\subsubsection{Computations of the PQL Predicates}
\label{app:PQL:predicates}

\definitionname~\ref{def:PQL:predicates} specifies the eight selected predicates over PQL tasks in terms of the corresponding predicates over transitions.
The reader can find techniques to compute the $\mathit{canConflict}$, $\mathit{canCooccur}$, and $\mathit{totalCausal}$ predicates over transitions in~\cite{PolyvyanyyWCRH14}. 
Using $\mathit{canConflict}$ and $\mathit{canCooccur}$, one can compute the $\mathit{conflict}$ and $\mathit{cooccur}$ predicates, refer to \definitionname~\ref{def:PQL:predicates}.
Next, we propose techniques that given a sound workflow system $S:=(P,T,F,\lambda,M)$ compute whether $t \in T$ \emph{can occur} in $S$, 
$t \in T$ \emph{always occurs} in $S$, and 
$t_1 \in T$ and $t_2 \in T$ are \emph{total concurrent} in $S$.
PQL performs these computations on systems obtained via unifications of sets of labels, refer to \sectionname~\ref{subsec:PQL:predicates:definitions} and~\cite{PolyvyanyyWCRH14} for details. 
Note that label unification in a sound workflow system often leads to a system that is not even a workflow system; due to the introduction of dead transitions, where a dead transition is a transition which is not part of any occurrence sequence of the system; see for example transitions $t_1$ and $t_2$ in \figurename~\ref{fig:unified}. 
Dead transitions are kept in the resulting systems as they may be required to perform subsequent transformations.
However, once all the transformations are applied, dead transitions can be removed from the system to result in a sound workflow system; this trivially follows from the definition of label unification. 
\reportornot{Note that in what follows, we restrict some discussions to sound workflow systems.}{}

\mypar{Can occur transition}
One can check whether a transition can occur in a workflow system by solving a reachability problem on a transformed version of the system.

\reportornot{\begin{define}{Can occur transition}{def:can:occur:transition}{\quad\\}}{\begin{define}{Can occur transition}{def:can:occur:transition}}
A transition $t \in T$ \emph{can occur} in a workflow system $S:=(P,T,F,\lambda,M)$ \ifaof there exists an execution $\sigma$ of $S$ such that $t \in \sigma$.
\end{define}

\noindent
Given a system and marking, the reachability problem consists of deciding if there is an occurrence sequence of the system that leads to the marking.
Note that the reachability problem is decidable~\cite{Hack75}.

\begin{lem}{Can occur transition}{lem:can:occur:transition}{\quad\\}
Let $S:=(P,T,F,\lambda,M)$ be a workflow system with a sink place $o \in P$.
A transition $t \in T$ can occur in $S$ \ifaof 
there exists an occurrence sequence $\sigma$ of 
$S':=(P \cup \{p'\}, T \cup \{t'\}, F \cup \{(p',t')\} \cup \set{(p,t')}{p \in \pre{t}} \cup \set{(t',p)}{p \in \post{t}}, \lambda \cup \{(t',\epsilon)\},M \cupplus [p'])$, 
where $p' \not\in P$ and $t' \not\in T$ are a fresh place and transition, respectively, such that
$t'$ is an element of $\sigma$, \ie $t' \in \sigma$, and 
$\sigma$ leads to $[o]$.
\end{lem}

\noindent
The proof of \lemmaname~\ref{lem:can:occur:transition} is straightforward due to the construction of $S'$.
If $t$ can occur in $S$, then there exists an execution $\gamma$ of $S$ that contains $t$ at some position $j$ of $\gamma$. 
A sequence of transitions obtained from $\gamma$ by replacing $t$ at position $j$ with $t'$ is an occurrence sequence of $S'$ that leads to $[o]$.
Let $\sigma$ be an occurrence sequence of $S'$ that leads to $[o]$ and $t' \in \sigma$.
Then, a sequence of transitions obtained from $\sigma$ by replacing $t'$ with $t$ and keeping the order of the remaining elements is an execution of $S$; note that $\sigma$ contains exactly one occurrence of $t'$.


\mypar{Always occurs transition}
Next, we present a technique for checking whether a transition always occurs in a sound workflow system.

\reportornot{\begin{define}{Always occurs transition}{def:always:occurs:transition}{\quad\\}}{\begin{define}{Always occurs transition}{def:always:occurs:transition}}
A transition $t \in T$ \emph{always occurs} in a sound workflow system $S:=(P,T,F,\lambda,M)$ \ifaof for every execution $\sigma$ of $S$ it holds that $t \in \sigma$.
\end{define}

\noindent
One can check whether a transition always occurs in a sound workflow system by solving a reachability problem on a transformed version of the system.

\begin{lem}{Always occurs transition}{lem:always:occurs:transition}{\quad\\}
Let $S:=(P,T,F,\lambda,M)$ be a workflow system with a sink place $o \in P$.
A transition $t \in T$ always occurs in $S$ \ifaof 
there exists no occurrence sequence $\sigma$ of 
$S':=(P \cup \{p'\},T,F \cup \{(p',t)\},\lambda, M \cupplus [p'])$, where $p' \not\in P$ is a fresh place, that leads to $[p',o]$.
\end{lem}

\noindent
The proof of \lemmaname~\ref{lem:always:occurs:transition} is straightforward due to the construction of $S'$.
If $t$ always occurs in $S$, then every execution of $S$ contains $t$. 
Let $\gamma$ be an execution of $S$ without $t$, then $\gamma$ is an occurrence sequence of $S'$ that leads to $[p',o]$.
Finally, let $\sigma$ be an occurrence sequence of $S'$ that leads to $[p',o]$. 
Then, clearly, $\sigma$ is an execution of $S$ without $t$.


\mypar{Total concurrent transitions}
Let $S:=(P,T,F,\lambda,M)$ be a workflow system.
By $\Xi_S(t_1,t_2)$, where $t_1,t_2 \in T$, we denote the set 
$\set{\pi \in \Pi_S}{\exists\, e_1,e_2 \in E_\pi : \rho_\pi(e_1)=t_1 \;\land\; \rho_\pi(e_2)=t_2}$.

\begin{figure}[t]
\vspace{-2mm}
\centering
\includegraphics[scale=.57]{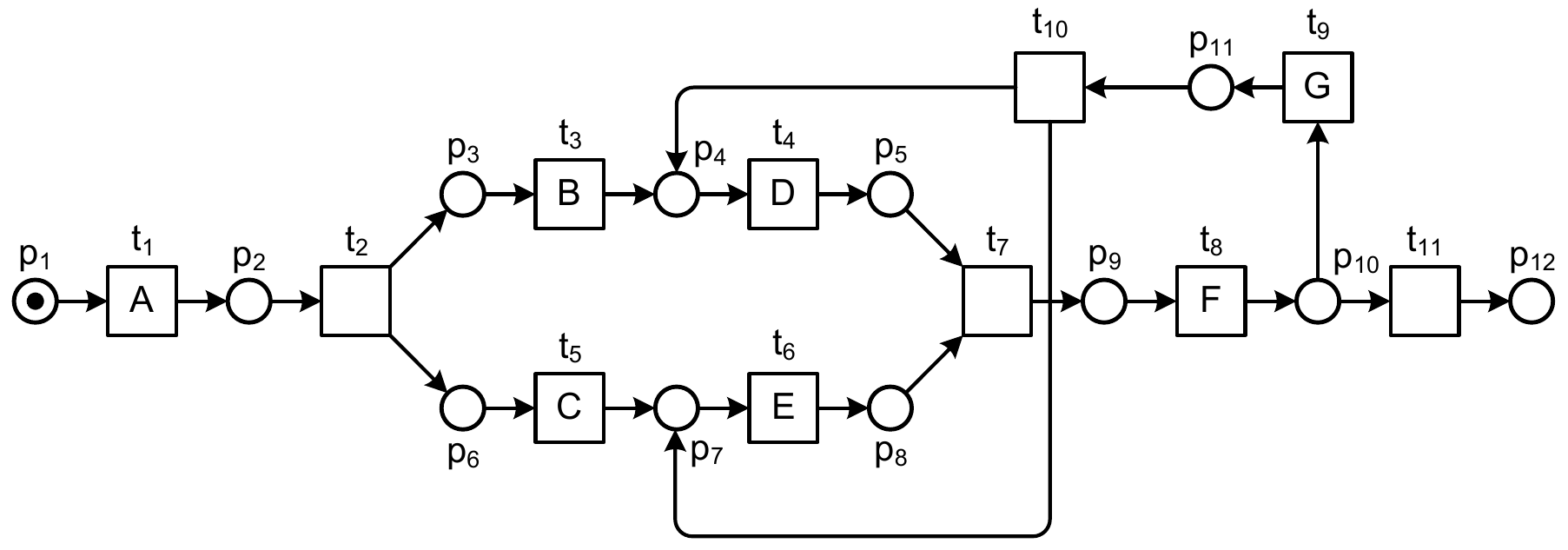}
\vspace{-3mm}
\caption{\small A workflow system.}\label{fig:to:unfold}
\vspace{-3mm}
\end{figure}

\begin{define}{Total concurrent transitions}{def:total:concurrent:transitions}{\quad\\}
Transitions $t_1 \in T$ and $t_2 \in T$ are \emph{total concurrent} in a sound workflow system $S:=(P,T,F,\lambda,M)$ \ifaof
$\forall\, \pi \in \Xi_S(t_1,t_2) \; \forall\, e_1 \in E_\pi \; \forall\, e_2 \in E_\pi : (e_1 \neq e_2 \;\land\; \rho_\pi(e_1)=t_1 \;\land\; \rho_\pi(e_2)=t_2) \Rightarrow e_1 \; ||_\pi \; e_2$.
\end{define}

\noindent
The \emph{unfolding} of a Petri net system is an acyclic graph that encodes all the processes of the system in a possibly infinite, tree-like structure~\cite{McMillan92,EsparzaRV02}. 
In~\cite{McMillan92}, McMillan proposed an algorithm for constructing a finite initial part of the unfolding, called a \emph{complete prefix} of the unfolding, which contains full information about the behavior of the system, \ie all the processes of the system. 

\figurename~\ref{fig:to:unfold} shows a workflow system $S$ obtained by translating the process model shown later in \figurename~\ref{fig:smodels} (model 9), refer to \sectionname~\ref{subsec:PQL:sample:queries}, into a Petri net system. 
\figurename~\ref{fig:unfolded} shows a complete prefix of the unfolding $U$ of $S$.
A complete prefix of the unfolding of a system $S:=(P,T,F,\lambda,M)$ is a 4-tuple $U:=(B,E,G,\rho)$, where $B$ and $E$ are disjoint sets of conditions and events, respectively, $G \subseteq (B \times E) \cup (E \times B)$ is the flow relation, such that $G^+$ is irreflexive, and $\rho : B \cup E \rightarrow P \cup T$ is a function that maps conditions to places and events to transitions.
\reportornot{Thus, a complete prefix of the unfolding is a process that allows forward conflicts~\cite{McMillan92}.}{}

\begin{figure}[t]
\vspace{-2mm}
\centering
\includegraphics[scale=.57]{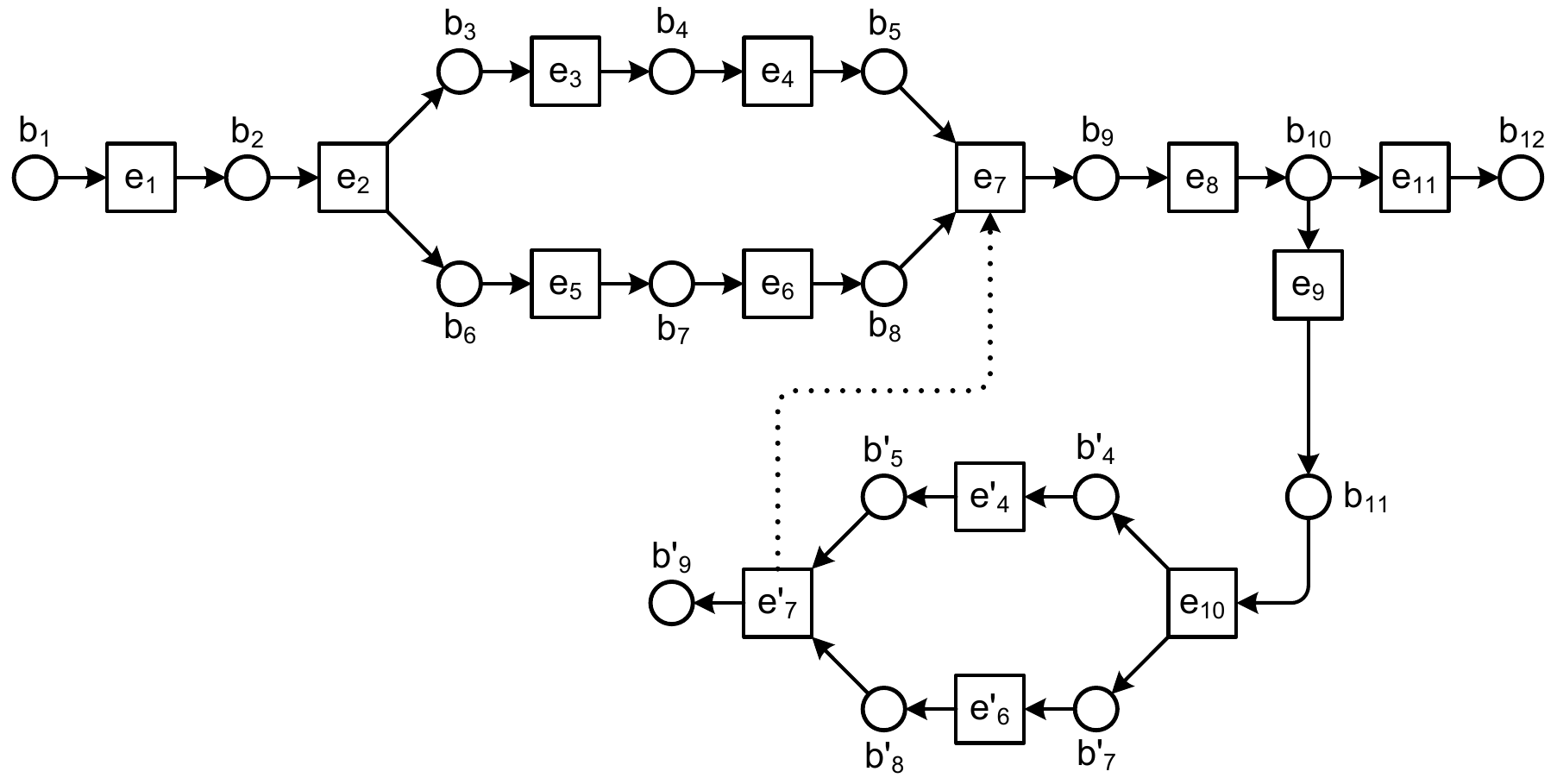}
\vspace{-2mm}
\caption{\small A complete prefix of the unfolding of the system in \figurename~\ref{fig:to:unfold}.}
\label{fig:unfolded}
\vspace{-2mm}
\end{figure}

In \figurename~\ref{fig:unfolded}, conditions and events are shown as circles and rectangles, respectively. 
Every condition $b_i, b'_i, \ldots$ represents one token at place $p_i$, $i \in \mathbb{N}$, refer to~\cite{PolyvyanyyROH15}.
Every event $e_j, e'_j, \ldots$ represents one occurrence of transition $t_j$, $j \in \mathbb{N}$.
McMillan proposes to associate every event $e$ of a compete prefix of the unfolding of a Petri net system with a marking that one reaches by firing all the transitions encoded by the events in the local configuration of $e$.
The \emph{local configuration} of an event $e$, denoted by $\left\lceil e \right\rceil$, is the set of events composed of $e$, all the events from which there is a directed path to $e$, and no other events.
For example, $\left\lceil e_4 \right\rceil = \{e_1,e_2,e_3,e_4\}$ and $\left\lceil e_7 \right\rceil = \{e_1,e_2,e_3,e_4,e_5,e_6,e_7\}$.
One reaches the marking $[p_5,p_6]$ by firing transitions represented by the events in $\left\lceil e_4 \right\rceil$, and the marking $[p_9]$ by firing transitions represented by the events in $\left\lceil e_7 \right\rceil$.
These markings are encoded by the sets of conditions $\{b_5,b_6\}$ and $\{b_9\}$ in \figurename~\ref{fig:unfolded}, respectively, and are denoted by $\mathit{Cut}(\left\lceil e_4 \right\rceil)$ and $\mathit{Cut}(\left\lceil e_7 \right\rceil)$, respectively.
Note that the marking $[p_9]$ can also be reached in the system in \figurename~\ref{fig:to:unfold} after occurrences of transitions encoded by the events in the local configuration of event $e'_7$.

Intuitively, if one continues constructing the prefix in \figurename~\ref{fig:unfolded}, the part that will follow $\mathit{Cut}(\left\lceil e'_7 \right\rceil)$ will be isomorphic to the part of the prefix that will follow $\mathit{Cut}(\left\lceil e_7 \right\rceil)$. 
Hence, McMillan proposed to stop the construction at $\mathit{Cut}(\left\lceil e'_7 \right\rceil)$ and refers to $e'_7$ and $e_7$ as a \emph{cutoff} and its \emph{corresponding} event, respectively.
We denote the set of all the cutoff events of a complete prefix of the unfolding $U$ by $\mathit{cutoffs}(U)$. 
Given a cutoff event $e$, by $\mathit{corr}(e)$, we denote the corresponding event of $e$. 
Thus, $\mathit{cutoffs}(U)$, where $U$ is a complete prefix of the unfolding in \figurename~\ref{fig:unfolded}, is equal to $\{e'_7\}$, and $\mathit{corr}(e'_7) = e_7$.
In the figure, the relation between the cutoff and its corresponding event is shown by the dotted arrow.
For further details on (complete prefixes of) the unfoldings please refer to~\cite{McMillan92,EsparzaRV02}.

Let $U:=(B,E,G,\rho)$ be a complete prefix of the unfolding of a system $S:=(P,T,F,\lambda,M)$.
Then, it holds that 
$\mathit{path}(U,e_1,e_2)$ 
\ifaof 
$(e_1,e_2) \in G^+$, or $(e_2,e_1) \in G^+$, or
there exists a sequence of  cutoffs $c \in \mathit{cutoffs}(U)^*$ such that 
(i) $\exists\, b \in \mathit{Cut}(\left\lceil c_1 \right\rceil) : (e_1,b) \in G^+$, 
(ii) $\exists\, b \in \mathit{Cut}(\left\lceil \mathit{corr}(c_{|c|}) \right\rceil) : (b,e_2) \in G^+$, and 
(iii) $\forall\, i \in [1..(|c|-1)] \exists\, b_1 \in \mathit{Cut}(\left\lceil \mathit{corr}(c_i) \right\rceil)\exists\, b_2 \in \mathit{Cut}(\left\lceil c_{i+1} \right\rceil): (b_1,b_2) \in G^*$.

\begin{lem}{Total concurrent transitions}{lem:total:concurrent:transitions}{\quad\\}
Let $S:=(P,T,F,\lambda,M)$ be a sound workflow system and let $U:=(B,E,G,\rho)$ be a complete prefix of the unfolding of $S$.
Transitions $t_1 \in T$ and $t_2 \in T$ are total concurrent in $S$ \ifaof 
$
\forall e_1 \!\in E\, \forall e_2 \!\in E :
(e_1 \neq e_2 \;\land\; \rho(e_1)=t_1 \;\land\; \rho(e_2)=t_2) 
\Rightarrow 
\neg\;\mathit{path}(U,e_1,e_2)
$.
\end{lem}

\noindent
Let us assume that $t_1$ and $t_2$ are total concurrent in $S$ and there exist two distinct events $e_1$ and $e_2$ that describe occurrences of $t_1$ and $t_2$, respectively, such that $\mathit{path}(U,e_1,e_2)$. 
Then, there is a process in $\Xi_S(t_1,t_2)$ that contains two distinct causal events that refer to $t_1$ and $t_2$. 
This follows immediately from the definition of a complete prefix of the unfolding, \seeCite{EsparzaRV02}, and the fact that every occurrence sequence of $S$ can be extended to an execution.
If for every two distinct events $e_1$ and $e_2$ that describe occurrences of $t_1$ and $t_2$ it holds that $\neg\,\mathit{path}(U,e_1,e_2)$, then $t_1$ and $t_2$ are total concurrent in $S$, as according to the definition of a complete prefix of the unfolding there exists no process in $\Xi_S(t_1,t_2)$ that evidences that some occurrences of $t_1$ and $t_2$ are causal.

Note that one can always construct a complete prefix of the unfolding of a bounded system~\cite{EsparzaRV02}, \ie a system with a finite number of reachable states, while a sound workflow system is guaranteed to be bounded~\cite{Aalst97ATPN}.
Using the prefix in \figurename~\ref{fig:unfolded} and \lemmaname~\ref{lem:total:concurrent:transitions} one can verify whether two given transitions are total concurrent in the system $S$ in \figurename~\ref{fig:to:unfold}.
For example, transitions $t_3$ and $t_5$ are total concurrent in $S$.
Transitions $t_3$ and $t_6$ are not total concurrent in $S$ because of events $e_3$ and $e'_6$ and the fact that $(e_3,e'_6) \in G^+$.
Transitions $t_9$ and $t_{11}$ are also not total concurrent in $S$ because of events $e_9$ and $e_{11}$ and the sequence of cutoff events $\sequence{e'_7}$; note that $(e_9,b'_9) \in G^+$ and $(b_9,e_{11}) \in G^+$.


%% file: tex/sample_queries.tex
\subsection{Sample PQL Queries}
\label{subsec:PQL:sample:queries}

\begin{figure}[htbp]
\centering
\includegraphics[width = 1.03\linewidth, trim = 5mm 0 0 0]{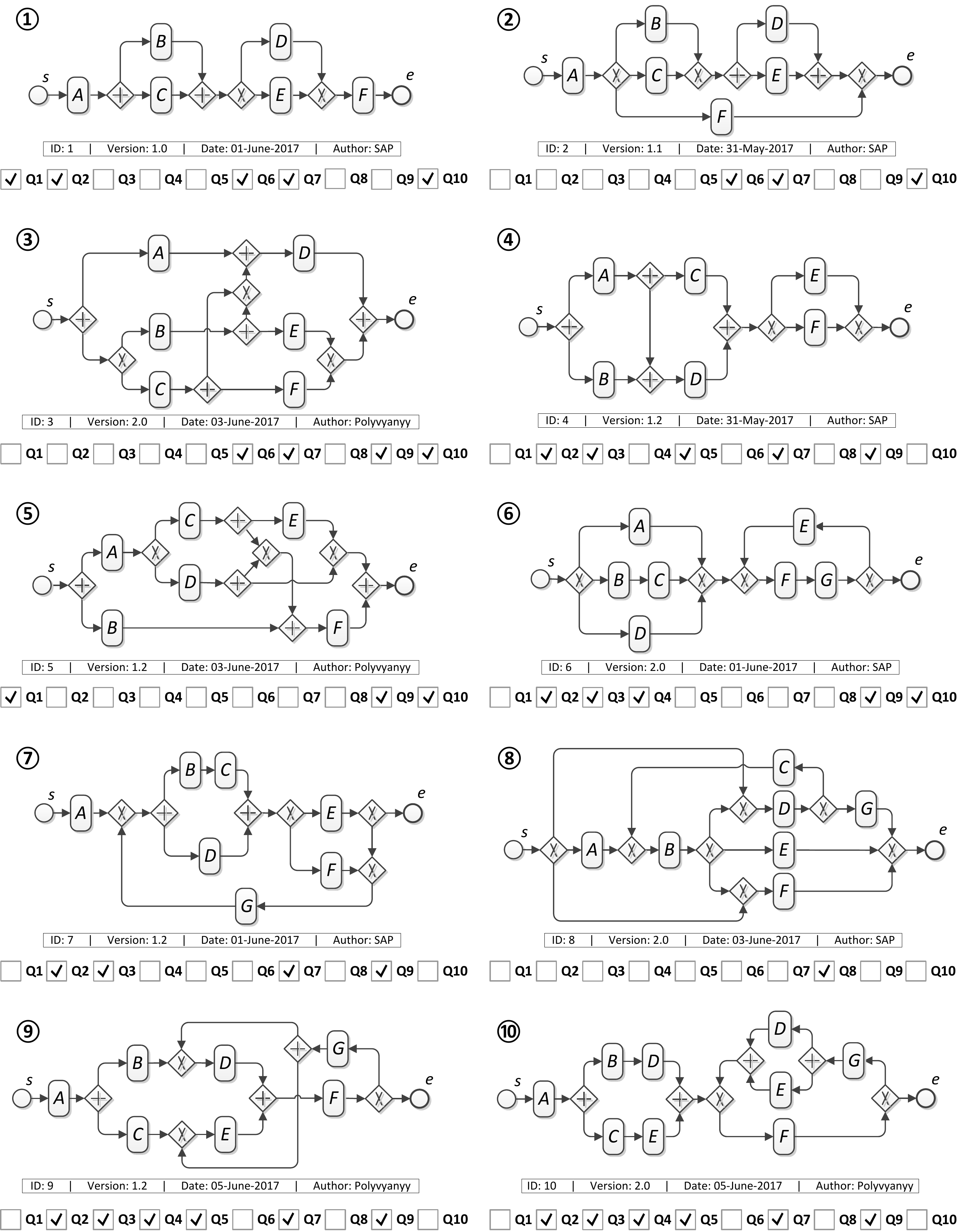}
\caption{\small A repository of ten process models (in BPMN) with their attributes information and results of evaluation of ten sample queries over each of these models.}
\label{fig:smodels}
\end{figure}

We now consider an example process repository consisting of ten process models that are sourced from the collection of the SAP R/3 reference model~\cite{Curran98} as well as from Polyvyanyy's PhD thesis~\cite{Polyvyanyy12}. 
\figurename~\ref{fig:smodels} depicts the ten process models using BPMN. 
In a BPMN model, activities are used to model tasks and are drawn as rectangles. 
Gateways are visualized as diamonds. 
Exclusive gateways use a marker which is shaped like ``$\times$'' inside
the diamond shape, wheres parallel gateways use a marker which is shaped like ``$+$'' inside the diamond shape.
Directed arcs encode control flow dependencies.
For simplicity, the models in \figurename~\ref{fig:smodels} use only abstract task labels (which are alphabet letters).  
In addition, the attributes information of each process model, including its ID, version, date created, and author, is shown in the text annotation under the model. 


The example process model repository discussed in this section can be formalized as the 6-tuple $(S,A,L,\mathit{val},\mathit{loc},\precsim)$, refer to \definitionname~\ref{def:repository}, where 
$S$ is the set of ten Petri net systems $\{s_1, \ldots, s_{10} \}$, $s_i$, $i \in \{1..10\}$, captures the behavior of model $i$ in \figurename~\ref{fig:smodels}; the corresponding net systems can be obtained from BPMN models using the approach from~\cite{DijkmanDO08},
$A:=$ \{ID, Version, Date, Author\} is the set of attribute names,
$L:=\{ /,\,\, /\texttt{Ten-Models-BPMN},\,\, /\texttt{SAP-R3-EPC-Repo} \}$ is the set of locations,
$\mathit{val}:=$ \{($s_1$, ID, 1), ($s_1$, Version, 1.0), ($s_1$, Date, 01-June-2017), $\ldots$\} is the attribute value assignment function,
$\mathit{loc}:= \cup_{i \in \{1..10\}} (s_i,/\texttt{Ten-Models-BPMN})$ is the location assignment function, and 
$\precsim$ is the location map such that $(l,l) \in\,\, \precsim$ for every $l \in L$, and $(/\texttt{Ten-Models-BPMN},/) \in\,\, \precsim$ and $(/\texttt{SAP-R3-EPC-Repo},/) \in\,\, \precsim$.

The process models in \figurename~\ref{fig:smodels} have various structural features and are sound. 
The first five models (models~1 to~5) capture acyclic processes and the other five (models~6 to~10) specify processes that contain cycles. 
Models~1, 2, 6, and~10 are well-structured, where a process model is \emph{well-structured} if and only if every node with multiple outgoing arcs (a split) has a corresponding node with multiple incoming arcs (a join), and vice versa, such that the set of nodes between the split and the join induces a SESE component~\cite{PolyvyanyyVV10}; otherwise the model is said to be \emph{unstructured}~\cite{Polyvyanyy12}. 
Models~3, 4, 5, 7, 8, and~9 in \figurename~\ref{fig:smodels} are unstructured. 
Model 7, 8, and~9 can be mapped to well-structured models. 
For example, model 10 is equivalent to model 9 and is well-structured.
Note that models 3 to 5 are inherently unstructured, \ie the behaviors they describe do not have equivalent well-structured representations (if the concurrency relations between process tasks must be preserved)~\cite{Polyvyanyy12}.
Among these models, model~4 has a typical Z-structure~\cite{KiepuszewskiHB00}, which is formed by the control flow dependencies between tasks $A$, $B$, $C$, and $D$. 

Next, we propose ten sample queries:

\smallskip
\begin{itemize}
	\item[\textit{Q1}.] \texttt{\SELECT~"Author" \FROM~"/Ten-Models-BPMN"}
							\item[]	\texttt{\WHERE~CanOccur("D")~\AND~Conflict("D","E");}\\
	\vspace*{-.75\baselineskip}
	\item[\textit{Q2}.] \texttt{\SELECT~"Version" \FROM~"/Ten-Models-BPMN"}
							\item[]	\texttt{\WHERE~AlwaysOccurs("C")~\OR~Cooccur("B","C");}\\
	\vspace*{-.75\baselineskip}
	\item[\textit{Q3}.] \texttt{\SELECT~"Date" \FROM~"/Ten-Models-BPMN"} 
							\item[]	\texttt{\WHERE~(CanOccur("G")~\AND~(\NOT~Conflict("E","G")))~\OR~}
							\item[] \texttt{\hspace*{3.1em}(TotalConcurrent("C","D")~\AND~AlwaysOccurs("D"));}\\
	\vspace*{-.75\baselineskip}
	\item[\textit{Q4}.] \texttt{\SELECT~"Author","Version" \FROM~"/Ten-Models-BPMN"}
							\item[]	\texttt{\WHERE~CanOccur(\{"F","G"\},\ALL)~\AND~AlwaysOccurs(\{"F","G"\},\ANY);}\\
	\vspace*{-.75\baselineskip}
	\item[\textit{Q5}.] \texttt{\SELECT~"Version","Date" \FROM~"/Ten-Models-BPMN"} 
							\item[]	\texttt{\WHERE~Cooccur("B",\{"C","D"\},\ALL)~\AND~TotalConcurrent("B",\{"C","D"\},\ANY);}\\
	\vspace*{-.75\baselineskip} 
	\item[\textit{Q6}.] \texttt{\SELECT~"Version","Author" \FROM~"/Ten-Models-BPMN"}
							\item[]	\texttt{\WHERE~Conflict(\{"A","B"\},\{"E","F"\},\ANY)~\OR~}
							\item[]	\texttt{\hspace*{3.1em}(Cooccur(\{"A","B"\},\{"E","F"\},\EACH)~\AND~}
							\item[]	\texttt{\hspace*{3.1em}TotalCausal(\{"A","B"\},\{"E","F"\},\ALL));}\\
	\vspace*{-.75\baselineskip}
	\item[\textit{Q7}.] \texttt{\SELECT~"Date","Author" \FROM~"/Ten-Models-BPMN"}
							\item[]	\texttt{\WHERE~"C"~\IN~(GetTasksAlwaysOccurs(\{"C"\})~\UNION~}	
							\item[]	\texttt{\hspace*{3.1em}GetTasksTotalCausal(\{"C"\},\{"B","D"\},\ALL));}\\
	\vspace*{-.75\baselineskip}
	\item[\textit{Q8}.] \texttt{\SELECT~"Date","Version" \FROM~"/Ten-Models-BPMN"}
							\item[]	\texttt{\WHERE~"G"~\IN~(GetTasksCanOccur(\{"G"\})~\INTERSECT~}	
							\item[]	\texttt{\hspace*{3.1em}GetTasksConflict(\{"G"\},\{"D","E","F"\},\ANY));}\\
	\vspace*{-.75\baselineskip}
	\item[\textit{Q9}.] \texttt{\SELECT~"Version","Date","Author" \FROM~"/Ten-Models-BPMN"}
							\item[]	\texttt{\WHERE~GetTasksCooccur(\{"A","B","C"\},\{"D","E"\},\ANY)~\NOT~\EQUALS~}	
							\item[]	\texttt{\hspace*{3.1em}GetTasksTotalConcurrent(\{"A","B","C"\},\{"D","E"\},\ANY);}\\
	\vspace*{-.75\baselineskip}
	\item[\textit{Q10}.]\texttt{\SELECT~* \FROM~"/Ten-Models-BPMN"~\WHERE~(\{"A","B","E","F"\}~\EXCEPT~}
							 \item[]\texttt{\hspace*{3.1em}GetTasksCooccur(\{"A","B","E","F"\},\{"C","D"\},\ALL))~\OVERLAPS~\WITH~}
							 \item[]\texttt{\hspace*{3.1em}GetTasksConflict(\{"A","B","E","F"\},\{"C","D"\},\ANY);}
\end{itemize}
\smallskip

\noindent
The PQL constructs used to specify these queries include the atomic behavioral predicates, the corresponding predicate macros, logical operations (over predicates and predicate macros), construction of task sets using predicates, set operations, and set comparison operations. 
Note that these sample queries (named \textit{Q1}--\textit{Q10}), are instantiations of ten query templates (out of a total of 150 query templates listed in \appendixname~\ref{app:PQL:queries}) to be used for evaluation of PQL in \sectionname~\ref{sec:dataset}. 


We executed the above ten queries on the repository of ten process models shown in \figurename~\ref{fig:smodels}. 
Each query was evaluated over each of these ten process models according to the semantics of PQL (defined in Sections~\ref{subsec:PQL:dynamic:semantics} and~\ref{subsec:PQL:predicates:semantics}). 
%
%
The evaluation result indicates whether a process model satisfies the condition specified in the \WHERE~clause of a specific query. 
In \figurename~\ref{fig:smodels}, this is depicted by a query name being ticked or not ticked under each process model. 
Note that since models~9 and~10 are behaviorally equivalent, every query evaluation over these models yields the same result. 

\begin{table}[t] 
\vspace{-2mm}
\footnotesize
  \centering
  \caption{\small Answers to ten sample queries over the repository of ten process models in \figurename~\ref{fig:smodels}.}
\vspace{-2mm}
		\begin{tabular}{| c | l |}
		\hline
		\textbf{Query} & \textbf{Query result} \\ 
		\hline 
		\hline 
		\textit{Q1} & \{($s_1$,\{(Author, SAP)\}), ($s_5$,\{(Author, Polyvyanyy)\})\} \\ 
		\hline
		\textit{Q2} & \{($s_1$,\{(Version, 1.0)\}), ($s_4$,\{(Version, 1.2)\}), ($s_6$,\{(Version, 2.0)\}), ($s_7$,\{(Version, 1.2)\}), \\ 
								& \ \ ($s_9$,\{(Version, 1.2)\}), ($s_{10}$,\{(Version, 2.0)\})\} \\ 
		\hline 
		\textit{Q3} & \{($s_4$,\{(Date, 31-May-2017)\}), ($s_6$, \{(Date, 01-June-2017)\}), ($s_7$,\{(Date, 01-June-2017)\}), \\
								& \ \ ($s_9$,\{(Date, 05-June-2017)\}), ($s_{10}$,\{(Date, 05-June-2017)\})\} \\ 
		\hline 
		\textit{Q4} & \{($s_6$,\{(Author, SAP), (Version, 2.0)\}), ($s_9$,\{(Author, Polyvyanyy), (Version, 1.2)\}), \\
								& \ \ ($s_{10}$,\{(Author, Polyvyanyy), (Version, 2.0)\})\} \\ 
		\hline 
		\textit{Q5} & \{($s_4$,\{(Version, 1.2), (Date, 31-May-2017)\}), ($s_9$,\{(Version, 1.2), (Date, 05-June-2017)\}), \\
								& \ \ ($s_{10}$,\{(Version, 2.0), (Date, 05-June-2017)\})\} \\ 
		\hline 
		\textit{Q6} & \{($s_1$,\{(Version, 1.0), (Author, SAP)\}), ($s_2$,\{(Version, 1.1), (Author, SAP)\}), \\
								& \ \ ($s_3$,\{(Version, 2.0), (Author, Polyvyanyy)\})\} \\ 
		\hline 
		\textit{Q7} & \{($s_1$,\{(Date, 01-June-2017), (Author, SAP)\}), ($s_2$,\{(Date, 31-May-2017), (Author, SAP)\}), \\
								& \ \ ($s_3$,\{(Date, 03-June-2017), (Author, Polyvyanyy)\}), ($s_4$,\{(Date, 31-May-2017), (Author, SAP)\}), \\ 
								& \ \ ($s_7$,\{(Date, 01-June-2017), (Author, SAP)\}), ($s_9$,\{(Date, 05-June-2017), (Author, Polyvyanyy)\}), \\ 
								& \ \ ($s_{10}$,\{(Date, 05-June-2017), (Author, Polyvyanyy)\})\} \\ 
		\hline 
		\textit{Q8} & \{($s_8$, \{(Date, 03-June-2017), (Version, 2.0)\})\} \\
		\hline 
		\textit{Q9} & \{($s_3$,\{(Version, 2.0), (Date, 03-June-2017), (Author, Polyvyanyy)\}), ($s_4$,\{(Version, 1.2),\\
								& \ \ (Date, 31-May-2017), (Author, SAP)\}), ($s_5$,\{(Version, 1.2), (Date, 03-June-2017), \\
								& \ \ (Author, Polyvyanyy)\}), ($s_6$,\{(Version, 2.0), (Date, 01-June-2017), (Author, SAP)\}), \\
								& \ \ ($s_7$,\{(Version, 1.2), (Date, 01-June-2017), (Author, SAP)\}), ($s_9$,\{(Version, 1.2),\\ 
								& \ \ (Date, 05-June-2017), (Author, Polyvyanyy)\}), ($s_{10}$,\{(Version, 2.0),\\ 
								& \ \ (Date, 05-June-2017), (Author, Polyvyanyy)\})\} \\ 
		\hline 
		\textit{Q10} & \{($s_1$,\{(ID, 1), (Version, 1.0), (Date, 01-June-2017), (Author, SAP)\}), ($s_2$,\{(ID, 2), (Version, 1.1),\\
								& \	\	(Date, 31-May-2017), (Author, SAP)\}), ($s_3$,\{(ID, 3), (Version, 2.0), (Date, 03-June-2017), \\
								& \	\	(Author, Polyvyanyy)\}), ($s_5$,\{(ID, 5), (Version, 1.2), (Date, 03-June-2017), (Author, Polyvyanyy)\}), \\
								& \ \ ($s_6$,\{(ID, 6), (Version, 2.0), (Date, 01-June-2017), (Author, SAP)\})\} \\
		\hline 
\end{tabular}
\label{tab:squeryrslt}
\vspace{-2mm}
\end{table}

\tablename~\ref{tab:squeryrslt} summarizes the results of executing the ten sample queries. 
For each query, only the corresponding systems (and their attributes) that satisfy the \WHERE~condition of the query are retrieved.
For example, Query~\textit{Q1} is designed to retrieve those process models where (i) task labeled~\texttt{"D"} (task~\texttt{D} for short) occurs in at least one process instance (\texttt{CanOccur("D")}) and (ii) there is no process instance in which tasks~\texttt{D} and~\texttt{E} both occur (\texttt{Conflict("D","E")}). 
The first condition holds in all models in \figurename~\ref{fig:smodels}, while the second condition holds in models~1 and~5 only. 
It is important to understand the behavior of a process model when evaluating the second condition. 
For example, in model~8, although tasks~\texttt{D} and~\texttt{E} are on two exclusive branches (subsequent to task~\texttt{B}), both tasks may occur in one process instance since task~\texttt{D} is part of a cycle. 
Another interesting example concerns query~\textit{Q3}, in particular, the condition \texttt{TotalConcurrent("C","D")} in the query.
This condition holds for a process model if and only if for every instance in which tasks~\texttt{C} and~\texttt{D} both occur every occurrence of task~\texttt{C} can be executed at the same time with every occurrence of task~\texttt{D}. 
For example, due to the characteristics of model~3, \ie it is unstructured, it is hard to evaluate this condition without knowledge of the model's behavior.



%% file: tex/implementation.tex
\section{Implementation}
\label{sec:implementation}

The proposed in the previous section querying method has been implemented and is publicly available.\footnote{\url{https://github.com/processquerying/PQL.git}}
The implementation exhibits a well-defined application programming interface (API) to facilitate its integration with other software products.
This API can be accessed by the users via command-line interfaces (CLIs) of two utilities: the PQL bot and the PQL tool.
We refer to them as \emph{PQL Tools} and, conjointly, they implement the \emph{PQL environment}.
The PQL bot is used to prepare models for querying, while the PQL tool is used to execute PQL queries over the indexed models.

The PQL environment is implemented using Java, ANTLR, and MySQL technologies. 
The Java language is chosen due to its ``architecture-neutral and portable'' principle. 
As a result, the environment can be deployed on various platforms running different operating systems.
The PQL tool uses an ANTLR~\cite{Parr2013} generated parser that can build and walk syntax trees of PQL queries. 
Given a context-free grammar expressed using extended Backus-Naur Form~\cite{HopcroftMU2003} as input, ANTLR automatically generates Java code of a parser of the grammar. 
The PQL grammar, which incorporates both the abstract and concrete syntax proposed in \sectionname~\ref{subsec:PQL:abstract:syntax} and \sectionname~\ref{subsec:PQL:concrete:syntax}, respectively, captured using ANTLR notation is listed in \appendixname~\ref{app:PQL:grammar}.
To improve execution times of PQL queries, the PQL tool relies on an index of behavioral relations---a special data structure that improves the computation speed of behavioral relations at the cost of time for its construction and space for its storage. 
The tool uses this index at runtime to avoid having to freshly compute PQL predicates every time a new query is issued. 
The index is stored in a MySQL relational database system.

The two PQL utilities load their configuration parameters from the global initialization file.
These parameters set values to configure database connections, a path to a model checker used by PQL bots, configuration parameters of an information retrieval engine for assessment of label similarities, the default and indexed label similarity thresholds, the maximum number of threads used by the PQL tool when executing queries, the time that PQL bots sleep (\ie stay idle) between two subsequent indexing jobs, and the maximal allowed time to index a single model. 
It is easy to configure the PQL environment to use any third-party tool for scoring label similarities, to fulfill the process querying requirements.
At the moment, one can configure the PQL environment to use one of the three integrated information retrieval engines to be used when scoring label similarities.
These are Apache Lucene\footnote{\url{https://lucene.apache.org/}}, Themis-IR~\cite{Polyvyanyy07}, and an implementation of the label similarity scoring approach based on the Levenshtein distance.
The Apache Lucene and Themis-IR engines are configured to use the vector space model to perform label similarity assessments, while 
the approach based on the Levenshtein distance is implemented based on the principles proposed in~\cite{YujianB07}.

\reportornot{
The PQL bot and the PQL tool are discussed in \sectionname~\ref{subsec:pql:bot} and \sectionname~\ref{subsec:pql:tool}, respectively.
\sectionname~\ref{subsec:apromore} is devoted to the discussion of the integration of the PQL environment with Apromore---an open source process model repository~\cite{RosaRADMDG11}.
}{}

\reportornot{
\subsection{The PQL Bot}
\label{subsec:pql:bot}

}{\mypar{The PQL Bot}}
The PQL bot is a standalone utility that can be used to systematically index models stored by the PQL tool.
Once a model is indexed, it can be matched to a query.
One can start multiple PQL bot instances simultaneously to index multiple models in parallel. 
To construct an index, a PQL bot instance computes all the behavioral predicates over all the PQL tasks of the model and stores them in the database.
A call to the PQL indexing routine takes a workflow system described in the Petri Net Markup Language (PNML) format as input. 
The PNML format is an XML-based syntax for high-level Petri nets, which has been designed as a standard interchange format aimed at enabling Petri net tools to exchange Petri net models~\cite{BillingtonCHKKPPSW03}. 
\reportornot{For many high-level process modeling languages, such as WS-BPEL, EPC, and BPMN, there exist mappings to the Petri net formalism. 
As a result, the PQL environment can work with models developed using a wide range of modeling tools captured using many mainstream notations.
For more information on Petri nets and workflow systems refer to \sectionname~\ref{sec:petri:net:systems}. 
This section also summarizes the technique for translating process models into workflow systems that was used in the reported implementation of the PQL bot; it is the technique for converting the EPC in \figurename~\ref{fig:EPC} into the workflow system in \figurename~\ref{fig:system}.}{}

\reportornot{
When initializing a PQL bot instance one can configure it via CLI. 
Some options of the PQL bot CLI are listed in \tablename~\ref{tab:PQL:bot:CLI:options}.
Every PQL bot instance has a unique name, which can be assigned using option \texttt{-n}.
If this option is not used, a random unique name is assigned.
Once started, a PQL bot instance indexes stored (but not yet indexed) models in succession.
One can use CLI options \texttt{-s} and \texttt{-i} to specify time to sleep, \ie to stay idle, between two successive indexing tasks, and the maximal time to attempt indexing of a model.
If these options are not used, the parameters get configured based on the values in the global configuration file.
If indexing of a model could not be completed within the given time frame, the model is marked as such that can not be indexed using this version of the bot, and the bot proceeds with indexing the next model.
The \texttt{-h} and \texttt{-v} CLI options, respectively, print the help message and get the version of the invoked PQL bot instance.

\begin{table}[htbp]
\footnotesize
  \centering
	\vspace{-2mm}
  \caption{\small CLI options of the PQL bot.}
	\vspace{-2mm}
\begin{tabular}{|l|c|c|l|}
  \hline
  \textbf{Option} & \textbf{Option (short)} & \textbf{Parameter} & \textbf{Description} \\
  \hline
  \hline
  \texttt{--help} & \texttt{-h} & & Print help message \\
  \hline
  \texttt{--index} & \texttt{-i} & \texttt{<number>} & Maximal indexing time (in seconds) \\
  \hline
  \texttt{--name} & \texttt{-n} & \texttt{<string>} & Name of this bot (maximum 36 characters) \\
  \hline
  \texttt{--sleep} & \texttt{-s} & \texttt{<number>} & Time to sleep between indexing jobs (in seconds) \\
  \hline
  \texttt{--version} & \texttt{-v} & & Get version of this bot \\
  \hline
\end{tabular}
\label{tab:PQL:bot:CLI:options}
\vspace{-1mm}
\end{table}

\noindent
Once started, a PQL bot instance run as a daemon, \ie a background process, until it is shut down.
An example of the command line output of a PQL bot instance is listed below.
}{}

\reportornot{
\medskip
{\footnotesize
\noindent
\texttt{>> java -jar PQL.BOT-1.0.jar -n=Brisbane -s=60 -i=86400}\\
\texttt{>> =======================================================================}\\
\texttt{>> Process Query Language (PQL) Bot ver. 1.0}\\
\texttt{>> =======================================================================}\\
\texttt{>> Name:               Brisbane}\\
\texttt{>> Sleep time:         60s}\\
\texttt{>> Max. index time:    86400s}\\
\texttt{>> =======================================================================}\\
\texttt{>> 10:45:18.487 Brisbane - There are no pending jobs}\\
\texttt{>> 10:45:18.487 Brisbane - Sent an alive message}\\
\texttt{>> 10:45:18.497 Brisbane - Going to sleep for 60 seconds}\\
\texttt{>> 10:46:18.505 Brisbane - Woke up}\\
\texttt{>> 10:46:18.525 Brisbane - Retrieved indexing job for the model with ID 1}\\
\texttt{>> 10:46:18.575 Brisbane - Start checking model with ID 1}\\
\texttt{>> 10:46:23.506 Brisbane - Finished checking model with ID 1}\\
\texttt{>> 10:46:23.506 Brisbane - Start indexing model with ID 1}\\
\texttt{>> 10:47:03.608 Brisbane - Finished indexing model with ID 1}\\
\texttt{>> 10:47:03.608 Brisbane - Going to sleep for 60 seconds}\\
\texttt{>> 10:48:03.613 Brisbane - Woke up}\\
\texttt{>> 10:48:03.623 Brisbane - Retrieved indexing job for the model with ID 2}\\
\texttt{>> 10:48:03.673 Brisbane - Start checking model with ID 2}\\
\texttt{>> 10:48:13.248 Brisbane - Finished checking model with ID 2}\\
\texttt{>> 10:48:13.249 Brisbane - Start indexing model with ID 2}\\
\texttt{>> 10:49:52.679 Brisbane - Finished indexing model with ID 2}\\
\texttt{>> 10:49:52.679 Brisbane - Going to sleep for 60 seconds}\\
\texttt{>> 10:50:52.704 Brisbane - Woke up}\\
\texttt{>> 10:50:52.704 Brisbane - There are no pending jobs}\\
\texttt{>> ...}
}

\medskip
\noindent
}{

}
The computation of a PQL predicate reduces to one of these three problems: 
the \emph{reachability problem}~\cite{Hack75}, 
the \emph{covering problem}~\cite{Rackoff78}, or 
the \emph{structural analysis over a complete prefix}~\cite{McMillan92,EsparzaRV02} of the \emph{unfolding}~\cite{NielsenPW81} of the system; refer to \sectionname~\ref{subsec:PQL:predicates:semantics} and~\cite{PolyvyanyyWCRH14}.
The PQL bot uses the solutions to the reachability and covering problems implemented in the LoLA tool version 2.0~\cite{Schmidt00}.\footnote{\url{http://service-technology.org/lola/}} 
LoLA uses state-of-the-art state space reduction techniques to check whether a Petri net system satisfies a given property.
When constructing finite complete prefixes of unfoldings, the PQL bot reuses the implementation of the algorithm from~\cite{EsparzaRV02} available as part of the \texttt{jBPT} initiative~\cite{PolyvyanyyW13}.

\reportornot{}{
\appendixname~\ref{app:implementation} lists some important CLI commands of the PQL bot and shows its sample output.
}

\reportornot{
\subsection{The PQL Tool}
\label{subsec:pql:tool}

}{\mypar{The PQL Tool}}
The PQL tool can be used to store, index, delete, and query process models.
\reportornot{
\tablename~\ref{tab:PQL:tool:CLI:options} lists some CLI options of the PQL tool. 
For example, the PQL tool allows a user to store a given model (option~\texttt{-s}), check if a model can be indexed (option~\texttt{-c}), index a model (option~\texttt{-i}), delete a model and its index (option~\texttt{-d}), visualize the parse tree of a given query (option~\texttt{-p}), execute a query (options~\texttt{-q}), and reset the PQL environment (option~\texttt{-r}). 
The CLI can be used to access help information (option~\texttt{-h}) and the version of the tool (option~\texttt{-v}).

\begin{table}[htbp]
\footnotesize
  \centering
	\vspace{-2mm}
  \caption{\small CLI options of the PQL tool.}
	\vspace{-2mm}
\begin{tabular}{|l|c|c|l|c|}
  \hline
  \textbf{Option} & \textbf{Option (short)} & \textbf{Parameter} &\textbf{Description} & \textbf{Requires option} \\
  \hline
  \hline
  \texttt{--check} & \texttt{-c} & & Check if model can be indexed & \texttt{-id} \\
  \hline  
  \texttt{--delete} & \texttt{-d} & & Delete model (and its index) & \texttt{-id} \\
  \hline
  \texttt{--help} & \texttt{-h} & & Print help message & \\
  \hline
  \texttt{--index} & \texttt{-i} & & Index model & \texttt{-id} \\
  \hline
  \texttt{--identifier} & \texttt{-id} & \texttt{<string>} & Model identifier & \texttt{-id} \\
  \hline
  \texttt{--parse} & \texttt{-p} & & Show PQL query parse tree & \texttt{-pql} \\
  \hline
  \texttt{--pnmlPath} & \texttt{-pnml} & \texttt{<path>} & PNML path &  \\
  \hline
  \texttt{--pqlPath} & \texttt{-pql} & \texttt{<path>} & PQL path &  \\
  \hline
  \texttt{--query} & \texttt{-q} & & Execute PQL query & \texttt{-pql} \\
  \hline
  \texttt{--reset} & \texttt{-r} & & Reset this PQL instance & \\
  \hline
  \texttt{--store} & \texttt{-s} & & Store model & \texttt{-pnml} (\texttt{-id}) \\
  \hline
  \texttt{--version} & \texttt{-v} & & Get version of this tool & \\
  \hline
\end{tabular}
\label{tab:PQL:tool:CLI:options}
\vspace{-1mm}
\end{table}

\noindent
To store models, the CLI option \texttt{-s} must be accompanied by option \texttt{-pnml} that specifies a path either to a single PNML file or to a directory that contains PNML files.
If a path to a PNML file is used, the call to the PQL tool must include option \texttt{-id} to specify a unique identifier to associate with the model.
Otherwise, models are attempted to be stored using their file names as unique identifiers.
Once stored, a model can be indexed by a PQL bot instance or by the PQL tool using the CLI option \texttt{-i} accompanied by option \texttt{-id} that specifies the unique identifier that was used to store the model.
When indexing a model, the PQL tool uses the same routines as the PQL bot.

Note that the dynamic semantics of PQL is implemented over sound workflow systems---a special class of Petri net systems, refer to \sectionname~\ref{subsec:PQL:predicates:semantics} for details.
One can check whether a given Petri net system is a sound workflow system by calling the PQL tool with option \texttt{-c}.
Note that every request to index a model in the PQL environment is automatically preceded by a soundness check of this model.
Alternatively, a user may delete a model using option \texttt{-d}.
By deleting a model, the user also deletes its index.
Both options \texttt{-c} and \texttt{-d} require option \texttt{-id} to uniquely identify a model to be checked and deleted, respectively.
To execute a PQL query, a user can use option \texttt{-q} together with option \texttt{-pql} that specifies a path to a file that contains a PQL query captured using the concrete syntax proposed in \sectionname~\ref{subsec:PQL:concrete:syntax}.
To visualize the parse tree of a PQL query, one can use option \texttt{-p} together with option \texttt{-pql}.
Finally, one can reset the PQL environment using option \texttt{-r}.
By resetting the environment, one deletes all stored models and indexes.

An example command line output of executing a PQL query that is discussed in detail in \sectname~\ref{subsec:PQL:predicates:semantics} using the PQL tool is shown below.

\smallskip
{\footnotesize
\noindent
\texttt{>> java -jar PQL.TOOL-1.0.jar -q -pql=query.pql}\\
\texttt{>> PQL query:  SELECT * FROM * WHERE AlwaysOccurs("process payment"[0.75]);}\\
\texttt{>> Attributes: [UNIVERSE]}\\
\texttt{>> Locations:  [UNIVERSE]}\\
\texttt{>> Task:       "process payment"[0.75] -> ["process payment by cash","process payment by check"]}\\
\texttt{>> Result:     [Fig.4.pnml]}

}
\smallskip
\noindent
}{}
The PQL tool supports multi-threaded querying. 
A user can configure the number of query threads to be used when evaluating queries in the global configuration file. 
As a result of executing a PQL query, the tool returns a collection of matching models and PQL tasks (sets of activity labels) that were used to retrieve the models. 
\reportornot{}{Again, we refer the reader to \appendixname~\ref{app:implementation} for a list of important CLI commands of the PQL tool and for its sample output.
}


\reportornot{
\subsection{Integration with Apromore}
\label{subsec:apromore}

}{\mypar{Integration with Apromore}}
PQL tools have been integrated with Apromore---an open source process model repository~\cite{RosaRADMDG11}, refer to~\cite{PolyvyanyyCCRLF15} for details. 
Apromore can store models developed using many of the commonly used notations. 
It allows users to edit and analyze the stored models.

\begin{figure}[t]
\centering
\vspace{-2mm}
\includegraphics[width = .75\linewidth]{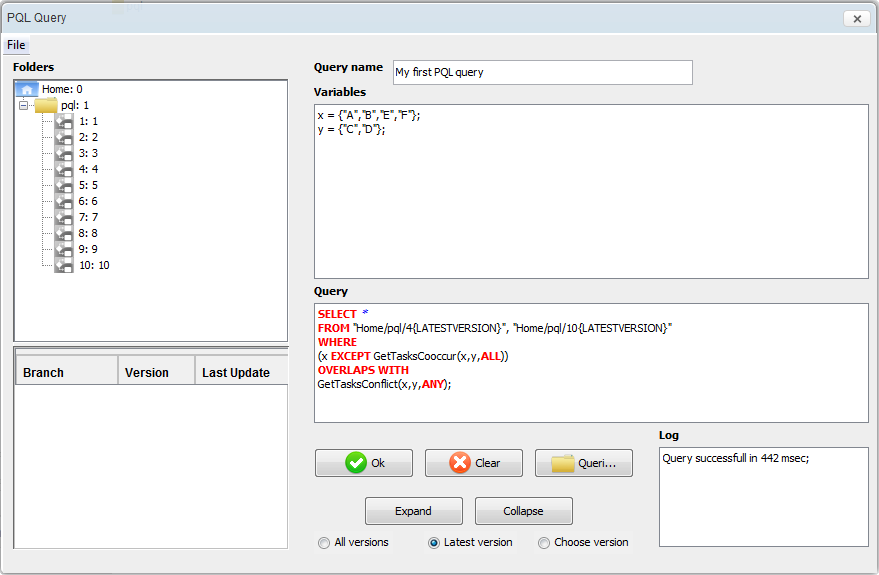}
\vspace{-2mm}
\caption{\small The PQL query editor in Apromore.}
\label{fig:query:editor}
\vspace{-4mm}
\end{figure}

Process models in the Apromore repository get converted to Petri net systems and then indexed by the PQL bots. 
The interface of Apromore has been extended to allow users to specify PQL queries. 
The interface is designed to improve user experience, for example, by suggesting standard PQL keywords and allowing to drag-and-drop process models to be considered as part of the \texttt{\FROM} clause.
A PQL query entered using the Apromore interface is passed to the PQL tool.
Based on the constructed indexes, the PQL tool evaluates the query. 
The result of the query, \ie a set of the unique identifiers of models that match the query, is channeled back to Apromore.
Apromore then displays to the user the list of models (and their attributes) that match the query. 
The user can examine the matching models using the Apromore's model editor.

The reader can take a look at a screencast that demonstrates the use of PQL in Apromore, cf. \url{https://www.youtube.com/watch?v=S_U6frTWd3M}.
\figurename~\ref{fig:query:editor} shows a screenshot of the Apromore's PQL query editor.
The editor specifies a query similar to query \textit{Q10} from \sectionname~\ref{subsec:PQL:sample:queries}.
In contrast to query \textit{Q10}, the query in \figurename~\ref{fig:query:editor} uses two variables \apqltext{x} and \apqltext{y}, and requests to be executed over the latest versions of two models with names \apqltext{4} and \apqltext{10} located in folder \apqltext{Home/pql/}.

\begin{figure}[t]
\centering
\vspace{-2mm}
\includegraphics[width = .75\linewidth]{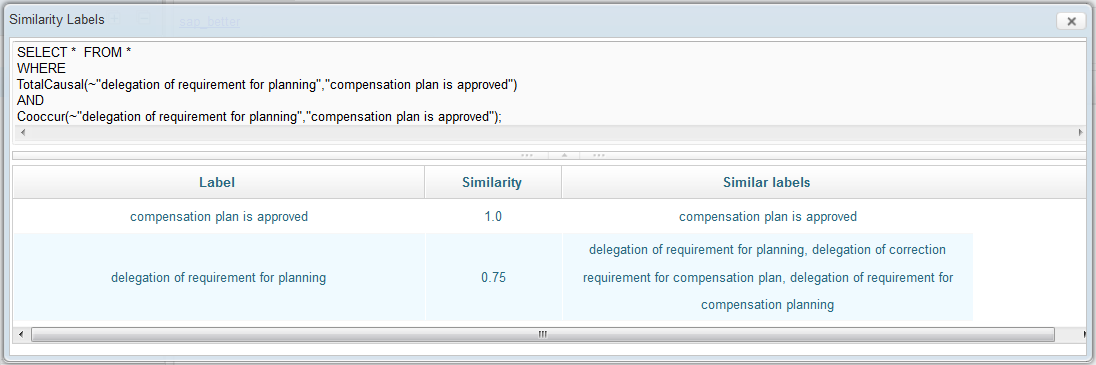}
\caption{\small The PQL query info window, borrowed from~\cite{PolyvyanyyCCRLF15}.}
\vspace{-2mm}
\label{fig:query:info}
\vspace{-2mm}
\end{figure}

\figurename~\ref{fig:query:info} shows another screenshot with information on the execution of yet another PQL query.
Concretely, the information displayed is the explanation of the label similarities used to execute the query.
As label ``compensation plan is approved'' is an instance of the \texttt{ExactTask} construct, according to the semantics of PQL, it resolves into the singleton that contains the label.
Note that label ``delegation of requirement for planning'' is an instance of the \texttt{DefSimTask} construct.
The PQL Tool used to execute this query was configured to use the default label similarity threshold of 0.75.
Based on the configured engine for scoring label similarities, and the threshold, the PQL tool has discovered three similar labels shown in the figure.

%% file: tex/evaluation.tex
\section{Evaluation}
\label{sec:evaluation}

Using the implementation from \sectionname~\ref{sec:implementation}, we conducted a series of experiments to assess the performance of PQL in terms of indexing and querying time. 
The experiments were performed on a computer with 8GB of RAM and 3.4GHz quad-core Intel Core CPU (8 logical processors), running Windows 7 and JVM~1.7. 
The results demonstrate the feasibility and scalability of using PQL in industrial settings. 
\reportornot{}{The detailed discussions of the results are in \appendixname~\ref{app:evaluation}.}
\reportornot{\sectionname~\ref{sec:dataset} proceeds with a presentation of the datasets used in the evaluation.}{}

\subsection{Datasets}
\label{sec:dataset}

\mypar{Process models} 
The study was conducted using 493 industrial and 1,000 synthetic process models.\footnote{The synthetic models are available at: \url{https://www.dropbox.com/s/plrlqe1ewsxwy3b/pnml-1000.zip?dl=0}.}
All the 1,493 models are sound workflow systems\reportornot{; the reader can refer to \sectionname~\ref{sec:preliminaries} for more information on sound workflow systems.}{.}
We obtained the industrial models from the SAP R/3 Reference Model~\cite{Curran98}, which is a collection of 604 EPCs in various domains such as sales, production, and procurement used to customize the SAP R/3 ERP system. 
We converted the EPCs to Petri net systems and then completed them to workflow systems. 
Next, we filtered out the unsound systems, resulting in 493 sound models.
We complemented this collection of real-life models with a collection of synthetic workflow systems, which we generated using the tool described in~\cite{YanDG15}. 
This tool takes a seed process model collection and generates models that share similar structural and label characteristics to the seed. 
We used the 604 EPCs in the SAP R/3 collection as a seed, with a multiplier of 50, to generate 30,200 artificial EPCs. 
We converted these EPCs into workflow systems and filtered out the unsound models, leading to 16,769 sound workflow systems. 
Finally, from these 16,769 systems, we randomly selected 1,000 systems, each with more than 10 nodes. 

\begin{table}[htbp]
\footnotesize
  \centering
	\vspace{-2mm}
  \caption{\small Structural characteristics of the industrial process models.} 
	\vspace{-2mm}
%
%
%
\begin{tabular}{|c|c|c|c|c|c|c|c|c|c|c|c|}
  \hline
   	& \textbf{\#P} & \textbf{\#T} & \textbf{\#F} & \textbf{\#OT} & \textbf{\#XS} & \textbf{\#XJ} & \textbf{\#AS} & \textbf{\#AJ} & \textbf{\#PG} & \textbf{\#B} & \textbf{\#R} \\
  \hline
  \hline
  \textbf{Average} & 17.85 & 15.91 & 35.7 & 8.57 & 0.64 & 0.65 & 1.13 & 1.13 & 5.73 & 1.78 & 0.002 \\ 
  \hline
  \textbf{Minimum} & 4 & 3 & 6 & 2 & 0 & 0 & 0 & 0 & 1 & 0 & 0  \\
  \hline
  \textbf{Maximum} & 74 & 86 & 180 & 35 & 6 & 6 & 6 & 6 & 31 & 9 & 1  \\
  \hline
  \textbf{Std. Dev.} & 12.73 & 11.89 & 27.61 & 6.21 & 0.89 & 0.91 & 1.21 & 1.22 & 4.93 & 1.66 & 0.05  \\
  \hline
\end{tabular}
\label{tab:collection:industrial}
\vspace{-2mm}
\end{table}

\noindent
Tables~\ref{tab:collection:industrial} and~\ref{tab:collection:synthetic} provide statistics on the structural properties of the two model collections.
The tables list basic statistics on the number of places (\#P), transitions (\#T), flow arcs (\#F), observable transitions (\#OT), XOR-splits (\#XS), XOR-joins (\#XJ), AND-splits (\#AS), AND-joins (\#AJ), polygons (\#PG), bonds (\#B), and rigids (\#R) in the models of both collections, which are the metrics for comparing the structural properties of model collections employed in~\cite{YanDG15}\reportornot{. By XOR-join and XOR-split we refer to a place with multiple input transitions and a place with multiple output transitions, respectively.
By AND-join and AND-split we refer to a transition with multiple input places and a transition with multiple output places, respectively.
Polygons, bonds and rigids are different types of SESE components in the WF-tree of a workflow system~\cite{WeidlichPMW11}, where a polygon is a sequence of SESE components in which every two subsequent components share one node, a bond is a collection of SESE components that share entry and exit nodes, and a rigid is an unstructured component~\cite{Polyvyanyy12}.}{; see \appendixname~\ref{app:evaluation} for details.}
The tables demonstrate that industrial and synthetic models have similar structural characteristics.


\begin{table}[htbp]
\footnotesize
  \centering
	\vspace{-2mm}
  \caption{\small Structural characteristics of the synthetic process models.}
	\vspace{-2mm}
%
%

\begin{tabular}{|c|c|c|c|c|c|c|c|c|c|c|c|}
  \hline
   	& \textbf{\#P} & \textbf{\#T} & \textbf{\#F} & \textbf{\#OT} & \textbf{\#XS} & \textbf{\#XJ} & \textbf{\#AS} & \textbf{\#AJ} & \textbf{\#PG} & \textbf{\#B} & \textbf{\#R} \\
  \hline
  \hline
  \textbf{Average} & 16.63 & 13.52 & 32.4 & 10.95 & 0.45 & 0.43 & 1.26 & 1.22 & 6.192 & 1.609 & 0.116 \\ 
  \hline
  \textbf{Minimum} & 5 & 5 & 10 & 3 & 0 & 0 & 0 & 0 & 1 & 0 & 0 \\
  \hline
  \textbf{Maximum} & 86 & 84 & 190 & 67 & 8 & 5 & 7 & 6 & 39 & 8 & 2 \\
  \hline
  \textbf{Std. Dev.} & 11.14 & 9.01 & 23.08 & 7.48 & 0.85 & 0.77 & 1.04 & 0.98 & 4.99 & 1.15 & 0.34 \\
  \hline
\end{tabular}
\label{tab:collection:synthetic}
\vspace{-2mm}
\end{table}

\mypar{Queries}
We designed 150 PQL query templates. 
Each template is a PQL query with placeholders for activity labels. 
During the experiments, the placeholders were instantiated with random labels.
The query templates were developed to exploit the various features of the PQL grammar. 
According to the PQL features they support, these query templates were divided into three categories and further subdivided into groups and subgroups. 
\tablename~\ref{tab:queries:cats} lists all the PQL query categories, groups, and subgroups, 
and provides numbers of query templates accordingly (see column ``\#~Templates''). 
All the PQL query templates used in the experiments are listed in \appendixname~\ref{app:PQL:queries}.



\begin{table}[h]
\vspace{-2mm}
\footnotesize
  \centering
  \caption{\small Categories, Groups, and Subgroups of PQL query templates.}
\vspace{-2mm}
\begin{tabular}{|c|l|c|}
  \hline
  \textbf{Category.Group.Subgroup} & \textbf{Name} & \textbf{\#~Templates} \\
  \hline
  \hline
  1.a & Unary atomic predicates & 2 \\ 
  \hline
  1.b & Binary atomic predicates & 4 \\
  \hline
  \hline
  2.a.1 & Negations of predicates (Same) & 6 \\
  \hline
  2.a.2 & Conjunctions of predicates (Same) & 18 \\
  \hline
  2.a.3 & Disjunctions of predicates (Same) & 18 \\
  \hline
  2.b.1 & Conjunctions of predicates (Mixed) & 3 \\
  \hline
  2.b.2 & Disjunctions of predicates (Mixed) & 3 \\
  \hline
  2.b.3 & Combinations of logical operations & 2 \\
  \hline
  \hline
  3.a.1 & Unary predicate macros & 12 \\
  \hline
  3.a.2 & Binary predicate macros (Task-Set) & 24 \\
  \hline
  3.a.3 & Binary predicate macros (Set-Set) & 36 \\
  \hline
  3.b.1 & Constructions via unary predicates & 2 \\
  \hline
  3.b.2 & Constructions via binary predicates & 8 \\
  \hline
  3.b.3 & Constructions with set operations & 5 \\
  \hline
  3.b.4 & Constructions with set comparisons & 7 \\
  \hline
\end{tabular}
\vspace{-2mm}
\label{tab:queries:cats}
\end{table}

\reportornot{
The first category contains six query templates capturing individual atomic behavioral predicates. These can be divided into two groups, one covering the two unary predicates (1.a), the other covering the four binary predicates (1.b).

The second category is a result of combining atomic predicates via logical operations. 
There are two groups, one for connecting the same predicates (2.a), the other for connecting mixed predicates (2.b). 
The former group can be further divided into three subgroups which capture, respectively, the negation of each predicate (2.a.1), the conjunctions of each predicate twice, three, or four times (2.a.2), and the disjunctions of each predicate twice, three, or four times  (2.a.3). 
The latter group also has three subgroups which capture, respectively, the conjunctions of any two, three, or four different predicates (2.b.1), the disjunctions of any two, three, or four different predicates (2.b.2), and a couple of different combinations of three logical operations between mixed predicates (2.b.3).

The third category captures predicate macros in one group (3.a) and construction of task sets using predicates in the other group (3.b). 
The first group has three subgroups. One applies each of the unary predicate macros over a task set of two, three, or four tasks in conjunction (\ALL) or in disjunction (\ANY) (3.a.1); one applies each of the binary predicate macros between a single task and a task set of two, three, or four tasks in conjunction or in disjunction (3.a.2); and one applies each of the binary predicate macros between two task sets each consisting of two, three, or four tasks (3.a.3). The second group has four subgroups. The first two subgroups apply the set predicate \texttt{TaskInSetOfTasks} to a task set that is constructed using unary behavioral predicates (3.b.1) or using binary predicates (3.b.2). The last two subgroups capture task set constructions using mixed behavioral predicates with set operations (3.b.3) or set comparisons (3.b.4).
}
{\noindent
The first category contains six query templates capturing individual atomic behavioral predicates.
The second category contains 50 query templates that result from combining atomic predicates via logical operations. 
Finally, the third category contains 94 query templates that use predicate macros and construction of task sets using predicates with set operations and comparisons. 
For details on the query templates used in the evaluation refer to \appendixname~\ref{app:evaluation}.}

\subsection{Indexing Performance}
\label{subsec:indexing:performance}
We conducted four experiments to measure the performance of PQL bots and the impact of different factors on indexing time. 
In what follows, for each of these four experiments, we detail its setup and discuss the obtained measurements.

\mypar{Experiment 1.1: Impact of PQL bots on indexing time}
The goal of this experiment was to measure the performance of PQL bots. 
We measured the time of indexing the industrial and synthetic models using different numbers of bots (from 1 to 8).
Each indexing exercise was repeated three times and we recorded the average indexing times of the three runs. 
In all the runs, the bots were configured to only index the label similarity threshold of 1.0. 
The procedure was repeated for different parts of the process model collections, \ie using 25\%, 50\%, 75\% and 100\% of models in each collection. 
Models for each part of each process model collection were selected randomly.

\begin{figure}[t]
\vspace{-2mm}
\begin{center}
   \subfigure[]{\label{fig:index:bots:industrial}\includegraphics[width = .48\textwidth, trim = 9mm 17mm 9mm 17mm]{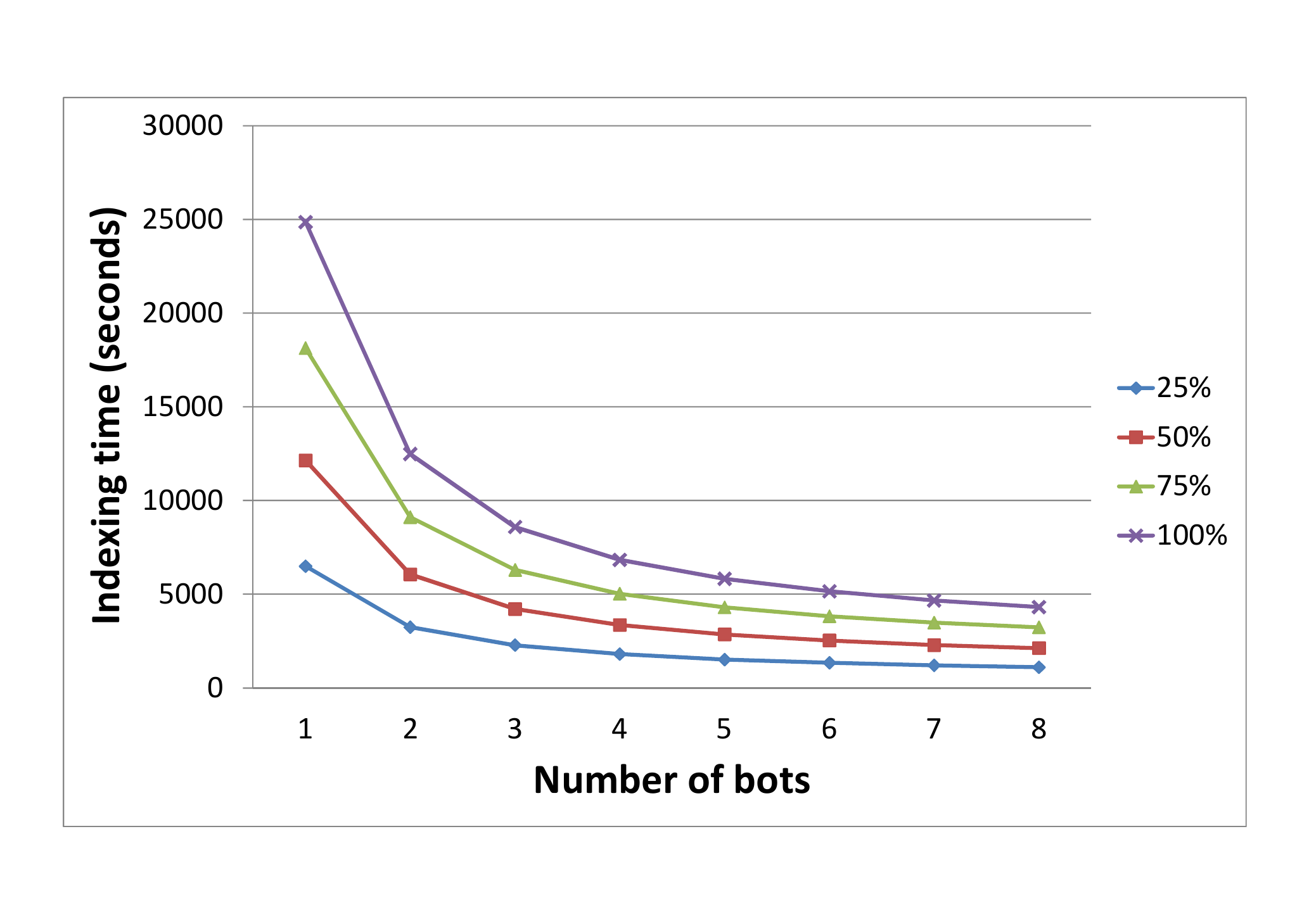}}
   \subfigure[]{\label{fig:index:bots:synthetic}\includegraphics[width = .48\textwidth, trim = 9mm 17mm 9mm 17mm]{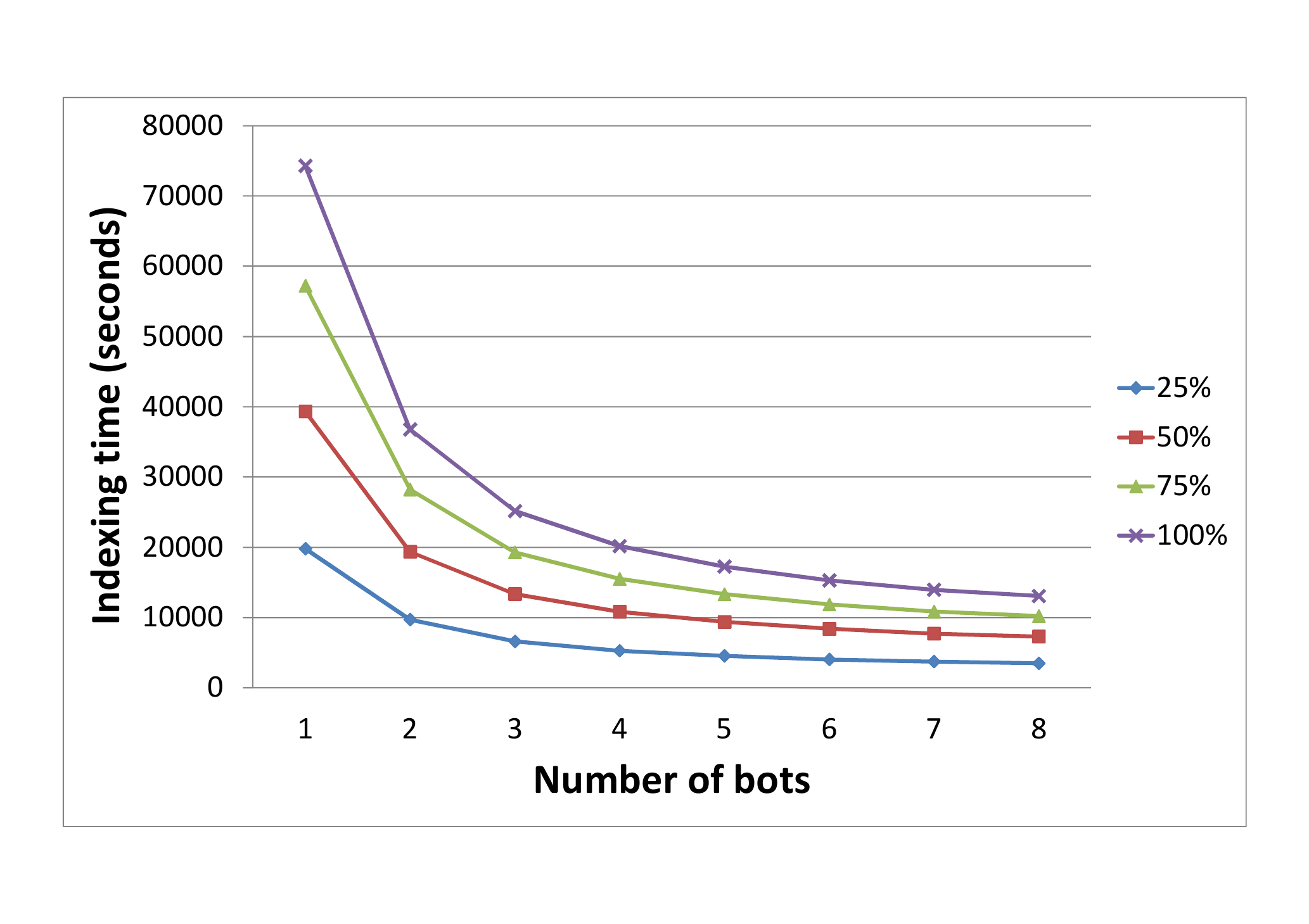}}
\end{center} 
\vspace{-3mm}
\caption{\small Impact of PQL bots on indexing time: (a) industrial models and (b) synthetic models.}
\label{fig:index:bots} 
\vspace{-4mm}
\end{figure}

\figurename~\ref{fig:index:bots} plots the indexing times (in seconds) for different parts of process model collections against different numbers of PQL bots for (a) the industrial models and (b) the synthetic models. 
The two plots demonstrate that adding bots decreases indexing time, though the decrease in indexing time gets less pronounced with the increase of the number of bots. 
\reportornot{Indexing of the whole collection of industrial models with one bot took 6 hours and 54 minutes. 
Two bots managed to index 493 systems in 3 hours and 28 minutes (approximately two times faster than with one bot). 
Note that eight bots spent 1 hour and 12 minutes indexing the same collection (5.8 times faster than with one bot). 
A similar trend can be observed for the synthetic models. 
The relationship between the indexing time and the number of bots is best described by the power function $y = t\times x^k$, where $x$ is the number of bots and $y$ is the indexing time. 
For the industrial process models, the estimated values for constant $t$ are 6,065.6, 11,241, 16,735, and 23,071, for 25\%, 50\%, 75\%, 100\% of models in the collection, respectively.
The estimated values for coefficient $k$ are -0.845, -0.834, -0.826, and -0.839, for 25\%, 50\%, 75\%, 100\% of models in the industrial collection, respectively.
For the synthetic process models, the estimated values for constant $t$ are 17,991, 35,494, 52,086, and 68,077, for 25\%, 50\%, 75\%, 100\% of models in the collection, respectively.
The estimated values for coefficient $k$ are -0.833, -0.806, -0.827, and -0.834, for 25\%, 50\%, 75\%, 100\% of models in the synthetic collection, respectively.
The coefficient of determination $R^2$ ranges from 0.9901 to 0.9936 for the industrial process models and from 0.9834 to 0.9884 for the synthetic process models, indicating that the fitted functions can explain most of the variance in the indexing time. 

This experiment also shows that the indexing time grows linearly with the size of a process model collection. 
Using one PQL bot, 25\% of models in the industrial collection were indexed in 1 hour and 48 minutes, 50\% were indexed in 3 hours and 22 minutes, 75\% in 5 hours and 2 minutes, and the whole collection was indexed in 6 hours and 54 minutes. 
This relationship between the indexing time and the size of the collection is best captured by the linear function $y = 49.549x + 158.65$, where $x$ is the number of models in the collection and $y$ is the indexing time.
The coefficient of determination $R^2$ for the above example is 0.9985.
The coefficients of determination $R^2$ for the fitted linear functions on four data points range from 0.9985 to 0.9997 (for different numbers of bots).
We observed the same trend for the synthetic collection, with $R^2$ values ranging from 0.9949 to 0.9992.
}{This experiment also confirmed that the indexing time grows linearly with the size of a process model collection. 
The detailed discussion of the results of this experiment can be found in \appendixname~\ref{app:evaluation}.}

\mypar{Experiment 1.2: Impact of model size on indexing time}
The goal of this experiment was to assess the impact of size of a model on its indexing time. 
The two model collections were indexed three times (using one bot) and for each model we recorded the average indexing time of the three runs. 
The bot was configured to only index the label similarity threshold of 1.0.

\begin{figure}[t]
\vspace{-2mm}
\begin{center}
   \subfigure[]{\label{fig:index:nodes:industrial}\includegraphics[width = .48\textwidth, trim = 9mm 17mm 9mm 17mm]{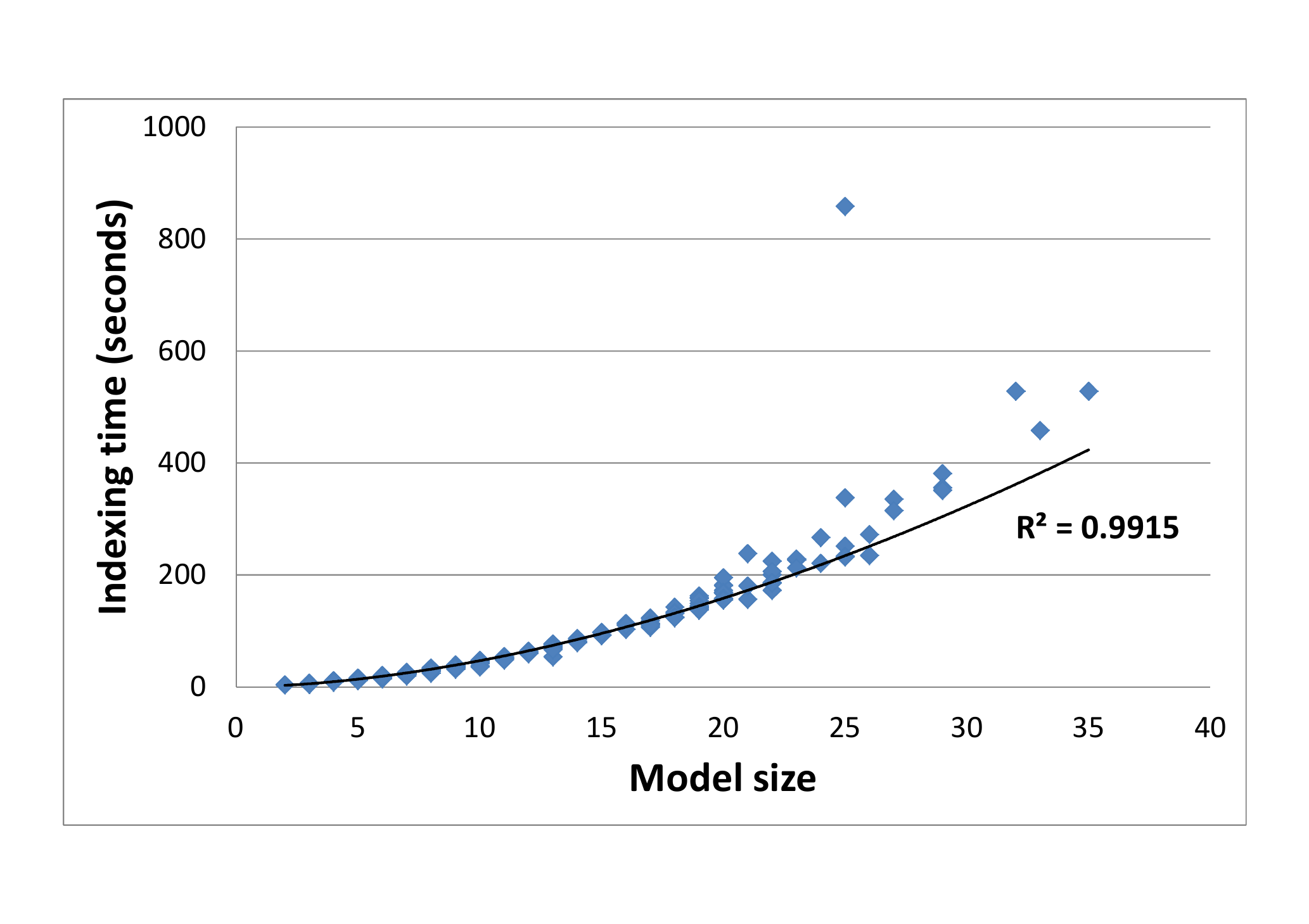}}
   \subfigure[]{\label{fig:index:nodes:synthetic}\includegraphics[width = .48\textwidth, trim = 9mm 17mm 9mm 17mm]{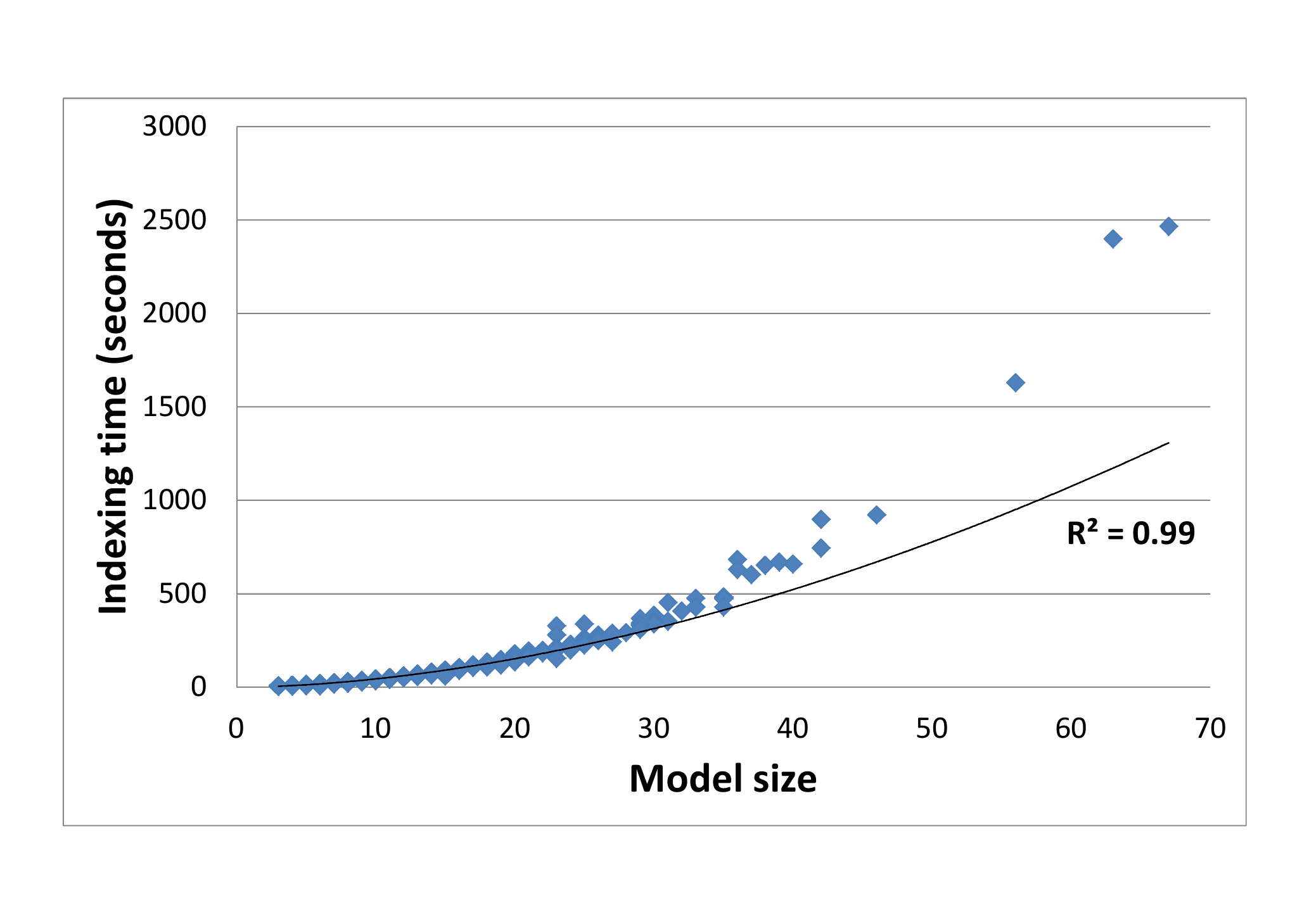}}
\end{center} 
\vspace{-3mm}
\caption{\small Impact of model size on indexing time: (a) industrial models and (b) synthetic models.}
\label{fig:index:nodes} 
\vspace{-4mm}
\end{figure}

\figurename~\ref{fig:index:nodes} plots the indexing times (in seconds) against different sizes of workflow systems for (a) the industrial models and (b) the synthetic models. 
In this experiment, the size of a workflow system is measured as the number of its observable transitions; the reason for this is that bots index behavioral relations over observable transitions. 
The observed average indexing time of a model in the industrial collection is 50.3 seconds, with a minimum of 4.0 seconds (for a model with 2 observable transitions) and a maximum of 858.7 seconds (for a model with 25 observable transitions). 
We have also observed that 95\% of models in the industrial collection can be indexed in less than 200 seconds. 
The average indexing time of a model in the synthetic collection is 74.0 seconds, with a minimum of 6.3 seconds (for a model with 3 observable transitions) and a maximum of 2,465.7 seconds (for a model with 67 observable transitions). 
The obtained measurements report that 95\% of models in the synthetic collection can be indexed in less than 250 seconds. 

\reportornot{
The relation between the indexing time and the model size in the industrial collection is best approximated by the power function $y = 0.8171 \times x^{1.7579}$, which results in a coefficient of determination $R^2$ of 0.9915. 
One obvious outlier workflow system in the industrial collection (a model with 25 observable transitions whose indexing took 858.7 seconds) can be explained by the much bigger size of its state space (2,097,422 reachable states, measured by the LoLA tool) compared to the sizes of state spaces of the other models in the collection (equal or less than 262,156 reachable states). 
For the synthetic models, the relation between the indexing time and the model size is best explained by the power function $y = 0.7423 \times x^{1.7775}$, which results in $R^2$ = 0.99. 

Given a system, either industrial or synthetic, we have noticed that it can be classified as an outlier using the value $f(x,y) = x^2 \times \mathit{log}(y)$, where $x$ is the number of observable transitions of the system and $y$ is the size of the state space of the system in terms of the number of its reachable states. 
The Pearson's correlation coefficient between the indexing times and the values of $f$ is $0.9426$ and $0.9692$ for the industrial and synthetic models, respectively.
The residual analysis has revealed that the four outliers in the industrial collection are among the six models with the highest values of $f$, while the seven outliers in the synthetic collection are among the eight models with the highest values of $f$.
The outliers were identified as models with the standardized residuals beyond $\pm3$ units. 
The four outliers in the industrial collection have 25 (5 minutes and 38 seconds to index), 25 (14 minutes and 19 seconds), 32 (8 minutes and 48 seconds) and 35 (8 minutes and 48 seconds) observable transitions.
Their state spaces comprise 114,717,\, 2,097,422,\, 390,\, and 882 reachable states, respectively.
The seven outliers in the synthetic collection have 36 (11 minutes and 23 seconds to index), 36 (10 minutes and 30 seconds), 42 (14 minutes and 58 seconds), 46 (15 minutes and 23 seconds), 56 (27 minutes and 10 seconds), 63 (39 minutes and 59 seconds), 67 (41 minutes and 6 seconds) observable transitions.
Their state spaces have 303,118,\, 131,108,\, 103,694,\, 1,602,\, 5,730,\, 311,306,\, and 6,818 reachable states, respectively.
}
{
On the employed datasets, the indexing times demonstrated polynomial dependency on the sizes of the indexed models.
For more details, and the analysis of outliers, refer to \appendixname~\ref{app:evaluation}.
}

\mypar{Experiment 1.3: Impact of label similarity threshold on indexing time}
The goal of this experiment was to assess the impact of different label similarity thresholds on the average time required to index a model. 
In this experiment, we varied the label similarity threshold (0.5, 0.6, 0.7, 0.8, 0.9 and 1.0) used to index the models. 
Both process model collections, \ie the industrial and the synthetic, were indexed with one bot three times (for each similarity threshold), and average indexing times for the three runs were recorded. 
The Lucene-VSM label comparison method was used in all the runs, refer to \sectionname~\ref{sec:implementation} for details. 
The measured average indexing times of an industrial model for the label similarity thresholds 0.5, 0.6, 0.7, 0.8, 0.9, and 1.0 are 49.73, 50.3, 50.44, 50.48, 50.3, 50.31 seconds, respectively, with the standard deviation in the range from 80.28 to 81.35 seconds.
The measured average indexing times of a synthetic model for the label similarity thresholds 0.5, 0.6, 0.7, 0.8, 0.9, and 1.0 are 74.53, 74.63, 74.62, 74.57, 74.61, 74.48 seconds, respectively, with the standard deviation in the range from 154.62 to 155.26 seconds.

There is no strong observed relation between the label similarity thresholds and the measured average indexing times, both for industrial and synthetic models.
We conclude that the impact of different label similarity thresholds on indexing times is negligible. 
This can be explained by the small number of unique labels in both collections (2,278 unique labels across the industrial models and 10,473 unique labels across the synthetic models) and the fact that the labels are short character strings (on average 4.84 and 6.08 words in a label of an industrial and a synthetic model, respectively), whereas modern information retrieval engines are known to be efficient on collections that comprise millions of natural language documents of much larger sizes~\cite{DBLPBaeza-YatesR2011}.

\mypar{Experiment 1.4: Impact of index size on indexing time}
The goal of this experiment was to measure the impact of index size on the average time required to index a model. 
To this end, we randomly split the industrial collection into four sets of (approximately) the same size (Sets 1--4). 
Each set was indexed four times: once when the index was empty, and then after 25\%, 50\% and 75\% of the collection was indexed. 
\reportornot{In total, we performed four indexing runs. 
In the first run, Set~1 was indexed first, followed by indexing of Set~2, followed by Set~3, and finally concluded by indexing Set~4.
The second run indexed the models in the order of Set 4, Set 1, Set 2, and finally Set 3.
The third run indexed the models in the order of Set 3, Set 4, Set 1, and finally Set 2.
The fourth run indexed the models in the order of Set 2, Set 3, Set 4, and finally Set 1.
In all the runs, the same label similarity threshold of 1.0 was used. 
All the four indexing runs were accomplished by one PQL bot.

For all the sets, we observed that the average indexing time is not significantly affected by the size of the index. 
For models in Set~1, the recorded average indexing times are 53.63 seconds, 53.65 seconds, 54.09 seconds, and 54.26 seconds in the first, second, third, and the fourth run, respectively.
For models in Set~2, the recorded average indexing times are 53.29 seconds, 53.46 seconds, 53.70 seconds, and 53.80 seconds in the first, second, third, and the fourth run, respectively.
For models in Set~3, the recorded average indexing times are 44.75 seconds, 44.79 seconds, 44.93 seconds, and 45.15 seconds in the first, second, third, and the fourth run, respectively.
Finally, for models in Set~4, the recorded average indexing times are 49.6 seconds, 49.52 seconds, 49.9 seconds, and 49.92 seconds in the first, second, third, and the fourth run, respectively.
Given any set out of the four sets of models, the relation between the index size (used as a starting point to index the given set) and the average indexing time of a model in the set is best approximated by a polynomial function.
The average indexing time for an industrial model in different sets at different positions in the indexing queue ranges from 44.75 to 54.26 seconds, with the difference between the maximum and minimum average indexing time for a given set at different indexing positions always being within 0.65 seconds. 
The measured coefficients of determination are equal to 0.9215, 0.9847, 0.9998, and 0.7366 for Set~1, Set~2, Set~3, and Set~4, respectively.
The experiment demonstrates a general trend of a negligible increase in the indexing time with the growth of the index size. 
This observation can be explained by the fact that the introduced overhead is due to write operations on the PQL index, which can be efficiently handled by modern database management systems. 

In the measurements of the average indexing times for the synthetic collection we observed a similar trend as in the measurements for the industrial collection.
In particular, given any set out of the four sets of models in the synthetic collection (note that each set of synthetic models is composed of 250 random models), the relation between the index size (used as a starting point to index the given set) and the average indexing time of a model in the set is best approximated by a polynomial function.
The average indexing time for a synthetic model in different sets at different positions in the indexing queue ranges from 61.7 to 86.74 seconds, with the difference between the maximum and minimum average indexing time for a given set at different indexing positions always being within 0.75 seconds. 
The measured coefficients of determination are equal to 0.8456, 0.9947, 0.9999, and 0.8545 for Set~1, Set~2, Set~3, and Set~4, respectively.}
{We observed that the average indexing time is not significantly affected by the size of the index.
The experiment revealed a general trend of a negligible increase in the indexing time with the growth of the index size. 
This observation can be explained by the fact that the introduced overhead is due to write operations on the PQL index, which can be efficiently handled by modern database management systems. 
We observed same trends in the measurements of the average indexing times for the synthetic collection.
For details, please refer to \appendixname~\ref{app:evaluation}.
}

\subsection{Querying Performance}
\label{subsec:querying:performance}
We conducted three experiments to assess the performance of executing different types of PQL queries. 
The queries were generated from the templates discussed in \sectionname~\ref{sec:dataset}. 
To conduct the evaluation, for each model collection and for each template we generated three queries by populating them with labels randomly selected from all the labels in the collection.

\mypar{Experiment 2.1: Impact of query threads on querying time}
This experiment measured the impact of the number of query threads and the size of a model collection on the querying time. 
We varied the number of query threads (from 1 to 8) and executed all the generated queries on different parts of the collections (25\%, 50\%, 75\%, and 100\% of models in each collection). 
For each part of each collection, models were randomly selected three times and the average querying times for the three runs were recorded. 
The models were indexed using the label similarity threshold of 1.0.

\begin{figure}[t]
\vspace{-2mm}
\begin{center}
   \subfigure[]{\label{fig:query:netsThreads:sap}\includegraphics[width = .48\textwidth, trim = 9mm 17mm 9mm 17mm]{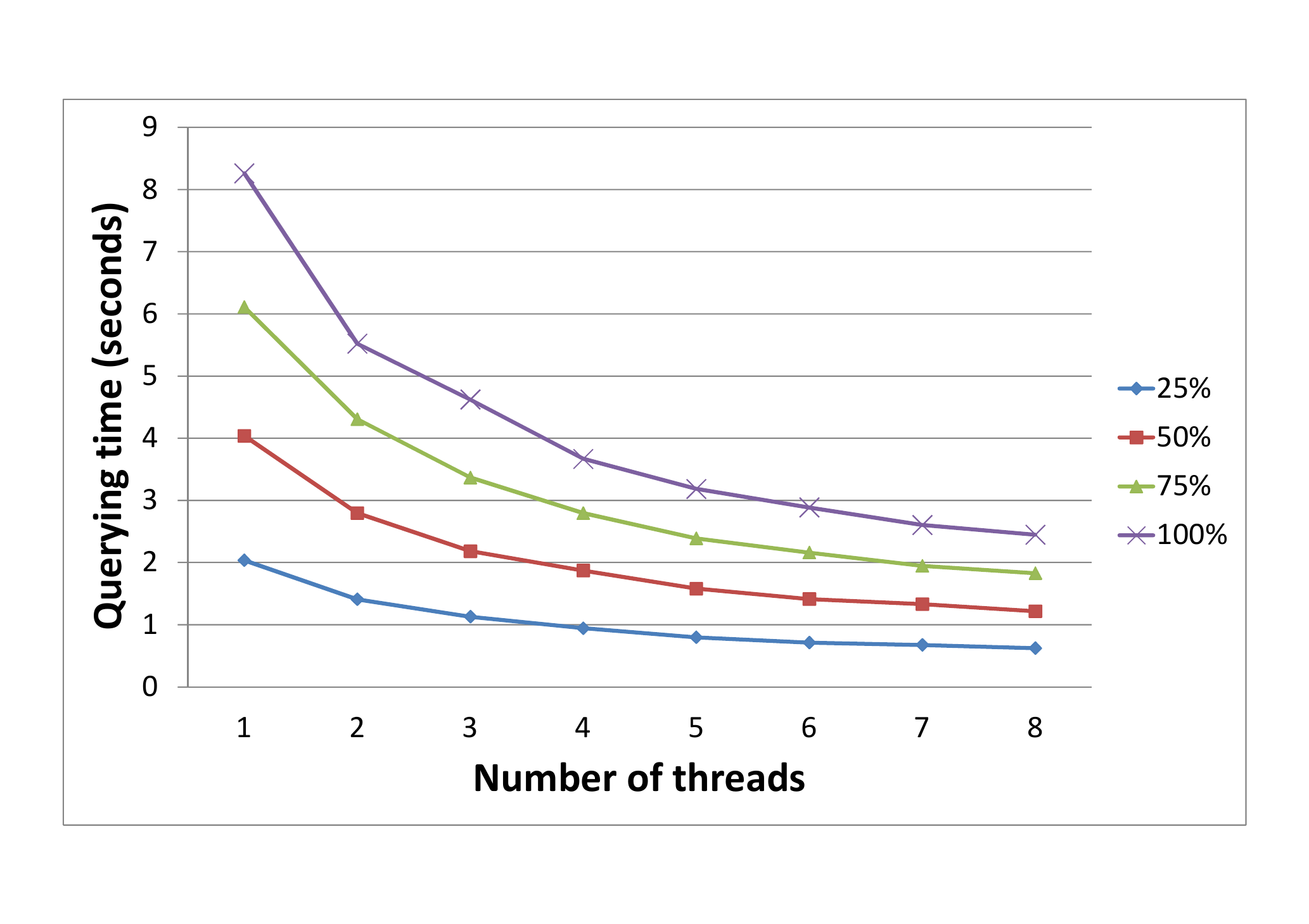}}
   \subfigure[]{\label{fig:query:netsThreads:synt}\includegraphics[width = .48\textwidth, trim = 9mm 17mm 9mm 17mm]{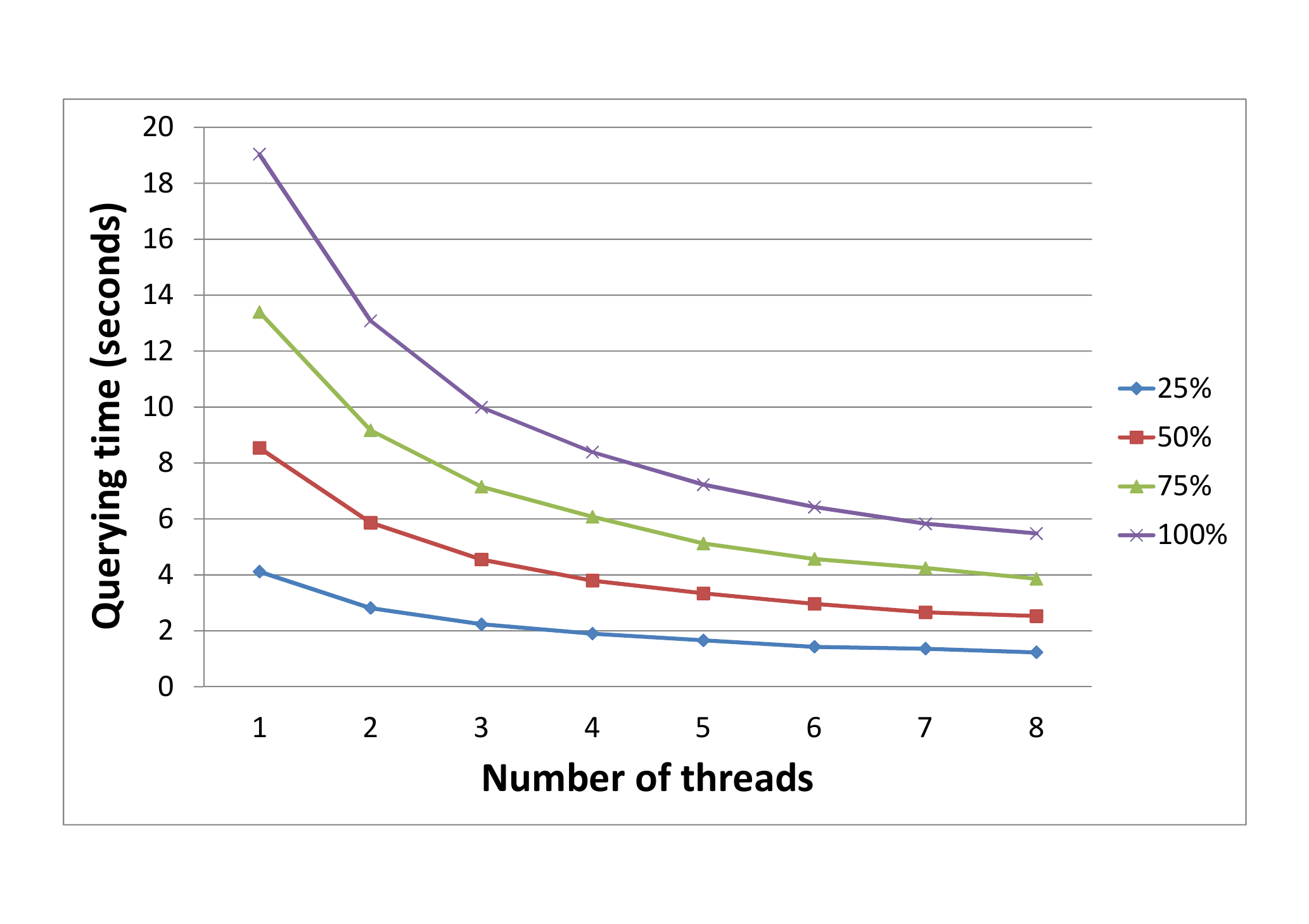}}
\end{center}
\vspace{-3mm}
\caption{\small Impact of query threads on querying time: (a) industrial models and (b) synthetic models.}
\label{fig:query:netThread} 
\vspace{-5mm}
\end{figure}

\figurename~\ref{fig:query:netThread} plots the querying times (in seconds) for different parts of process model collections against different numbers of query threads for (a) the industrial models and (b) the synthetic models. 
It demonstrates that additional query threads decrease querying time. 
As in Experiment 1.1, the gain in performance gets less pronounced with the increase of the number of query threads. 
For example, querying of the industrial models with one thread took on average 8.259 seconds, two threads managed to accomplish queries over 493 systems in 6.109 seconds (1.35 times faster than using one thread), while eight thread used 2.037 seconds to execute a query over the whole collection (four times faster than with one thread). 
A similar trend can be observed for the synthetic models.

\reportornot{
The relation between the querying time and the number of threads is best captured by the power function $y = t\times x^k$, where $y$ is the querying time and $x$ is the number of query threads. 
For the industrial process models, the estimated values for constant $t$ are 2.0793, 4.1112, 6.3066, and 8.3917, for 25\%, 50\%, 75\%, 100\% of models in the collection, respectively.
The estimated values for coefficient $k$ are -0.582, -0.584, -0.595, and -0.594, for 25\%, 50\%, 75\%, 100\% of models in the industrial collection, respectively.
For the synthetic process models, the estimated values for constant $t$ are 4.1811, 8.6902, 13.693, and 19.444, for 25\%, 50\%, 75\%, 100\% of models in the collection, respectively.
While the estimated values for coefficient $k$ are -0.581, -0.597, -0.604, and -0.612, for 25\%, 50\%, 75\%, 100\% of models in the synthetic collection, respectively.
The coefficient of determination $R^2$ ranges from 0.9969 to 0.9982 for the industrial models and from 0.9983 to 0.9989 for the synthetic models, indicating that the fitted models can explain most of the variance in the querying time.
%
%

The measurements obtained in this experiment can be used to show that the querying time grows linearly with the size of a process model collection. 
For example, the measured average querying time with one query thread over 25\% of the industrial models is 2.037 seconds, over 50\% is 4.037 seconds, over 75\% is 6.109 seconds, and over the whole collection is 8.259 seconds.
This trend is best described by the linear relation $y = 0.0168 \times x - 0.0658$, where $x$ is the number of models in the collection and $y$ is the querying time.
The coefficient of determination $R^2$ for the above example is 0.9998.
The coefficients of determination $R^2$ for the fitted linear functions on four data points range from 0.9988 to 1.0 (for different numbers of query threads).
We observed the same trend for the synthetic collection, with $R^2$ values ranging from 0.9967 to 0.9998.
}{
The relation between the querying time and the number of threads is best captured by power functions with negative exponents.
Also, the collected measurements reveal that the querying time grows linearly with the size of a process model collection. 
For details refer to \appendixname~\ref{app:evaluation}.
}

\mypar{Experiment 2.2: Impact of query types on querying time}
\begin{figure}[!thpb]
\vspace{-2mm}
\begin{center}
   \subfigure[]{\label{fig:query:cat12:nets:sap}\includegraphics[width = .47\textwidth, trim = 9mm 17mm 9mm 17mm]{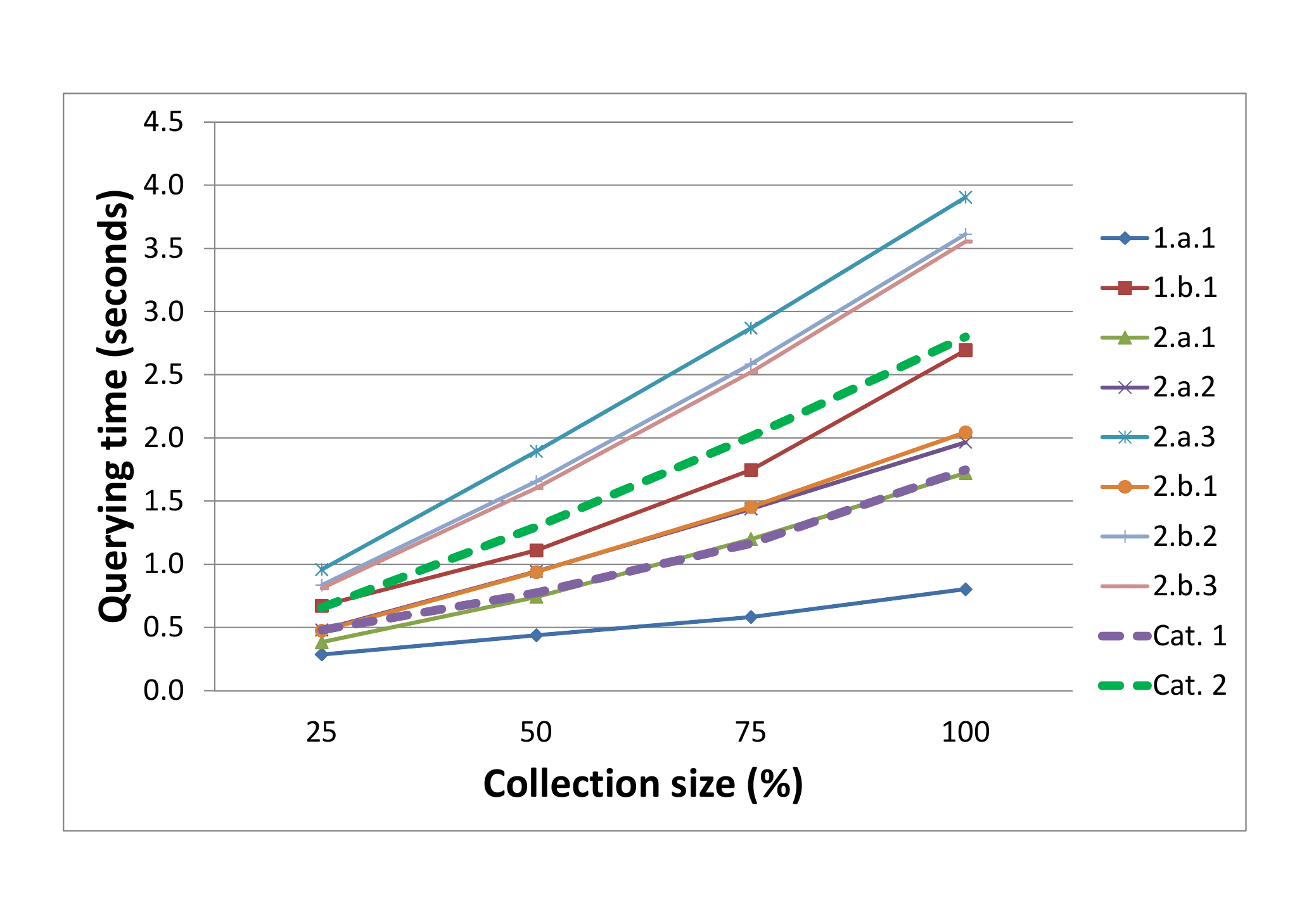}}
   \subfigure[]{\label{fig:query:cat12:threads:sap}\includegraphics[width = .47\textwidth, trim = 9mm 17mm 9mm 17mm]{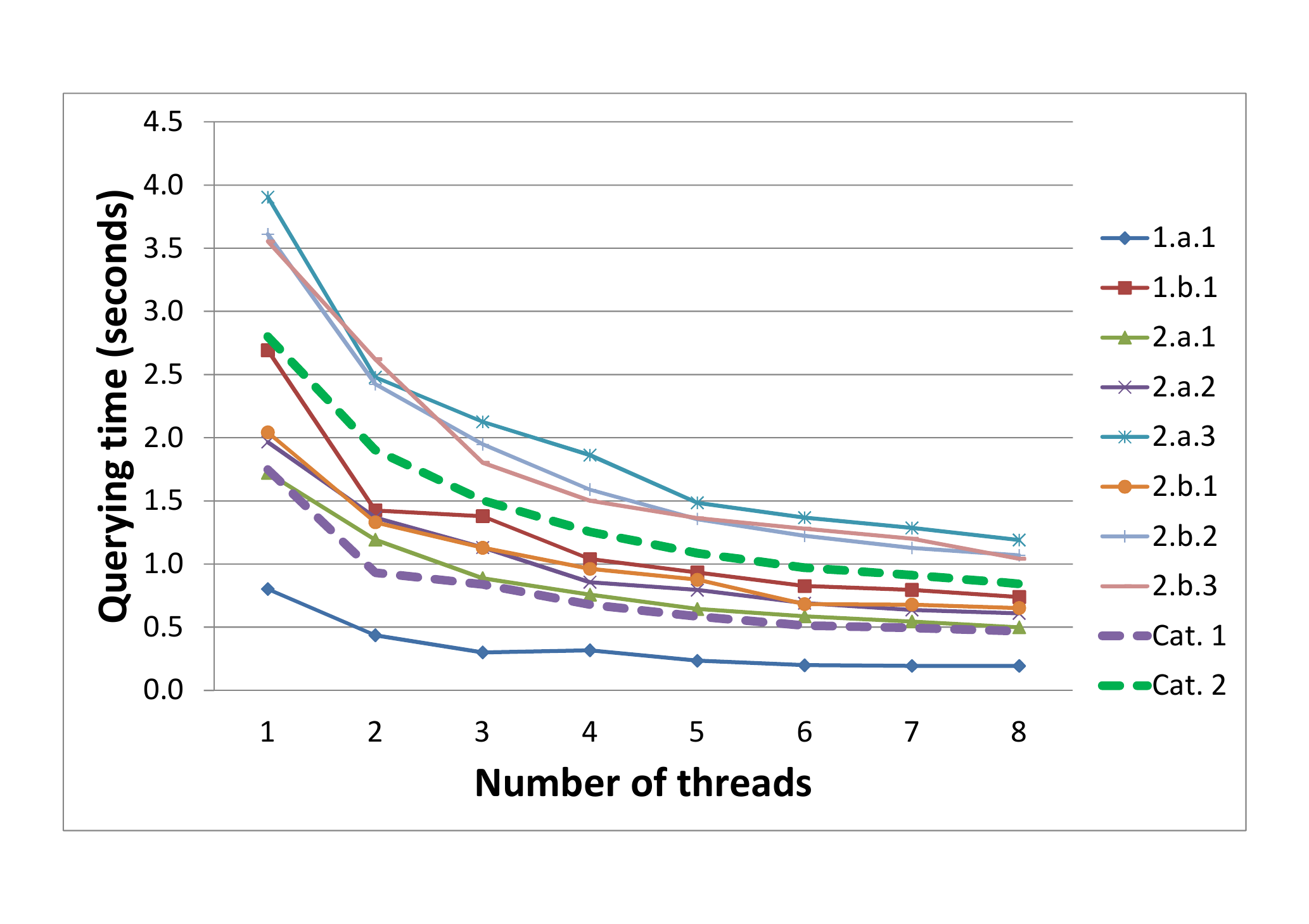}}
   \subfigure[]{\label{fig:query:cat3:nets:sap}\includegraphics[width = .47\textwidth, trim = 9mm 17mm 9mm 17mm]{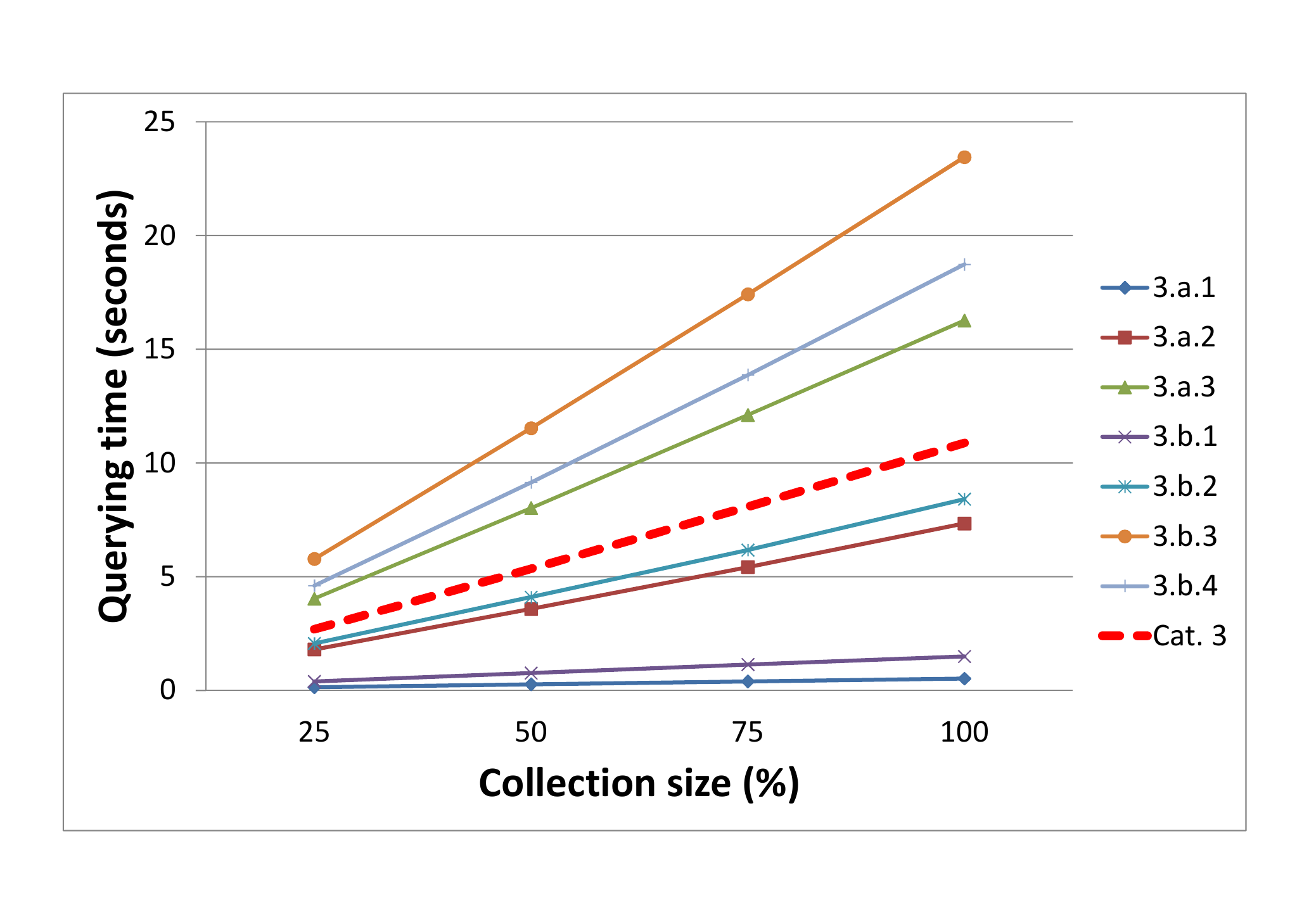}}
   \subfigure[]{\label{fig:query:cat3:threads:sap}\includegraphics[width = .47\textwidth, trim = 9mm 17mm 9mm 17mm]{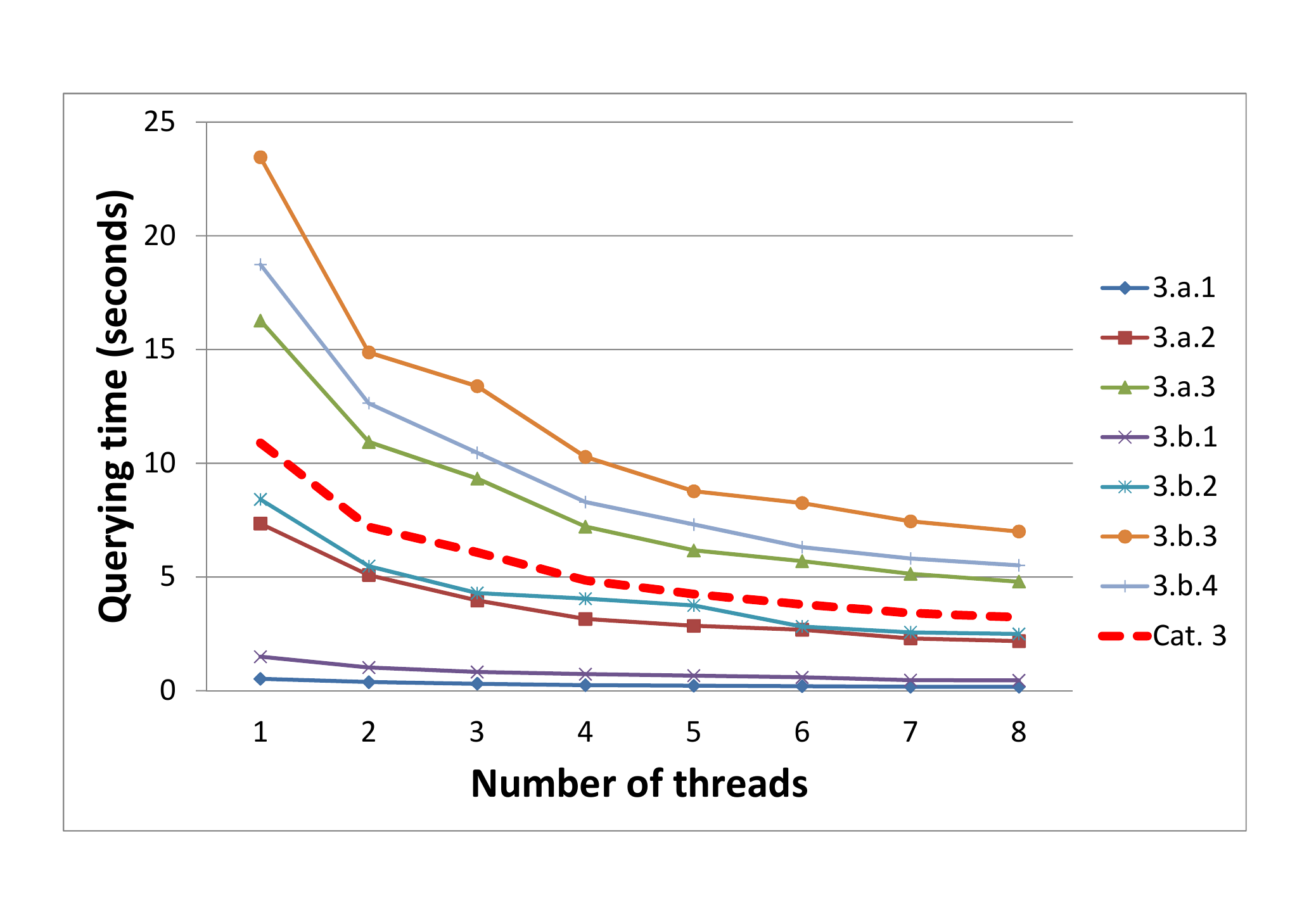}}
   \subfigure[]{\label{fig:query:cat12:nets:synt}\includegraphics[width = .47\textwidth, trim = 9mm 17mm 9mm 17mm]{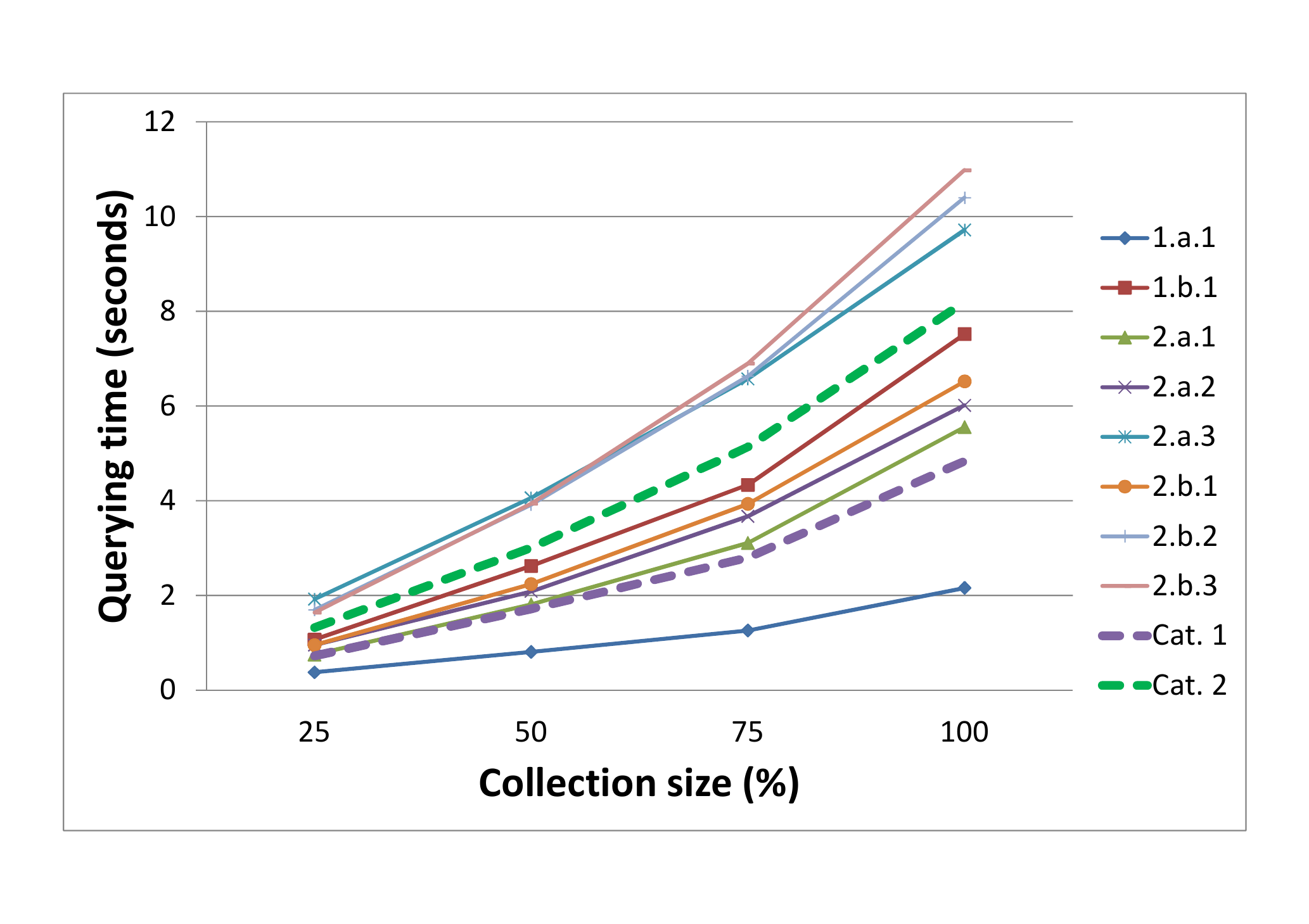}}
   \subfigure[]{\label{fig:query:cat12:threads:synt}\includegraphics[width = .47\textwidth, trim = 9mm 17mm 9mm 17mm]{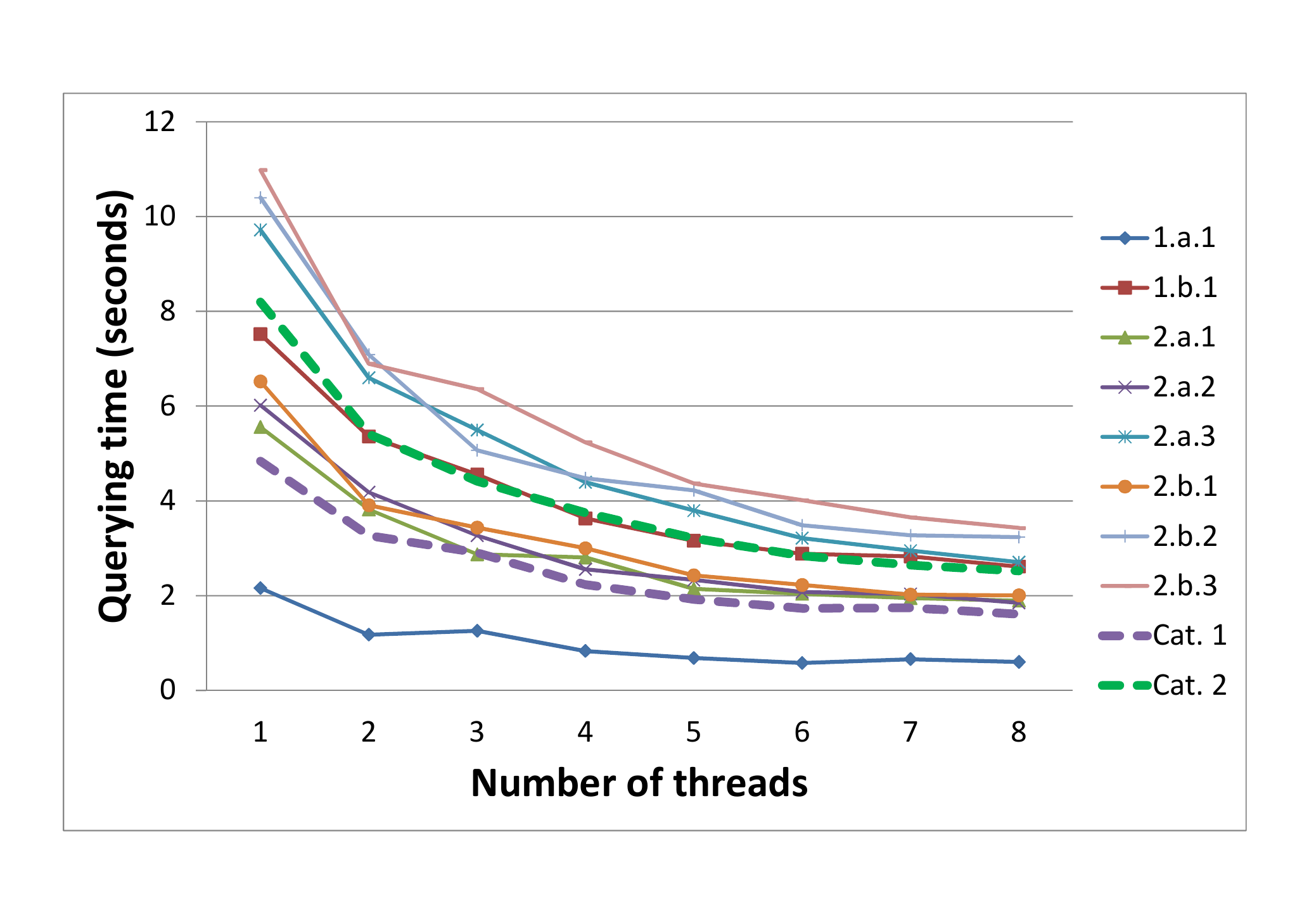}}
   \subfigure[]{\label{fig:query:cat3:nets:synt}\includegraphics[width = .47\textwidth, trim = 9mm 17mm 9mm 17mm]{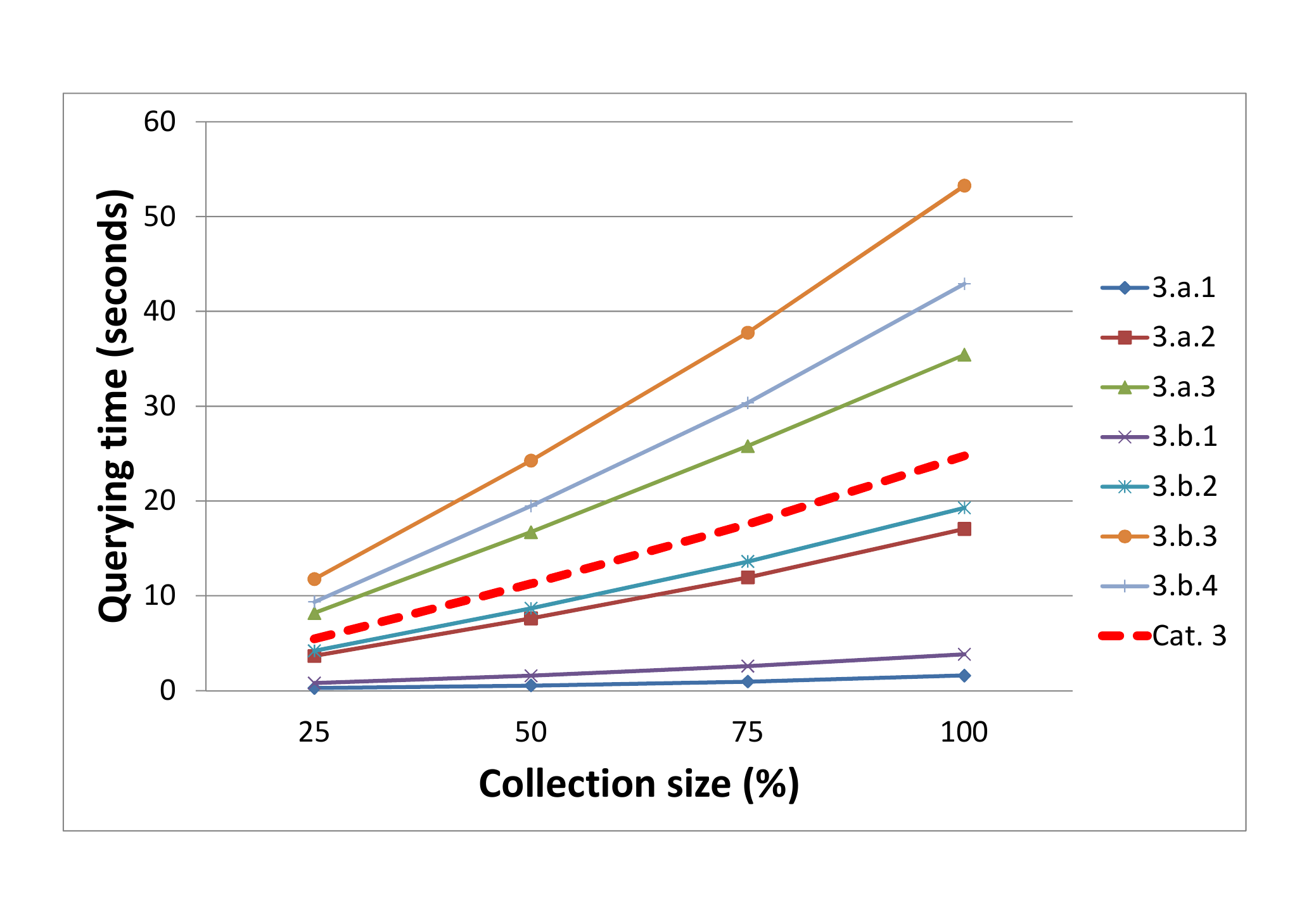}}
   \subfigure[]{\label{fig:query:cat3:threads:synt}\includegraphics[width = .47\textwidth, trim = 9mm 17mm 9mm 17mm]{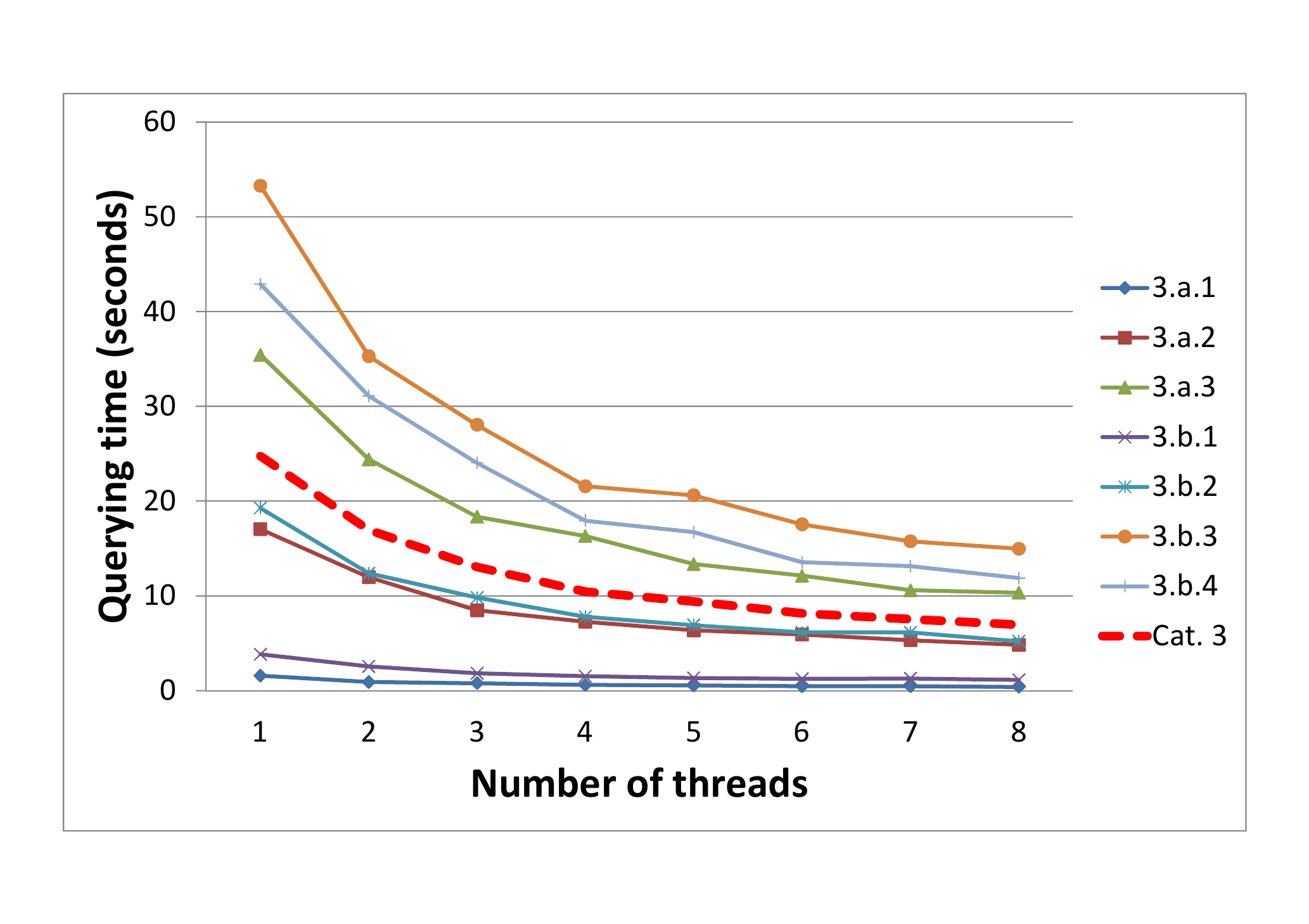}}
\end{center} 
\vspace{-4mm}
\caption{\small 
The average querying times for different model collection sizes (a,c,e,g) and for different numbers of query threads (b,d,f,h); PQL queries in Cats. 1 and 2 for the industrial models (a,b) and for the synthetic models (e,f), and PQL queries in Cat. 3 for the industrial models (c,d) and for the synthetic models (g,h).}
\label{fig:querying:time} 
\vspace{-3mm}
\end{figure}
This experiment aimed to assess the impact of different query types on querying time.
It uses the same setup as Experiment~2.1. 
In this experiment, we measured querying times for different query groups discussed in Section~\ref{sec:dataset}. 

\reportornot{
\figurenames~\ref{fig:query:cat12:nets:sap} and~\ref{fig:query:cat12:threads:sap} show the average querying times for the collection of industrial models for queries in Categories~1 and~2, while \figurenames~\ref{fig:query:cat3:nets:sap} and~\ref{fig:query:cat3:threads:sap} show the average querying times for queries in Category~3. 
The average querying times for the synthetic collection are shown in \figurenames~\ref{fig:query:cat12:nets:synt} and~\ref{fig:query:cat12:threads:synt} (Categories~1 and~2) and \figurenames~\ref{fig:query:cat3:nets:synt} and~\ref{fig:query:cat3:threads:synt} (Category~3). 
Queries in Category~1 only comprise atomic predicates and, thus, are the fastest.
The measured average querying time for the Category~1 queries using one query thread is 1.75 seconds for the 493 models in the industrial collection (approximately 3.5ms per one model-query check) and 4.84 seconds for the 1,000 models in the synthetic collection (approximately 4.8ms per one model-query check).
With eight query threads, the Category~1 queries were on average accomplished in 0.47 seconds for the 493 models in the industrial collection and 1.61 seconds for the 1,000 models in the synthetic collection.
Queries in Category~2 comprise atomic predicates and logical connectives. 
Thus, they require a bit more time to accomplish than queries in Category 1. 
The measured average querying time for Category~2 queries with one thread is 2.8 seconds for the industrial models and 8.2 seconds for the synthetic models.
With eight query threads, it is 0.84 and 2.52 seconds for the industrial and synthetic models, respectively.
Queries in Category~3 comprise macros and, hence, are the lengthiest.
The measured average querying time for Category 3 queries with one query thread is 10.89 seconds for the industrial models (approximately 11ms per one model-query check) and 24.76 seconds for the synthetic models (approximately 25ms per one model-query check). 
With eight query threads, it is 3.23 and 6.96 seconds for the industrial and synthetic models, respectively.

\figurenames~\ref{fig:query:cat12:nets:sap},~\ref{fig:query:cat3:nets:sap},~\ref{fig:query:cat12:nets:synt} and~\ref{fig:query:cat3:nets:synt} demonstrate the linear dependency between the number of models in a collection and querying time for different query types. 
The coefficients of determination $R^2$ for the fitted linear functions for different query subgroups range from 0.9718 to 1.0 for the industrial models and from 0.9592 to 0.9992 for the synthetic models.
Finally, \figurenames~\ref{fig:query:cat12:threads:sap},~\ref{fig:query:cat3:threads:sap},~\ref{fig:query:cat12:threads:synt}, and~\ref{fig:query:cat3:threads:synt} show that for all the query subgroups the relation between the querying time and the number of query threads follows the trend observed in Experiment~2.1. 
The coefficients of determination $R^2$ for the fitted power functions for different query subgroups range from 0.9724 to 0.998 for the industrial models and from 0.9329 to 0.9955 for the synthetic models.
}{
\figurename~\ref{fig:querying:time} shows the average querying times for different model collection sizes and different numbers
of query threads.
\figurenames~\ref{fig:query:cat12:nets:sap},~\ref{fig:query:cat3:nets:sap},~\ref{fig:query:cat12:nets:synt} and~\ref{fig:query:cat3:nets:synt} show the linear dependency between the number of models in a collection and querying time for different query types, while \figurenames~\ref{fig:query:cat12:threads:sap},~\ref{fig:query:cat3:threads:sap},~\ref{fig:query:cat12:threads:synt}, and~\ref{fig:query:cat3:threads:synt} demonstrate a trend that is similar to the one observed in Experiment~2.1. 
In general, the results revealed the feasibility of using PQL in industrial settings.
For exmaple, with eight query threads, the Category~1 queries were on average accomplished in 0.47 seconds for the 493 models in the industrial collection and 1.61 seconds for the 1,000 models in the synthetic collection.
The detailed discussions of the results can be found in \appendixname~\ref{app:evaluation}.
}

\mypar{Experiment 2.3: Impact of label similarity on querying time}
The aim of this experiment was to measure the impact of label similarity on querying time. 
To this end, we repeated Experiment~2.1 with the following modifications: 
(i) the bots were configured to index label similarity thresholds of 0.75 and 1.0, (ii) all the query templates were augmented to include the ``tilde'' symbol immediately before every activity label (thus, similar labels were considered during querying), and (iii) the tool was configured to use a default label similarity threshold of 0.75. 

We compared the measured querying times with the querying times obtained in Experiment~2.1. 
For the industrial process models, the querying times (using one query thread) on average are 10.9\% higher for queries that account for similar labels. 
For the synthetic process models, the querying times (again, using one query thread) on average are 9.47\% higher for queries that account for similar labels. 
This small overhead can be explained by the fact that the PQL tool indexes behavioral relations at the level of PQL tasks and, thus, at runtime, the overhead is only due to the additional time required to retrieve information on the indexed PQL tasks.



%% file: tex/related.tex
\section{Related Work}
\label{sec:related}

Recently, we conducted a systematic literature review of the state of the art methods for querying process repositories~\cite{PolyvyanyyOBA17}.
As part of that study, we designed a framework for developing process querying methods.
The framework is an abstract system in which components can be selectively replaced to result in a new process querying method.
According to this framework, PQL addresses querying of \emph{formal} process models using a query language with a \emph{formal} semantics that implements the \emph{read} querying intent, \ie is designed to retrieve models from repositories.
Two recent surveys of process querying methods~\cite{WangJWW14,PolyvyanyyOBA17} confirm that query languages with these characteristics constitute a gap in the area of process querying. 
In~\cite{Polyvyanyy2018}, we categorized and summarized the existing process querying methods. 
According to that classification, two prominent existing works for behavioral querying of process models that share similar characteristics with PQL are the works on ``Behavior Query Language'' (BQL)~\cite{JinWW11} and ``A Process-model Query Language'' (APQL)~\cite{HofstedeORS0P13}.
Next, we briefly review APQL and BQL, summarize some important results on behavioral predicates, and discuss several related techniques from the areas outside process querying.

\medskip
\noindent
\textbf{Behavioral querying.} 
PQL is inspired by our work on APQL and redesigns APQL in many respects.
PQL differs from APQL in that it is grounded in a different and empirically justified collection of behavioral predicates, extends the abstract syntax of APQL, proposes a concrete syntax, has an implementation that demonstrates its feasibility, and operates on the level of observable behaviors, \ie incorporates mechanisms for interpreting two different tasks as such that have the same meaning for the purpose of evaluating the queries. 
If compared with BQL, PQL is rigorous since both the syntax and semantics of the language are formally defined.
BQL is grounded in three behavioral predicates.
All these predicates belong to the 4C spectrum and were included in our empirical study.
These three predicates are \texttt{Conflict} (called ``exclude'' in~\cite{JinWW11}), \texttt{ExistCausal} (``precede''), and \texttt{ExistConcurrent} (``parallel with''), refer to \sectionname~\ref{sec:predicates} for more details. 
As a result of our empirical evaluation, only one of these three predicates, namely \texttt{Conflict}, was selected to be included in the set of eight core PQL predicates.
As to \texttt{ExistCausal} and \texttt{ExistConcurrent}, the stakeholders have given their preference to ``stronger'' predicates of \texttt{TotalCausal} and \texttt{TotalConcurrent}, which ensure that the respective behavioral relation holds for all (rather than only for some) occurrences of the tasks.
Finally, the fundamental difference between PQL and BQL is that BQL predicates are defined over occurrences of tasks, while PQL predicates are defined over aggregations (derived using quantification) over the occurrences of tasks.


\medskip
\noindent
\textbf{Model Checking.} 
Model checking studies problems that can verify various properties of process models~\cite{ChanABBMNR98,BaierK08}.
A model checking problem is a problem that, given a formal specification of a property, usually captured using some \emph{specification language}, and a process model, answers whether the property holds in the given model.
Often, in order to solve the problem, a model checking technique proceeds by constructing an alternative representation of the model that indicates whether the given property holds or not in the model.
Model checking techniques usually use \emph{temporal logics} as property specification languages, \eg linear temporal logic (LTL) and computational tree logic (CTL).
Model checking techniques can be employed for process querying to retrieve process models that fulfill a given property~\cite{GurfinkelCD03}.

As recently demonstrated by Wolf~\cite{Wolf2018}, computations of most of the 4C behavioral relations can be reduced, via non-trivial transformations that require exponential space, to classical interleaving-based model checking problems. 
Only one problem remains completely unsolved, whereas several problems were solved in the absence of auto-concurrency.
In fact, based on the results reported in~\cite{PolyvyanyyWCRH14,Wolf2018}, we know that all the predicates selected for inclusion into PQL, refer to \sectionname~\ref{subsec:empirical:evaluation:results}, can be reduced to model checking problems.
However, one cannot directly apply the proposed solutions for process querying.
Note that the proposed in~\cite{Wolf2018} LTL and CTL properties for computing the 4C relations are formulated over the transformed models and, hence, do not convey the meanings of the properties to be computed, which makes them unsuitable for the user interpretations.
Also, the performance of the approach proposed in~\cite{Wolf2018} has not been evaluated; note that model checking on infinite-state systems is undecidable and is PSPACE-complete on finite state systems~\cite{EsparzaN94}.
Such an evaluation, and development of new efficient techniques for computing the 4C relations, may contribute to the development of PQL.

\medskip
\noindent
\textbf{Behavioral predicates.} 
In~\cite{Dwyer1999}, the authors report the results of a survey of property specifications (a.k.a. behavioral predicates) captured in \emph{temporal logics}, e.g., LTL or CTL.
The authors collected and classified 555 properties, most of which are specified in LTL, from a wide range of domains, including hardware protocols, communication protocols, avionics, operating systems, and database systems.
Interestingly, the authors conclude that ``even with significant expertise, dealing with the complexity of such a specification [a temporal logic property] can be daunting'' and suggest that often ``complexity is addressed by the definition and use of abstraction''. 
PQL implements such an abstraction. 
There are no translations from PQL predicates to temporal logic properties over concepts of the original process model, see above.
However, it is easy to see that several properties surveyed in~\cite{Dwyer1999} can be expressed as logical expressions over the 4C spectrum relations.
A comprehensive study of such translations is an interesting direction to explore.

To the best of our knowledge, there are no existing works that study the relevance of behavioral predicates for the purpose of querying (business) process models.
Future empirical studies will aim at gaining a better understanding of the suitability of behavioral predicates for process querying, as per the process querying compromise between decidable, efficiently computable, and suitable process querying methods~\cite{PolyvyanyyOBA17}.

\medskip
\noindent
\textbf{Verification of software systems.}
In~\cite{Dwyer2004}, the authors proposed FLAVERS---a finite-state verification technique to analyze whether a concurrent software system satisfies a given user-specified property. 
To perform the verification, FLAVERS constructs an abstract representation of the system.
This abstraction step comes at a price of precision of the analysis results. 
In our work, we suggest using PQL to query repositories of business process models. 
However, one can explore PQL for querying, testing, and verification of software systems.
Unlike FLAVERS, PQL is precise, \ie the result of every PQL query is free from false positive and false negative errors.
FLAVERS and PQL differ in how they interpret models of analyzed models, according to the interleaving and noninterleaving semantics of concurrent systems~\cite{Sassone1996}, respectively.
Thus, PQL can be used to express properties that address potential simultaneous execution of instructions of software systems.

In software engineering, well-established finite-state verification techniques, like the methods based on classical model checking or the FLAVERS technique, are used to inform continuous improvement of processes~\cite{Clarke2008,Osterweil2017}. 
PQL can enrich this repertoire of verification techniques for detecting errors in semantically rich process models with user interpretable and relevant properties.

\medskip
\noindent
\textbf{Declarative process discovery.}
Process mining assists business analysts in dealing with the emerging complexities and uncertainties introduced by business processes, as it aims to discover, monitor and improve processes observed in the real world using the knowledge accumulated in event logs produced by modern information systems~\cite{Aalst2016}.
An event log is a collection of traces, each representing events executed by 
a business process.
The process discovery problem consists of obtaining a good process model that describes the behavior recorded in a given event log.
Declarative process discovery aims to construct such process models as collections of declarative constraints over possible executions of business activities.
These declarative constraints are often expressed as behavioral predicates, similar to those used in PQL.

In~\cite{Ciccio2015}, the authors proposed an algorithm for declarative process discovery, called MINERful.
The algorithm performs the statistical analysis of the input event log, and then uses the derived knowledge to compose the declarative constraints which collectively describe the traces recorded in the event log.
In particular, MINERful discovers Declare constraints~\cite{Aalst2009}, which is a particular repertoire of 20 LTL templates.
In \cite{Westergaard2013}, the authors propose another algorithm capable of discovering Declare constraints from event logs, called UnconstrainedMiner.
To address the problem of LTL semantics over finite traces, in that work, Declare templates are translated to regular expressions.

Next generation declarative process discovery algorithms can target discovery of the 4C spectrum constraints, in general, and PQL predicates, in particular.
This will allow overcoming two limitations of the existing techniques: the inability of encoding noninterleaving semantics~\cite{Sassone1996} and lack of empirical justification of the discovered constraints.

%% file: tex/conclusions.tex
\section{Conclusion}
\label{sec:conclusion}

This paper proposed a query language for retrieving models from process model repositories. 
The language has an \emph{SQL-like syntax} to facilitate ease of use and its basic predicates are \emph{empirically grounded}.
Noteworthy is also the fact that the language targets process \emph{behavior} rather than structural aspects of process models. 
This focus on the behavior seems more natural than the usual way of querying the representations of the behavior.
The language supports label similarity, \ie labels are assessed as to whether they are sufficiently similar (which is determined with respect to a given threshold) to another label. 
Support for this feature is essential as variations on task names can easily emerge when a range of people are involved in the creation of process models. 

The proposed query language has been implemented and its runtime performance evaluated using real-life and synthetic process models. 
The conducted experiments indicate the feasibility of computing the basic PQL predicates and demonstrate that the implemented PQL Tools can execute PQL queries in almost real-time.

There are several limitations of this work that should be acknowledged. 
Some of these limitations naturally give rise to possible avenues for future work. 

First of all, the expressiveness of the proposed query language is constrained by the choice of basic predicates. 
As it turns out though, this is a fundamental limitation that no semantic process model query language grounded in behavioral predicates can escape~\cite{PolyvyanyyADG16}. Expressiveness can be extended, for instance, by allowing the use of example executions in queries, in particular executions that contain wildcards to express search intents like ``find all process models that allow the execution $\sequence{\texttt{a},\texttt{b},*,\texttt{b},\texttt{b},*,\texttt{c},*}$''. 
Through the use of wildcards one can search for those process models in which \emph{any} trace that fits the template can be executed.

Second, the proposed design of PQL, and the design of its subsequent versions, should be justified by new empirical evidence. For example, it is important to study the ratio between the queries that users want to express and PQL can support and to improve the language accordingly. One can conduct empirical studies to assess how simple it is for the users to specify complex behavioral queries, and use the obtained insights to simplify the language. As another example, one can empirically assess the usefulness of the proposed exploratory search principle.

Third, the proposed query language is only concerned with control-flow aspects of process models and does not take the data and resource perspectives into account. 
The potential use of the query language can be widened by allowing queries that refer to aspects other than control-flow. 

Fourth, PQL can be extended with functionality that is auxiliary to behavioral querying, e.g., aggregate functions over retrieved processes.
PQL predicates operate over the global process scope, \ie over processes that represent completed executions of process models.
Similar to some properties surveyed by Dwyer et al.~\cite{Dwyer1999}, PQL predicates can be generalized to operate over other scopes, \eg between given process conditions. However, it is worth mentioning that 80\% of the surveyed in~\cite{Dwyer1999} properties address the global scope.

Fifth, PQL can be extended with manipulation statements (think INSERT, DELETE, UPDATE). 
Through the use of these statements, it should be possible to add, delete, and modify process models. 

Another limitation concerns the requirement that the class of process models that the query language can deal with is the class of sound workflow systems. 
In practice, process models may not always be sound, for example as models may not yet have been completed, and it is worthwhile to explore how the current work could be adapted to allow for a wider class of process models.

Despite the benefits of a query language based on the behavior described in models, there is a drawback that the behavioral aspects can be complex and consequently query results may not always be intuitive to stakeholders. 
To facilitate query understanding, process models can be annotated to explain the reasons for their inclusion in the query results. 

\medskip
\textbf{Acknowledgements}

\smallskip
\noindent
We would like to thank Jan Recker for his input in the design of the experiment on assessing practical relevance of using behavioral predicates for querying process repositories.


%
%


%% file: tex/appendix_interview.tex
\section{Interview Instrument}
\label{app:PQL:interview}

This appendix contains the instrument used to conduct the empirical study reported in \sectname~\ref{sec:behavioral:relations}.




\begin{center}
\vspace{15mm}
\includegraphics[page=1,scale=.7,angle=90,trim=25mm 25mm 25mm 25mm]{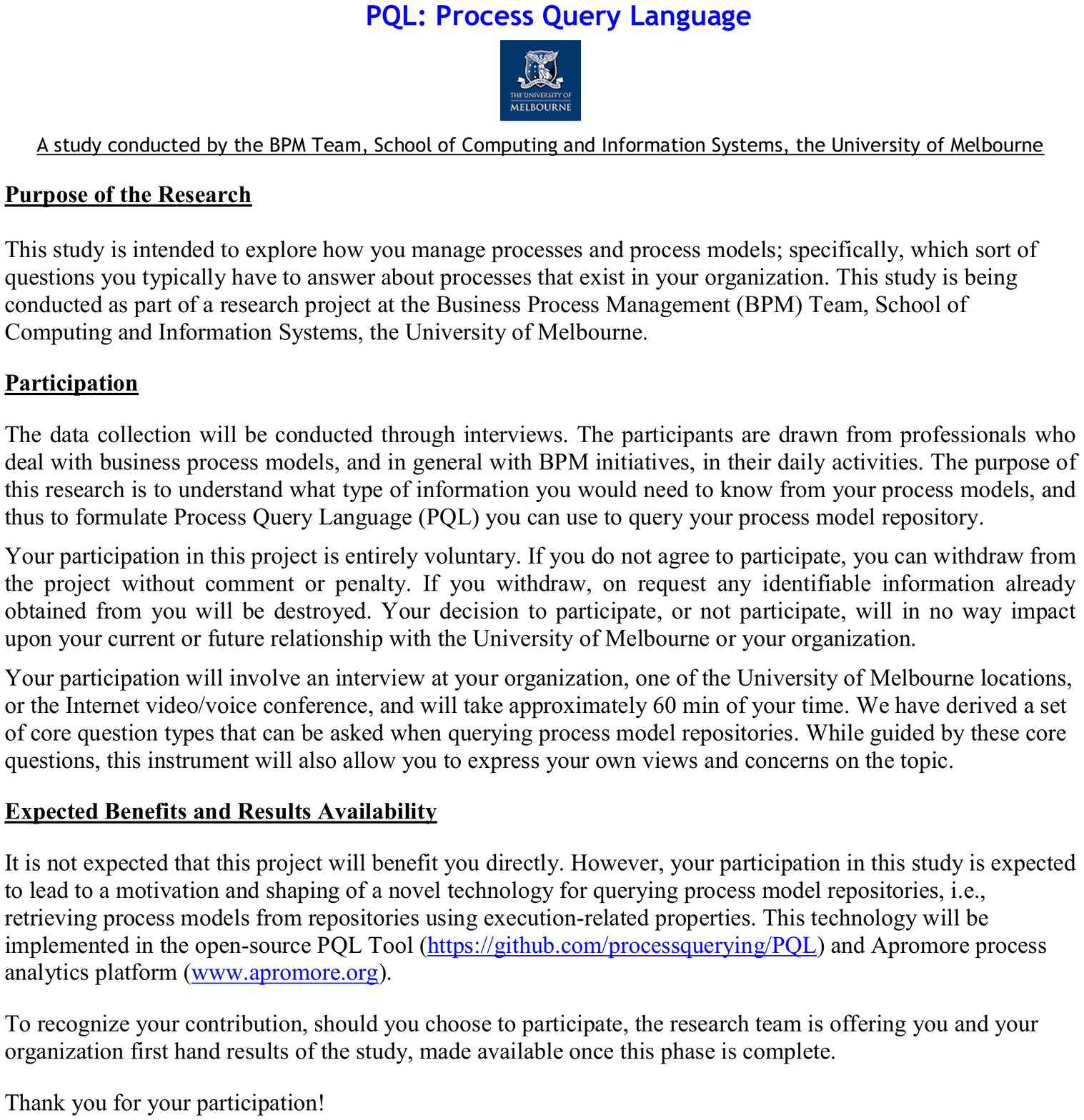}
\end{center}

\newpage
\begin{center}
\vspace{10mm}
\includegraphics[page=2,scale=.7,angle=90,trim=25mm 25mm 0mm 25mm]{./interview/PQL_FINAL_UoM}
\end{center}

\newpage
\begin{center}
\vspace{10mm}
\includegraphics[page=3,scale=.7,angle=90,trim=25mm 65mm 0mm 25mm]{./interview/PQL_FINAL_UoM}
\end{center}

\newpage
\begin{center}
\vspace{10mm}
\includegraphics[page=4,scale=.7,angle=90,trim=25mm 65mm 0mm 25mm]{./interview/PQL_FINAL_UoM}
\end{center}

\newpage
\begin{center}
\vspace{10mm}
\includegraphics[page=5,scale=.7,angle=90,trim=25mm 25mm 0mm 25mm]{./interview/PQL_FINAL_UoM}
\end{center}

\newpage
\begin{center}
\vspace{10mm}
\includegraphics[page=6,scale=.7,angle=90,trim=25mm 25mm 0mm 25mm]{./interview/PQL_FINAL_UoM}
\end{center}

\newpage
\begin{center}
\vspace{10mm}
\includegraphics[page=7,scale=.7,angle=90,trim=25mm 25mm 0mm 25mm]{./interview/PQL_FINAL_UoM}
\end{center}


%% file: tex/appendix_grammar.tex
\newpage
\section{PQL Grammar}
\label{app:PQL:grammar}

This appendix specifies the complete grammar of the PQL language captured in the ANTLR notation.
ANTLR (ANother Tool for Language Recognition) is a parser generator for reading and translating structured text or binary files~\cite{ParrQ95,Parr2013}. 
ANTLR can take a grammar of a language as input and generate source code for a parser that can build and walk syntax trees~\cite{Meyer1990}. 
The language must be specified using a context-free grammar which is expressed using extended Backus-Naur Form~\cite{HopcroftMU2003}.


\newenvironment{code}
{\par\runninglinenumbers
\modulolinenumbers[1]
\linenumbersep=-.7em
\footnotesize
\def\linenumberfont
{\normalfont\scriptsize\textit}}

\begin{multicols}{2}
\begin{code}
\leftskip=4em
\begin{verbatim}
   // PQL version 1.0 grammar for ANTLR v4   
   // [The "BSD licence"]
   // Copyright (c) 2014-2016 Artem Polyvyanyy
   // All rights reserved.

   grammar PQL;
                
   query      : variables 
              SELECT attributes 
              FROM locations 
              (WHERE predicate)? EOS ;

   variables  : variable* ;
   variable   : varName ASSIGN 
                setOfTasks EOS ;
   
   varName    : VARIABLE_NAME ;

   attributes : attribute (SEP attribute)* ;
   attribute  : universe
              | attributeName ;
           
   locations  : location (SEP location)* ;
   location   : universe
              | locationPath ;
              
   universe           : UNIVERSE ;
   attributeName      : STRING ;
   locationPath       : STRING ;

   setOfTasks : tasks
              | union
              | intersection
              | difference ;

   tasks      : varName  
              | setOfAllTasks
              | setOfTasksLiteral
              | setOfTasksConstruction
              | setOfTasksParentheses ;

   setOfAllTasks      :
              GET_TASKS LP RP;
              
   setOfTasksLiteral  : 
              LB (task (SEP task)*)? RB ;
   
   task       : approximate label 
              | label (LSB similarity RSB)? ; 
   
   approximate: TILDE ;
   label      : STRING ;
   similarity : SIMILARITY ;
   
   setOfTasksConstruction      :
                unaryPredicateConstruction
              | binaryPredicateConstruction ;

   unaryPredicateConstruction  :
                (GET_TASKS)unaryPredicateName 
                LP setOfTasks RP ;
   
   binaryPredicateConstruction :
                (GET_TASKS)binaryPredicateName 
                LP setOfTasks SEP setOfTasks 
                SEP anyAll RP ;

   anyAll     : ANY | ALL ;
   
   unaryPredicateName : CAN_OCCUR
              | ALWAYS_OCCURS;

   binaryPredicateName: CAN_CONFLICT
              | CAN_COOCCUR 
              | CONFLICT
              | COOCCUR
              | TOTAL_CAUSAL
              | TOTAL_CONCUR ;

   predicate  : proposition
              | conjunction
              | disjunction
              | logicalTest ;

   proposition: unaryPredicate
              | binaryPredicate
              | unaryPredicateMacro
              | binaryPredicateMacro
              | setPredicate
              | truthValue
              | parentheses
              | negation ;

   unaryPredicate      : unaryPredicateName 
                LP task RP ;

   binaryPredicate     : binaryPredicateName
                LP task SEP task RP ;  			
              
   unaryPredicateMacro : unaryPredicateName
                LP setOfTasks SEP anyAll RP ;
		   
   binaryPredicateMacro: 
                binaryPredicateMacroTaskSet
              | binaryPredicateMacroSetSet ;    

   binaryPredicateMacroTaskSet :
                binaryPredicateName LP task 
                SEP setOfTasks SEP anyAll RP ;
		   
   binaryPredicateMacroSetSet  : 
              binaryPredicateName 
              LP setOfTasks SEP setOfTasks 
              SEP anyEachAll RP ;

   anySomeEachAll  : ANY | SOME | EACH | ALL ;
    
   setPredicate: taskInSetOfTasks
              | setComparison ;
                     
   taskInSetOfTasks : task IN setOfTasks ;

   setComparison    : setOfTasks 
              setComparisonOperator 
              setOfTasks ;

   setComparisonOperator : identical
              | different
              | overlapsWith
              | subsetOf
              | properSubsetOf ;              

   truthValue : TRUE
              | FALSE ;

   logicalTest: isTrue
              | isNotTrue
              | isFalse
              | isNotFalse ;
              
   union      : (tasks | difference | 
                intersection) UNION (tasks | 
                difference | intersection)
                (UNION (tasks | difference 
                | intersection))* ;

   intersection : (tasks | difference) 
                  INTERSECTION 
                  (tasks | difference)
                  (INTERSECTION (tasks 
                  | difference))* ;
   
   difference   : tasks DIFFERENCE tasks
                | tasks DIFFERENCE 
                  difference ; 
 
   negation     : NOT proposition ;

   isTrue       : proposition IS TRUE ;
   isNotTrue    : proposition IS NOT TRUE ;
   isFalse      : proposition IS FALSE ;
   isNotFalse   : proposition IS NOT FALSE ;

   disjunction  : (proposition | logicalTest | 
            conjunction) OR (proposition | 
            logicalTest | conjunction) (OR 
            (proposition | logicalTest 
            | conjunction))* ;
		   
   conjunction  : (proposition | logicalTest) 
            AND (proposition | logicalTest)
            (AND (proposition 
            | logicalTest))* ;

   parentheses  : LP proposition RP 
           | LP conjunction RP 
           | LP disjunction RP 
           | LP logicalTest RP ;

   setOfTasksParentheses : LP varName RP
           | LP universe RP
           | LP setOfTasksLiteral RP
           | LP setOfTasksConstruction RP
           | LP union RP
           | LP difference RP
           | LP intersection RP
           | LP setOfTasksParentheses RP ;

   UNIVERSE     : '*' ;
   
   STRING       : DQ ( ESC_SEQ 
                | ~('\\'|'"') )* DQ ;
   VARIABLE_NAME: ('a'..'z'|'_') 
                  ('a'..'z'|'0'..'9'|'_')*;
   SIMILARITY   : '1' | '0' ('.' '0'..'9'+)? 
                | '.' '0'..'9'+ ;

   LP          : '(' ;
   RP          : ')' ;
   LB          : '{' ;
   RB          : '}' ;
   LSB         : '[' ;
   RSB         : ']' ;
   DQ          : '"' ;
   EOS         : ';' ; 
   SEP         : ',' ;
   ASSIGN      : '=' ;
   TILDE       : '~' ;
   
   ESC_SEQ     : '\\' ('\"'|'\\'|'/'|'b'|
               'f'|'n'|'r'|'t') 
               | UNICODE_ESC ;
   UNICODE_ESC : '\\' 'u' HEX_DIGIT 
               HEX_DIGIT HEX_DIGIT HEX_DIGIT ;
   HEX_DIGIT   : ('0'..'9'|
               'a'..'f'|'A'..'F') ;
   WS          : [ \r\t\n]+ -> skip ;
   LINE_COMMENT: '--' ~[\r\n]* -> skip ;

   SELECT      : 'SELECT' ;
   FROM        : 'FROM' ;
   WHERE       : 'WHERE' ;
  
   EQUALS      : 'EQUALS' ;
   OVERLAPS    : 'OVERLAPS' ;
   WITH        : 'WITH' ;
   SUBSET      : 'SUBSET' ;
   PROPER      : 'PROPER' ;
   GET_TASKS   : 'GetTasks' ;

   NOT         : 'NOT' ;
   AND         : 'AND' ;
   OR          : 'OR' ;

   ANY         : 'ANY' ;
   SOME        : 'SOME' ;
   EACH        : 'EACH' ;
   ALL         : 'ALL' ;

   IN          : 'IN' ;
   IS          : 'IS' ;
   OF          : 'OF' ;
   
   TRUE        : 'TRUE' ; 
   FALSE       : 'FALSE' ;

   identical       : EQUALS ;
   different       : NOT EQUALS ;
   overlapsWith    : OVERLAPS WITH ;
   subsetOf        : IS SUBSET OF ;
   properSubsetOf  : IS PROPER SUBSET OF ;

   UNION           : 'UNION' ;
   INTERSECTION    : 'INTERSECT' ;
   DIFFERENCE      : 'EXCEPT' ;

   CAN_OCCUR       : 'CanOccur' ;
   ALWAYS_OCCURS   : 'AlwaysOccurs' ;
   CAN_CONFLICT    : 'CanConflict' ;
   CAN_COOCCUR     : 'CanCooccur' ;
   CONFLICT        : 'Conflict' ;
   COOCCUR         : 'Cooccur' ;
   TOTAL_CAUSAL    : 'TotalCausal' ;
   TOTAL_CONCUR    : 'TotalConcurrent' ;
\end{verbatim}
\end{code}
\end{multicols}


%% file: tex/appendix_predicate.tex
\section{Denotations of Four Predicate Alternatives}
\label{app:predicate}

This appendix lists mathematical denotations of four alternatives of the \texttt{Predicate} constrcut.

\noindent
Every specimen of \texttt{SetComparison} denotes a set comparison operation.\footnote{As the meaning of a \texttt{SetOfTasks} construct is a set composed of the meanings of all and only its \texttt{Task} elements, each interpreted as a set of character strings, the identity of elements is defined as set equality over sets of character strings.}
\addtolength{\jot}{-.2em}
\small
\begin{eqnarray*}
\begin{split}
M_{\texttt{SetComparison}}\\ [p:\texttt{SetComparison},\\s:\mathbb{S},\mathit{vals}:\powerset{(\mathbb{V} \times \powerset{(\mathbb{T})})}] 
\end{split}
& \triangleq & 
\begin{cases} 
\;  \begin{split}
    M_{\texttt{SetOfTasks}}(p.\mathit{tasks}_1,s,\mathit{vals}) =\;\;\;\;\;\;\;\;\;\;\\ M_{\texttt{SetOfTasks}}(p.\mathit{tasks}_2,s,\mathit{vals})\;\;\;\;\;\;\;\;\;\;
    \end{split}
    & 
    \begin{split}
    p.\mathit{oper}\,\,\,\textbf{\texttt{is}}\;\;\;\\\texttt{Identical}
    \end{split}
    \\[1em] 
\;  \begin{split}
    M_{\texttt{SetOfTasks}}(p.\mathit{tasks}_1,s,\mathit{vals}) \neq\\ M_{\texttt{SetOfTasks}}(p.\mathit{tasks}_2,s,\mathit{vals}) 
    \end{split}
    & 
    \begin{split}
    p.\mathit{oper}\,\,\,\textbf{\texttt{is}}\;\;\;\\\texttt{Different}
    \end{split}
    \\[1em] 
\;  \begin{split}
    M_{\texttt{SetOfTasks}}(p.\mathit{tasks}_1,s,\mathit{vals})\,\, \cap\\ M_{\texttt{SetOfTasks}}(p.\mathit{tasks}_2,s,\mathit{vals}) \neq \emptyset 
    \end{split}
    & 
    \begin{split}
    p.\mathit{oper}\,\,\,\textbf{\texttt{is}}\;\;\;\;\;\;\;\;\;\;\;\\\texttt{OverlapsWith}
    \end{split}
    \\[1em] 
\;  \begin{split}
    M_{\texttt{SetOfTasks}}(p.\mathit{tasks}_1,s,\mathit{vals})\,\, \subseteq\\ M_{\texttt{SetOfTasks}}(p.\mathit{tasks}_2,s,\mathit{vals})
    \end{split}
    & 
    \begin{split}
    p.\mathit{oper}\,\,\,\textbf{\texttt{is}}\;\\\texttt{SubsetOf}
    \end{split}
    \\[1em] 
\;  \begin{split}
    M_{\texttt{SetOfTasks}}(p.\mathit{tasks}_1,s,\mathit{vals})\,\, \subset\\ M_{\texttt{SetOfTasks}}(p.\mathit{tasks}_2,s,\mathit{vals})
    \end{split}
    & 
    \begin{split}
    p.\mathit{oper}\,\,\,\textbf{\texttt{is}}\;\;\;\;\;\;\;\;\;\;\;\;\;\;\;\;\\\texttt{ProperSubsetOf}
    \end{split}
    \\ 
\end{cases}
\end{eqnarray*}
\normalsize
\noindent
The denotation of a specimen of \texttt{UnaryPredicate} is a unary behavioral relation.
\small
\begin{eqnarray*}
\begin{split}
M_{\texttt{UnaryPredicate}}[p:\texttt{UnaryPredicate},\\s:\mathbb{S},\mathit{vals}:\powerset{(\mathbb{V} \times \powerset{(\mathbb{T})})}]
\end{split}
& \triangleq &
\begin{cases} 
    \;\mathit{canOccur(s,M_\texttt{Task}(p.\mathit{task}))} & p.\mathit{name}\,\,\,\textbf{\texttt{is}}\,\,\,\texttt{CanOccur}\\
    \;\mathit{alwaysOccurs(s,M_\texttt{Task}(p.\mathit{task}))} & p.\mathit{name}\,\,\,\textbf{\texttt{is}}\,\,\,\texttt{AlwaysOccurs}
\end{cases}
\end{eqnarray*}
\normalsize
\noindent
Note that the respective behavioral relation is computed for the system that is being matched to the query.
Similarly, the meaning of a specimen of \texttt{BinaryPredicate} is a binary behavioral relation.
\small
\begin{eqnarray*}
\begin{split}
M_{\texttt{BinaryPredicate}}\\ [p:\texttt{BinaryPredicate},\\s:\mathbb{S},\mathit{vals}:\powerset{(\mathbb{V} \times \powerset{(\mathbb{T})})}]
\end{split}
& \triangleq &
\begin{cases}
\;  \begin{split}
    \mathit{canConflict}(s,M_\texttt{Task}(p.\mathit{task}_1),\;\;\;\;\;\;\;\;\;\;\;\;\;\;\;\;\; \\M_\texttt{Task}(p.\mathit{task}_2))\;\;\;\;\;\;\;\;\;
    \end{split}
    & 
    \begin{split}
    p.\mathit{name}\,\,\,\textbf{\texttt{is}}\;\;\;\;\;\;\;\\\texttt{CanConflict}
    \end{split}
    \\[1em] 
\;  \begin{split}
    \mathit{canCooccur}(s,M_\texttt{Task}(p.\mathit{task}_1),\;\;\;\;\;\;\, \\M_\texttt{Task}(p.\mathit{task}_2))
    \end{split}
    & 
    \begin{split}
    p.\mathit{name}\,\,\,\textbf{\texttt{is}}\;\;\;\;\\\texttt{CanCooccur}
    \end{split}
    \\[1em] 
\;  \begin{split}
    \mathit{conflict}(s,M_\texttt{Task}(p.\mathit{task}_1),\;\;\;\;\;\;\;\;\;\;\;\;\;\;\;\;\, \\M_\texttt{Task}(p.\mathit{task}_2))
    \end{split}
    & 
    \begin{split}
    p.\mathit{name}\,\,\,\textbf{\texttt{is}}\!\!\\\texttt{Conflict}
    \end{split}
    \\[1em] 
\;  \begin{split}
    \mathit{cooccur}(s,M_\texttt{Task}(p.\mathit{task}_1),\;\;\;\;\;\;\;\;\;\;\;\;\;\;\; \\M_\texttt{Task}(p.\mathit{task}_2))
    \end{split}
    & 
    \begin{split}
    p.\mathit{name}\,\,\,\textbf{\texttt{is}}\!\!\!\!\!\!\!\\\texttt{Cooccur}
    \end{split}
    \\[1em] 
\;  \begin{split}
    \mathit{totalCausal}(s,M_\texttt{Task}(p.\mathit{task}_1),\;\;\;\;\;\;\;\;\, \\M_\texttt{Task}(p.\mathit{task}_2))
    \end{split}
    & 
    \begin{split}
    p.\mathit{name}\,\,\,\textbf{\texttt{is}}\;\;\;\;\;\;\;\\\texttt{TotalCausal}
    \end{split}
    \\[1em] 
\;  \begin{split}
    \mathit{totalConcurrent}(s,M_\texttt{Task}(p.\mathit{task}_1),\\M_\texttt{Task}(p.\mathit{task}_2))
    \end{split}
    & 
    \begin{split}
    p.\mathit{name}\,\,\,\textbf{\texttt{is}}\;\;\;\;\;\;\;\;\;\;\;\;\;\;\;\;\;\\\texttt{TotalConcurrent}
    \end{split}
    \\ 
\end{cases}\\
\end{eqnarray*}
\normalsize
\noindent
As of today, PQL offers two unary and six binary behavioral predicates, refer to \sectionname~\ref{subsec:empirical:evaluation:results}. 
We expect that future studies will justify the use of other behavioral predicates for process querying, which consequently will result in the introduction of new predicates in PQL, ultimately improving its expressiveness. 
The reader can refer to \sectionname~\ref{subsec:PQL:predicates:semantics} to learn about methods for computing the PQL predicates (both unary and binary predicates).

As already mentioned, PQL utilizes a well-known mechanism of macros to combine results of several unary or binary relations.
A specimen of \texttt{UnaryPredicateMacro} combines several results of unary behavioral relations using existential and universal quantifiers as follows.
\small
\begin{eqnarray*}
\begin{split}
M_{\texttt{UnaryPredicateMacro}}\\ [p:\texttt{UnaryPredicateMacro},\\s:\mathbb{S},\mathit{vals}:\powerset{(\mathbb{V} \times \powerset{(\mathbb{T})})}] 
\end{split}
& \triangleq & 
\begin{cases}
    \begin{split}
    \;\exists\,\, t \in M_\texttt{SetOfTasks}(p.\mathit{tasks},s,\mathit{vals}) :\\ \mathit{canOccur}(s,t) 
    \end{split}
    & 
    \begin{split}
    (p.\mathit{name}\,\,\,\textbf{\texttt{is}}\,\,\,\texttt{CanOccur}) \,\land\\ (p.q\,\,\,\textbf{\texttt{is}}\,\,\,\texttt{Any})
    \end{split}
    \\[1em]
    \begin{split}
    \;\forall\, t \in M_\texttt{SetOfTasks}(p.\mathit{tasks},s,\mathit{vals}) :\\ \mathit{canOccur}(s,t) 
    \end{split}
    & 
    \begin{split}
    (p.\mathit{name}\,\,\,\textbf{\texttt{is}}\,\,\,\texttt{CanOccur}) \,\land\\ (p.q\,\,\,\textbf{\texttt{is}}\,\,\,\texttt{All})
    \end{split}
    \\[1em]
    \begin{split}
    \;\exists\,\, t \in M_\texttt{SetOfTasks}(p.\mathit{tasks},s,\mathit{vals}) :\\ \mathit{alwaysOccurs}(s,t) 
    \end{split}
    & 
    \begin{split}
    (p.\mathit{name}\,\,\,\textbf{\texttt{is}}\,\,\,\texttt{AlwaysOccurs}) \,\land\\ (p.q\,\,\,\textbf{\texttt{is}}\,\,\,\texttt{Any})
    \end{split}
    \\[1em]
    \begin{split}
    \;\forall\, t \in M_\texttt{SetOfTasks}(p.\mathit{tasks},s,\mathit{vals}) :\\ \mathit{alwaysOccurs}(s,t) 
    \end{split}
    & 
    \begin{split}
    (p.\mathit{name}\,\,\,\textbf{\texttt{is}}\,\,\,\texttt{AlwaysOccurs}) \,\land\\ (p.q\,\,\,\textbf{\texttt{is}}\,\,\,\texttt{All})
    \end{split}
    \\[1em]
\end{cases}
\end{eqnarray*}
\normalsize
\noindent
For example, a specimen $p$ of \texttt{UnaryPredicateMacro} such that $p.\mathit{name}$ is of type \texttt{AlwaysOccurs} and $p.q$ is of type \texttt{All} denotes the truth value of \emph{true} in the context of system $s$ \ifaof for every task $t$ in the set of tasks defined by $p.\mathit{tasks}$ it holds that $t$ occurs in every instance of $s$.

PQL offers two types of binary predicate macros: macros that relate a task to a set of tasks, and macros that relate two sets of tasks.
One can use a specimen of the \texttt{BinaryPredicateMacroTaskSet} construct to test if a task is in a binary behavioral relation with some (or all) tasks in a given set.

\small
\begin{eqnarray*}
\begin{split}
M_{\texttt{BinaryPredicateMacroTaskSet}}\\ [p:\texttt{BinaryPredicateMacroTaskSet},\\s:\mathbb{S},\mathit{vals}:\powerset{(\mathbb{V} \times \powerset{(\mathbb{T})})}] 
\end{split}
& \triangleq & 
\begin{cases}
    \begin{split}
    \;\exists\,\, t \in M_\texttt{SetOfTasks}(p.\mathit{tasks},s,\mathit{vals}) :\;\;\;\;\;\;\;\\ \mathit{canConflict}(s,M_\texttt{Task}(p.\mathit{task}),t)\;\;\;\;\;\;\;
    \end{split}
    & 
    \begin{split}
    (p.\mathit{name}\,\,\,\textbf{\texttt{is}}\\\texttt{CanConflict}) \,\land\\ (p.q\,\,\,\textbf{\texttt{is}}\,\,\,\texttt{Any})
    \end{split}
    \\[2.5em]
    \begin{split}
    \;\forall\, t \in M_\texttt{SetOfTasks}(p.\mathit{tasks},s,\mathit{vals}) :\\ \mathit{canConflict}(s,M_\texttt{Task}(p.\mathit{task}),t) 
    \end{split}
    & 
    \begin{split}
    (p.\mathit{name}\,\,\,\textbf{\texttt{is}}\\\texttt{CanConflict}) \,\land\\ (p.q\,\,\,\textbf{\texttt{is}}\,\,\,\texttt{All})
    \end{split}\\
    \;\;\;\ldots &
\end{cases}
\end{eqnarray*}
\normalsize
\noindent
We only specify the denotation of a specimen $p$ for which it holds that $p.\mathit{name}$ is of type $\texttt{CanConflict}$. 
Note that the denotations of specimens for other binary predicate names are defined analogously.
E.g., one can use a specimen $p$ of \texttt{BinaryPredicateMacroTaskSet} such that $p.\mathit{name}$ is of type \texttt{TotalCausal} and $p.q$ is of type \texttt{All} to check if the task defined by $p.\mathit{task}$ is in the \texttt{TotalCausal} behavioral relation with every task in the set of tasks defined by $p.\mathit{tasks}$, etc.

Finally, the denotation of a specimen of \texttt{BinaryPredicateMacroSetSet} is as follows.
\small
\begin{eqnarray*}
\begin{split}
M_{\texttt{BinaryPredicateMacroSetSet}}\\ [p:\texttt{BinaryPredicateMacroSetSet},\\s:\mathbb{S},\mathit{vals}:\powerset{(\mathbb{V} \times \powerset{(\mathbb{T})})}] 
\end{split}
& \triangleq & 
\begin{cases}
    \begin{split}
    \;\exists\, t_1 \in M_\texttt{SetOfTasks}(p.\mathit{tasks}_1,s,\mathit{vs})\;\;\;\;\;\;\;\\ \exists\, t_2 \in M_\texttt{SetOfTasks}(p.\mathit{tasks}_2,s,\mathit{vs}):\;\;\;\;\;\;\;\\ \mathit{canConflict}(s,t_1,t_2)\;\;\;\;\;\;\;
    \end{split}
    & 
    \begin{split}
    (p.\mathit{name}\,\,\,\textbf{\texttt{is}}\\\texttt{CanConflict}) \,\land\\ (p.q\,\,\,\textbf{\texttt{is}}\,\,\,\texttt{Any})
    \end{split}
    \\[2.5em]
    \begin{split}
    \;\exists\, t_1 \in M_\texttt{SetOfTasks}(p.\mathit{tasks}_1,s,\mathit{vs})\;\;\;\;\;\;\;\\ \forall\, t_2 \in M_\texttt{SetOfTasks}(p.\mathit{tasks}_2,s,\mathit{vs}):\;\;\;\;\;\;\;\\ \mathit{canConflict}(s,t_1,t_2)\;\;\;\;\;\;\;
    \end{split}
    & 
    \begin{split}
    (p.\mathit{name}\,\,\,\textbf{\texttt{is}}\\\texttt{CanConflict}) \,\land\\ (p.q\,\,\,\textbf{\texttt{is}}\,\,\,\texttt{Some})
    \end{split}
    \\[2.5em]
    \begin{split}
    \;\forall\, t_1 \in M_\texttt{SetOfTasks}(p.\mathit{tasks}_1,s,\mathit{vs})\\ \exists\, t_2 \in M_\texttt{SetOfTasks}(p.\mathit{tasks}_2,s,\mathit{vs}):\\ \mathit{canConflict}(s,t_1,t_2) 
    \end{split}
    & 
    \begin{split}
    (p.\mathit{name}\,\,\,\textbf{\texttt{is}}\\\texttt{CanConflict}) \,\land\\ (p.q\,\,\,\textbf{\texttt{is}}\,\,\,\texttt{Each})
    \end{split}
    \\[2.5em]
    \begin{split}
    \;\forall\, t_1 \in M_\texttt{SetOfTasks}(p.\mathit{tasks}_1,s,\mathit{vs})\\ \forall\, t_2 \in M_\texttt{SetOfTasks}(p.\mathit{tasks}_2,s,\mathit{vs}):\\ \mathit{canConflict}(s,t_1,t_2) 
    \end{split}
    & 
    \begin{split}
    (p.\mathit{name}\,\,\,\textbf{\texttt{is}}\\\texttt{CanConflict}) \,\land\\ (p.q\,\,\,\textbf{\texttt{is}}\,\,\,\texttt{All})\\
    \end{split}\\
    \;\;\ldots &
\end{cases}
\end{eqnarray*}
\normalsize
\noindent
For space considerations, we omit the precise definitions of denotations of all specimens of the \texttt{BinaryPredicateMacroSetSet} construct. 
Note that the missing definitions are similar to those shown above for a specimen $p$ for which $p.\mathit{name}$ is of type \texttt{CanConflict}.
For example, one can use a specimen $p$ of \texttt{BinaryPredicateMacroSetSet} such that $p.\mathit{name}$ is of type \texttt{conflict} and $p.q$ is of type \texttt{Any} to check if some task in the set of tasks defined by $p.\mathit{tasks}_1$ is in the \texttt{Conflict} behavioral relation with some task in the set of tasks defined by $p.\mathit{tasks}_2$, etc.


%% file: tex/appendix_implementation.tex
\section{Implementation Details}
\label{app:implementation}

This appendix contains details on PQL tools discussed in \sectname~\ref{sec:implementation}.

\mypar{The PQL Bot}
When initializing a PQL bot instance one can configure it via CLI. 
Some options of the PQL bot CLI are listed in \tablename~\ref{tab:PQL:bot:CLI:options}.
Every PQL bot instance has a unique name, which can be assigned using option \texttt{-n}.
If this option is not used, a random unique name is assigned.
Once started, a PQL bot instance indexes stored (but not yet indexed) models in succession.
One can use CLI options \texttt{-s} and \texttt{-i} to specify time to sleep, \ie to stay idle, between two successive indexing tasks, and the maximal time to attempt indexing of a model.
If these options are not used, the parameters get configured based on the values in the global configuration file.
If indexing of a model could not be completed within the given time frame, the model is marked as such that can not be indexed using this version of the bot, and the bot proceeds with indexing the next model.
The \texttt{-h} and \texttt{-v} CLI options, respectively, print the help message and get the version of the invoked PQL bot instance.

\begin{table}[htbp]
\footnotesize
  \centering
	\vspace{-2mm}
  \caption{\small CLI options of the PQL bot.}
	\vspace{-2mm}
\begin{tabular}{|l|c|c|l|}
  \hline
  \textbf{Option} & \textbf{Option (short)} & \textbf{Parameter} & \textbf{Description} \\
  \hline
  \hline
  \texttt{--help} & \texttt{-h} & & Print help message \\
  \hline
  \texttt{--index} & \texttt{-i} & \texttt{<number>} & Maximal indexing time (in seconds) \\
  \hline
  \texttt{--name} & \texttt{-n} & \texttt{<string>} & Name of this bot (maximum 36 characters) \\
  \hline
  \texttt{--sleep} & \texttt{-s} & \texttt{<number>} & Time to sleep between indexing jobs (in seconds) \\
  \hline
  \texttt{--version} & \texttt{-v} & & Get version of this bot \\
  \hline
\end{tabular}
\label{tab:PQL:bot:CLI:options}
\vspace{-1mm}
\end{table}

\noindent
Once started, a PQL bot instance run as a daemon, \ie a background process, until it is shut down.
An example of the command line output of a PQL bot instance is listed below.

{
\footnotesize
\begin{verbatim}
>> java -jar PQL.BOT-1.0.jar -n=Brisbane -s=60 -i=86400
>> =======================================================================
>> Process Query Language (PQL) Bot ver. 1.0
>> =======================================================================
>> Name:               Brisbane
>> Sleep time:         60s
>> Max. index time:    86400s
>> =======================================================================
>> 10:45:18.487 Brisbane - There are no pending jobs
>> 10:45:18.487 Brisbane - Sent an alive message
>> 10:45:18.497 Brisbane - Going to sleep for 60 seconds
>> 10:46:18.505 Brisbane - Woke up
>> 10:46:18.525 Brisbane - Retrieved indexing job for the model with ID 1
>> 10:46:18.575 Brisbane - Start checking model with ID 1
>> 10:46:23.506 Brisbane - Finished checking model with ID 1
>> 10:46:23.506 Brisbane - Start indexing model with ID 1
>> 10:47:03.608 Brisbane - Finished indexing model with ID 1
>> 10:47:03.608 Brisbane - Going to sleep for 60 seconds
>> 10:48:03.613 Brisbane - Woke up
>> 10:48:03.623 Brisbane - Retrieved indexing job for the model with ID 2
>> 10:48:03.673 Brisbane - Start checking model with ID 2
>> 10:48:13.248 Brisbane - Finished checking model with ID 2
>> 10:48:13.249 Brisbane - Start indexing model with ID 2
>> 10:49:52.679 Brisbane - Finished indexing model with ID 2
>> 10:49:52.679 Brisbane - Going to sleep for 60 seconds
>> 10:50:52.704 Brisbane - Woke up
>> 10:50:52.704 Brisbane - There are no pending jobs
>> ...
\end{verbatim}
}

\mypar{The PQL Tool}
\tablename~\ref{tab:PQL:tool:CLI:options} lists some CLI options of the PQL tool. 
For example, the PQL tool allows a user to store a given model (option~\texttt{-s}), check if a model can be indexed (option~\texttt{-c}), index a model (option~\texttt{-i}), delete a model and its index (option~\texttt{-d}), visualize the parse tree of a given query (option~\texttt{-p}), execute a query (options~\texttt{-q}), and reset the PQL environment (option~\texttt{-r}). 
The CLI can be used to access help information (option~\texttt{-h}) and the version of the tool (option~\texttt{-v}).

\begin{table}[htbp]
\footnotesize
  \centering
	\vspace{-2mm}
  \caption{\small CLI options of the PQL tool.}
	\vspace{-2mm}
\begin{tabular}{|l|c|c|l|c|}
  \hline
  \textbf{Option} & \textbf{Option (short)} & \textbf{Parameter} &\textbf{Description} & \textbf{Requires option} \\
  \hline
  \hline
  \texttt{--check} & \texttt{-c} & & Check if model can be indexed & \texttt{-id} \\
  \hline  
  \texttt{--delete} & \texttt{-d} & & Delete model (and its index) & \texttt{-id} \\
  \hline
  \texttt{--help} & \texttt{-h} & & Print help message & \\
  \hline
  \texttt{--index} & \texttt{-i} & & Index model & \texttt{-id} \\
  \hline
  \texttt{--identifier} & \texttt{-id} & \texttt{<string>} & Model identifier & \texttt{-id} \\
  \hline
  \texttt{--parse} & \texttt{-p} & & Show PQL query parse tree & \texttt{-pql} \\
  \hline
  \texttt{--pnmlPath} & \texttt{-pnml} & \texttt{<path>} & PNML path &  \\
  \hline
  \texttt{--pqlPath} & \texttt{-pql} & \texttt{<path>} & PQL path &  \\
  \hline
  \texttt{--query} & \texttt{-q} & & Execute PQL query & \texttt{-pql} \\
  \hline
  \texttt{--reset} & \texttt{-r} & & Reset this PQL instance & \\
  \hline
  \texttt{--store} & \texttt{-s} & & Store model & \texttt{-pnml} (\texttt{-id}) \\
  \hline
  \texttt{--version} & \texttt{-v} & & Get version of this tool & \\
  \hline
\end{tabular}
\label{tab:PQL:tool:CLI:options}
\vspace{-1mm}
\end{table}

\noindent
To store models, the CLI option \texttt{-s} must be accompanied by option \texttt{-pnml} that specifies a path either to a single PNML file or to a directory that contains PNML files.
If a path to a PNML file is used, the call to the PQL tool must include option \texttt{-id} to specify a unique identifier to associate with the model.
Otherwise, models are attempted to be stored using their file names as unique identifiers.
Once stored, a model can be indexed by a PQL bot instance or by the PQL tool using the CLI option \texttt{-i} accompanied by option \texttt{-id} that specifies the unique identifier that was used to store the model.
When indexing a model, the PQL tool uses the same routines as the PQL bot.

Note that the dynamic semantics of PQL is implemented over sound workflow systems---a special class of Petri net systems, refer to \sectionname~\ref{subsec:PQL:predicates:semantics} for details.
One can check whether a given Petri net system is a sound workflow system by calling the PQL tool with option \texttt{-c}.
Note that every request to index a model in the PQL environment is automatically preceded by a soundness check of this model.
Alternatively, a user may delete a model using option \texttt{-d}.
By deleting a model, the user also deletes its index.
Both options \texttt{-c} and \texttt{-d} require option \texttt{-id} to uniquely identify a model to be checked and deleted, respectively.
To execute a PQL query, a user can use option \texttt{-q} together with option \texttt{-pql} that specifies a path to a file that contains a PQL query captured using the concrete syntax proposed in \sectionname~\ref{subsec:PQL:concrete:syntax}.
To visualize the parse tree of a PQL query, one can use option \texttt{-p} together with option \texttt{-pql}.
Finally, one can reset the PQL environment using option \texttt{-r}.
By resetting the environment, one deletes all stored models and indexes.

An example command line output of executing a PQL query that is discussed in detail in \sectname~\ref{subsec:PQL:predicates:semantics} using the PQL tool is shown below.

{\footnotesize
\begin{verbatim}
>> java -jar PQL.TOOL-1.0.jar -q -pql=query.pql
>> PQL query:  SELECT * FROM * WHERE AlwaysOccurs("process payment"[0.75]);
>> Attributes: [UNIVERSE]
>> Locations:  [UNIVERSE]
>> Task:       "process payment"[0.75] -> ["process payment by cash",
>>             "process payment by check"]
>> Result:     [Fig.4.pnml]
\end{verbatim}
}


%% file: tex/appendix_evaluation.tex
\section{Discussion of Evaluation Results}
\label{app:evaluation}

This appendix contains detailed discussions of the results reported in \sectname~\ref{sec:evaluation}.

\mypar{Datasets} 
Tables~\ref{tab:collection:industrial} and~\ref{tab:collection:synthetic} report statistics on the structural properties of the two model collections used in the evaluation.
By XOR-join and XOR-split we refer to a place with multiple input transitions and a place with multiple output transitions, respectively.
By AND-join and AND-split we refer to a transition with multiple input places and a transition with multiple output places, respectively.
Polygons, bonds and rigids are different types of SESE components in the WF-tree of a workflow system~\cite{WeidlichPMW11}, where a polygon is a sequence of SESE components in which every two subsequent components share one node, a bond is a collection of SESE components that share entry and exit nodes, and a rigid is an unstructured component~\cite{Polyvyanyy12}.


\mypar{Queries}
\tablename~\ref{tab:queries:cats} lists all the PQL query categories, groups, and subgroups, and provides numbers of query templates used in the evaluation. 
The first category contains six query templates capturing individual atomic behavioral predicates. These can be divided into two groups, one covering the two unary predicates (1.a), the other covering the four binary predicates (1.b).

The second category is a result of combining atomic predicates via logical operations. 
There are two groups, one for connecting the same predicates (2.a), the other for connecting mixed predicates (2.b). 
The former group can be further divided into three subgroups which capture, respectively, the negation of each predicate (2.a.1), the conjunctions of each predicate twice, three, or four times (2.a.2), and the disjunctions of each predicate twice, three, or four times  (2.a.3). 
The latter group also has three subgroups which capture, respectively, the conjunctions of any two, three, or four different predicates (2.b.1), the disjunctions of any two, three, or four different predicates (2.b.2), and a couple of different combinations of three logical operations between mixed predicates (2.b.3).

The third category captures predicate macros in one group (3.a) and construction of task sets using predicates in the other group (3.b). 
The first group has three subgroups. One applies each of the unary predicate macros over a task set of two, three, or four tasks in conjunction (\ALL) or in disjunction (\ANY) (3.a.1); one applies each of the binary predicate macros between a single task and a task set of two, three, or four tasks in conjunction or in disjunction (3.a.2); and one applies each of the binary predicate macros between two task sets each consisting of two, three, or four tasks (3.a.3). The second group has four subgroups. The first two subgroups apply the set predicate \texttt{TaskInSetOfTasks} to a task set that is constructed using unary behavioral predicates (3.b.1) or using binary predicates (3.b.2). The last two subgroups capture task set constructions using mixed behavioral predicates with set operations (3.b.3) or set comparisons (3.b.4).


\mypar{Experiment 1.1: Impact of PQL bots on indexing time}
\figurename~\ref{fig:index:bots} plots the indexing times (in seconds) for different parts of process model collections against different numbers of PQL bots for (a) the industrial models and (b) the synthetic models. 

Indexing of the whole collection of industrial models with one bot took 6 hours and 54 minutes. 
Two bots managed to index 493 systems in 3 hours and 28 minutes (approximately two times faster than with one bot). 
Note that eight bots spent 1 hour and 12 minutes indexing the same collection (5.8 times faster than with one bot). 
A similar trend can be observed for the synthetic models. 
The relationship between the indexing time and the number of bots is best described by the power function $y = t\times x^k$, where $x$ is the number of bots and $y$ is the indexing time. 
For the industrial process models, the estimated values for constant $t$ are 6,065.6, 11,241, 16,735, and 23,071, for 25\%, 50\%, 75\%, 100\% of models in the collection, respectively.
The estimated values for coefficient $k$ are -0.845, -0.834, -0.826, and -0.839, for 25\%, 50\%, 75\%, 100\% of models in the industrial collection, respectively.
For the synthetic process models, the estimated values for constant $t$ are 17,991, 35,494, 52,086, and 68,077, for 25\%, 50\%, 75\%, 100\% of models in the collection, respectively.
The estimated values for coefficient $k$ are -0.833, -0.806, -0.827, and -0.834, for 25\%, 50\%, 75\%, 100\% of models in the synthetic collection, respectively.
The coefficient of determination $R^2$ ranges from 0.9901 to 0.9936 for the industrial process models and from 0.9834 to 0.9884 for the synthetic process models, indicating that the fitted functions can explain most of the variance in the indexing time. 

This experiment also shows that the indexing time grows linearly with the size of a process model collection. 
Using one PQL bot, 25\% of models in the industrial collection were indexed in 1 hour and 48 minutes, 50\% were indexed in 3 hours and 22 minutes, 75\% in 5 hours and 2 minutes, and the whole collection was indexed in 6 hours and 54 minutes. 
This relationship between the indexing time and the size of the collection is best captured by the linear function $y = 49.549x + 158.65$, where $x$ is the number of models in the collection and $y$ is the indexing time.
The coefficient of determination $R^2$ for the above example is 0.9985.
The coefficients of determination $R^2$ for the fitted linear functions on four data points range from 0.9985 to 0.9997 (for different numbers of bots).
We observed the same trend for the synthetic collection, with $R^2$ values ranging from 0.9949 to 0.9992.


\mypar{Experiment 1.2: Impact of model size on indexing time}
\figurename~\ref{fig:index:nodes} plots the indexing times (in seconds) against different sizes of workflow systems for the industrial and synthetic models. 

The relation between the indexing time and the model size in the industrial collection is best approximated by the power function $y = 0.8171 \times x^{1.7579}$, which results in a coefficient of determination $R^2$ of 0.9915. 
One obvious outlier workflow system in the industrial collection (a model with 25 observable transitions whose indexing took 858.7 seconds) can be explained by the much bigger size of its state space (2,097,422 reachable states, measured by the LoLA tool) compared to the sizes of state spaces of the other models in the collection (equal or less than 262,156 reachable states). 
For the synthetic models, the relation between the indexing time and the model size is best explained by the power function $y = 0.7423 \times x^{1.7775}$, which results in $R^2$ = 0.99. 

Given a system, either industrial or synthetic, we have noticed that it can be classified as an outlier using the value $f(x,y) = x^2 \times \mathit{log}(y)$, where $x$ is the number of observable transitions of the system and $y$ is the size of the state space of the system in terms of the number of its reachable states. 
The Pearson's correlation coefficient between the indexing times and the values of $f$ is $0.9426$ and $0.9692$ for the industrial and synthetic models, respectively.
The residual analysis has revealed that the four outliers in the industrial collection are among the six models with the highest values of $f$, while the seven outliers in the synthetic collection are among the eight models with the highest values of $f$.
The outliers were identified as models with the standardized residuals beyond $\pm3$ units. 
The four outliers in the industrial collection have 25 (5 minutes and 38 seconds to index), 25 (14 minutes and 19 seconds), 32 (8 minutes and 48 seconds) and 35 (8 minutes and 48 seconds) observable transitions.
Their state spaces comprise 114,717,\, 2,097,422,\, 390,\, and 882 reachable states, respectively.
The seven outliers in the synthetic collection have 36 (11 minutes and 23 seconds to index), 36 (10 minutes and 30 seconds), 42 (14 minutes and 58 seconds), 46 (15 minutes and 23 seconds), 56 (27 minutes and 10 seconds), 63 (39 minutes and 59 seconds), 67 (41 minutes and 6 seconds) observable transitions.
Their state spaces have 303,118,\, 131,108,\, 103,694,\, 1,602,\, 5,730,\, 311,306,\, and 6,818 reachable states, respectively.


\mypar{Experiment 1.4: Impact of index size on indexing time}
In total, we performed four indexing runs to observed the impact on the indexing time. 
In the first run, Set~1 was indexed first, followed by indexing of Set~2, followed by Set~3, and finally concluded by indexing Set~4.
The second run indexed the models in the order of Set 4, Set 1, Set 2, and finally Set 3.
The third run indexed the models in the order of Set 3, Set 4, Set 1, and finally Set 2.
The fourth run indexed the models in the order of Set 2, Set 3, Set 4, and finally Set 1.
In all the runs, the same label similarity threshold of 1.0 was used. 
All the four indexing runs were accomplished by one PQL bot.

For all the sets, we observed that the average indexing time is not significantly affected by the size of the index. 
For models in Set~1, the recorded average indexing times are 53.63 seconds, 53.65 seconds, 54.09 seconds, and 54.26 seconds in the first, second, third, and the fourth run, respectively.
For models in Set~2, the recorded average indexing times are 53.29 seconds, 53.46 seconds, 53.70 seconds, and 53.80 seconds in the first, second, third, and the fourth run, respectively.
For models in Set~3, the recorded average indexing times are 44.75 seconds, 44.79 seconds, 44.93 seconds, and 45.15 seconds in the first, second, third, and the fourth run, respectively.
Finally, for models in Set~4, the recorded average indexing times are 49.6 seconds, 49.52 seconds, 49.9 seconds, and 49.92 seconds in the first, second, third, and the fourth run, respectively.
Given any set out of the four sets of models, the relation between the index size (used as a starting point to index the given set) and the average indexing time of a model in the set is best approximated by a polynomial function.
The average indexing time for an industrial model in different sets at different positions in the indexing queue ranges from 44.75 to 54.26 seconds, with the difference between the maximum and minimum average indexing time for a given set at different indexing positions always being within 0.65 seconds. 
The measured coefficients of determination are equal to 0.9215, 0.9847, 0.9998, and 0.7366 for Set~1, Set~2, Set~3, and Set~4, respectively.
The experiment demonstrates a general trend of a negligible increase in the indexing time with the growth of the index size. 
This observation can be explained by the fact that the introduced overhead is due to write operations on the PQL index, which can be efficiently handled by modern database management systems. 

In the measurements of the average indexing times for the synthetic collection we observed a similar trend as in the measurements for the industrial collection.
In particular, given any set out of the four sets of models in the synthetic collection (note that each set of synthetic models is composed of 250 random models), the relation between the index size (used as a starting point to index the given set) and the average indexing time of a model in the set is best approximated by a polynomial function.
The average indexing time for a synthetic model in different sets at different positions in the indexing queue ranges from 61.7 to 86.74 seconds, with the difference between the maximum and minimum average indexing time for a given set at different indexing positions always being within 0.75 seconds. 
The measured coefficients of determination are equal to 0.8456, 0.9947, 0.9999, and 0.8545 for Set~1, Set~2, Set~3, and Set~4, respectively.


\mypar{Experiment 2.1: Impact of query threads on querying time}
\figurename~\ref{fig:query:netThread} plots the querying times (in seconds) for different parts of process model collections against different numbers of query threads. 
The relation between the querying time and the number of threads is best captured by the power function $y = t\times x^k$, where $y$ is the querying time and $x$ is the number of query threads. 
For the industrial process models, the estimated values for constant $t$ are 2.0793, 4.1112, 6.3066, and 8.3917, for 25\%, 50\%, 75\%, 100\% of models in the collection, respectively.
The estimated values for coefficient $k$ are -0.582, -0.584, -0.595, and -0.594, for 25\%, 50\%, 75\%, 100\% of models in the industrial collection, respectively.
For the synthetic process models, the estimated values for constant $t$ are 4.1811, 8.6902, 13.693, and 19.444, for 25\%, 50\%, 75\%, 100\% of models in the collection, respectively.
While the estimated values for coefficient $k$ are -0.581, -0.597, -0.604, and -0.612, for 25\%, 50\%, 75\%, 100\% of models in the synthetic collection, respectively.
The coefficient of determination $R^2$ ranges from 0.9969 to 0.9982 for the industrial models and from 0.9983 to 0.9989 for the synthetic models, indicating that the fitted models can explain most of the variance in the querying time.
%
%

The measurements obtained in this experiment can be used to show that the querying time grows linearly with the size of a process model collection. 
For example, the measured average querying time with one query thread over 25\% of the industrial models is 2.037 seconds, over 50\% is 4.037 seconds, over 75\% is 6.109 seconds, and over the whole collection is 8.259 seconds.
This trend is best described by the linear relation $y = 0.0168 \times x - 0.0658$, where $x$ is the number of models in the collection and $y$ is the querying time.
The coefficient of determination $R^2$ for the above example is 0.9998.
The coefficients of determination $R^2$ for the fitted linear functions on four data points range from 0.9988 to 1.0 (for different numbers of query threads).
We observed the same trend for the synthetic collection, with $R^2$ values ranging from 0.9967 to 0.9998.


\mypar{Experiment 2.2: Impact of query types on querying time}
\figurenames~\ref{fig:query:cat12:nets:sap} and~\ref{fig:query:cat12:threads:sap} show the average querying times for the collection of industrial models for queries in Categories~1 and~2, while \figurenames~\ref{fig:query:cat3:nets:sap} and~\ref{fig:query:cat3:threads:sap} show the average querying times for queries in Category~3. 
The average querying times for the synthetic collection are shown in \figurenames~\ref{fig:query:cat12:nets:synt} and~\ref{fig:query:cat12:threads:synt} (Categories~1 and~2) and \figurenames~\ref{fig:query:cat3:nets:synt} and~\ref{fig:query:cat3:threads:synt} (Category~3). 
Queries in Category~1 only comprise atomic predicates and, thus, are the fastest.
The measured average querying time for the Category~1 queries using one query thread is 1.75 seconds for the 493 models in the industrial collection (approximately 3.5ms per one model-query check) and 4.84 seconds for the 1,000 models in the synthetic collection (approximately 4.8ms per one model-query check).
With eight query threads, the Category~1 queries were on average accomplished in 0.47 seconds for the 493 models in the industrial collection and 1.61 seconds for the 1,000 models in the synthetic collection.
Queries in Category~2 comprise atomic predicates and logical connectives. 
Thus, they require a bit more time to accomplish than queries in Category 1. 
The measured average querying time for Category~2 queries with one thread is 2.8 seconds for the industrial models and 8.2 seconds for the synthetic models.
With eight query threads, it is 0.84 and 2.52 seconds for the industrial and synthetic models, respectively.
Queries in Category~3 comprise macros and, hence, are the lengthiest.
The measured average querying time for Category 3 queries with one query thread is 10.89 seconds for the industrial models (approximately 11ms per one model-query check) and 24.76 seconds for the synthetic models (approximately 25ms per one model-query check). 
With eight query threads, it is 3.23 and 6.96 seconds for the industrial and synthetic models, respectively.

\figurenames~\ref{fig:query:cat12:nets:sap},~\ref{fig:query:cat3:nets:sap},~\ref{fig:query:cat12:nets:synt} and~\ref{fig:query:cat3:nets:synt} demonstrate the linear dependency between the number of models in a collection and querying time for different query types. 
The coefficients of determination $R^2$ for the fitted linear functions for different query subgroups range from 0.9718 to 1.0 for the industrial models and from 0.9592 to 0.9992 for the synthetic models.
Finally, \figurenames~\ref{fig:query:cat12:threads:sap},~\ref{fig:query:cat3:threads:sap},~\ref{fig:query:cat12:threads:synt}, and~\ref{fig:query:cat3:threads:synt} show that for all the query subgroups the relation between the querying time and the number of query threads follows the trend observed in Experiment~2.1. 
The coefficients of determination $R^2$ for the fitted power functions for different query subgroups range from 0.9724 to 0.998 for the industrial models and from 0.9329 to 0.9955 for the synthetic models.


%% file: tex/appendix_queries.tex
\section{PQL Queries}
\label{app:PQL:queries}

\newcommand{\pql}[1]{\texttt{#1}}
\newcommand{\specialcellc}[2][c]{\begin{tabular}[#1]{@{}c@{}}#2\end{tabular}}
\newcommand{\specialcell}[2][c]{\begin{tabular}[#1]{@{}l@{}}#2\end{tabular}}
\newcommand{\rbr}{\scaleleftright[2ex]{.}{\rule[-.35ex]{0ex}{2ex}}{\rbrace}}
\newcommand{\lbr}{\scaleleftright[2ex]{.}{\rule[-.35ex]{0ex}{2ex}}{\lbrace}}


This appendix contains 150 PQL query templates used in the evaluation reported in \sectionname~\ref{sec:evaluation}. 
Each template is a PQL query with placeholders for activity labels.
The templates can be instantiated with specific labels and used as a benchmark to evaluate performance of PQL tools.
The query templates were developed to exploit the various features of the PQL grammar. 
According to the PQL features they support, these query templates are divided into three categories and further subdivided into groups and subgroups, refer to \sectionname~\ref{sec:dataset} for details.

\small
\setlength\extrarowheight{2pt}
\begin{longtable}{| p{.04\textwidth} | p{.075\textwidth} | p{.815\textwidth} |}
\caption{PQL query templates.}
\label{tab:queries} \\
\hline
\textbf{\small No.} & \textbf{\small ID} & \textbf{\small PQL template} \\
\hline
\hline
1 & 1.a.1 & \pql{\SELECT~* \FROM~* \WHERE~CanOccur("L1");} \\ \hline
2 & 1.a.2 & \pql{\SELECT~* \FROM~* \WHERE~AlwaysOccurs("L1");} \\ \hline\hline
3 & 1.b.1	& \pql{\SELECT~* \FROM~* \WHERE~Cooccur("L1","L2");} \\ \hline
4 & 1.b.2	& \pql{\SELECT~* \FROM~* \WHERE~Conflict("L1","L2");} \\ \hline
5 & 1.b.3	& \pql{\SELECT~* \FROM~* \WHERE~TotalCausal("L1","L2");} \\ \hline
6 & 1.b.4	& \pql{\SELECT~* \FROM~* \WHERE~TotalConcurrent("L1","L2");} \\ \hline\hline
7  & 2.a.1.1	& \pql{\SELECT~* \FROM~* \WHERE~\NOT~CanOccur("L1");} \\ \hline
8  & 2.a.1.2	& \pql{\SELECT~* \FROM~* \WHERE~\NOT~AlwaysOccurs("L1");} \\ \hline
9  & 2.a.1.3	& \pql{\SELECT~* \FROM~* \WHERE~\NOT~Cooccur("L1","L2");} \\ \hline
10 & 2.a.1.4	& \pql{\SELECT~* \FROM~* \WHERE~\NOT~Conflict("L1","L2");} \\ \hline
11 & 2.a.1.5	& \pql{\SELECT~* \FROM~* \WHERE~\NOT~TotalCausal("L1","L2");} \\ \hline
12 & 2.a.1.6	& \pql{\SELECT~* \FROM~* \WHERE~\NOT~TotalConcurrent("L1","L2");} \\ \hline\hline
13 & 2.a.2.1	& \pql{\SELECT~* \FROM~* \WHERE~CanOccur("L1") \AND~CanOccur("L2");} \\ \hline
14 & 2.a.2.2	& \specialcell{\pql{\SELECT~* \FROM~* \WHERE~CanOccur("L1") \AND}\\ \pql{CanOccur("L2") \AND~CanOccur("L3");}} \\ \hline
15 & 2.a.2.3	& \specialcell{\pql{\SELECT~* \FROM~* \WHERE~CanOccur("L1") \AND}\\\pql{CanOccur("L2") \AND~CanOccur("L3") \AND~CanOccur("L4");}} \\ \hline
16 & 2.a.2.4	& \pql{\SELECT~* \FROM~* \WHERE~AlwaysOccurs("L1") \AND~AlwaysOccurs("L2");} \\ \hline
17 & 2.a.2.5	& \specialcell{\pql{\SELECT~* \FROM~* \WHERE~AlwaysOccurs("L1") \AND}\\ \pql{AlwaysOccurs("L2") \AND~AlwaysOccurs("L3");}} \\ \hline
18 & 2.a.2.6	& \specialcell{\pql{\SELECT~* \FROM~* \WHERE~AlwaysOccurs("L1") \AND}\\ \pql{AlwaysOccurs("L2") \AND~AlwaysOccurs("L3") \AND~AlwaysOccurs("L4");}} \\ \hline
19 & 2.a.2.7	& \pql{\SELECT~* \FROM~* \WHERE~Cooccur("L1","L2") \AND~Cooccur("L3","L4");} \\ \hline
20 & 2.a.2.8	& \specialcell{\pql{\SELECT~* \FROM~* \WHERE~Cooccur("L1","L2") \AND}\\ \pql{Cooccur("L3","L4") \AND~Cooccur("L5","L6");}} \\ \hline
21 & 2.a.2.9	& \specialcell{\pql{\SELECT~* \FROM~* \WHERE~Cooccur("L1","L2") \AND}\\ \pql{Cooccur("L3","L4") \AND~Cooccur("L5","L6") \AND~Cooccur("L7","L8");}} \\ \hline
22 & 2.a.2.10	& \pql{\SELECT~* \FROM~* \WHERE~Conflict("L1","L2") \AND~Conflict("L3","L4");} \\ \hline
23 & 2.a.2.11	& \specialcell{\pql{\SELECT~* \FROM~* \WHERE~Conflict("L1","L2") \AND}\\ \pql{Conflict("L3","L4") \AND~Conflict("L5","L6");}} \\ \hline
24 & 2.a.2.12	& \specialcell{\pql{\SELECT~* \FROM~* \WHERE~Conflict("L1","L2") \AND}\\ \pql{Conflict("L3","L4") \AND~Conflict("L5","L6") \AND~Conflict("L7","L8");}} \\ \hline
25 & 2.a.2.13	& \specialcell{\pql{\SELECT~* \FROM~* \WHERE~TotalCausal("L1","L2") \AND}\\ \pql{TotalCausal("L3","L4");}} \\ \hline
26 & 2.a.2.14	& \specialcell{\pql{\SELECT~* \FROM~* \WHERE~TotalCausal("L1","L2") \AND}\\ \pql{TotalCausal("L3","L4") \AND~TotalCausal("L5","L6");}} \\ \hline
27 & 2.a.2.15	& \specialcell{\pql{\SELECT~* \FROM~* \WHERE~TotalCausal("L1","L2") \AND}\\ \pql{TotalCausal("L3","L4") \AND~TotalCausal("L5","L6") \AND}\\ \pql{TotalCausal("L7","L8");}} \\ \hline
28 & 2.a.2.16	& \specialcell{\pql{\SELECT~* \FROM~* \WHERE~TotalConcurrent("L1","L2") \AND}\\ \pql{TotalConcurrent("L3","L4");}} \\ \hline
29 & 2.a.2.17	& \specialcell{\pql{\SELECT~* \FROM~* \WHERE~TotalConcurrent("L1","L2") \AND}\\ \pql{TotalConcurrent("L3","L4") \AND~TotalConcurrent("L5","L6");}} \\ \hline
30 & 2.a.2.18	& \specialcell{\pql{\SELECT~* \FROM~* \WHERE~TotalConcurrent("L1","L2") \AND}\\ \pql{TotalConcurrent("L3","L4") \AND~TotalConcurrent("L5","L6") \AND}\\ \pql{TotalConcurrent("L7","L8");}} \\ \hline\hline
31 & 2.a.3.1	& \pql{\SELECT~* \FROM~* \WHERE~CanOccur("L1") \OR~CanOccur("L2");} \\ \hline
32 & 2.a.3.2	& \specialcell{\pql{\SELECT~* \FROM~* \WHERE~CanOccur("L1") \OR~CanOccur("L2") \OR}\\ \pql{CanOccur("L3");}} \\ \hline
33 & 2.a.3.3	& \specialcell{\pql{\SELECT~* \FROM~* \WHERE~CanOccur("L1") \OR~CanOccur("L2") \OR}\\ \pql{CanOccur("L3") \OR~CanOccur("L4");}} \\ \hline
34 & 2.a.3.4	& \pql{\SELECT~* \FROM~* \WHERE~AlwaysOccurs("L1") \OR~AlwaysOccurs("L2");} \\ \hline
35 & 2.a.3.5	& \specialcell{\pql{\SELECT~* \FROM~* \WHERE~AlwaysOccurs("L1") \OR~AlwaysOccurs("L2") \OR}\\ \pql{AlwaysOccurs("L3");}} \\ \hline
36 & 2.a.3.6	& \specialcell{\pql{\SELECT~* \FROM~* \WHERE~AlwaysOccurs("L1") \OR~AlwaysOccurs("L2") \OR}\\ \pql{AlwaysOccurs("L3") \OR~AlwaysOccurs("L4");}} \\ \hline
37 & 2.a.3.7	& \pql{\SELECT~* \FROM~* \WHERE~Cooccur("L1","L2") \OR~Cooccur("L3","L4");} \\ \hline
38 & 2.a.3.8	& \specialcell{\pql{\SELECT~* \FROM~* \WHERE~Cooccur("L1","L2") \OR~Cooccur("L3","L4") \OR}\\ \pql{Cooccur("L5","L6");}} \\ \hline
39 & 2.a.3.9	& \specialcell{\pql{\SELECT~* \FROM~* \WHERE~Cooccur("L1","L2") \OR~Cooccur("L3","L4") \OR}\\ \pql{Cooccur("L5","L6") \OR~Cooccur("L7","L8");}} \\ \hline
40 & 2.a.3.10	& \pql{\SELECT~* \FROM~* \WHERE~Conflict("L1","L2") \OR~Conflict("L3","L4");} \\ \hline
41 & 2.a.3.11	& \specialcell{\pql{\SELECT~* \FROM~* \WHERE~Conflict("L1","L2") \OR~Conflict("L3","L4") \OR}\\ \pql{Conflict("L5","L6");}} \\ \hline
42 & 2.a.3.12	& \specialcell{\pql{\SELECT~* \FROM~* \WHERE~Conflict("L1","L2") \OR~Conflict("L3","L4") \OR}\\ \pql{Conflict("L5","L6") \OR~Conflict("L7","L8");}} \\ \hline
43 & 2.a.3.13	& \specialcell{\pql{\SELECT~* \FROM~* \WHERE~TotalCausal("L1","L2") \OR}\\ \pql{TotalCausal("L3","L4");}} \\ \hline
44 & 2.a.3.14	& \specialcell{\pql{\SELECT~* \FROM~* \WHERE~TotalCausal("L1","L2") \OR}\\ \pql{TotalCausal("L3","L4") \OR~TotalCausal("L5","L6");}} \\ \hline
45 & 2.a.3.15	& \specialcell{\pql{\SELECT~* \FROM~* \WHERE~TotalCausal("L1","L2") \OR}\\ \pql{TotalCausal("L3","L4") \OR~TotalCausal("L5","L6") \OR}\\ \pql{TotalCausal("L7","L8");}} \\ \hline
46 & 2.a.3.16	& \specialcell{\pql{\SELECT~* \FROM~* \WHERE~TotalConcurrent("L1","L2") \OR}\\ \pql{TotalConcurrent("L3","L4");}} \\ \hline
47 & 2.a.3.17	& \specialcell{\pql{\SELECT~* \FROM~* \WHERE~TotalConcurrent("L1","L2") \OR}\\ \pql{TotalConcurrent("L3","L4") \OR~TotalConcurrent("L5","L6");}} \\ \hline
48 & 2.a.3.18	& \specialcell{\pql{\SELECT~* \FROM~* \WHERE~TotalConcurrent("L1","L2") \OR}\\ \pql{TotalConcurrent("L3","L4") \OR~TotalConcurrent("L5","L6") \OR}\\ \pql{TotalConcurrent("L7","L8");}} \\ \hline\hline
49 & 2.b.1.1	& \pql{\SELECT~* \FROM~* \WHERE~CanOccur("L1") \AND~Conflict("L2","L3");} \\ \hline
50 & 2.b.1.2	& \specialcell{\pql{\SELECT~* \FROM~* \WHERE~AlwaysOccurs("L1") \AND}\\ \pql{Cooccur("L2","L3") \AND~TotalConcurrent("L4","L5");}} \\ \hline
51 & 2.b.1.3	& \specialcell{\pql{\SELECT~* \FROM~* \WHERE~Cooccur("L1","L2") \AND}\\ \pql{TotalCausal("L3","L4") \AND~TotalConcurrent("L5","L6");}} \\ \hline\hline
52 & 2.b.2.1	& \pql{\SELECT~* \FROM~* \WHERE~AlwaysOccurs("L1") \OR~Cooccur("L2","L3");} \\ \hline
53 & 2.b.2.2	& \specialcell{\pql{\SELECT~* \FROM~* \WHERE~CanOccur("L1") \OR~AlwaysOccurs("L2") \OR}\\ \pql{Conflict("L3","L4");}} \\ \hline
54 & 2.b.2.3	& \specialcell{\pql{\SELECT~* \FROM~* \WHERE~Conflict("L1","L2") \OR}\\ \pql{TotalCausal("L3","L4") \OR~TotalConcurrent("L5","L6");}} \\ \hline\hline
55 & 2.b.3.1	& \specialcell{\pql{\SELECT~* \FROM~* \WHERE~AlwaysOccurs("L1") \OR}\\ \pql{(Cooccur("L2","L3") \AND~(\NOT~TotalCausal("L4","L5")));}} \\ \hline
56 & 2.b.3.2	& \specialcell{\pql{\SELECT~* \FROM~* \WHERE}\\ \pql{(CanOccur("L1") \AND~(\NOT~Conflict("L2","L3"))) \OR}\\ \pql{(TotalConcurrent("L4","L5") \AND~AlwaysOccurs("L6"));}} \\ \hline\hline
57 & 3.a.1.1	& \pql{\SELECT~* \FROM~* \WHERE~CanOccur(\lbr"L1","L2"\rbr,\ALL);} \\ \hline
58 & 3.a.1.2	& \pql{\SELECT~* \FROM~* \WHERE~CanOccur(\lbr"L1","L2","L3"\rbr,\ALL);} \\ \hline
59 & 3.a.1.3	& \pql{\SELECT~* \FROM~* \WHERE~CanOccur(\lbr"L1","L2","L3","L4"\rbr,\ALL);} \\ \hline
60 & 3.a.1.4	& \pql{\SELECT~* \FROM~* \WHERE~CanOccur(\lbr"L1","L2"\rbr,\ANY);} \\ \hline
61 & 3.a.1.5	& \pql{\SELECT~* \FROM~* \WHERE~CanOccur(\lbr"L1","L2","L3"\rbr,\ANY);} \\ \hline
62 & 3.a.1.6	& \pql{\SELECT~* \FROM~* \WHERE~CanOccur(\lbr"L1","L2","L3","L4"\rbr,\ANY);} \\ \hline
63 & 3.a.1.7	& \pql{\SELECT~* \FROM~* \WHERE~AlwaysOccurs(\lbr"L1","L2"\rbr,\ALL);} \\ \hline
64 & 3.a.1.8	& \pql{\SELECT~* \FROM~* \WHERE~AlwaysOccurs(\lbr"L1","L2","L3"\rbr,\ALL);} \\ \hline
65 & 3.a.1.9	& \pql{\SELECT~* \FROM~* \WHERE~AlwaysOccurs(\lbr"L1","L2","L3","L4"\rbr,\ALL);} \\ \hline
66 & 3.a.1.10	& \pql{\SELECT~* \FROM~* \WHERE~AlwaysOccurs(\lbr"L1","L2"\rbr,\ANY);} \\ \hline
67 & 3.a.1.11	& \pql{\SELECT~* \FROM~* \WHERE~AlwaysOccurs(\lbr"L1","L2","L3"\rbr,\ANY);} \\ \hline
68 & 3.a.1.12	& \pql{\SELECT~* \FROM~* \WHERE~AlwaysOccurs(\lbr"L1","L2","L3","L4"\rbr,\ANY);} \\ \hline\hline
69 & 3.a.2.1	& \pql{\SELECT~* \FROM~* \WHERE~Cooccur("L1",\lbr"L2","L3"\rbr,\ALL);} \\ \hline
70 & 3.a.2.2	& \pql{\SELECT~* \FROM~* \WHERE~Cooccur("L1",\lbr"L2","L3","L4"\rbr,\ALL);} \\ \hline
71 & 3.a.2.3	& \pql{\SELECT~* \FROM~* \WHERE~Cooccur("L1",\lbr"L2","L3","L4","L5"\rbr,\ALL);} \\ \hline
72 & 3.a.2.4	& \pql{\SELECT~* \FROM~* \WHERE~Cooccur("L1",\lbr"L2","L3"\rbr,\ANY);} \\ \hline
73 & 3.a.2.5	& \pql{\SELECT~* \FROM~* \WHERE~Cooccur("L1",\lbr"L2","L3","L4"\rbr,\ANY);} \\ \hline
74 & 3.a.2.6	& \pql{\SELECT~* \FROM~* \WHERE~Cooccur("L1",\lbr"L2","L3","L4","L5"\rbr,\ANY);} \\ \hline
75 & 3.a.2.7	& \pql{\SELECT~* \FROM~* \WHERE~Conflict("L1",\lbr"L2","L3"\rbr,\ALL);} \\ \hline
76 & 3.a.2.8	& \pql{\SELECT~* \FROM~* \WHERE~Conflict("L1",\lbr"L2","L3","L4"\rbr,\ALL);} \\ \hline
77 & 3.a.2.9	& \pql{\SELECT~* \FROM~* \WHERE~Conflict("L1",\lbr"L2","L3","L4","L5"\rbr,\ALL);} \\ \hline
78 & 3.a.2.10	& \pql{\SELECT~* \FROM~* \WHERE~Conflict("L1",\lbr"L2","L3"\rbr,\ANY);} \\ \hline
79 & 3.a.2.11	& \pql{\SELECT~* \FROM~* \WHERE~Conflict("L1",\lbr"L2","L3","L4"\rbr,\ANY);} \\ \hline
80 & 3.a.2.12	& \pql{\SELECT~* \FROM~* \WHERE~Conflict("L1",\lbr"L2","L3","L4","L5"\rbr,\ANY);} \\ \hline
81 & 3.a.2.13	& \pql{\SELECT~* \FROM~* \WHERE~TotalCausal("L1",\lbr"L2","L3"\rbr,\ALL);} \\ \hline
82 & 3.a.2.14	& \pql{\SELECT~* \FROM~* \WHERE~TotalCausal("L1",\lbr"L2","L3","L4"\rbr,\ALL);} \\ \hline
83 & 3.a.2.15	& \pql{\SELECT~* \FROM~* \WHERE~TotalCausal("L1",\lbr"L2","L3","L4","L5"\rbr,\ALL);} \\ \hline
84 & 3.a.2.16	& \pql{\SELECT~* \FROM~* \WHERE~TotalCausal("L1",\lbr"L2","L3"\rbr,\ANY);} \\ \hline
85 & 3.a.2.17	& \pql{\SELECT~* \FROM~* \WHERE~TotalCausal("L1",\lbr"L2","L3","L4"\rbr,\ANY);} \\ \hline
86 & 3.a.2.18	& \pql{\SELECT~* \FROM~* \WHERE~TotalCausal("L1",\lbr"L2","L3","L4","L5"\rbr,\ANY);} \\ \hline
87 & 3.a.2.19	& \pql{\SELECT~* \FROM~* \WHERE~TotalConcurrent("L1",\lbr"L2","L3"\rbr,\ALL);} \\ \hline
88 & 3.a.2.20	& \pql{\SELECT~* \FROM~* \WHERE~TotalConcurrent("L1",\lbr"L2","L3","L4"\rbr,\ALL);} \\ \hline
89 & 3.a.2.21	& \specialcell{\pql{\SELECT~* \FROM~* \WHERE}\\ \pql{TotalConcurrent("L1",\lbr"L2","L3","L4","L5"\rbr,\ALL);}} \\ \hline
90 & 3.a.2.22	& \pql{\SELECT~* \FROM~* \WHERE~TotalConcurrent("L1",\lbr"L2","L3"\rbr,\ANY);} \\ \hline
91 & 3.a.2.23	& \pql{\SELECT~* \FROM~* \WHERE~TotalConcurrent("L1",\lbr"L2","L3","L4"\rbr,\ANY);} \\ \hline
92 & 3.a.2.24	& \specialcell{\pql{\SELECT~* \FROM~* \WHERE}\\ \pql{TotalConcurrent("L1",\lbr"L2","L3","L4","L5"\rbr,\ANY);}} \\ \hline\hline
93 & 3.a.3.1	& \pql{\SELECT~* \FROM~* \WHERE~Cooccur(\lbr"L1","L2"\rbr,\lbr"L3","L4"\rbr,\ALL);} \\ \hline
94 & 3.a.3.2	& \specialcell{\pql{\SELECT~* \FROM~* \WHERE}\\ \pql{Cooccur(\lbr"L1","L2","L3"\rbr,\lbr"L4","L5","L6"\rbr,\ALL);}} \\ \hline
95 & 3.a.3.3	& \specialcell{\pql{\SELECT~* \FROM~* \WHERE}\\ \pql{Cooccur(\lbr"L1","L2","L3","L4"\rbr,\lbr"L5","L6","L7","L8"\rbr,\ALL);}} \\ \hline
96 & 3.a.3.4	& \pql{\SELECT~* \FROM~* \WHERE~Cooccur(\lbr"L1","L2"\rbr,\lbr"L3","L4"\rbr,\ANY);} \\ \hline
97 & 3.a.3.5	& \specialcell{\pql{\SELECT~* \FROM~* \WHERE}\\ \pql{Cooccur(\lbr"L1","L2","L3"\rbr,\lbr"L4","L5","L6"\rbr,\ANY);}} \\ \hline
98 & 3.a.3.6	& \specialcell{\pql{\SELECT~* \FROM~* \WHERE}\\ \pql{Cooccur(\lbr"L1","L2","L3","L4"\rbr,\lbr"L5","L6","L7","L8"\rbr,\ANY);}} \\ \hline
99 & 3.a.3.7	& \pql{\SELECT~* \FROM~* \WHERE~Cooccur(\lbr"L1","L2"\rbr,\lbr"L3","L4"\rbr,\EACH);} \\ \hline
100 & 3.a.3.8	& \specialcell{\pql{\SELECT~* \FROM~* \WHERE}\\ \pql{Cooccur(\lbr"L1","L2","L3"\rbr,\lbr"L4","L5","L6"\rbr,\EACH);}} \\ \hline
101 & 3.a.3.9	& \specialcell{\pql{\SELECT~* \FROM~* \WHERE}\\ \pql{Cooccur(\lbr"L1","L2","L3","L4"\rbr,\lbr"L5","L6","L7","L8"\rbr,\EACH);}} \\ \hline
102 & 3.a.3.10	& \pql{\SELECT~* \FROM~* \WHERE~Conflict(\lbr"L1","L2"\rbr,\lbr"L3","L4"\rbr,\ALL);} \\ \hline
103 & 3.a.3.11	& \specialcell{\pql{\SELECT~* \FROM~* \WHERE}\\ \pql{Conflict(\lbr"L1","L2","L3"\rbr,\lbr"L4","L5","L6"\rbr,\ALL);}} \\ \hline
104 & 3.a.3.12	& \specialcell{\pql{\SELECT~* \FROM~* \WHERE}\\ \pql{Conflict(\lbr"L1","L2","L3","L4"\rbr,\lbr"L5","L6","L7","L8"\rbr,\ALL);}} \\ \hline
105 & 3.a.3.13	& \pql{\SELECT~* \FROM~* \WHERE~Conflict(\lbr"L1","L2"\rbr,\lbr"L3","L4"\rbr,\ANY);} \\ \hline
106 & 3.a.3.14	& \specialcell{\pql{\SELECT~* \FROM~* \WHERE}\\ \pql{Conflict(\lbr"L1","L2","L3"\rbr,\lbr"L4","L5","L6"\rbr,\ANY);}} \\ \hline
107 & 3.a.3.15	& \specialcell{\pql{\SELECT~* \FROM~* \WHERE}\\ \pql{Conflict(\lbr"L1","L2","L3","L4"\rbr,\lbr"L5","L6","L7","L8"\rbr,\ANY);}} \\ \hline
108 & 3.a.3.16	& \pql{\SELECT~* \FROM~* \WHERE~Conflict(\lbr"L1","L2"\rbr,\lbr"L3","L4"\rbr,\EACH);} \\ \hline
109 & 3.a.3.17	& \specialcell{\pql{\SELECT~* \FROM~* \WHERE}\\ \pql{Conflict(\lbr"L1","L2","L3"\rbr,\lbr"L4","L5","L6"\rbr,\EACH);}} \\ \hline
110 & 3.a.3.18	& \specialcell{\pql{\SELECT~* \FROM~* \WHERE}\\ \pql{Conflict(\lbr"L1","L2","L3","L4"\rbr,\lbr"L5","L6","L7","L8"\rbr,\EACH);}} \\ \hline
111 & 3.a.3.19	& \pql{\SELECT~* \FROM~* \WHERE~TotalCausal(\lbr"L1","L2"\rbr,\lbr"L3","L4"\rbr,\ALL);} \\ \hline
112 & 3.a.3.20	& \specialcell{\pql{\SELECT~* \FROM~* \WHERE}\\ \pql{TotalCausal(\lbr"L1","L2","L3"\rbr,\lbr"L4","L5","L6"\rbr,\ALL);}} \\ \hline
113 & 3.a.3.21	& \specialcell{\pql{\SELECT~* \FROM~* \WHERE}\\ \pql{TotalCausal(\lbr"L1","L2","L3","L4"\rbr,\lbr"L5","L6","L7","L8"\rbr,\ALL);}} \\ \hline
114 & 3.a.3.22	& \pql{\SELECT~* \FROM~* \WHERE~TotalCausal(\lbr"L1","L2"\rbr,\lbr"L3","L4"\rbr,\ANY);} \\ \hline
115 & 3.a.3.23	& \specialcell{\pql{\SELECT~* \FROM~* \WHERE}\\ \pql{TotalCausal(\lbr"L1","L2","L3"\rbr,\lbr"L4","L5","L6"\rbr,\ANY);}} \\ \hline
116 & 3.a.3.24	& \specialcell{\pql{\SELECT~* \FROM~* \WHERE}\\ \pql{TotalCausal(\lbr"L1","L2","L3","L4"\rbr,\lbr"L5","L6","L7","L8"\rbr,\ANY);}} \\ \hline
117 & 3.a.3.25	& \pql{\SELECT~* \FROM~* \WHERE~TotalCausal(\lbr"L1","L2"\rbr,\lbr"L3","L4"\rbr,\EACH);} \\ \hline
118 & 3.a.3.26	& \specialcell{\pql{\SELECT~* \FROM~* \WHERE}\\ \pql{TotalCausal(\lbr"L1","L2","L3"\rbr,\lbr"L4","L5","L6"\rbr,\EACH);}} \\ \hline
119 & 3.a.3.27	& \specialcell{\pql{\SELECT~* \FROM~* \WHERE}\\ \pql{TotalCausal(\lbr"L1","L2","L3","L4"\rbr,\lbr"L5","L6","L7","L8"\rbr,\EACH);}} \\ \hline
120 & 3.a.3.28	& \pql{\SELECT~* \FROM~* \WHERE~TotalConcurrent(\lbr"L1","L2"\rbr,\lbr"L3","L4"\rbr,\ALL);} \\ \hline
121 & 3.a.3.29	& \specialcell{\pql{\SELECT~* \FROM~* \WHERE}\\ \pql{TotalConcurrent(\lbr"L1","L2","L3"\rbr,\lbr"L4","L5","L6"\rbr,\ALL);}} \\ \hline
122 & 3.a.3.30	& \specialcell{\pql{\SELECT~* \FROM~* \WHERE}\\ \pql{TotalConcurrent(\lbr"L1","L2","L3","L4"\rbr,\lbr"L5","L6","L7","L8"\rbr,\ALL);}} \\ \hline
123 & 3.a.3.31	& \pql{\SELECT~* \FROM~* \WHERE~TotalConcurrent(\lbr"L1","L2"\rbr,\lbr"L3","L4"\rbr,\ANY);} \\ \hline
124 & 3.a.3.32	& \specialcell{\pql{\SELECT~* \FROM~* \WHERE}\\ \pql{TotalConcurrent(\lbr"L1","L2","L3"\rbr,\lbr"L4","L5","L6"\rbr,\ANY);}} \\ \hline
125 & 3.a.3.33	& \specialcell{\pql{\SELECT~* \FROM~* \WHERE}\\ \pql{TotalConcurrent(\lbr"L1","L2","L3","L4"\rbr,\lbr"L5","L6","L7","L8"\rbr,\ANY);}} \\ \hline
126 & 3.a.3.34	& \pql{\SELECT~* \FROM~* \WHERE~TotalConcurrent(\lbr"L1","L2"\rbr,\lbr"L3","L4"\rbr,\EACH);} \\ \hline
127 & 3.a.3.35	& \specialcell{\pql{\SELECT~* \FROM~* \WHERE}\\ \pql{TotalConcurrent(\lbr"L1","L2","L3"\rbr,\lbr"L4","L5","L6"\rbr,\EACH);}} \\ \hline
128 & 3.a.3.36	& \specialcell{\pql{\SELECT~* \FROM~* \WHERE}\\ \pql{TotalConcurrent(\lbr"L1","L2","L3","L4"\rbr,\lbr"L5","L6","L7","L8"\rbr,\EACH);}} \\ \hline\hline
129 & 3.b.1.1	& \specialcell{\pql{\SELECT~* \FROM~* \WHERE}\\ \pql{"L1" \IN~GetTasksCanOccur(\lbr"L2","L3","L4"\rbr);}} \\ \hline
130 & 3.b.1.2	& \specialcell{\pql{\SELECT~* \FROM~* \WHERE}\\ \pql{"L1" \IN~GetTasksAlwaysOccurs(\lbr"L2","L3","L4"\rbr);}} \\ \hline\hline
131 & 3.b.2.1	& \specialcell{\pql{\SELECT~* \FROM~* \WHERE}\\ \pql{"L1" \IN~GetTasksCooccur(\lbr"L2","L3"\rbr,\lbr"L4","L5","L6"\rbr,\ALL);}} \\ \hline
132 & 3.b.2.2	& \specialcell{\pql{\SELECT~* \FROM~* \WHERE}\\ \pql{"L1" \IN~GetTasksCooccur(\lbr"L2","L3"\rbr,\lbr"L4","L5","L6"\rbr,\ANY);}} \\ \hline
133 & 3.b.2.3	& \specialcell{\pql{\SELECT~* \FROM~* \WHERE}\\ \pql{"L1" \IN~GetTasksConflict(\lbr"L2","L3"\rbr,\lbr"L4","L5","L6"\rbr,\ALL);}} \\ \hline
134 & 3.b.2.4	& \specialcell{\pql{\SELECT~* \FROM~* \WHERE}\\ \pql{"L1" \IN~GetTasksConflict(\lbr"L2","L3"\rbr,\lbr"L4","L5","L6"\rbr,\ANY);}} \\ \hline
135 & 3.b.2.5	& \specialcell{\pql{\SELECT~* \FROM~* \WHERE}\\ \pql{"L1" \IN~GetTasksTotalCausal(\lbr"L2","L3"\rbr,\lbr"L4","L5","L6"\rbr,\ALL);}} \\ \hline
136 & 3.b.2.6	& \specialcell{\pql{\SELECT~* \FROM~* \WHERE}\\ \pql{"L1" \IN~GetTasksTotalCausal(\lbr"L2","L3"\rbr,\lbr"L4","L5","L6"\rbr,\ANY);}} \\ \hline
137 & 3.b.2.7	& \specialcell{\pql{\SELECT~* \FROM~* \WHERE}\\ \pql{"L1" \IN~GetTasksTotalConcurrent(\lbr"L2","L3"\rbr,\lbr"L4","L5","L6"\rbr,\ALL);}} \\ \hline
138 & 3.b.2.8	& \specialcell{\pql{\SELECT~* \FROM~* \WHERE}\\ \pql{"L1" \IN~GetTasksTotalConcurrent(\lbr"L2","L3"\rbr,\lbr"L4","L5","L6"\rbr,\ANY);}} \\ \hline\hline
139 & 3.b.3.1& \specialcell{\pql{\SELECT~* \FROM~* \WHERE}\\ \pql{"L1" \IN~GetTasksAlwaysOccurs(\lbr"L2","L3","L4"\rbr) \UNION}\\ \pql{GetTasksTotalCausal(\lbr"L5","L6"\rbr,\lbr"L7","L8","L9"\rbr,\ALL);}} \\ \hline
140 & 3.b.3.2	& \specialcell{\pql{\SELECT~* \FROM~* \WHERE}\\ \pql{"L1" \IN~GetTasksCanOccur(\lbr"L2","L3","L4"\rbr) \INTERSECT}\\ \pql{GetTasksConflict(\lbr"L5","L6"\rbr,\lbr"L7","L8","L9"\rbr,\ANY);}} \\ \hline
141 & 3.b.3.3	& \specialcell{\pql{\SELECT~* \FROM~* \WHERE}\\ \pql{"L1" \IN~GetTasksCooccur(\lbr"L2","L3"\rbr,\lbr"L4","L5","L6"\rbr,\ANY) \EXCEPT}\\ \pql{GetTasksTotalConcurrent(\lbr"L7","L8"\rbr,\lbr"L9","L10","L11"\rbr,\ANY);}} \\ \hline
142 & 3.b.3.4	& \specialcell{\pql{\SELECT~* \FROM~* \WHERE}\\ \pql{"L1" \IN~(GetTasksCooccur(\lbr"L2","L3"\rbr,\lbr"L4","L5","L6"\rbr,\ANY) \EXCEPT}\\ \pql{GetTasksTotalConcurrent(\lbr"L7","L8"\rbr,\lbr"L9","L10","L11"\rbr,\ANY)) \UNION}\\ \pql{(GetTasksCanOccur(\lbr"L12","L13","L14"\rbr) \INTERSECT}\\ \pql{GetTasksConflict(\lbr"L15","L16"\rbr,\lbr"L17","L18","L19"\rbr,\ALL));}} \\ \hline
143 & 3.b.3.5	& \specialcell{\pql{\SELECT~* \FROM~* \WHERE}\\ \pql{"L1" \IN~(GetTasksCooccur(\lbr"L2","L3"\rbr,\lbr"L4","L5","L6"\rbr,\ANY) \UNION}\\ \pql{GetTasksTotalConcurrent(\lbr"L7","L8"\rbr,\lbr"L9","L10","L11"\rbr,\ALL))}\\ \pql{\INTERSECT~(GetTasksCanOccur(\lbr"L12","L13","L14"\rbr) \EXCEPT}\\ \pql{GetTasksConflict(\lbr"L15","L16"\rbr,\lbr"L17","L18","L19"\rbr,\ALL));}} \\ \hline\hline
144 & 3.b.4.1	& \specialcell{\pql{\SELECT~* \FROM~* \WHERE~(\lbr"L1","L2","L3"\rbr~\EXCEPT}\\ \pql{GetTasksConflict(\lbr"L4","L5"\rbr,\lbr"L6","L7","L8"\rbr,\ALL)) \EQUALS}\\ \pql{GetTasksTotalCausal(\lbr"L9","L10"\rbr,\lbr"L11","L12","L13"\rbr,\ALL);}} \\ \hline
145 & 3.b.4.2 & \specialcell{\pql{\SELECT~* \FROM~* \WHERE}\\ \pql{GetTasksCooccur(\lbr"L1","L2"\rbr,\lbr"L3","L4","L5"\rbr,\ANY) \NOT~\EQUALS}\\ \pql{GetTasksTotalConcurrent(\lbr"L6","L7"\rbr,\lbr"L8","L9","L10"\rbr,\ANY);}} \\ \hline
146 & 3.b.4.3	& \specialcell{\pql{\SELECT~* \FROM~* \WHERE~(\lbr"L1","L2","L3"\rbr~\EXCEPT}\\ \pql{GetTasksCooccur(\lbr"L4","L5"\rbr,\lbr"L6","L7","L8"\rbr,\ALL)) \OVERLAPS~\WITH}\\ \pql{GetTasksConflict(\lbr"L9","L10"\rbr,\lbr"L11","L12","L13"\rbr,\ANY);}} \\ \hline
147 & 3.b.4.4	& \specialcell{\pql{\SELECT~* \FROM~* \WHERE~(\lbr"L1","L2","L3"\rbr~\EXCEPT}\\ \pql{GetTasksAlwaysOccurs(\lbr"L4","L5","L6"\rbr) \IS~\SUBSET~\OF}\\ \pql{GetTasksCanOccur(\lbr"L7","L8","L9"\rbr));}} \\ \hline
148 & 3.b.4.5	& \specialcell{\pql{\SELECT~* \FROM~* \WHERE}\\ \pql{GetTasksTotalCausal(\lbr"L1","L2"\rbr,\lbr"L3","L4","L5"\rbr,\ALL) \IS~\PROPER}\\ \pql{\SUBSET~\OF~(\lbr"L6","L7","L8"\rbr~\EXCEPT}\\ \pql{GetTasksTotalConcurrent(\lbr"L9","L10"\rbr,\lbr"L11","L12","L13"\rbr,\ALL));}} \\ \hline
149 & 3.b.4.6	& \specialcell{\pql{\SELECT~* \FROM~* \WHERE}\\ \pql{(GetTasksCooccur(\lbr"L1","L2"\rbr,\lbr"L3","L4","L5"\rbr,\ALL)) \EQUALS}\\ \pql{GetTasksTotalConcurrent(\lbr"L6","L7"\rbr,\lbr"L8","L9","L10"\rbr,\ALL) \OR}\\ \pql{((GetTasksCanOccur(\lbr"L11","L12","L13"\rbr) \OVERLAPS~\WITH}\\ \pql{GetTasksConflict(\lbr"L14","L15"\rbr,\lbr"L16","L17","L18"\rbr,\ALL)) \AND}\\ \pql{((\lbr"L19","L20","L21"\rbr~\EXCEPT}\\ \pql{GetTasksTotalCausal(\lbr"L22","L23"\rbr,\lbr"L24","L25","L26"\rbr,\ANY))}\\ \pql{\IS~\SUBSET~\OF~GetTasksAlwaysOccurs(\lbr"L27","L28","L29"\rbr)));}} \\
\hline
150 & 3.b.4.7	& \specialcell{\pql{\SELECT~* \FROM~* \WHERE}\\ \pql{(\NOT~(GetTasksCooccur(\lbr"L1","L2"\rbr,\lbr"L3","L4","L5"\rbr,\ANY) \OVERLAPS}\\ \pql{\WITH~GetTasksTotalConcurrent(\lbr"L6","L7"\rbr,\lbr"L8","L9","L10"\rbr,\ANY)))}\\ \pql{\AND~((GetTasksConflict(\lbr"L11","L12"\rbr,\lbr"L13","L14","L15"\rbr,\ALL)}\\ \pql{\IS~\PROPER~\SUBSET~\OF~GetTasksCanOccur(\lbr"L16","L17","L18"\rbr))}\\ \pql{\AND~((\lbr"L19","L20","L21"\rbr~\EXCEPT}\\ \pql{GetTasksTotalCausal(\lbr"L22","L23"\rbr,\lbr"L24","L25","L26"\rbr,\ALL))}\\ \pql{\NOT~\EQUALS~GetTasksAlwaysOccurs(\lbr"L27","L28","L29"\rbr)));}}
\\* \hline 
\end{longtable}
\normalsize